\documentclass{elsarticle}
\usepackage{graphicx,amssymb}
\usepackage{subfigure}
\usepackage[%
  breaklinks=false,%
  colorlinks=true,%
  linkcolor=blue,anchorcolor=blue,%
  citecolor=blue,filecolor=blue,%
  menucolor=blue,%
  urlcolor=blue]{hyperref}
\usepackage[show]{ed}
\usepackage{amsmath
}        
\usepackage{axodraw}
\usepackage{subfigure}
\usepackage{ifpdf}
\numberwithin{equation}{section}
\numberwithin{table}{section}
\numberwithin{figure}{section}
\newcount\Farbe
\ifnum\Farbe=1
        
\else
        
\fi
\def\e{\epsilon}
\def\sud{s_{12}}
\def\sdt{s_{23}}

\def\zb{\bar{z}}
\def\abs#1{|#1|}
\def\spab#1.#2.#3{\langle\mskip-1mu{#1} 
                  | #2 | {#3}\mskip-1mu]}
\def\spba#1.#2.#3{[\mskip-1mu{#1} 
                  | #2 | {#3}\mskip-1mu\rangle}
\def\spa#1.#2{\langle#1\,#2\rangle}
\def\spb#1.#2{[#1\,#2]}
\def\cm{{\cal M}}
\def\ep{\epsilon}
\def\beq{\begin{equation}}
\def\eeq{\end{equation}}
\def\la{\langle}
\def\ra{\rangle}
\def\cg{c_\Gamma}
\def\bom#1{{\mbox{\boldmath $#1$}}}

\def\Eq{Eq.~}
\def\Eqs{Eqs.~}
\def\tb{\tilde b}
\def\bb{\bar b}
\def\slsh{\rlap{$\;\!\!\not$}}     
\def\vNV{van Neerven - Vermaseren}
\def\Nc{N_c}
\newcommand\fverb{\setbox\fverbbox=\hbox\bgroup\verb}
\newcommand\fverbdo{\egroup\medskip\noindent%
			\fbox{\unhbox\fverbbox}\ }
\newcommand\fverbit{\egroup\item[\fbox{\unhbox\fverbbox}]}
\newbox\fverbbox

\def\ib{\bar{\imath}}
\def\jb{\bar{\jmath}}
\def\qb{{\bar q}}
\def\Qb{{\bar Q}}

\newcommand{\eps}{\epsilon}
\newcommand{\nn}{\nonumber \\}
\newcommand{\be}{\begin{equation}}
\newcommand{\ee}{\end{equation}}
\newcommand{\ba}{\begin{eqnarray}}
\newcommand{\ea}{\end{eqnarray}}
\newcommand{\beqa}{\begin{eqnarray}}
\newcommand{\eeqa}{\end{eqnarray}}

\def\mo{m_1}
\def\mt{m_2}
\def\nh{\hat{n}}
\def\ne{n_{\epsilon}}
\def\beq{\begin{equation}}
\def\eeq{\end{equation}}
\def\beqn{\begin{eqnarray}}
\def\eeqn{\end{eqnarray}}
\def\lt{l_{(2)}}
\def\nn{\nonumber}

\newbox\charbox
\newbox\slabox

\def\s#1{{      
        \setbox\charbox=\hbox{$#1$}
        \setbox\slabox=\hbox{$/$}
        \dimen\charbox=\ht\slabox
        \advance\dimen\charbox by -\dp\slabox
        \advance\dimen\charbox by -\ht\charbox
        \advance\dimen\charbox by \dp\charbox
        \divide\dimen\charbox by 2
        \raise-\dimen\charbox\hbox to \wd\charbox{\hss/\hss}
        \llap{$#1$} }}

\def\tree{{\rm tree}}
\def\oneloop{{1 \mbox{-} \rm loop}}

\def\A{{\cal A}}
\def\B{{\cal B}}
\def\la{\langle}
\def\ra{\rangle}

\def\si{\sigma}

\def\ket#1{|#1]}
\def\aket#1{|#1\rangle}
\newcommand{\bra}[1]{\langle{#1}|}
\newcommand{\cut}[2]{ \!{\vphantom{\Big|}}_{#1}{\Big|}_{#2}\! }

\newcommand{\cO}{{\cal O}}

\newfont{\liste}{pzdr scaled 1100}
\newfont{\grfett}{cmmib10 scaled 1100}

\DeclareMathSymbol{\varPhi}{\mathalpha}{operators}{"08}
\DeclareMathSymbol{\varOmega}{\mathalpha}{operators}{"0A}

\makeatletter

\def\myfootnotesize{\@setsize\footnotesize{10pt}\ixpt\@ixpt} 

\@ifundefined{square}{}{\let\Box\square}
\makeatother

\begin{document}
\title{One-loop calculations in quantum field theory:
from Feynman diagrams to unitarity cuts}

\author[Fermilab]{R. Keith Ellis}
\ead{ellis@fnal.gov}
\address[Fermilab]{Fermilab, Batavia, IL 60510, USA }
\author[Zurich]{Zoltan Kunszt}
\ead{kunszt@itp.phys.ethz.ch}
\address[Zurich]{Institute for Theoretical Physics, ETH, Zurich,
       CH-8093 Zurich, Switzerland}
\author[Baltimore]{Kirill Melnikov}
\ead{melnikov@pha.jhu.edu}
\author[Oxford]{Giulia Zanderighi}
\ead{g.zanderighi1@physics.ox.ac.uk}
\address[Baltimore]{Department of Physics, Johns Hopkins University,
Baltimore,USA} 
\address[Oxford]{Rudolf Peierls Centre for Theoretical Physics, 1 Keble Road, University of Oxford, UK}

\begin{abstract}
  The success of the experimental program at the Tevatron re-inforced
  the idea that precision physics at hadron colliders is desirable
  and, indeed, possible. The Tevatron data strongly suggests that
  one-loop computations in QCD describe hard scattering well.
  Extrapolating this observation to the LHC, we conclude that
  knowledge of many short-distance processes at next-to-leading order
  may be required to describe the physics of hard scattering.  While
  the field of one-loop computations is quite mature, parton
  multiplicities in hard LHC events are so high that traditional
  computational techniques become inefficient.  Recently new
  approaches based on unitarity have been developed for calculating
  one-loop scattering amplitudes in quantum field theory.  These
  methods are especially suitable for the description of
  multi-particle processes in QCD and are amenable to numerical
  implementations.  We present a systematic pedagogical description of
  both conceptual and technical aspects of the new methods.
\end{abstract}


\begin{flushright}
FERMILAB-PUB-11-195-T\\
OUTP-11-39P
\end{flushright}
\maketitle

\newpage
\tableofcontents
\newpage
\section{Introduction}

Perturbation theory is one of the few rigorous ways to connect Quantum
Chromodynamics (QCD) to observations. It is particularly important because,
during the next decade, our understanding of particle physics will be
challenged by experiments at the Large Hadron Collider (LHC) at CERN.
Hadron collisions at large momentum transfer are important for the
direct observation of new forces and new forms of matter.  Thanks to
the phenomenon of asymptotic freedom in QCD, the strong coupling
constant becomes small at large momentum transfer, making it possible
to describe hard hadron collisions in QCD perturbation theory.

During the past decade, both Tevatron experiments -- CDF and D0 --
have accumulated an enormous luminosity, of the order of ten inverse
femtobarns.  This high luminosity enabled detailed and careful
investigation of Standard Model processes, including jets and
electroweak gauge boson production as well as studies of the top quark
and the Higgs boson.  In all the cases considered, the comparison of
observables that are calculable in perturbative QCD with experimental
results improved if next-to-leading order (NLO) QCD computations were
used.  This fact establishes perturbative QCD as a systematic
framework to describe the physics of hard hadronic collisions.  It
also suggests that, ideally, the theoretical toolkit for the LHC
should contain next-to-leading computations for a large variety of
processes.

This idea was formalized by the so-called wishlist which puts together
a collection of processes whose computation through NLO QCD is thought
to be most useful \cite{Bern:2008ef}. The list includes large number
of processes with three or four particles in the final state; those
final-state particles include QCD partons, heavy quarks and
electroweak gauge bosons.  The wishlist was originally compiled back
in 2003 and, at that time, the computation of even the simplest
processes on the wishlist was considered to be very challenging.

The situation changed dramatically in the past four to five years
since new techniques for one-loop computations lead to the explosion
of new results in the field. In the past three years, a large number
of one-loop computations with four (massive and massless) particles in
the final state were completed
\cite{Bredenstein:2009aj,Bredenstein:2010rs,
Bevilacqua:2009zn,Berger:2009zg,KeithEllis:2009bu,Berger:2009ep,Melnikov:2009wh,
  Berger:2010vm,Frederix:2010ne,Melia:2010bm,Denner:2010jp,Bevilacqua:2010ve,
  Bevilacqua:2010qb,Melia:2011dw,Greiner:2011mp} and even 
the first $2 \to 5$ results -- the NLO
QCD corrections to $pp \to W(Z)+4j$ \cite{Berger:2010zx,Ita:2011wn} -- was
obtained.\footnote{As usual, we do not include 
  decay products of heavy particles
in this counting.}  Some of this progress is to be
attributed to spectacular improvements of the existing
Feynman-diagrammatic algorithms for one-loop calculations, whose
efficiency and stability was boosted beyond expectations.
Furthermore, the past three years have witnessed a full development of
the idea that one-loop amplitudes can be reconstructed from their
unitarity cuts. This idea was put forward by Bern, Dixon and Kosower
in the 1990s and was used in a number of phenomenological calculations
\cite{Bern:1997sc}.  It was revived in recent years when Britto,
Cachazo and Feng observed that the coefficients of the four-point
functions, obtained when one-loop amplitudes are reduced to scalar
integrals, are products of on-shell tree scattering amplitudes
evaluated at complex momenta \cite{Britto:2004nc}, and when Ossola,
Papadopoulos and Pittau (OPP) discovered a simple algebraic method for
reducing tensor integrals to scalar master integrals
\cite{Ossola:2006us}.  

In this article we describe these developments
using the technique developed by some of us, in collaboration with
W.~Giele, that we will refer to as the generalized $D$-dimensional
unitarity \cite{Ellis:2007br,Giele:2008ve,Ellis:2008ir}. In contrast
to other unitarity-related techniques, this is the only method known
today that delivers complete one-loop scattering amplitudes in
renormalizable quantum field theories, regardless of whether one-loop
amplitudes involve massless or massive particles.

The goal of this review is to present ideas and techniques of
generalized $D$-dimensional unitarity in a pedagogical
manner. Although a fair number of advanced one-loop calculations has
been already performed using this method, the subject will benefit
from a detailed and critical review that will make it accessible also
to a non-expert audience.  While we believe that essential details of
the unitarity methods are well-established by now, further algorithmic
improvements are not to be excluded.  Our goal therefore is  to
provide detailed information about the method of generalized
$D$-dimensional unitarity with the hope that it can be used as a
foundation for further development of this approach. We also describe
some aspects of other unitarity-based techniques to one-loop
computations. A more detailed  description of these alternative 
approaches  
and many references to original publications 
can be found in recent reviews \cite{Berger:2009zb,Britto:2010xq}.
The remainder of this article is organized as follows.  

In Section~\ref{sec:tradoneloop}, we summarize the traditional
approaches to one-loop computations. Such traditional approaches often
employ the Passarino-Veltman procedure to express tensor one-loop
integrals through scalar integrals.  For reference purposes, we
present full details of the Passarino-Veltman reduction procedure in
\ref{app:PV}.

In Section~\ref{sec3} we introduce the van Neerven - Vermaseren basis
which is important for the explicit implementation of the unitarity
technique.
Subsequent sections refer back to and use various results obtained in
Section~\ref{sec3}.

In Section~\ref{sec:red2d} we discuss the reduction algorithm in
detail by considering two-dimensional examples. Working in
two-dimensional space-time offers clear advantages since it allows us
to derive concise analytic results.
We make use of the van Neerven - Vermaseren basis to reduce a scalar
triangle to a sum of scalar two-point functions. This derivation
generalizes to higher dimensions leading to the  important result
that any $N$-point function in $D$ dimensions, with $N>D$, can
be written as a combination of $D$-point functions.
We also discuss the reduction of the rank two tensor two-point
function in two dimensions in two different ways. The first method
employs a decomposition of the loop momentum using the van Neerven -
Vermaseren basis. The second method shows that the reduction can be
performed if the integrand is known at special values of the loop
momentum, namely the ones for which the inverse Feynman propagators
vanish. We use this example to introduce the connection between
reduction procedures and the OPP/unitarity ideas \cite{Ossola:2006us}.
We finish this Section with an example where we compute the so-called
rational part -- a remnant of the ultraviolet regularization -- of the
photon vacuum polarization in two-dimensional QED. This contribution
is responsible for the dynamical generation of the photon mass in
two-dimensional QED -- a remarkable phenomenon first pointed out by
J.~Schwinger \cite{Schwinger:1962tp}.

In Section~\ref{sect5} we describe the explicit construction of the
OPP reduction procedure in $D$-dimensions. We give full details about
the parametrization of the integrand, we describe how discrete Fourier
transforms can be used to extract reduction coefficients, and we show
how specific situations, that could give rise to numerical
instabilities, can be handled. Finally, we discuss the rational part,
explain its ultraviolet origin and point out that ultraviolet finite
integrals can have a rational part. In~\ref{app:RT} we give
the explicit expression for the rational part of specific tensor
integrals used here.

In Section~\ref{sect6} we discuss the color decomposition of one-loop
amplitudes and introduce the concept of color-ordered amplitudes. We
discuss specific examples involving only gluons, one quark-pair and
many gluons, as well as the case of amplitudes involving multiple
quark pairs. 

In Section~\ref{sec:7new} we explain how the OPP procedure can be
related to unitarity and describe various ingredients that are
important for a practical implementation of the computational
algorithm. We present the construction of polarization states and
spinors in higher dimensions. We explain how tree-level amplitudes can
be computed using recursive Berends-Giele relations
\cite{Berends:1987cv} and present examples of the recursive equations
for amplitudes with up to four fermions and an arbitrary number of
gluons.  We also show how Britto-Cachazo-Feng-Witten (BCFW) relations
between scattering amplitudes can be proven using Berends-Giele
recursions, and that the BCFW relation is independent of the number of
dimensions.  Finally, we discuss subtleties related to the
implementation of $D$-dimensional unitarity methods for calculating
scattering amplitudes with massive particles.

In Section~\ref{analytic} we describe several methods of more analytic
nature that are closely related to the OPP method and generalized
unitarity. We give an extensive discussion of the method suggested by
Forde which allows the direct computation of the cut-constructible
reduction coefficients \cite{Forde:2007mi}.  We describe a
generalization of this method, suggested by Badger
\cite{Badger:2008cm}, that enables the calculation of the rational
part.  We also discuss a technique suggested by Mastrolia for the
computation of the double-cut reduction coefficient
\cite{Mastrolia:2009dr}.  One aspect of our discussion that makes it
different from much of the literature is that we avoid the extensive
use of spinor-helicity decomposition of the loop momentum and show
that the analytic approaches described in
Refs.~\cite{Forde:2007mi,Badger:2008cm,Mastrolia:2009dr} can be
understood using a simple parametrization of the loop momentum
phase-space in terms of polar and azimuthal angles.

In Section~\ref{sec:examples} we present three examples that emphasize
the anomalous nature of the rational part.  We discuss the decay of
the Higgs boson to two photons through a loop of massless scalars.  We
demonstrate that the rational part of the photon-photon scattering
amplitude is independent of the mass of the virtual particle that
mediates the photon-photon scattering, and discuss the absence of the
rational part in $n$-photon scattering amplitudes for $n \ge 6$,
following Ref.~\cite{Badger:2008rn}.  Thereafter, we show that the
rational part of the one-loop triangle amplitude gives the anomalous
part of the divergence of the axial current. We conclude this Section
with a simple example of the analytic calculation of a one-loop
diagram using the spinor helicity method. A brief introduction to
the spinor helicity method is given in \ref{App:SH}. 

In Section~\ref{sec:num} we describe how generalized unitarity can be
implemented in a computer code.  We contrast the application of the 
OPP reduction technique to Feynman diagrams with the
unitarity-based implementation. We explain a convenient method to
handle cuts in parent diagrams systematically. We discuss issues
related to having colorless (unordered) particles in the scattering
process. We also present the standard checks that are done on numerical
calculations and issues related to numerical instabilities. Finally,
we show that the computational time for one-loop amplitudes depends on
the number of external particles in a polynomial way and give a few
examples.  

We conclude in Section~\ref{sec:conclu}.  As a final remark we note
that other approaches to one-loop calculations beyond traditional 
Passarino-Veltman reduction and generalized unitarity have 
have been studied in the literature \cite{Soper:2001hu,Passarino:2001wv,
  Ferroglia:2002mz,Nagy:2003qn,Anastasiou:2007qb,
  Lazopoulos:2007ix,Nagy:2006xy,
  Gong:2008ww,Kilian:2009wy,Becker:2010ng,Becker:2011vg,
Catani:2008xa}, 
but we will not discuss them in this review.
 
\section{One loop diagrams: the traditional approach}
\label{sec:tradoneloop}
\subsection{Preliminary remarks} 

In this Section, we review traditional approaches to the computation
of one-loop integrals. While the discussion below is general, it is
useful to have in mind the mathematical structure of the Standard
Model of particle physics.  One-loop computations in the Standard
Model require calculating integrals of the following form
\begin{equation}
I_N \sim  \int \frac{{\rm d}^4 l}{(2\pi)^4} 
\frac{{\cal N}(l)}{((l+q_0)^2 - m_1^2)((l+q_1)^2 - m_2^2)....
((l+q_{N-1})^2 - m_N^2)},
\label{sect2_eq1}
\end{equation}
\begin{figure}[t]
\begin{center}
\includegraphics[scale=0.55]{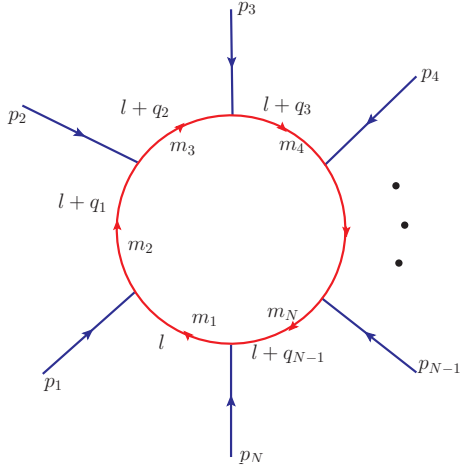}
\end{center}
\caption{Generic diagram at one-loop with $N$ external momenta.}
\label{fig:oneloopdiag}
\end{figure}
where $N$ is the number of external particles, $q_j=\sum_{k=1}^{j}p_k$
and $p_1 + p_2 + ....p_N = 0$ thanks to the momentum conservation, see
Fig.~\ref{fig:oneloopdiag}.
The special case where ${\cal N}(l)=1$ is referred to as a scalar
integral.  In general ${\cal N}(l)$ is a polynomial function of the
loop momentum $l$ as well as external momenta $p_i$, external
polarization vectors, spinors etc.  The goal is to compute $I_N$ in an
efficient way.

Before discussing how this can be done, we point out two things in
connection with the structure of $I_N$. First, we note that if ${\cal
  N}(l)$ contains the loop momentum $l$ in a high enough power, the
integral $I_N$ will be ultraviolet (UV) divergent. From simple power
counting, it is clear that for an $N$-point integral, the UV
divergence appears if ${\cal N}(l)$ contains tensor integrals of rank
$r$ higher than $r \ge 2N - 4$.  In particular, only one-point and
two-point scalar integrals are UV divergent. The highest rank of an
$N$-point one-loop diagram, occurring in renormalizable quantum field
theories, such as the Standard Model or QCD, is $r=N$. Hence, only
one-, two-, three- and four-point one-loop integrals can be divergent
in the ultraviolet region while five- and higher-point one-loop
integrals are {\it ultraviolet finite}.  In the presence of
ultraviolet divergences the integrals require regularization. It is
conventional to employ dimensional
regularization~\cite{tHooft:1972fi}, where the dimensionality of
space-time is set to $D = 4-2\epsilon$, and the limit $\epsilon \to 0$ 
is taken at the end of the calculation. 
As a result, the loop momentum $l$ becomes a
$D$-dimensional vector and the integration measure in
\Eq(\ref{sect2_eq1}) is changed to
\begin{equation}
\frac{{\rm d}^4 l}{(2\pi)^4}  \to \frac{{\rm d}^D l}{(2\pi)^D}.
\label{sect2_eq3}
\end{equation}

It is easy to see that this modification of the integration measure in 
\Eq(\ref{sect2_eq3}) regularizes ultraviolet divergences. Indeed, 
consider as an example $I_4$ with ${\cal N}(l) = l_\mu l_\nu l_\rho l_\delta$.
Power counting suggests that the divergence is logarithmic; as a consequence
it is insensitive to external kinematic parameters. The divergence 
can be isolated by considering 
\begin{eqnarray}
I_4 \to 
\int \frac{{\rm d}^Dl }{(2\pi)^D }
\frac{l_\mu l_\nu l_\rho l_\delta}{[d(l)]^4}
= \frac{( g_{\mu \nu} g_{\rho \delta} + g_{\mu \rho} g_{\nu \delta}
+ g_{\mu \delta} g_{\nu \rho})}{D(D+2)} 
\int \frac{{\rm d}^Dl }{(2\pi)^D } \frac{l^4}{[d(l)]^4}, 
\end{eqnarray}
where $d(l) = l^2 - \mu^2$ and 
$\mu$ is some kinematic invariant that we keep to regulate 
potential divergences 
at small values of $l^2$.  The integral is calculated with the help of 
the following equation 
\begin{eqnarray} \label{x20}
&& \int \frac{ d^D l}{i \pi^{D/2}} \frac{(l^2)^r}
{\Big ( l^2-\mu^2 \Big ) ^m}  =
\frac{\Omega_D}{\pi^{D/2}}
(-1)^{r-m} \mu^{D+2r-2m}\int_0^\infty dx\frac{x^{2r+D-1}}{(1+x^2)^m}
\nonumber \\ && =
(-1)^{r-m} 
\mu^{D+2 r-2 m}  
\frac{\Gamma(r+D/2) \Gamma(m-r-D/2)
}{\Gamma(D/2) \Gamma(m)},
\end{eqnarray}
where  $\Omega_D = 2 \pi^{D/2}/\Gamma(D/2)$ is 
the solid angle in $D$ dimensions. 
\Eq(\ref{x20}) gives a finite result for $D \neq 4$ for $m=4$ and $r=2$. 
We therefore conclude that dimensional regularization does indeed
regularize UV divergences in one-loop computations in quantum field
theories.

There is a second type of potential problem related to one-loop
integrals $I_N$. These problems appear when a sufficient number of
propagators in the integrand can go on the mass-shell simultaneously,
introducing potentially non-integrable singularities.  The general
theory of such singularities is provided by the Landau
rules~\cite{Landau:1959fi}.  The two most important examples of these
singularities are the soft and collinear ones, related to the presence
of massless particles~\cite{Kinoshita:1962ur}.  Those singularities
can also be regularized dimensionally~\cite{Marciano:1975de}.

Dimensional regularization therefore provides an economical tool to
{\it define} one-loop integrals in UV-divergent, renormalizable
theories that contain massless particles since, with a single
parameter $\epsilon = (4-D)/2$, we are able to regulate both types of
divergences and make the integrals finite.  From now on, we assume
that dimensional regularization is always applied to loop integrals
and all the loop integrals that we have to calculate, need to be
understood in that framework.

It turns out that, in the limit $D \to 4$, any integral $I_N$ can be
written as a linear combination of one-loop scalar integrals, that
include four-, three-, two- and one-point functions and a remnant of
the dimensional regularization procedure that is called {\it the rational part} 
${\cal R}$
\be
I_N = c_{4;j} I_{4;j} +c_{3;j} I_{3;j} +c_{2;j} I_{2;j} +c_{1;j} I_{1;j} + {\cal R} + {\cal O}(D-4). 
\label{eq5}
\ee 
In \Eq(\ref{eq5}) the coefficients $c_{N,j}$ ($N=1,\dots4$) are
evaluated in $D=4$, i.\ e.\ they do not have any dependence on
$\epsilon$, and $I_{L;j}$ stands for an $L$-point one-loop scalar
integral of the type $j$.  The type $j$ specifies, cryptically, which
combinations of the external momenta $p_i$ build up the $q_i$ that
enter the (master) integrals in the right hand side of \Eq(\ref{eq5}).
The existence of this decomposition is one of the most important
results for one-loop calculations; its origin relies on simple Lorentz
invariance which allows a decomposition of tensor integrals to
invariant form factors and on the four-dimensional nature of space
time which allows scalar higher point integrals to be reduced to sums
of boxes.  Thus a scalar pentagon in $D$ dimensions can be written as
a sum of the five box diagrams obtainable by removing one propagator
if we neglect terms of order
$\epsilon$~\cite{Melrose:1965kb,vanNeerven:1983vr,Bern:1992em}.  The
general one loop $N$-point integral in $D = 4 - 2\epsilon$ dimensions
for $N\geq 6$ can be recursively obtained as a linear combination of
pentagon integrals~\cite{Melrose:1965kb,vanNeerven:1983vr}, provided
that the external momenta are restricted to four dimensions.  This
will be discussed in detail in the later sections.
The significance of \Eq(\ref{eq5}) is that once scalar one-loop
integrals are tabulated for $N \leq 4$, 
{\it any} one-loop calculation is reduced to
the determination of both the coefficients $c_{L;j}$ and the rational
part ${\cal R}$.  As we shall demonstrate in this review, the
reduction coefficients and the rational part, can be obtained by
efficient numerical methods based on an algebraic understanding of the
structure of the integrand.

\subsection{One-loop scalar integrals}

We now discuss the set of scalar integrals that appear on the right
hand side in \Eq(\ref{eq5}). These integrals $I_N$ -- traditionally
referred to as tadpoles, bubbles, triangles and boxes -- are defined
as
\begin{equation}
\label{Scalarintegrals}
\begin{split} 
& I_1(m_1^2) =
 \frac{\mu^{4-D}}{i \pi^{\frac{D}{2}}r_{\Gamma}}\int  \frac{d^D l}{d_1},  \\
&I_2(p_1^2;m_1^2,m_2^2)  =
\frac{\mu^{4-D}}{i \pi^{\frac{D}{2}}r_{\Gamma}}\int  \;
\frac{d^D l}{d_1 d_2},   \\
&I_3(p_1^2,p_2^2,p_3^2;m_1^2,m_2^2,m_3^2) =
\frac{\mu^{4-D}}{i \pi^{\frac{D}{2}}r_{\Gamma}}
\int \; \frac{d^D l}{d_1 d_2 d_3}, \;   \\
&I_4(p_1^2,p_2^2,p_3^2,p_4^2;s_{12},s_{23};m_1^2,m_2^2,m_3^2,m_4^2)= 
\frac{\mu^{4-D}}{i \pi^{\frac{D}{2}}r_{\Gamma}}
\int  \frac{d^D l}{d_1 d_2 d_3 d_4}\;,
\end{split}
\end{equation}
where $d_{i} = (l+q_{i-1})^2 - m_{i}^2 + i \varepsilon$,
$q_n\equiv \sum_{i=1}^{n} p_i$, $q_0 = 0$, $s_{ij}=
(p_i+p_j)^2$ and  $r_{\Gamma} = \Gamma^2(1-\ep)
\Gamma(1+\ep)/\Gamma(1-2\ep)$.

For real masses $m_i$ the results for these four types of integrals
are given in Ref.~\cite{tHooft:1978xw}. A further simplification of the
general box integral can be found in Ref.~\cite{Denner:1991qq}.  In the
case where some of the masses vanish, leading to infrared and
collinear divergences, the dimensionally regularized results are given
in Ref.~\cite{Ellis:2007qk}. All scalar one-loop integrals for $N \leq
4$ with vanishing or real masses can be obtained from the Fortran 77
program QCDLoop~\cite{qcdloop}.  The extension of these results to
cases where some of the internal masses are complex -- relevant for
calculations with unstable particles -- was given in
Refs.~\cite{Denner:2010tr,vanHameren:2010cp}.  As a consequence of
these papers, the problem of the analytic calculation of one-loop
integrals and of their numerical evaluation can be considered
completely solved.

\subsection{One-loop tensor integrals and form factor expansion}
\label{Onelooptensorintegrals}

In the calculation of a general one-loop amplitude, individual Feynman
diagrams will give rise to tensor integrals containing powers of the
loop momentum in the numerator. In a renormalizable theory the number
of powers of the loop momentum, $r$ will be limited such that $r \leq
N$ where $N$ is the number of external legs. We shall define $r$ to be
the rank of the tensor integral. The calculation of these tensor
integrals is simple but tedious so it is expedient to reduce the
tensor integrals to the scalar integrals presented in the previous
Section. This was first proposed by Passarino and
Veltman~\cite{Passarino:1978jh}, and we give an explanation of their
method in this Section.  We first define the tensor integrals.  For
reasons of convenience, we have here adopted an alternative notation,
similar to the notation of Passarino and Veltman, for the scalar
integrals, $A_0,B_0,C_0,D_0$ which correspond to scalar tadpole,
bubble, triangle, and box integrals. We write
\begin{equation}
A_0(m_1)= 
\frac{1}{i \pi^{D/2}}\int \; d^D l  \; \frac{1} {d_1},
\end{equation}
\begin{equation}
B_0;B^\mu;B^{\mu\nu}(p_1,m_1,m_2)= 
\frac{1}{i \pi^{D/2}}\int \; d^D l \;  \frac{1;l^\mu; l^\mu l^\nu} {d_1 d_2},
\end{equation}

\begin{equation}
\label{eq:Cint}
C_0;C^\mu;C^{\mu\nu};C^{\mu\nu\alpha}(p_1,p_2,m_1,m_2,m_3) = 
\frac{1}{i \pi^{D/2}}\int \; d^D l \;  \frac{1;l^\mu; l^\mu l^\nu; l^\mu l^\nu l^\alpha}{d_1 d_2 d_3 },
\end{equation}
\begin{eqnarray}
&&D_0;D^\mu;D^{\mu\nu};D^{\mu\nu\alpha};D^{\mu\nu\alpha\beta}(p_1,p_2,p_3,
m_1,m_2,m_3,m_4) = \nonumber \\
&&\frac{1}{i \pi^{D/2}}\int \; d^D l  \; 
\frac{1;l^\mu; l^\mu l^\nu; l^\mu l^\nu l^\alpha; l^\mu l^\nu l^\alpha l^\beta}{d_1 d_2 d_3 d_4 },
\end{eqnarray}
where denominators are given by 
$d_i = (l + \sum \limits_{k=1}^{i-1}p_k)^2 - m_i^2$, $i=1,..,4$.
We give a complete description of the Passarino-Veltman reduction
in~\ref{app:PV}. In the current section we shall present the details
of the reduction of rank-one and rank-two tensor triangle functions to
scalar integrals. This will be sufficient to illustrate the pattern of
the reduction and to make clear what are the potential shortcomings of
the method.  We note that the rank three tensor triangle integrals,
which also occur in a renormalizable theory, are treated
in~\ref{app:PV}.  As a consequence of Lorentz invariance, we may write
\begin{eqnarray}
C^\mu &=& p_{1}^{\mu} C_1 +p_{2}^{\mu} C_2 \,, 
\label{Crank1text} \\
C^{\mu \nu} 
            &=& g^{\mu\nu} C_{00} + \sum_{i,j=1}^2 p_i^\mu p_j^\nu C_{ij}\,, \;\;
\mbox{where}~C_{21}=C_{12}. 
\label{Crank2text}
\end{eqnarray}
We shall refer to the coefficients $C_{i},C_{00},C_{ij}, i,j=1,2$ as
form factors.  The dependence of these form factors on the Lorentz
invariants of the problem,
$p_1^2,p_2^2,(p_1+p_2)^2,m_1^2,m_2^2,m_3^2$ has been
suppressed. 
We contract both sides of \Eq(\ref{Crank1text}) with $p_1$ and
$p_2$. In the numerator of the left-hand side we obtain the following
dot products which may be expressed in terms of the denominators,
\begin{eqnarray}
l \cdot p_1 &=&  
\frac{1}{2} ( f_1+d_2 -d_1),\;\;\; f_1 = m_2^2-m_1^2-p_1^2, 
\label{ldotp1}  \\
l \cdot p_2 &=&  \frac{1}{2} ( f_2+d_3 -d_2),\;\;\; 
f_2 = m_3^2-m_2^2-p_2^2-2 p_1\cdot p_2\,. 
\label{ldotp2}
\end{eqnarray}
We use \Eqs(\ref{Crank1text},\ref{ldotp1},\ref{ldotp2}) 
and obtain a system of equations for the coefficients $C_1,C_2$.
It reads 
\begin{equation}
G_2 \left( \begin{array}{c} C_{1}\\
                            C_{2} \end{array}\right) =
\left( \begin{array}{c} \langle l \cdot p_1\rangle \\
                        \langle l \cdot p_2\rangle \end{array}\right) =
\left( \begin{array}{c} R^{[c]}_1\\
                        R^{[c]}_2 \end{array}\right),
\label{Cformfactoreq0}
\end{equation}
where $G_2$ is the $2\times 2$ Gram matrix
\begin{equation}
G_2= \left( \begin{array}{cc} p_1\cdot p_1 & p_1 \cdot p_2  \\
                              p_1\cdot p_2 &  p_2 \cdot p_2 \\
                         \end{array}\right)\, ,
\end{equation}
and we have introduced the notation
\be
\langle l \cdot p_{j} \rangle  = 
\int \;  \frac{d^D l}{i \pi^{D/2}} 
\; \frac{l \cdot p_{j}}{d_1 d_2 d_3},\;\;\mbox{for}~j=1,2 \; .
\ee
We use \Eqs(\ref{ldotp1},\ref{ldotp2}) to find  explicit 
expressions for the $ R^{[c]}_{1,2}$. They are 
\begin{equation}
\begin{split}
& R^{[c]}_{1} = \frac{1}{2} (f_1 C_0(1,2,3)+B_0(1,3)-B_0(2,3)), 
\\
& R^{[c]}_{2} = \frac{1}{2} (f_2 C_0(1,2,3)+B_0(1,2)-B_0(1,3)) \,. 
\label{Rdef}
\end{split}
\end{equation}
To express our results in \Eq(\ref{Rdef}) we have introduced a
compact notation which labels the form factors by the denominators
they contain.  Thus, for example, in \Eq(\ref{Rdef}), $B_{0}(2,3)$ is
defined as the integral
\beq
B_{0}(2,3) \equiv B_{0}(p_2,m_2,m_3)=
\int \; 
\frac{d^D l}{i \pi^{D/2}}
  \frac{1} {(l^2-m_2^2) ((l+p_2)^2-m_3^2)}\,.
\label{CompactNotation1}
\eeq
Note that in \Eq(\ref{CompactNotation1}) the loop momentum $l$ has
been shifted with respect to the defining equation for the triangle
integrals because $d_1$ has been cancelled.  Finally, solving the
system of equations (\ref{Cformfactoreq0}) we obtain,
\begin{equation}
\left( \begin{array}{c} C_{1}\\
C_{2} \end{array}\right) =G_2^{-1} \left( \begin{array}{c} R^{[c]}_1\\
 R^{[c]}_2 \end{array}\right)\,.
\label{Cformfactoreq1}
\end{equation}

A similar contraction procedure can also be applied to the rank-two
tensor triangle integral, \Eq(\ref{Crank2text}), or to higher rank
tensors as described in~\ref{app:PV}.  For example, contracting
\Eq(\ref{Crank2text}) with $p_1$ and $p_2$ we obtain
\begin{equation}
\begin{split}
& p_{1\;\mu} C^{\mu \nu} = p_1^\nu (p_1\cdot p_1 C_{11} +p_1\cdot p_2 C_{12}+C_{00}) 
+p_2^\nu (p_1\cdot p_1 C_{12} +p_1\cdot p_2 C_{22}),  \\ 
& p_{2\;\mu} C^{\mu \nu} = p_1^\nu (p_1\cdot p_2 C_{11} +p_2\cdot p_2 C_{12}) 
+p_2^\nu (p_1\cdot p_2 C_{12} +p_2\cdot p_2 C_{22}+C_{00})\,.  
\label{Rank2}
\end{split}
\end{equation}
Using \Eqs(\ref{ldotp1},\ref{ldotp2}) we can derive the following 
two equations
\begin{equation}
G_2 \left( \begin{array}{c} C_{11}\\
                            C_{12} \end{array}\right) =
\left( \begin{array}{c} R^{[c1]}_1\\
                        R^{[c1]}_2 \end{array}\right),
\;\;\;\;\;
G_2 \left( \begin{array}{c} C_{12}\\
                            C_{22} \end{array}\right) =
\left( \begin{array}{c} R^{[c2]}_1\\
                        R^{[c2]}_2 \end{array}\right),
\label{Cformfactoreq3}
\end{equation}
where 
\begin{equation}
\begin{split}
& R^{[c1]}_{1} 
=\frac{1}{2} (f_1 C_{1}(1,2,3)+B_1(1,3)+B_0(2,3)-2 C_{00}(1,2,3)), \\ 
& R^{[c1]}_{2} =\frac{1}{2} (f_2 C_{1}(1,2,3)+B_1(1,2)-B_1(1,3)),  
\label{Rc1}
\end{split}
\end{equation}
and 
\begin{equation}
\begin{split}
& R^{[c2]}_{1}=\frac{1}{2} (f_1 C_{2}(1,2,3)+B_1(1,3)-B_1(2,3)), \\ 
& R^{[c2]}_{2}=\frac{1}{2} (f_2 C_{2}(1,2,3)-B_1(1,3)-2 C_{00}(1,2,3))\; . 
\label{Rc2}
\end{split}
\end{equation}
In this way we obtain a ladder of relations which allow us to express
a rank $r$ triangle form-factors in terms of rank $r-1$ triangle
form-factors and sums of bubble form-factors, with rank $r-1$ or less.
Thus, as a general rule, the Passarino-Veltman procedure relates rank
$r$ form-factors of a Feynman integral with $N$ denominators to rank
$r-1$ form-factors of Feynman integrals with $N-1$ denominators plus
other terms which are less ultraviolet singular.  The full pattern of
reduction to scalar integrals is given in
Table~\ref{PVreductionpaths}.
\begin{table}[t]
\begin{center}
\begin{tabular}{|ll|}
\hline
$D_{ijkl}$& $\to\quad D_{00ij},D_{ijk},C_{ijk},C_{ij},C_{i},C_{0}$ \\
$D_{00ij}$& $\to\quad D_{ijk},D_{ij},C_{ij},C_{i}$ \\
$D_{0000}$& $\to\quad D_{00i},D_{00},C_{00}$ \\
$D_{ijk}$& $\to\quad D_{00i},D_{ij},C_{ij},C{_i}$ \\
$D_{00i}$& $\to\quad D_{ij},D_{i},C_{i},C_{0}$ \\
$D_{ij}$& $\to\quad D_{00},D_{i},C_{i},C_{0}$ \\
$D_{00}$& $\to\quad D_{i},D_{0},C_{0}$ \\
$D_{i}$&  $\to\quad D_{0}, C_{0}$\\
\hline
$C_{ijk}$& $ \to\quad C_{00i},C_{ij},B_{ij},B{_i}$ \\
$C_{00i}$& $ \to\quad C_{ii},C{_i},B_{i},B_{0}$ \\
$C_{ij}$& $\to\quad C_{00},C_{i},B_{i},B_{0}$ \\
$C_{00}$& $\to\quad C_{i},C_{0},B_{0}$ \\
$C_{i}$&  $\to\quad C_{0}, B_{0}$\\
\hline
$B_{ii}$& $\to\quad B_{00},B_{i},A_{0}$ \\
$B_{00}$& $\to\quad B_{i},B_{0},A_{0}$ \\
$B_{i}$&  $\to\quad B_{0}, A_{0}$\\
\hline
\end{tabular}
\end{center}
\caption{Reduction chains for Passarino-Veltman procedure, see \ref{app:PV} for a definition of all coefficients.}
\label{PVreductionpaths}
\end{table}

The exception to this rule is the $C_{00}$ term that we treat below.  We also note
that, since the external vectors are purely four-dimensional, the
contraction procedure that we just described does not introduce
an explicit dependence on the dimensionality of space-time in the
reduction equations.

To find the $C_{00}$ coefficient, we note that a further relation can
be obtained by contracting the rank-two tensor integral in
\Eq(\ref{Crank2text}) with the metric tensor $g^{\mu \nu}$.  We find
\begin{eqnarray}
\langle l^2
-m_1^2 \rangle = D\;C_{00} + R^{[c1]}_{1} +R^{[c2]}_{2} -m_1^2 C_{0}.
\end{eqnarray}
Inserting the explicit forms from \Eqs(\ref{Rc1},\ref{Rc2}) we find,
\begin{eqnarray}
C_{00}(1,2,3)&=&\frac{1}{2 (D-2)}
   (2 m_1^2 C_0(1,2,3)-f_2 C_{2}(1,2,3)-f_1 C_{1}(1,2,3)  \nonumber \\
&+&B_0(2,3)). 
\label{C00reduction} 
\end{eqnarray}
Therefore, we see that a pattern of reduction appears
\begin{eqnarray}
C_{ij} &\to& C_{00},C_{i},B_i,(B_0)\,, \nonumber \\
C_{00} &\to& C_{i}, (C_0,B_0)\,,\nonumber \\
C_{i} &\to& (C_0,B_0) .
\end{eqnarray}
The scalar integrals in the reduction path are shown in brackets. 

A simple generalization of the above procedure accomplishes the
reduction of all tensor integrals to scalar integrals. A reduction of
six- and higher-point functions requires additional input, since the
external momenta are not linearly independent, but the basic
principles remain intact. It appears therefore, that the reduction
procedure is a well-established technique that can be applied to any
process of interest in a straightforward way.

However, there are three primary reasons for why it is non-trivial to
perform such a reduction in practice for complicated collider
processes. First, the number of Feynman diagrams grows dramatically
with the number of external particles. For the LHC processes of
interest, the number of diagrams can easily reach a few thousand.
Second, the number of terms generated during the reduction of tensor
integrals grows rapidly with the number of external particles and with
the rank of the integral.  Third, in cases with degenerate kinematics,
the traditional reduction procedure may lead to numerical
instabilities which we consider in Section~\ref{2.4}.

\subsection{Singular regions}
\label{2.4}

As an example, we consider the reduction of the rank one triangle. The
form factors for this integral can be found by solving
\Eq(\ref{Cformfactoreq0}). We obtain
\begin{equation}
\left(\begin{array}{c} C_1\\
                       C_2  \end{array}\right)
= {\boldmath G_2^{-1}} \left(\begin{array}{c} \langle  l \cdot p_{1} \rangle  \\
                                               \langle  l \cdot p_{2} \rangle  
                                               \end{array}\right),
\label{eq347}
\end{equation}
where the inverse of the Gram matrix is given by 
\begin{equation}
G_2^{-1}= \frac{\left( \begin{array}{cc}  p_2\cdot p_2 & - p_1 \cdot p_2 \\
                           - p_1\cdot p_2 &  p_1 \cdot p_1 \\
                         \end{array}\right)}{\Delta_2(p_1,p_2)},
\end{equation}
and 
\be
\Delta_2(p_1,p_2)
={\rm det}[ G_2 ] = p_1^2 p_2^2 -(p_1 \cdot p_2)^2,
\ee
is the determinant of the Gram matrix, the so-called 
{\it Gram determinant}.  

We now investigate the solution \Eq(\ref{eq347}) in the limit $p_1 ||
p_2$ with $p_1^2 \neq 0$.  In this limit, the Gram determinant
$\Delta_2$ vanishes, so that the inverse matrix $G_2^{-1}$ needed for
the construction of the solution \Eq(\ref{eq347}) cannot be
obtained. On the other hand, the original integral, $C^\mu$,
\Eq(\ref{eq:Cint}),
is well-defined in that limit. Therefore, the
problem appears because we attempt to treat the two momenta
$p_1^{\mu}$ and $p_2^{\mu}$ as independent in the form factor
expansion \Eq(\ref{Crank1text}) even in a situation when they are linearly
dependent.

It is easy to remedy this situation, at least in this simple case, by 
using linearly-independent  momenta. To this end, 
we introduce 
\begin{equation} 
\tilde p_2^{\mu} = p_2^\mu  
- \frac{p_1 \cdot p_2}{p_1^2} p_1^{\mu},
\end{equation} 
and write 
\begin{equation}
C^{\mu} 
= p_1^\mu {\tilde C}_1 + \tilde p_2^{\mu} {\tilde C}_2,
\label{eq457}
\end{equation}
instead of decomposition shown in \Eq(\ref{Crank1text}).

Since $p_1 \cdot {\tilde p}_2 = 0$, it is easy to solve for ${\tilde C}_1$ 
and ${\tilde C}_2$. By contracting the integral with $p_{1,2}$ we obtain the 
set of equations 
\begin{equation} 
\begin{split} 
 \langle l \cdot p_1 \rangle   = p_1^2 {\tilde C_1},\;\;\;\; 
 \langle l \cdot p_2 \rangle   = p_2 \cdot p_1 {\tilde C_1} 
+ \frac{\Delta_2(p_1,p_2)}{p_1^2}  {\tilde C}_2,
\end{split} 
\end{equation} 
and find 
\begin{equation} 
\begin{split} 
{\tilde C}_1 = \frac{\langle l \cdot p_1 \rangle }{p_1^2},
\;\;\;
{\tilde C}_2 = 
\frac{p_1^2 \langle l \cdot \tilde p_2 \rangle}{\Delta_2(p_1,p_2)}.  
\label{tildec2}
\end{split} 
\end{equation} 

We can now analyze the limit $p_1 || p_2$.  We write $p_2^\mu = \kappa
p_1^\mu + \delta n^\mu$, $n \cdot p_1 = 0$, $n^2 = 1$, $\delta \ll 1$.
It follows that $\Delta(p_1,p_2)= \delta^2 p_1^2$ and ${\tilde p}_2 =
\delta n^\mu$.  It is easy to see from \Eq(\ref{tildec2}) that 
$p_1^\mu \tilde C_1$ is finite in the limit $p_1 || p_2$ and, because the
integral $C^\mu$ is also finite in that  limit,
$\tilde{p}_2^\mu \tilde C_2$ must be finite as well.  However this
finiteness occurs because there is a cancellation between numerator
and denominator in the expression for $C^\mu$ in \Eq(\ref{eq457}).  In
detail, $\tilde{p}_2^\mu$ is $O(\delta)$ and $\tilde C_2$ is
$O(1/\delta)$.  These features -- and the ensuing cancellations --
become obscured if $\tilde C_2$ is rewritten in terms of master
integrals.  In this case the finiteness of $\tilde C_2$ is achieved
through the cancellation of a number of ${\cal O}(\delta^{-2})$ terms,
including non-trivial relations between three- and two- point
integrals which, in the limit $p_2 \to \kappa p_1$, become linearly
dependent.

The situation described here generalizes to more complicated cases: 
a brute-force application of the Passarino-Veltman reduction procedure
can lead to numerical instabilities due to the vanishing of Gram
determinant at so called ``exceptional points'', despite the fact that
no singularity is present in the original integral. We will discuss in
Sections~\ref{sec:advanced} and \ref{sec:num} how to rescue these
exceptional points.

\subsection{Advanced diagrammatic methods}
\label{sec:advanced}

The computational algorithms for tree and one-loop calculations that
employ Feynman diagrams are suitable for automation. There are public
codes for generating Feynman
diagrams~\cite{Nogueira:1991ex,Hahn:2000kx} which, in conjunction with
algebraic manipulation codes such as Form, Maple and Mathematica,
allow to automatically generate Fortran or C computer codes that
numerically compute scattering amplitudes or cross-sections.  This
type of approach is used in many applications to calculate physical
observables at leading and next-to-leading order accuracy.

Unfortunately, computational approaches based on Feynman diagrams
experience worse than factorial scaling with the number of the
external particles.  As the number of external particles grows, 
high-rank tensor integrals appear in virtual diagrams.
The
Passarino-Veltman reduction generates a multitude of terms and the
number of these terms grows faster than exponentially with the rank of
the tensor.  Furthermore, as we already mentioned, in the numerical
evaluation of one-loop amplitudes one needs to address the issue of
numerical stability related to the vanishing of Gram determinants.

The main subject of this review is to describe an alternative to
Feynman-diagrammatic methods. Still, we emphasize that the question of
when the practical limit on Feynman-diagrammatic one-loop calculations
is reached remains open.  For a long time it was believed that it was
extremely hard, if not altogether impossible, for Feynman-diagrammatic
computations to pass the $2 \to 3$ threshold and deliver physical
results for $2 \to 4$ processes. Yet, in 2009 this threshold was
successfully passed: as the result of additional technical
improvements, techniques based on Feynman diagrams were successfully used
to describe $pp \to t\bar{t}b\bar{b}$ and $pp \to W^+W^-b \bar b$
processes at next-to-leading in perturbative QCD
\cite{Bredenstein:2009aj,Denner:2010jp}.  Similar ideas are
implemented in the GOLEM program~\cite{Binoth:2008uq,Cullen:2010hz} 
and first results on the next-to-leading order computation of 
processes with six external particles have been presented by the 
GOLEM collaboration.

Taking as an example the process $pp \to t\bar{t}b\bar{b}$, we note
that at next-to-leading order in perturbative QCD, it involves two
partonic channels $q\bar{q} \to b\bar{b}t\bar{t}$ and $gg \to b
\bar{b}t\bar{t}$.  There are respectively 188 and 1003 loop diagrams
that contribute to the one-loop amplitudes for the two channels
\cite{Bredenstein:2009aj}.  The computational cost of having to deal
with a large number of Feynman diagrams is compensated by a careful
organization of the computation.  The key idea is to decouple, to the
extent possible, the reduction of tensor loop integrals and other
operations such as multiplication of gamma matrices and spinors,
summations of colors and helicities, etc. from each other.  This is
achieved by paying careful attention to the following issues
\cite{Bredenstein:2009aj}.

First, the interference of the leading (LO) and next-to-leading order (NLO) 
matrix elements is decomposed into the individual contributions
of loop diagrams $\Gamma$. The sum over helicities and colors are  performed
for each loop diagram separately
\be
\begin{split}
\sum_{\rm col}   \sum_{\rm hel} 
 {\cal M}^{\rm (NLO)} \left ( {\cal M}^{\rm (LO)}
\right)^{*}
= 
\sum_{\Gamma}\left[\sum_{\rm col}\sum_{\rm hel}
{\cal M}^{(\Gamma)} \left ( {\cal M}^{\rm (LO)} \right)^{*}\right]\,.
\end{split}
\ee

Second, each individual loop diagram has a color factor that can be 
expanded in a compact color basis.  The leading order amplitude is treated as vector
in this color basis and it is contracted against the color factor of the loop
diagram.

Third, similar to the treatment of color, the spin-dependent parts of
all diagrams can be expanded in a compact spin basis for a given
channel.  For example, a representative term in the spin basis needed
to describe the four-quark, two-gluon channel has the form \be
\begin{split}
\hat{S } =Q^{\mu_1\mu_2\rho_1\ldots \rho_l}
\eps_{\mu_1}(p_1)
\eps_{\mu_2}(p_2)
&
\left[\bar{v}(p_3) \gamma_{\rho_1}
\ldots \gamma_{\rho_m}u(p_4)\right]
\\
&
\times \left[\bar{v(p_5) }\gamma_{\rho_{m+1}}\ldots
\gamma_{\rho_l}u(p_6)\right],
\end{split}
\ee
where $Q^{\mu_1\mu_2\rho_1\ldots \rho_l}$ is a tensor of the
appropriate rank that is composed of metric tensors and external
momenta.  The use of a compact spin basis enables very efficient
helicity summations.

Fourth, clever techniques are used for tensor integral reductions.
Those techniques minimize difficulties related to inverse Gram
determinants and employ a cache system to recycle tensor integrals
with common sub-topologies and avoid the computation of relevant scalar
integrals more than once.  For tensor $N$-point integrals with $N \ge
5$, a procedure can be used that reduces the number of propagators and
the rank of the corresponding integral at the same time
\cite{Denner:2005nn}.  Such a procedure does not introduce inverse
Gram determinants.  In the case of three- and four-point functions,
where the reduction does introduce small Gram determinants, an
expansion procedure around the limit of vanishing Gram determinants 
and other relevant kinematical structures is
applied~\cite{Denner:2005nn,Giele:2004ub}. 

Fifth, careful identification of terms that need to be treated in $D$
dimensions and terms that can be treated in four dimensions, plays an
important role in achieving high computational efficiency.  Finally,
it was observed \cite{Bredenstein:2009aj} that it is possible to give
up on the optimization of the spin basis.  Indeed, in earlier work it
was considered crucial that a minimal, absolutely independent number
of Lorentz structures was used in the parametrization of the one-loop
amplitude.  Instead, it was found in Ref.~\cite{Bredenstein:2009aj}
that one does not lose computational efficiency by only employing
generic four-dimensional identities to reduce the number of
independent terms in the spin basis.  This feature is important since
it minimizes the amount of human intervention in the simplification of
spinor chains.

\section{Van Neerven - Vermaseren basis}
\label{sec3}

\subsection{The Van Neerven - Vermaseren basis}

On-shell scattering amplitudes in gauge field theories are
gauge-invariant. A practical version of this statement is that an
on-shell scattering amplitude must vanish, if evaluated replacing the
polarization vector of a particular massless gauge boson by its four-momentum,
provided all the other gauge bosons have physical polarizations.  This
provides both a constraint on the form of the amplitude and a powerful
check on the computation.  However, it is well-known that in
complicated cases involving higher-point scattering amplitudes,
the analytic demonstration of this cancellation is non-trivial.  One
reason why such complications arise is the four-dimensional nature of
space-time, since it implies that for high-point amplitudes the external
momenta are not linearly independent.  In four dimensions, the
``dimensionality constraint'' can be stated in the form of the
Schouten identity
\begin{equation}
\label{Schouten}
l^{\lambda}\ep^{\mu_1\mu_2\mu_3\mu_4} = 
l^{\mu_1}\ep^{\lambda\mu_2\mu_3\mu_4}
+l^{\mu_2}\ep^{\mu_1\lambda\mu_3\mu_4}
+ l^{\mu_3}\ep^{\mu_1\mu_2\lambda\mu_4} 
+ l^{\mu_4}\ep^{\mu_1\mu_2\mu_3\lambda}\,,
\end{equation}
which follows from the vanishing of the totally antisymmetric
rank-five tensor in four dimensions.  Since these constraints are not
implemented in the Passarino-Veltman procedure, it is usually not easy
to demonstrate gauge cancellations in that framework.

The Schouten identities also provide us with a way to introduce a
particular reference frame that, as we will see, is very useful for
reducing tensor one-loop integrals to scalar integrals.  To motivate
this choice, consider a two-dimensional vector space spanned by two
{\it non-orthogonal} two-dimensional vectors
$q_1^{\mu_1},q_2^{\mu_2}$.  Any two-dimensional vector from that
vector space can be written as a linear combination of these vectors
$l^{\alpha}=c_1q_1^{\alpha}+c_2q_2^{\alpha}$.  However, since
$q_1,q_2$ are not orthogonal, $c_i \neq l \cdot q_i$.  A standard way
to introduce the orthonormal basis is the Gram-Schmidt
orthogonalization procedure, but we will not pursue it here. Instead
we will choose a convenient basis starting from Schouten identity in
two dimensions
\begin{equation}
\label{Schouten2}
l^{\lambda}\ep^{\mu_1\mu_2} = 
l^{\mu_1}\ep^{\lambda\mu_2}
+l^{\mu_2}\ep^{\mu_1\lambda}\,.
\end{equation}
Contracting both sides of this equation 
with $q_1^{\mu_1}q_2^{\mu_2}$ we obtain
\be
l^{\lambda} \ep^{q_1q_2}=(l \cdot q_1)\ep^{\lambda q_2}
+(l \cdot q_2)\ep^{q_1\lambda},
\ee
where $\quad e^{\lambda q_2}=
\ep^{\lambda \mu_2} q_{2,\mu_2}$ and $\ep^{q_1q_2} = 
\epsilon^{\mu_1 \mu_2} q_{1,\mu_1} q_{2,\mu_2}.$
We divide both sides by $\ep^{q_1q_2}$, introduce
vectors~\footnote{The case $\ep^{q_1q_2}\to 0$ requires special care,
  see for instance the discussion in Sect.~\ref{2.4}.} 
\be
v_1^{\lambda}=\frac{ \ep^{\lambda q_2} }{\ep^{q_1 q_2}},
\;\;\;\;\;\;\;\; 
v_2^{\lambda}=\frac{\ep^{q_1\lambda} }{\ep^{q_1 q_2}},
\;\;\;\;\;\; v_i \cdot  q_j=\delta_{i j},
\ee
and write 
\be
\label{Schouten3}
l^{\lambda} =(l \cdot q_1)v_1^{\lambda}+(l \cdot q_2) v_2^{\lambda}.
\ee
The vector sets  $\{ v_i \}$ and $\{ q_j \}$ are orthogonal, 
but  vectors $v_i$ are not orthonormal, $v_i \cdot v_j \neq  \delta_{ij}$. 
Nevertheless, they are useful because they define
a coordinate system where
the $v_i$-coordinate of an arbitrary vector $l$ 
is the projection of this vector on 
the vector $q_i$ . If  we identify 
$l+q_i$ with momenta that appear in propagators of 
one-loop diagrams (see Fig.~\ref{generic}), 
the  scalar product  $ l \cdot q_i$
can be replaced by differences of denominators
\be
l \cdot q_i=\frac{1}{2}\left[
((l+q_i)^2-m_i^2)-(l^2-m_0^2)+m_i^2-m^2_0-q_i^2\right].
\ee
This strategy has already been used in the Passarino-Veltman reduction, 
c.f.\ \Eqs(\ref{ldotp1},\ref{ldotp2}), and 
as we explain later in the review, this feature will be 
used to develop a systematic procedure to determine the 
parametric form of one-loop integrands.

The two-dimensional example, however,  is not 
sufficient since we have to deal with higher-dimensional
vector spaces.  We therefore need to extend the considerations 
described above. 
To this end, we write 
\be
v_1^\mu = \frac{\epsilon_{q_1 q_2} \epsilon^{\mu q_2}}{
\epsilon_{q_1 q_2} \epsilon^{q_1 q_2}},\;\;\;\;
v_2^\mu = \frac{\epsilon_{q_1 q_2} \epsilon^{q_1 \mu}}{
\epsilon_{q_1 q_2} \epsilon^{q_1 q_2}},
\ee
and use 
\be
\ep^{\mu_1\mu_2}\ep_{\nu_1\nu_2}=
\delta^{\mu_1}_{\nu_1}\delta^{\mu_2}_{\nu_2}
 - \delta^{\mu_1}_{\nu_2}\delta^{\mu_2}_{\nu_1}
= \det | \delta^{\mu}_{\nu} |
\equiv \delta^{\mu_1\mu_2}_{\nu_1\nu_2},
\ee
to write  vectors $v_{1,2}$ using the  basis of generalized 
Kronecker delta-symbols
\be
\label{Schouten4}
v_1^\mu=\frac{\delta^{\mu q_2}_{q_1 q_2}}{\Delta_2} \,,\quad
v_2^\mu=\frac{\delta^{q_1 \mu}_{q_1 q_2}}{\Delta_2} 
\,,\quad \Delta_2=\delta^{q_1 q_2}_{q_1 q_2}=q_1^2q_2^2-(q_1 \cdot q_2)^2.
\ee
We recognize that, in contrast to Levi-Civita tensors, the generalized
Kronecker deltas can be introduced for vector spaces of arbitrary
dimensions, allowing us to define the basis of dual vectors $v_i$ that
can be used in four-dimensional calculations.  Such basis is called
the van Neerven - Vermaseren basis \cite{vanNeerven:1983vr} .

The van Neerven - Vermaseren basis has proved to be very useful to
understand a number of important results concerning the reduction
of tensor integrals and the applicability of the generalized
unitarity.  First, in four dimensions, simple algorithms were derived
for the reduction of tensor integrals to the linear combination of
box, triangle, bubble and tadpole scalar integrals.  The number of
terms generated in this process is smaller than in
the standard Passarino-Veltman reduction procedure.  Some illustrative
results are presented in Section 4. Second, using the van Neerven -
Vermaseren basis, it is straightforward to show that in four
dimensions the scalar five-point Feynman integral is given by a linear
combination of scalar box
integrals~\cite{Melrose:1965kb,vanNeerven:1983vr}.  Third, using the
van Neerven - Vermaseren basis it is easy to understand that in four
dimensions the {\it integrand} of any one-loop Feynman diagram in any
renormalizable theory is given by a linear combination of 
rational functions containing products of four, three, two  or one Feynman denominators
and with the numerators
of a very restrictive form, see Section~\ref{sect5}.  Finally,
employing the van Neerven - Vermaseren decomposition, it is
straightforward to find the loop momenta that satisfy quadruple,
triple-, double-, and single-cut on-shell conditions, see
Section~\ref{sect5}.  These features of the van Neerven - Vermaseren
basis make it important for the construction of the
generalized $D$-dimensional unitarity technique. We therefore devote 
the following subsection to a detailed explanation of 
the van Neerven - Vermaseren basis.

\subsection{Physical  and  transverse space}

\begin{figure}[t]
\begin{center}
\includegraphics[scale=0.8]{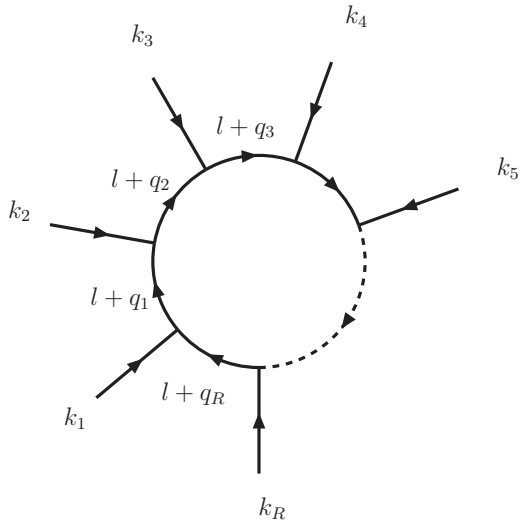}
\end{center}
\caption{Generic diagram with $R$ external momenta at one-loop.}
\label{generic}
\end{figure}

We consider a $N$-particle scattering amplitude in a renormalizable
quantum field theory in $D$-dimensional space-time.  Such an amplitude
can be computed from the relevant Feynman diagrams, each given by an
integral over the loop momentum $l$ of an integrand function.  We
study one of these Feynman diagrams and imagine that it has $R$
loop-momentum-dependent propagators, see Fig.~\ref{generic}.  The
integrand ${\cal I}_N$ is a rational function of the loop momentum $l$
given by the product of $R$, $l$-dependent scalar inverse propagators
$d_i$ and a polynomial in $l$ of rank $r_l \leq R$.
\beq
{\cal I}_{N}(p_1,p_2,\ldots,p_N|l)=
\frac{{\cal N_{\cal I}}(p_1,p_2,\ldots,p_N;l)}{d_1 d_2\cdots d_R} \; .
\eeq
The amplitude  has   a set of $R$ inflow momenta, $k_1,\ldots,k_R$. 
The inflow momenta are either 
equal to the external momenta $p_i$, or are given 
by their linear combinations 
\beq
d_i=(l+q_i)^2-m_i^2\,, \quad
k_i = q_{i} - q_{i-1}, \;\; 
k_i=\sum_{j=1}^{N} \alpha_{ij} p_j\,,\quad \sum_{i=1}^R k_i=0,
\eeq
where $\alpha_{ij}=0,1$ are diagram-specific numbers. Sometimes we refer 
to the $q_i$ vectors as the  ``propagator momenta'' .
We call the vector space spanned by the inflow momenta the {\it
  physical space}.  We emphasize that the dimensionality $D_P$ of the
physical space changes from diagram to diagram. Accounting for 
momentum conservation $\sum \limits_{i=1}^R k_i=0$, we obtain
\be
D_P=
 \min(D,R-1),
\ee 
which implies that for $R \le D$, the dimensionality of the physical
space is smaller than the dimensionality of space-time.  The authors
of Ref.~\cite{vanNeerven:1983vr} advocate 
the use of a non-orthogonal coordinate system in the physical space.
This coordinate system is dual to the non-orthogonal coordinate system of the inflow momenta.
By contrast, in the
$D_T$-dimensional transverse space we can use a standard
ortho-normal coordinate system.  The dimensionalities of various spaces
satisfy obvious constraints
\beq
D=D_P+D_T,\;\; D_P=\min{(D,R-1)},\;\; D_T=\max{(0,D-R+1)}.
\eeq
If $R>D$, the transverse space is zero-dimensional.

To define the van Neerven - Vermaseren basis we introduce the generalized 
Kronecker symbol~\cite{vanOldenborgh:1989wn}
\footnote{
This notation is closely related to the asymmetric Gram determinant 
notation of Ref.~\cite{Kajantie},
\[ G\left(\begin{array}{ccc} k_1 & \cdots & k_R \\ q_1 & \cdots & q_R \end{array}\right)=
\delta^{k_1k_2\cdots k_R}_{q_1q_2\cdots q_R}\ .\]}
\beq
\delta^{\mu_1\mu_2\cdots\mu_R}_{\nu_1\nu_2\cdots\nu_R}=\left| \begin{array}{cccc} 
\delta_{\nu_1}^{\mu_1} & \delta_{\nu_2}^{\mu_1} & \dots & \delta_{\nu_R}^{\mu_1} \\
\delta_{\nu_1}^{\mu_2} & \delta_{\nu_2}^{\mu_2} & \dots & \delta_{\nu_R}^{\mu_2} \\
\vdots & \vdots & &\vdots \\ 
\delta_{\nu_1}^{\mu_R} & \delta_{\nu_2}^{\mu_R} & \dots & \delta_{\nu_R}^{\mu_R}
\end{array}\right|\ ,
\eeq
a compact notation for the Kronecker symbol contracted with momenta,
\beq
\delta^{k \mu_2\cdots\mu_R}_{\nu_1k\cdots\nu_R}\equiv
\delta^{\mu_1\mu_2\cdots\mu_R}_{\nu_1\nu_2\cdots\nu_R}k_{\mu_1}k^{\nu_2}\ ,
\eeq
and the $R$-particle Gram determinant
\beq
\Delta(k_1,k_2,\cdots,k_{R})=\delta^{k_1k_2\cdots k_{R}}_{k_1k_2\cdots k_{R}}\ .
\eeq
Note that for $R\geq D+1$ the generalized Kronecker delta vanishes.
For the special case $D=R$ the Kronecker delta factorizes into the
product of two Levi-Civita tensors $\delta^{\mu_1\mu_2\cdots\mu_
  R}_{\nu_1\nu_2\cdots\nu_R}=\varepsilon^{\mu_1\mu_2\cdots\mu_R}\varepsilon_{\nu_1\nu_2\cdots\nu_R}$.
The generalized Kronecker-deltas are determinants of $R$-dimensional
matrices.  For example for $R=2$ and $R=3$ we have the expressions
\begin{equation}
\begin{split}
 \delta^{k_1k_2}_{k_1\mu} &=
k_1\cdot k_1\,\delta^{k_2}_{\mu}-{k_1}_{\mu}\delta^{k_2}_{k_1}, 
= k_1\cdot k_1\,{k_2}_{\mu}-k_2\cdot k_1\, {k_1}_{\mu},
\\
 \delta^{k_1k_2k_3}_{k_1k_2k_3} & =k_1\cdot k_1\,\delta^{k_2k_3}_{k_2k_3}
-k_1\cdot k_2\,\delta^{k_2k_3}_{k_1k_3}
+k_1\cdot k_3\,\delta^{k_2k_3}_{k_1k_2} 
\\
& =k_1\cdot k_1\,(k_2\cdot k_2\,k_3\cdot k_3-k_2\cdot k_3\, k_3\cdot k_2) 
\\
& -k_1\cdot k_2\,(k_2\cdot k_1\,k_3\cdot k_3-k_2\cdot k_3\, k_3\cdot k_1) 
\\
& +k_1\cdot k_3\,(k_2\cdot k_1\,k_3\cdot k_2-k_2\cdot k_2\, k_3\cdot k_1). 
\end{split} 
\end{equation}

We can use the Kronecker $\delta$-symbols to construct the 
van  Neerven - Vermaseren basis vectors  for the physical 
space $D_P$
\beq
v_i^\mu (k_1,\ldots,k_{D_P}) \equiv  
 \frac{\delta^{k_1\ldots k_{i-1}\mu k_{i+1}\ldots k_{D_P}}_{k_1\ldots 
k_{i-1}k_ik_{i+1}\ldots k_{D_P}}}{\Delta (k_1,\ldots,k_{D_P})}\ ,
\label{def:v}
\eeq
The basis vectors satisfy orthogonality and normalization properties
\be
\label{vinorm} 
v_i\cdot k_j=\delta_{ij}\,, \quad {\rm for}\quad  j\leq D_P. 
\ee
When $R\leq D$ it is convenient  to define  also the projection operator 
onto the transverse space 
\beq
{w_\mu}^\nu (k_1,  \ldots ,k_{R-1}) \equiv  
\frac{\delta^{k_1\cdots k_{R-1}\nu}_{k_1\ldots k_{R-1}\mu}}{\Delta(k_1,\ldots,k_{R-1})}\ .
\label{def:w}
\eeq
It fulfills  the properties of a projection operator 
\be
{w_\mu}^\mu=D_T=D+1-R\,,\; 
k_i^\mu \, w_{\mu \nu} =0\,,\; 
 {w^{\mu}}_{\alpha}w^{\alpha\nu}=w^{\mu\nu}.
\ee
We denote the orthonormal unit vectors of the transverse space by
$n_r^{\mu}\,, r=1\ldots D_T$. They satisfy the standard orthogonality
and normalization requirements
\be
\begin{split}
n_r\cdot n_s=\delta _{r s}\,,\; &k_i\cdot n_r=0\,,\; 
v_i\cdot n_r=0,\;\;\;\;
w^{\mu\nu}=\sum_{r=1}^{D_T}n_r^{\mu}n_r^{\nu},
\end{split}
\label{metricdecomp}
\ee
where $i=1\ldots D_P$,~ $r,s=1\ldots D_T$ and $w^{\mu \nu}$ is the
metric tensor of the transverse subspace.  The tensor decomposition of
the full metric tensor is given by the expression
\beq\label{fullmetricdecomp}
g^{\mu\nu}=\sum_{i=1}^{D_P} k_i^{\mu}v_i^{\nu}+w^{\mu\nu}
=\sum_{i=1}^{D_P} k_i^{\mu}v_i^{\nu}+\sum_{i=1}^{D_T} n_i^{\mu}n_i^{\nu}\ .
\eeq
Note that the right hand side of this equation is,  actually, 
a symmetric tensor since, by explicitly writing  the generalized Kronecker 
delta-function using  $k_i$ vectors, one can show that the following 
equation holds 
\be
\sum \limits_{i=1}^{D_P} k_i^\mu v_i^\nu = 
\sum \limits_{i=1}^{D_P} k_i^\nu v_i^\mu.
\ee
For the case $D=R$, the transverse space is one-dimensional and the
unit vector $n_1$ is proportional to a Levi-Civita tensor.  For the
cases $R < D$ it is a simple task to construct explicitly the
$n_r^{\mu}$ basis vectors that fulfill the requirements given in
\Eq(\ref{metricdecomp}).  As an example, if $D=4$ and $R=4$, we get
\be
\begin{split}
& v_1^{\mu}(k_1,k_2,k_3)=\frac{\delta^{\mu k_2k_3}_{k_1k_2k_3}}{\Delta (k_1,k_2,k_3)},\
v_2^{\mu}(k_1,k_2,k_3)=\frac{\delta^{k_1\mu k_3}_{k_1k_2k_3}}{\Delta (k_1,k_2,k_3)},\
\\
& v_3^{\mu}(k_1,k_2,k_3)
=\frac{\delta^{k_1k_2\mu }_{k_1k_2k_3}}{\Delta (k_1,k_2,k_3)};
\\
& {w_{\mu}}^{\nu}(k_1,k_2,k_3)
=\frac{\delta^{k_1k_2k_3\nu}_{k_1k_2k_3\mu}}{\Delta(k_1,k_2,k_3)}
={n_1}_{\mu}{n_1}^{\nu}
=\frac{\varepsilon_{k_1k_2k_3\mu}\varepsilon^{k_1k_2k_3\nu}}{\Delta(k_1,k_2,k_3)}\ .
\end{split} 
\ee
In applications to one loop calculations, it is often needed to write
the loop momentum $l$ as a linear combination of the van Neerven -
Vermaseren basis vectors for a particular graph with the denominator
factors $d_1,d_2,...d_R$.  The denominators are given by
$d_i=(l+q_i)^2-m_i^2$ and the inflow momenta read $k_i=q_i-q_{i-1}$.
The decomposition is obtained by contracting the loop momentum with
the metric tensor given in \Eq(\ref{fullmetricdecomp})
\beq
l^{\mu}=\sum_{i=1}^{D_P} 
( l\cdot k_i )\,v_i^{\mu}
+\sum_{i=1}^{D_T} ( l \cdot n_i )\ n_i^{\mu}\ .
\eeq
Using the identity
\beq
\label{qiIdentity}
l\cdot k_i=\frac{1}{2}\left[d_i-d_{i-1}-\left(q_i^2-m_i^2\right)+\left(q_{i-1}^2-m_{i-1}^2\right)\right]\ ,
\eeq
we find
\beq
l^\mu = V_R^{\mu}
+\frac{1}{2} \sum_{i=1}^{D_P} (d_i-d_{i-1})\, 
v_i^{\mu}+\sum_{i=1}^{D_T}(l \cdot n_i)\, n_i^\mu\ ,
\label{eq:l3} 
\eeq
where $d_0=d_R$, $m_0=m_R$ and
\beq
V_R^\mu=-\frac{1}{2}\sum_{i=1}^{D_P} \Big((q_i^2-m_i^2)-(q_{i-1}^2-m_{i-1}^2)\Big)\, v_i^\mu\ .
\eeq

As an illustration of this procedure, we explicitly give the
loop-momentum decomposition in two cases. The first example concerns
the five-point function in four dimensions, so that $D=4$ and $R=5$.
We derive
\begin{equation} 
\begin{split} 
l^\mu &=V_5^{\mu}+\frac{1}{2}(d_1-d_5)\, v_1^{\mu}+\frac{1}{2}(d_2-d_1)\, v_2^{\mu}
\\
& +\frac{1}{2}(d_3-d_2)\, v_3^{\mu}
+\frac{1}{2}(d_4-d_3)\, v_4^{\mu}, 
\\
V_5^{\mu} &=-\frac{1}{2}(q_1^2-q_5^2-m_1^2+m_5^2)\, 
v_1^{\mu}-\frac{1}{2}(q_2^2-q_1^2-m_2^2+m_1^2)\, v_2^{\mu}
\\
&-\frac{1}{2}(q_3^2-q_2^2-m_3^2+m_2^2)\, 
v_3^{\mu}-\frac{1}{2}(q_4^2-q_3^2-m_4^2+m_3^2)\, v_4^{\mu} \; .
\end{split} 
\end{equation} 
Similarly, for a three-point function in 
four dimensions $D=4$ and $R=3$. We obtain 
\begin{equation} 
\begin{split} 
l^\mu & =V_3^{\mu}+\frac{1}{2}(d_1-d_3)\, 
v_1^{\mu}+\frac{1}{2}(d_2-d_1)\, v_2^{\mu}
+(l \cdot n_1) \, n_1^{\mu}+(l \cdot n_2) \, n_2^{\mu}, 
\\
V_3^{\mu}& =-\frac{1}{2}(q_1^2-q_3^2-m_1^2+m_3^2)\, v_1^{\mu}
-\frac{1}{2}(q_2^2-q_1^2-m_2^2+m_1^2)\, v_2^{\mu} \; .
\end{split}
\end{equation} 

We conclude this Section with a few comments.  We note that if the
number of inflow momenta $R$ exceeds the dimensionality of space-time
$D$, the decomposition of the loop momentum into the van Neerven -
Vermaseren basis may be used to prove that the $D+m$ point functions
$m \ge 1$ can all be written as linear combinations of the $D$-point
functions.  We will show an example of this in the next Section.  We
also remark that if we set $q_0=0$ we can choose to parametrize the
loop momenta using the coordinate system of the $q_i$ vectors with its
associated dual coordinate basis $v_i(q_1,q_2,\ldots,q_{D_P}) $ as
given by \Eqs(\ref{def:v},\ref{def:w}).  We note that
$\Delta(k_1,k_2,\ldots, k_{D_P})=\Delta(q_1,q_2,\ldots,_{D_P})$.
Similarly the projection operator onto the transverse space remains
the same. The identity \Eq(\ref{qiIdentity}) and all those relations
that depend on this identity will be modified accordingly. We conclude
that it is possible to change from one basis, to another linearly
dependent one at essentially no cost. This fact can be useful in
numerical applications. Finally, we emphasize that both versions of
the van Neerven - Vermaseren basis allow us to include the unitarity
constraints without resorting to the spinor-helicity formalism, which is
most often used in analytic calculations with massless particles. By
avoiding the spinor-helicity formalism, the method can be used in
computations with massive internal particles, where the mass can be
either real or complex-valued.

\section{Reduction at the integrand level in two dimensions}
\label{sec:red2d}

Analytic calculations in four dimensions require significant algebraic
effort, that often obscures the conceptual aspects of reduction
techniques using the van Neerven-Vermaseren basis. The amount of
algebra can be kept to a minimum by working in two-dimensional
space-time.  This section presents a number of two dimensional
examples.  In what follows we first show that, in two dimensions, the
three-point function can be always written as a linear combination of
two-point functions. After that, we express a rank-two, two-point
function in terms of scalar two-point functions and tadpoles using
unitarity-based ideas.  Finally, we discuss a physical example in
two-dimensional space-time where the rational part plays an important
role.

\subsection{Reduction of a scalar triangle}

We consider a scalar three-point function in two-dimensional space-time, 
\beq
I_3 = \int \frac{{\rm d}^2 l}{(2\pi)^2}\; {\cal I}_3\;\;,
\eeq
and focus specifically on the integrand given by,
\beq
{\cal I}_3=\frac{1}{d_0d_1d_2},
\eeq 
where 
$l$ is the loop momentum, $d_i = (l+q_i)^2 - m_i^2$, $i \in [0,1,2]$  
and $q_0 = 0$. 
To show that, in two dimensions, the three-point 
function is given by the linear combination 
of the two-point functions, we use the fact that 
the loop momentum can be written as a linear combination of the vectors 
$q_{1,2}$.  It is convenient to 
employ  the van Neerven  - Vermaseren basis for this purpose.  For the 
case of a three-point function in two dimensions, 
the dimensionality of space-time and the dimensionality 
of the physical space coincide. Therefore, we write 
\beq
\label{lSchouten}
l^\mu=v_1^\mu (l \cdot q_1)+v_2^\mu (l \cdot q_2)\,, \quad 
v_1^\mu=\frac{\delta^{\mu q_2}_{q_1q_2}}{\Delta_2},\;\;
v_2^\mu=\frac{\delta^{q_1 \mu}_{q_1q_2}}{\Delta_2},
\eeq
where, as usual $v_i \cdot q_j = \delta_{ij}$ and $\Delta_2$ is 
the two-dimensional Gram determinant 
\beq
\Delta_2= q_1^2 q_2^2 - (q_1 \cdot q_2)^2.
\eeq
We can eliminate the scalar products $l \cdot q_{1,2}$ 
from \Eq(\ref{lSchouten}) using the following equations
\beq
l\cdot q_i =\frac{1}{2} (d_i-d_0-r_i)\,, \quad 
r_i=q_i^2-m_i^2+m_0^2\,, \quad i=1,2.
\label{z150}
\eeq
To perform this elimination we contract \Eq(\ref{lSchouten})  with $l$, use 
\Eq(\ref{z150}) and $l^2 = d_0 +m_0^2$, 
and write 
\beqa
\label{scalarSH}
2(d_0+m_0^2)=\sum \limits_{i=1}^{2}(l\cdot v_i)(d_i-d_0) 
-\sum \limits_{i=1}^{2}(l\cdot v_i) r_i .
\eeqa
We can use the identity 
\be
l^\mu = g^{\mu \nu} l_\nu 
= \sum \limits_{j=1}^{2} v_j^\mu \, (q_j \cdot l),
\ee
to rewrite the last term in \Eq(\ref{scalarSH})
in the following form
\beq
\label{lvident}
\sum \limits_{i=1}^{2} (l \cdot v_i)r_i 
= \sum \limits_{j=1}^{2} ( w \cdot v_j) (l \cdot q_j) 
=\frac{1}{2} \sum \limits_{j=1}^{2} ( w\cdot v_j)  \; 
(d_j-d_0-r_j),
\eeq 
where the vector $w^\mu$ is defined as 
\beq
w^\mu=\sum \limits_{i=1}^{2}  v_i^\mu  r_i  \,.
\eeq
We use \Eq(\ref{lvident}) to simplify the last term in \Eq(\ref{scalarSH}),
and get 
\beq 
2 d_0 + 2 m_0^2 = \frac{1}{2}\left(\sum_i \left[(2l \cdot v_i)-(w\cdot v_i)\right]d_i
- \sum_i (2l - w)\cdot v_i d_0 +
\sum_i r_i (w\cdot v_i)\right)\,.
\eeq
Dividing this equation by $d_0 d_1 d_2$, using $w^2 = \sum_i
r_i (w\cdot v_i)$, 
and collecting similar terms,
we obtain the reduction formula for the integrand of the 
three-point 
function in two-dimensional space-time 
\be 
\label{2DimIntegrandreduction}
\begin{split} 
{\cal I}_3  =\frac{1}{(4 m_0^2-w^2)}
\Bigg \{& \frac{2(l\cdot v_1)-(w \cdot v_1)}{d_0d_2} 
    +  \frac{2(l \cdot v_2)-(w \cdot v_2)}{d_0d_1}
\\
& 
 - \frac{   4+ (2l-w)  \cdot  ( v_1+v_2)   } {  d_1 d_2  }
\Bigg \}.
\end{split} 
\ee

Later on, we will  discuss in detail 
the parametrization of the integrand due to Ossola, Pittau 
and Papadopoulos (OPP),   
but \Eq(\ref{2DimIntegrandreduction})
provides a first example of the OPP parametrization applied to the 
case where the dimensionality of space-time and the dimensionality of 
the physical space coincide, $R=D=2$.
Indeed, the numerator of each 
of the three terms on the right hand side of \Eq(\ref{2DimIntegrandreduction}) 
is a rank-one tensor of the form 
$
b_0 + b_1 (n \cdot l). 
$
This tensor is special since the loop momentum $l$ appears multiplied 
by a vector $n$, that is orthogonal to the reference vector in the denominator 
of each individual term. This is evident for the first two terms 
on the right-hand side of \Eq(\ref{2DimIntegrandreduction})
since 
$q_2 \cdot v_1 = q_1 \cdot v_2 = 0$, but it  is also true 
in the third term.
Indeed,  shifting the loop momentum in the 
third term in \Eq(\ref{2DimIntegrandreduction})
$
l \to \tilde l = l - q_1, 
$
we observe that the reference momentum in the denominator 
becomes $q_2 - q_1$. Since 
$ (q_2 - q_1) \cdot (v_2 + v_1) = 0$, we have demonstrated our assertion.

To see why the special form of the tensor 
is important, we compute the three-point 
function by performing the integral over the loop momentum 
\be
I_3 = \int \frac{{\rm d}^2 l}{(2\pi)^2}\; {\cal I}_3.
\label{z345}
\ee
In spite of the fact that the right hand side 
of \Eq(\ref{z345}) contains an integral of a rank-one 
tensor, cf. \Eq(\ref{2DimIntegrandreduction}), 
the integration is trivial. Indeed, the loop momentum 
is always contracted with the basis vector of the transverse space 
and the corresponding angular integrals vanish by symmetry. 
We obtain\footnote{It is convenient to shift of the loop momentum $l
  \to l -q_1$ in the last term of \Eq(\ref{2DimIntegrandreduction}), so
  that the remaining vector integral vanishes by symmetry.}
\be 
I_3 = \frac{(-1)}{(4
  m_0^2-w^2)} \left\{ ( w \cdot v_1) I_{02} + ( w \cdot v_2 ) I_{01}
  +(2 - w \cdot ( v_1+v_2)) I_{12} \right\},
\label{z346}
\ee
where  $\displaystyle 
I_{ij} = \int \frac{{\rm d}^2l}{(2\pi)^2} \frac{1}{d_i d_j} $ is a two-point 
function. 
\Eq(\ref{z346}) completes our proof that in the two-dimensional 
space-time, the three-point function is given by a linear combination 
of two-point functions. It is important to realize that 
this result generalizes. Indeed, any $N$-point function, for 
$N > D$, where $D$ is the dimensionality of space-time, can be written 
as a linear combination of the $D$-point functions.  Finally, 
we will see in what follows that 
the two principal ideas behind the reduction process just outlined -- 
the use of the van Neerven - Vermaseren basis and the reduction 
by an algebraic integration over the transverse space after 
establishing the parametric form of the integrand can be easily extended 
to higher-dimensional spaces and higher-point functions, making 
it a powerful tool for one-loop computations. 

\subsection{Reduction of a rank-two two-point function}

Our next two-dimensional example concerns 
the reduction of a rank-two two-point function using 
van Neerven - Vermaseren basis. Consider an integrand given by
\beq
{\cal I} (k,\mo,\mt) = \frac{(\nh \cdot l)^2 }{d_1 d_2},
\label{eq121}
\eeq
where $d_1 = l^2-\mo^2$, $d_2 = (l+k)^2-\mt^2$, $\nh \cdot k = 0, k^2 \neq 0$ 
and $\nh^2 = 1$.  Because of the projection onto $\nh$, 
the momentum $l$ in the numerator in \Eq(\ref{eq121}) 
lies in the transverse space.
We want to express this integral in terms of scalar integrals. 

Note that in contrast to 
the three-point function considered in the preceding 
Section, the rank-two two-point function in two dimensions 
has an ultraviolet divergence. We regularize this divergence by continuing 
the loop momentum to $D = 2-2\ep$ dimensions 
and begin by constructing the van Neerven - Vermaseren 
basis. As the basis vector of the physical space, we take 
\beq
n^\mu = \frac{k^\mu}{\sqrt{k^2}},\;\;\; n^2 =1. 
\eeq
We choose  $\nh$ to be the basis vector of the transverse space, which 
is allowed since $n$ and $\nh$ are orthogonal, 
$n \cdot \nh  = 0$.  As the consequence 
of the completeness relation, the two vectors satisfy 
\be
n^\mu  n^\nu  + \nh^\mu  \nh^\nu  =g_{(2)}^{\mu \nu}, 
\ee
where $g_{(2)}^{\mu \nu}$ is the two-dimensional metric tensor. 
Contracting this equation with the loop momentum, we obtain 
\beq
(\nh \cdot l)^2=\lt^2 - (n \cdot l)^2 = \lt^2 
- \frac{(l \cdot k)^2}{k^2}.
\label{eq59}
\eeq
Because $l$ is a $d$-dimensional vector, we can decompose it as 
\beq
l^\mu = (l \cdot n) n^\mu+(l \cdot \nh) \nh^\mu + \ne^\mu (l\cdot \ne),
\eeq
where $\ne $ is the unit vector that parametrizes the $(D-2)$-dimensional 
vector space. It follows that the square of the $d$-dimensional loop momentum 
can be written as 
\beq
l^2 = l_{(2)}^2+(\ne \cdot l)^2 = l_{(2)}^2+\mu^2,
\label{eq62}
\eeq
where $\mu^2 = (\ne \cdot l)^2$ is introduced.  
To proceed further,  we express various scalar products through 
inverse Feynman propagators $d_{1,2}$ 
\be
l_{(2)}^2 = d_1 + m_1^2-\mu^2,\;\;\; 2 \, l\cdot k = d_2 - d_1 - r_1^2,
\label{eq632}
\ee
and use \Eqs(\ref{eq59},\ref{eq62}) to obtain 
\beq \label{x1}
\frac{(\nh \cdot l)^2 }{d_1 d_2} = -\frac{(\lambda^2+\mu^2)}{d_1 d_2} 
       + \frac{1}{4 k^2} \Bigg[
           \frac{r_1^2 - 2 \, l \cdot k}{ d_1}
          + \frac{r_2^2 + 2 \, l \cdot k+2 k^2}{ d_2} \Bigg].
\eeq
In \Eqs(\ref{eq632},\ref{x1}), 
we use the following short-hand notations
\be
\begin{split}
& r_1^2 = k^2+\mo^2-\mt^2,\;\;\; r_2^2 = k^2+\mt^2-\mo^2, \\
& \lambda^2 = \frac{k^4-2 k^2 (\mo^2+\mt^2)+(\mo^2-\mt^2)^2}{4 k^2}.
\end{split}
\label{eq63}
\ee

Even if we did not know the result displayed in \Eq(\ref{x1}), we could 
still argue on general grounds that the integrand can be written 
as 
\be
\label{x2}
\begin{split}
\frac{(\nh \cdot l)^2 }{d_1 d_2} = &
\frac{b_0+b_1 (\nh \cdot l) + b_2 (\ne \cdot l)^2 }{d_1 d_2}  
+\frac{a_{1,0}+a_{1,1} (n \cdot l) + a_{1,2} (\nh \cdot l) }{d_1 } 
 \\
& 
+\frac{a_{2,0}+a_{2,1} (n \cdot l) + a_{2,2} (\nh \cdot l) }{d_2 }. 
\end{split} 
\ee
We will explain in Section~\ref{sect5} where this parametrization 
comes from. Here, we compare 
terms in \Eq(\ref{x1}) and \Eq(\ref{x2}) and obtain 
\be
\label{x3}
\begin{split} 
& b_0  = -\lambda^2,\;\;
b_1  = 0 ,\;\;
b_2  = -1, \\
& a_{1,0} = \frac{r_1^2}{4 k^2},\;\;
a_{1,1}=-\frac{1}{2 \sqrt{k^2}},\;\;
a_{1,2}=0,  \\
& a_{2,0} = \frac{r_2^2}{4 k^2} +\frac{1}{2},\;\;
a_{2,1}=\frac{1}{2 \sqrt{k^2}},\;\;
a_{2,2}=0.
\end{split} 
\ee

It is instructive to rederive  \Eq(\ref{x3}) 
using an alternative procedure. 
This procedure is important because it generalizes to four-dimensions, 
without modification, and  because it shows how  the reduction techniques 
are connected to unitarity.  We begin 
by multiplying both sides of \Eq(\ref{x2}) by $d_1,d_2$ 
and obtain 
\be
\label{x4}
\begin{split}
 (\nh \cdot l)^2  & = 
\left [ b_0+b_1 (\nh \cdot l) + b_2 (\ne \cdot l)^2 \right ] 
+\left [ a_{1,0}+a_{1,1} (n \cdot l) + a_{1,2} (\nh \cdot l) \right] d_2
\\
& 
+\left [a_{2,0}+a_{2,1} (n \cdot l) + a_{2,2} (\nh \cdot l ) \right ] d_1.
\end{split} 
\ee
We would like to use \Eq(\ref{x4}) to find all the $b$- and $a$-coefficients.
Since there are nine unknowns, we can evaluate \Eq(\ref{x4}) for nine 
different values of the loop momentum $l$, invert the nine-by-nine 
matrix and find the coefficients. While this procedure does, indeed, provide 
a solution to the problem, it requires inverting a large matrix and is 
therefore impractical. A better algorithm exploits the fact that, 
under special choices of the loop momentum $l$ in \Eq(\ref{x4}), 
the matrix to invert becomes block-diagonal.

To see how this works, we first describe  a procedure to compute 
the $b$-coefficients {\it only}. To project the right hand side of 
\Eq(\ref{x4})  onto $b$-coefficients, we choose the loop momentum $l$ 
to satisfy $d_1(l) = d_2(l) = 0$.   For the moment, consider 
the loop momentum $l$ that satisfies those constraints and, simultaneously, 
has zero projection on the $d$-dimensional space, $\ne \cdot l$ = 0.
We find that there are just two loop momenta $l$ that satisfy those 
constraints; they can be written as 
\be
\label{eqlm}
l_c^{\pm} = \alpha_c n \pm i \beta_c \nh,
\ee
where 
\be
\alpha_c = -\frac{r_1^2}{2\sqrt{k^2}},\;\;\;
\beta_c = \lambda.
\label{eq69}
\ee
The parameters $r_1$ and $\lambda$ are shown in 
\Eq(\ref{eq63}).  
We substitute these two solutions into \Eq(\ref{x4}) and obtain two equations 
for the coefficients $b_{0,1}$ 
\be
\begin{split} 
 b_0 + b_1 \nh \cdot l_c^+ = -\lambda^2,\;\;\;
 b_0 +  b_1 \nh \cdot l_c^- = -\lambda^2.
\end{split} 
\ee
It follows that $b_0 = -\lambda^2$ and $b_1 = 0$, in agreement with 
\Eq(\ref{x3}).

To find $b_2$ we proceed along similar lines but we require that the  
scalar product $ l \cdot \ne $ does not vanish.  Since the 
conditions $d_1 = 0, d_2 = 0$ are equivalent to $ 2 l \cdot k + r_1^2 = 0$, 
$l^2 = m_1^2$, the loop momentum that satisfies those constraints 
is the same as in \Eq(\ref{eqlm}), up to a change  
$\nh \to \ne$,
\be
l^\pm = \alpha_c n \pm i \beta_c \ne.
\ee
Substituting $l^\pm$ into \Eq(\ref{x4}) and using $b_0 = -\lambda^2$,
$b_1 = 0$, we obtain 
\be
0 = (1 + b_2) \lambda^2,
\ee
which implies that $b_2 = -1$, in agreement with \Eq(\ref{x3}).

The next step is to identify the coefficients of the tadpoles 
in \Eq(\ref{x4}). We will focus on a set $a_{1,0},a_{1,1},a_{1,2}$.
We can project \Eq(\ref{x4}) on these coefficients by choosing 
the loop momentum for which $d_1$ vanishes but $d_2$ is different 
from zero. Note that no $l \cdot \ne$ terms are needed to find the 
$a$-coefficients.  As a consequence, we can work with a two-dimensional 
loop momentum 
\beq
l_1 = \gamma_1 n + \gamma_2 \nh .
\eeq
The equation  $d_1(l_1)=0 $ implies $\gamma_1^2+ \gamma_2^2 = \mo^2$, 
so that $\gamma_1,\gamma_2$ lie on a circle of a radius $\mo$.
Substituting $l_1$ into \Eq(\ref{x4}), we find  
\be
\gamma_2^2 = -\lambda^2 + (2 \sqrt{k^2} \gamma_1 + r_1^2 ) (a_{1,0} 
+ a_{1,1}\gamma_1 + a_{1,2} \gamma_2 ).
\label{x5}
\ee
To solve \Eq(\ref{x5}), we choose $\gamma_1 = 0, \gamma_2 = \pm m_1$ 
and obtain two equations 
\be
a_{1,0} \pm  a_{1,2} m_1 = \frac{m_1^2 + \lambda^2}{r_1^2}
= \frac{r_1^2}{4 k^2}.
\ee
Hence, it follows that $a_{1,0} = r_1^2/(4k^2)$ and 
$a_{1,2} = 0$, in agreement with \Eq(\ref{x3}). To find 
$a_{1,1}$, we choose $\gamma_2 = 0, \gamma_1 = m_1$, solve 
\Eq(\ref{x5}) and obtain $a_{1,1} = -(4k^2)^{-1/2}$.

We can determine coefficients $a_{2,0},a_{2,1},a_{2,2}$ in the same
manner, by choosing the loop momentum that satisfies $d_2(l) = 0$.
The calculation is similar to the one performed above and for this
reason we do not present it here. We emphasize that the procedure that
we just explained implies that, for the reduction of one-loop
integrals to a set of scalar integrals, we need to know integrands at
{\it special values of the loop momenta}, for which at least one of
the inverse Feynman propagators that contributes to a particular
diagram, vanishes.  Since zeros of Feynman denominators correspond to
situations when virtual particles go on their mass shells, the
connection between the reduction procedure and the ideas of unitarity
begins to emerge.

\subsection{The photon mass in the Schwinger model}

We will conclude  our discussion of the two-dimensional 
physics by emphasizing the role that remnants 
of ultraviolet regularization play in  one-loop calculations. 
Such  terms are known as the {\it rational part}; an example 
is the $ (l \cdot \ne)^2/(d_1 d_2)$ term in 
\Eq(\ref{x2}). Note that those terms cannot be found 
by studying the integrand as a function of the 
four-dimensional loop momentum; the analytic 
continuation of the loop momentum to $d$-dimensions 
is crucial for the identification of those terms.

We will  discuss the rational part extensively in the following 
sections. However, the importance 
of those terms can be illustrated by considering 
two-dimensional QED -- the Schwinger model \cite{Schwinger:1962tp}. 
The Schwinger model is exactly solvable; it is often 
used to illustrate a variety of interesting phenomena 
in quantum field theory. Among these phenomena is the well-known 
result that  the photon acquires a dynamical mass. As we now 
show, this mass is generated by the rational part of 
the photon vacuum polarization at one loop.
To this end, consider the photon vacuum polarization 
function due to a loop of massless fermions,
\be
\Pi_{12} = -e^2 \int \frac{{\rm d}^D l}{(2\pi)^D}
\frac{{\rm Tr}(\hat \epsilon_1 \hat l \hat \epsilon_2  ( \hat l + \hat k) ) 
}{l^2 ( l+k)^2},
\label{q1}
\ee
where $\epsilon_1$ and $\epsilon_2$ are the two-dimensional {\it
  polarization vectors} of the virtual photon, $D = 2-2\ep$ and $\hat
p \equiv \gamma_\mu p^\mu$. The reason we introduce the polarization
vectors in the above equation is to simplify the bookkeeping of the
two- and $d$-dimensional Lorentz indices in what follows.

It is clear from the discussion in the previous Section, that 
$\Pi_{12}$ can be written as a linear combination of the scalar 
one- and two-point functions. The one-point functions with massless 
fermion propagators vanish  in dimensional regularization. Therefore, 
the non-vanishing contribution 
to $\Pi_{12}$ can only come from the scalar two-point function. 
According to the previous Section, a reduction coefficient in this 
case can be obtained by studying the integrand in \Eq(\ref{q1}) 
for values of the loop momentum $l$ such that the two inverse 
Feynman propagators vanish
\be
l^2 = 0,\;\;\;(l+k)^2 = 0.
\ee

Since such contribution corresponds to both intermediate particles in
the loop being on the mass shell, we refer to such a contribution as
the ``double cut''.  We can parametrize the loop-momentum using the
van Neerven - Vermaseren basis introduced in Sect.\ref{sec3}
\be
l^\mu = (l \cdot k) v_1^\mu + l_\perp^\mu + (l \cdot \ne) \ne^\mu,\;\;\;
\label{q2}
\ee where $v_1^\mu = k^\mu/k^2$, $l_\perp^\mu = (l\cdot n) n^\mu$ 
is a two-dimensional vector orthogonal to $k$ and
$\ne^\mu$ denotes a unit vector that parametrizes the $(D-2)$ dimensions.
Solving the two constraints, we find
\be
l^\mu = -\frac{1}{2} k^\mu + l_\perp^\mu + (l \cdot \ne) \ne^\mu,\;\;\;
l_\perp^2 + (l \cdot \ne)^2 = - \frac{k^2}{4} \, ,
\label{qq2}
\ee
with $ k\cdot l_\perp=\ne \cdot k=\ne\cdot l_\perp=0$.
Calculating the trace in \Eq(\ref{q1}), substituting solutions from 
\Eq(\ref{qq2}) and disregarding terms that are linear in $l_\perp$ 
or $\ne$, we obtain 
\be
\begin{split} 
\label{q3}
\Pi_{12}  = -2 e^2 & 
\int \frac{{\rm d}^D l}{(2\pi)^D}
\frac{1}{l^2 (l+k)^2} \Bigg  [   
2 (l_\perp \cdot \epsilon_1) (l_\perp \cdot \epsilon_2) 
\\
& + \frac{k^2}{2} \left ( \epsilon_1 \cdot \epsilon_2 
- \frac{(\epsilon_1 \cdot k) (\epsilon_2 \cdot k)}{k^2} \right )
\Bigg ].
\end{split} 
\ee 
Because $l^2 = l_\perp^2 + ...$ and 
$(l+k)^2 = l_\perp^2 + ...$, the 
angular integration over the direction of $l_\perp$ is trivial and 
 we can write 
\be
\begin{split}
(l_\perp \cdot \epsilon_1) (l_\perp \cdot \epsilon_2) 
 \to &  \; l_\perp^2 \left ( \epsilon_1 \cdot \epsilon_2 
- \frac{(\epsilon_1 \cdot k) (\epsilon_2 \cdot k)}{k^2} \right )
\\
 = &\left ( -\frac{k^2}{4} - (l \cdot \ne)^2 \right )
\left ( \epsilon_1 \cdot \epsilon_2 
- \frac{(\epsilon_1 \cdot k) (\epsilon_2 \cdot k)}{k^2} \right ).
\end{split}
\ee
Using this result in \Eq(\ref{q3}), we derive a simple formula
for the polarization operator 
\be
\Pi_{12}  = 4 e^2 \left ( \epsilon_1 \cdot \epsilon_2 
- \frac{(\epsilon_1 \cdot k) (\epsilon_2 \cdot k)}{k^2} \right )\;
\int \frac{{\rm d}^D l}{(2\pi)^D}
\frac{(l \cdot \ne)^2 }{l^2 (l+k)^2}. 
\label{eq146}
\ee
\Eq(\ref{eq146})
shows that the photon vacuum polarization in two-dimensional 
QED is non-vanishing only because of a remnant of the ultraviolet 
regularization. Using modern jargon, we say that it is purely rational.
We compute the 
integral in the limit $D \to 2$,
\be
\int \frac{{\rm d}^Dl}{(2\pi)^D}
\frac{(l \cdot \ne)^2}{l^2 ( l+k)^2} = \;\frac{i}{4\pi}\,,
\ee
and obtain 
\be
\Pi_{12} = i \frac{e^2}{\pi} 
\left ( \epsilon_1 \cdot \epsilon_2 
- \frac{(\epsilon_1 \cdot k) (\epsilon_2 \cdot k)}{k^2} \right ).
\label{q4}
\ee
A resummation of the vacuum polarization contributions shown 
in \Eq(\ref{q4}) gives  the well-known {\it massive}
photon propagator in the Schwinger  model 
\cite{Schwinger:1962tp}
\be
\Pi_{\mu \nu} = \frac{-i}{k^2 - m_\gamma^2} 
\left ( g_{\mu \nu} - \frac{k_\mu k_\nu}{k^2} \right ),
\;\;\; m_\gamma^2 = \frac{e^2}{\pi}. 
\ee
Hence, the photon mass $m_\gamma^2 = e^2/\pi$ comes 
entirely from the rational part of the photon vacuum polarization 
 diagram in two dimensions. 

\section{Reduction at the integrand level in $D$-dimensions}
\label{sect5}

In this Section, we generalize the two-dimensional reduction procedure
described in the previous Section to $D$-dimensional space-time. We
are ultimately interested in the limit $D \to 4$.
\subsection{Parametrization of the integrand}
\label{sec_param}

We begin with the observation that, in any renormalizable quantum
field theory, the rank of the one-loop tensor integrals that appear
does not exceed the number of external lines. Therefore, we will only
be concerned with the reduction of one-loop integrals of restricted
rank, e.g. the rank-five or less for five-point functions, rank-four
or less for four-point functions and so on.

We would like to establish a simple parametrization of  one-loop 
integrands, following a seminal suggestion 
by  Ossola, Papadopoulos and Pittau
\cite{Ossola:2006us}. 
It reads 
\be
\begin{split}
I_N  & = \int \frac{{\rm d}^Dl}{(2\pi)^D}
\frac{{\rm Num}(l)}{\prod_i d_i(l)}
= \int \frac{{\rm d}^D l}{(2\pi)^D} 
\frac{1}{\prod_i d_i(l)} \times \Bigg \{
\\
& 
\sum_{i_1,i_2,i_3,i_4,i_5} {\tilde e}_{i_1,i_2,i_3,i_4,i_5}(l)  
\prod \limits_{j \neq [i_1,i_2,i_3,i_4,i_5]}^{}  d_{j}(l) 
\\
& + \sum_{i_1,i_2,i_3,i_4} {\tilde d}_{i_1,i_2,i_3,i_4}(l)  
\prod \limits_{j \neq [i_1,i_2,i_3,i_4]}^{} d_{j}(l)
\\
& + \sum_{i_1,i_2,i_3} {\tilde c}_{i_1,i_2,i_3}(l)  
\prod \limits_{j \neq [i_1,i_2,i_3]}^{} d_{j}(l)
\\
& + \sum_{i_1,i_2} {\tilde b}_{i_1,i_2}(l)  
\prod \limits_{j \neq [i_1,i_2]}^{} d_{j}(l)
+ \sum_{i_1} {\tilde a}_{i_1}(l)  \prod \limits_{j \neq i_1} d_j(l)
\Bigg \}. 
\end{split}
\label{x11}
\ee
The index $i$ runs over all possible inverse Feynman propagators $d_i$. Similarly,
the index $j$ runs over all inverse Feynman propagators, except those explicitly excluded.
The important feature of this parametrization is that all inverse
propagators $d_i(l)$ on the right hand side appear in the first power,
i.e.\ there are no terms of the form $d_i^2(l)$ for any $i$.  In the
spirit of the previous section, this allows us to project on different
${\tilde e},{\tilde d},{\tilde c},{\tilde b}$ and ${\tilde
  a}$-functions, by considering loop momenta that nullify different
sets of inverse propagators.

We will discuss first the reduction of a rank-five five-point
function; the general case then easily follows.  To this end, we
consider $d_i(l) = (l+q_i)^2 - m_i^2$, $i=0, \dots ,4$, $q_0 = 0$ and
assume that the numerator function reads
\be
{\rm Num}(l)= N_5(l) = \prod \limits_{i=1}^{5} u_i \cdot l, 
\label{eqnum}
\ee
where $u_{i}$  are  some external four-dimensional vectors.   

As the first step in the reduction procedure, we find the reduction
coefficients of the five-point function, ${\tilde e}_{01234}$.  To
accomplish this, we construct the \vNV~basis out of four vectors $q_i$
and decompose the loop momentum

\be
l^\mu = \sum \limits_{i=1}^{4} (l \cdot q_i) v_i^\mu + (l \cdot \ne) \ne^\mu.
\ee
The scalar products $l \cdot q_i$ are expressed in terms of inverse 
Feynman propagators 
\be
l \cdot q_i = \frac{1}{2} \left ( d_i - d_0 - (q_i^2 - m_i^2 + m_0^2 ) 
\right ).
\ee
Since $u_5 \cdot \ne =0$, we can rewrite \Eq(\ref{eqnum}) as  
\be
\begin{split} 
& N_5(l) = \left ( \prod \limits_{i}^{4} u_i \cdot l \right ) 
( u_5 \cdot l ) = 
\frac{1}{2} \sum \limits_{j=1}^{4} (u_5 \cdot v_j)
\left ( \prod \limits_{i}^{4} u_i \cdot l \right ) 
\left ( d_j - d_0 \right )  
\\
& 
- \frac{1}{2} \sum \limits_{j=1}^{4} (u_5 \cdot v_j)
\left ( \prod \limits_{i}^{4} u_i \cdot l \right ) 
( q_j^2 - m_j^2 + m_0^2).
\label{Numerator(l)}
\end{split} 
\ee
Upon dividing the numerator function by the product of inverse Feynman
propagators $d_0 d_1 d_2 d_3 d_4$, we find that the first term on the
right-hand-side of \Eq(\ref{Numerator(l)}), produces a collection of
rank-four four-point functions and the second term -- a rank-four
five-point function.  We now repeat the same procedure with the
rank-four five-point function and conclude that it can be expressed
through a combination of rank-three four-point functions and the
rank-three five point function.
Whenever, as a result of these manipulations, the propagator $d_0$
cancels, it is possible to shift the loop-momentum to bring the
integrand to the standard form.
We can clearly continue this procedure until we are left with a {\it
  scalar} five-point function and a collection of four-point functions
of the ranks from zero (scalar) to four (maximal).  Hence, we
have established that the function ${\tilde e}_{01234}(l) $ in
\Eq(\ref{x11}) is $l$-independent \be {\tilde e}_{01234}(l) = {\tilde e}^{(0)}_{01234}.
\ee

In the course of the procedure described above, the highest rank
integral left unreduced is the rank-four four-point function.  We now
discuss how it can be reduced.  For definiteness, we consider the
four-point function with four propagators $d_0, d_1, d_2, d_3$, but
our discussion can be applied to any other four-point function, by the
appropriate re-definition of the propagator momenta and masses.  We
construct \vNV~basis vectors out of the three momenta $q_1,q_2,q_3$.
The physical space in this case is three-dimensional and the
transverse space is one-dimensional.  We parametrize the transverse
space by the unit vector $n_4$.

The decomposition in terms of \vNV~basis then reads
\be
l^\mu =  \sum \limits_{i=1}^{3} v_i^\mu (l \cdot q_i)
 + ( l\cdot n_4) n_4^\mu 
 + ( l \cdot \ne ) \ne^\mu. 
\label{eq:vnv4}
\ee
Using this parametrization we can write the numerator of the four-point 
function as, 
\be
\label{x12}
\begin{split} 
& N_{4} (l) = 
\left ( \prod \limits_{i}^{3} u_i \cdot l \right ) 
( u_4 \cdot l ) = 
\frac{1}{2} \sum \limits_{j=1}^{3} u_4 \cdot v_j
\left ( \prod \limits_{i}^{3} u_i \cdot l \right ) 
\left ( d_j - d_0 \right )  
\\
& 
- \frac{1}{2} \sum \limits_{j=1}^{3} u_4 \cdot v_j
( q_j^2 - m_j^2 + m_0^2)  \prod \limits_{i}^{3} u_i \cdot l 
+  (l \cdot n_4) (u_4 \cdot n_4) \prod \limits_{i}^{3} u_i \cdot l.
\end{split} 
\ee
The first two terms on the right-hand side are considered ``reduced'',
since they are rank-three three-point and four-point functions. The
last term, however, is a rank-four four-point function, and so it does
not appear that we gained anything. To demonstrate that we, actually,
did gain something, we take the last term in \Eq(\ref{x12}) and
repeat the reduction procedure described above.  It is clear that a
variety of terms will be produced, most of lower-point or lower-rank
type, and the only term that we should consider as ``not-reduced''
reads
\be
\left ( \prod \limits_{i}^{3} u_i \cdot l \right ) (l \cdot n_4)  
\to 
\left ( \prod \limits_{i}^{2} u_i \cdot l \right ) (l \cdot n_4)^2.
\ee
We simplify it by examining the square of the loop momentum $l$. 
%
Using the decomposition in terms of \vNV~basis, \Eq(\ref{eq:vnv4})
and the relations $ 2 l \cdot q_i = (d_i - d_0 -q_i^2+m_i^2-m_0^2)$
and $l^2 = d_0 + m_0^2$, we find
\be
(l \cdot n_4)^2 = - (l \cdot \ne )^2 + {\rm constant~terms} 
+ {\cal O}(d_0,d_1,d_2,d_3).
\label{eq159}
\ee
Terms dubbed ``constant'' in the above formula contribute (after
multiplication by $(u_1 \cdot l) (u_2 \cdot l)$ to rank-two
four-point functions while terms that contain at least one inverse
Feynman propagator, contribute to three-point functions.  The
``not-reduced'' part of the rank-four four-point function therefore
reads
\be
\begin{split}
\prod \limits_{i}^{4} u_i \cdot l  \to
\left ( \prod \limits_{i}^{2} u_i \cdot l \right ) (l \cdot n_4)^2
\to \left ( \prod \limits_{i}^{2} u_i \cdot l \right ) (l \cdot \ne)^2.
\end{split}
\ee
It is clear that if we repeat the reduction process, we express any
tensor four-point function integral (of rank not higher than four),
through the following numerator function
\be
{\tilde d}_{0123}(l) = {\tilde d}_0 + {\tilde d}_1 (l \cdot n_4) 
+ {\tilde d}_2 (l \cdot \ne)^2 
+ {\tilde d}_3 (l \cdot \ne)^2 (l \cdot n_4) + {\tilde d}_4 ( l \cdot \ne)^4,
\label{x0123}
\ee
where the $l$-dependence is shown explicitly.  We note that the degree
of the polynomial on the right hand side of \Eq(\ref{x0123}) is the
direct consequence of the fact that the highest rank tensor four-point
functions that we consider is four.  This restriction works well if we
deal with  renormalizable quantum field theories but it might not be
general enough if one-loop calculations with effective field theories
are attempted. The extension of the algorithm to those cases is
straightforward since the required parametrization of a numerator
function of, say, a four-point function will be an extension of
\Eq(\ref{x0123}) to higher rank tensors.  It is straightforward to
figure out the required extension, following the line of reasoning
explained above. Interestingly, such extensions are very economical;
for example, we mention that to achieve reduction of a {\it rank-five}
four-point function, we need only include one additional term
${\tilde d}_5 (l \cdot \ne)^4 (l \cdot n_4)$ in the parametrization 
of ${\tilde d}_{0123}$ in  \Eq(\ref{x0123}).

We now turn our attention to the three-point functions that are
obtained in the course of the reduction of the four-point
functions. The highest tensor rank we have to care about is three.
The physical space is two-dimensional and the transverse space is
two-dimensional as well.  The loop momentum reads
\be
l^\mu = \sum_{i=1}^{2} v_i^\mu (l \cdot q_i) 
+ (l \cdot n_3) n_3^\mu + (l \cdot n_4) n_4^\mu + ( l \cdot \ne) \ne^\mu.
\label{x134}
\ee
We follow the same  procedure as already described in the context of 
five- and four-point functions. The reduced terms will be at most 
rank-two two-point functions. The irreducible structures read 
\be
\label{x135}
\begin{split}
& \prod \limits_{i=3}^{4} (l \cdot u_i)  \to  
\sum \limits_{i=3}^{4} c_{1i} (l \cdot n_i)
+
\sum \limits_{i=3}^{4} c_{2i} (l \cdot n_i)^2
+\sum \limits_{i=3}^{4} c_{3i} (l \cdot n_i)^3
\\
& 
+ c_4 (l \cdot n_4)(l \cdot n_3)  
+ c_5 (l \cdot n_3)^2(l \cdot n_4)
+ c_6 (l \cdot n_3) (l \cdot n_4)^2.
\end{split}
\ee
Similar to the case of the four-point function,
not all the terms 
in \Eq(\ref{x135}) are independent in the four-dimensional case. 
To make this dependence explicit, 
we square $l^\mu$ in \Eq(\ref{x134}), use  $l^2 = d_0 + m_0^2$ and  
find 
\be
\label{eqcon}
(l \cdot n_3)^2 + (l \cdot n_4)^2 
+ (l \cdot \ne )^2 = {\rm constant~terms} + {\cal O}(d_0,d_1,d_2).
\ee
We use this constraint in \Eq(\ref{x135}), to trade 
$(l \cdot n_3)^2(l \cdot n_4),\;\;
(l \cdot n_4)^2(l \cdot n_3)$ for 
$(l \cdot \ne)^2 (l \cdot n_4)$ and 
$(l \cdot \ne)^2 (l \cdot n_3)$. Also, given \Eq(\ref{eqcon}),
we can use $(l \cdot \ne)^2$ and $(l \cdot n_3)^2 - (l \cdot n_4)^2$ 
as two independent structures, instead of $(l \cdot n_3)^2$ and 
$ (l \cdot n_4)^2$.  Hence, the parametrization of the function 
${\tilde c}_{012}$ becomes 
\be
\label{eq_c_coeff}
 \begin{split} 
& {\tilde c}_{012}(l) = {\tilde c}_0 + {\tilde c}_1 (l \cdot n_3) + {\tilde c}_2 (l \cdot n_4) 
+ {\tilde c}_3 ( ( l  \cdot n_3)^2 -  (l \cdot n_4)^2 )
\\
&  
+ {\tilde c}_4 ( l  \cdot n_3) (l \cdot n_4) 
+ {\tilde c}_5  ( l  \cdot n_3)^3 
+ {\tilde c}_6  (l \cdot n_4)^3
\\
&  + {\tilde c}_7  (l \cdot \ne)^2 
 + {\tilde c}_8  (l \cdot \ne)^2  ( l  \cdot n_3) 
 + {\tilde c}_9  (l \cdot \ne)^2 (l \cdot n_4) .
\end{split} 
\ee 
The advantage of this parametrization, compared to \Eq(\ref{x135}),
is that in four dimensions only $\tilde c_0$ gives a non-vanishing
contribution after integration.

Similar considerations can be used to derive the general
parametrization of the two-point and one-point functions. Recall that
the highest tensor rank of the two-point function that we consider is
two; the highest tensor rank of the one-point function is one.  We
will not discuss the derivation and just give the results for the
numerator functions.  The numerator of the two-point function can be
written as
\beqa {\tilde b}_{01}(l) &=& {\tilde b}_0 + {\tilde b}_1 (l \cdot n_2) +
{\tilde b}_2 (l \cdot n_3) + {\tilde b}_3 (l \cdot n_4)
\nonumber \\
& +&{\tilde b}_4 ( (l \cdot n_2)^2 - (l \cdot n_4)^2) +{\tilde b}_5 (
(l \cdot n_3)^2 - (l \cdot n_4)^2) +{\tilde b}_6 (l \cdot n_2) (l
\cdot n_3)
\nonumber \\
& + &{\tilde b}_7 (l \cdot n_3) (l \cdot n_4) + {\tilde b}_8 (l \cdot
n_2) (l \cdot n_4) + {\tilde b}_9 (l \cdot \ne)^2,
\label{eq987}
\eeqa
while the general parametrization of the numerator of the 
one-point function for propagator $d_i$ is
\be
{\tilde a_i}(l) = {\tilde a}_0 
+ {\tilde a}_1 (l \cdot n_1)
+ {\tilde a}_2 (l \cdot n_2) 
+ {\tilde a}_3 (l \cdot n_3) 
+ {\tilde a}_4 (l \cdot n_4).  
\label{eqtadpole}
\ee
In \Eq(\ref{eqtadpole}) $\tilde a_0$ is the relevant reduction
coefficient since all other terms integrate to zero.

\subsection{Computation of the reduction coefficients} 
\label{sec5.2}

In the previous Section we showed how an integrand of a general
one-loop integral in a renormalizable quantum field theory can be
parametrized. An important feature of this parametrization is that
all $l$-dependent four-dimensional tensors that are present in the
coefficients $\tilde d^{}_{i_1..i_4},..,\tilde a^{}_{i_1}$ 
vanish if 
angular integration in the transverse space of the respective
reduced integral is performed.  We will refer to such tensors as ``traceless''.
This feature is very  important since it allows us to 
rewrite \Eq(\ref{x11}) in a simplified,
fully reduced form
\be
\begin{split}
I_N  & = \int \frac{{\rm d}^D l}{(2\pi)^D}
\frac{{\rm Num}(l)}{\prod_i d_i(l)}
= \sum_{i_1,i_2,i_3,i_4,i_5} 
{\tilde e}_{i_1,i_2,i_3,i_4,i_5}^{(0)} I_{i_1 i_2 i_3 i_4 i_5}  
\\
& + \sum_{i_1,i_2,i_3,i_4} {\tilde d}_{i_1,i_2,i_3,i_4}^{(0)} I_{i_1 i_2 i_3 i_4}   
+ \sum_{i_1,i_2,i_3} {\tilde c}_{i_1,i_2,i_3}^{(0)} I_{i_1 i_2 i_3}   
\\
& + \sum_{i_1,i_2} {\tilde b}_{i_1,i_2}^{(0)} I_{i_1 i_2}   
+ \sum_{i_1} {\tilde a}_{i_1}^{(0)} I_{i_1} + {\cal R}.
\end{split}
\label{y11}
\ee 
%
 \Eq(\ref{y11}) proves the reduction formula stated in \Eq(\ref{eq5}).
The right hand side of \Eq(\ref{y11}) contains scalar integrals
multiplied by $l$-independent contributions of the reduction
coefficients ${\tilde e}^{(0)}, {\tilde d}^{(0)}, {\tilde
  b}^{(0)},~{\rm etc.}$ and the ``rational'' term ${\cal R}$ which
originates from the integration over the loop momentum of tensorial
structures involving $(l \cdot \ne)$. 
The rational part  ${\cal R}$ is given explicitly by the following equation
\be
{\cal R}=
-\sum_{i_1,i_2,i_3,i_4}
\frac{\tilde{d}^{(4)}_{i_1i_2i_3i_4}}{6}
+\sum_{i_1,i_2,i_3} \frac{\tilde{c}^{(7)}_{i_1i_2i_3}}{2} +
\sum_{i_1,i_2}\left[\frac{m_{i_1}^2 + m_{i_2}^2}{2}
- \frac{\left(q_{i_1}-q_{i_2}\right)^2}{6}\right]
\tilde{b}^{(9)}_{i_1i_2}.
\label{eq_rational}
\ee

We note that, in order to  arrive at \Eq(\ref{y11}), we have integrated over 
some directions  of the loop momentum.
The integration over the loop momentum is so simple 
because the projection on the transverse space is always given in
terms of traceless tensors.  To illustrate this point, consider a
contribution of a general two-point function to the right-hand side of
\Eq(\ref{x11}).  It reads
\be
\begin{split} 
& \int \frac{{\rm d}^D l}{(2\pi)^D}
\frac{1}{(l^2 - m_0^2) ( l^2 + 2l \cdot q + q^2 - m_1^2)}
\Bigg \{
{\tilde b}_0 
+ {\tilde b}_1 (l \cdot  n_2) 
\\
& 
+ {\tilde b}_2 (l \cdot  n_3) 
+ {\tilde b}_3 (l \cdot  n_4) 
+{\tilde b}_4 ( (l \cdot  n_2)^2 - (l \cdot  n_4)^2) 
\\
& 
+{\tilde b}_5 ( (l \cdot  n_3)^2 - (l \cdot  n_4)^2) 
+{\tilde b}_6 (l \cdot n_2) (l \cdot  n_3) 
 + {\tilde b}_7 (l \cdot n_3) (l \cdot  n_4) 
\\
& + {\tilde b}_8 (l \cdot n_2) (l \cdot n_4) 
+ {\tilde b}_9 (l \cdot  \ne)^2
\Bigg \}.
\end{split}
\label{y12}
\ee
Because $q \cdot \ne=0, q \cdot n_i = 0$, $i=2,3,4$, the integration 
over the directions 
of the transverse space 
$l_\perp = 
n_2 (l \cdot n_2) 
+ n_3 (l \cdot n_3) 
+ n_4 (l \cdot n_4)
+ \ne (l \cdot \ne)
$ 
is  straightforward. We obtain  
\be
\label{eqm123}
\begin{split} 
& \int {\rm d}^{D_1} l_\perp \delta (l_\perp^2 - \mu_0^2)
\left ( l_\perp^\mu, l_\perp^\mu l_\perp^{\nu} \right ) = 
\int {\rm d}^{D_1} l_\perp \delta (l_\perp^2 - \mu_0^2)
\left ( 0,\; \frac{g_{\perp}^{\mu \nu}}{D_1}\; l_\perp^2 \right ),
\end{split} 
\ee
where $D_1 = D-1$.

Using this result in \Eq(\ref{y12}) together with the orthonormality
property of the transverse space basis vectors $n_i n_j =
\delta_{ij}$, we conclude that only two terms -- ${\tilde b}_0$ and
${\tilde b}_9$ contribute after the integration over the loop momentum
is performed. The term with ${\tilde b}_9$ contributes to the rational
part ${\cal R}$ in \Eq(\ref{y11}), while ${\tilde b}_0$ is the
reduction coefficient of the relevant two-point master integral.
Similar considerations apply to all other reduction coefficients,
leading to \Eq(\ref{y11}).  Clearly, we are interested in the
calculation of quantities that are integrated over the loop
momentum. It follows from \Eq(\ref{y11}) that, in addition to the
rational part, we only require a modest number of the reduction
coefficients ${\tilde e}^{(0)}, {\tilde d}^{(0)}, {\tilde c}^{(0)},..$
etc. The question that we address now is how to find those
coefficients efficiently.

In the course of the discussion of the two-dimensional case, we have
seen that a powerful way to find coefficients $\tilde
e^{}_{i_1..i_5},...,\tilde a^{}_{i_1}$ involves calculations of both
sides of \Eq(\ref{x11}) for special values of the loop momentum $l$,
where a chosen subset of inverse Feynman propagators
$d_{0},d_{1},...,d_{N}$ vanish. We now discuss this procedure in
detail, pointing out some subtleties that appear once we 
implement it.

We begin with the five-point function contribution. We choose five
inverse propagators, say $d_0,d_1,...d_4$ and find the loop momentum
for which {\it all} of these inverse propagators vanish.  This requires
the momentum $l$ to span more than four dimensions, so, for definiteness,
we make the minimal choice and take $l$ to be five-dimensional. Clearly, the
only term in the right hand side of \Eq(\ref{x11}) that is non-zero
is the term that does not contain any of the five propagators.  This
is ${\tilde e}_{01234}$ -- the term that we would like to find.  We
argued previously that this term is constant, so computing the left
hand side of \Eq(\ref{x11}) with the momentum $l^*$ such that
$d_0(l^*) = 0, d_1(l^*) =0, ...d_4(l^*)=0$, gives us $\tilde e_{01234}$.

While this procedure is correct, it often becomes impractical since it
treats the scalar five-point function as a master integral.
This would have been fully justified if we were interested in a
five-dimensional calculation, but, in practical computations, we
eventually take the limit $D \to 4$. In this limit, the five-point
function becomes a linear combination of five four-point functions.  We
would therefore like to eliminate the five-point integral from the
set of master integrals right away, avoiding large cancellations between four-
and five-point functions in the $D \to 4$ limit.  To see how this can
be done note that the loop momentum in the five-point function can be
written as
\be
l^\mu = \frac{1}{2} \sum \limits_{i=1}^{4} v_i^{\mu} \left ( d_i - d_0
-(q_i^2 - m_i^2 + m_0^2) \right ) + (l \cdot \ne ) \ne^\mu.  
\ee 
Squaring the two sides of this equation and using $l^2 = d_0 + m_0^2$, 
we see that for a loop momentum that satisfies $d_0 = 0, d_1= 0,\dots
d_4 = 0$, we have  
\be
( l \cdot \ne)^2 = -\frac{1}{4}\sum \limits_{ij} ( v_i \cdot  v_j ) 
(q_i^2 - m_i^2 + m_0^2)(q_j^2-m_j^2+m_0^2) + m_0^2\,.
\ee
It follows that we can either choose a scalar five-point function as
the master integral or the integral with additional $(l \cdot \ne)^2$
in the numerator. However, because
\be
\lim_{D \to 4} \int \frac{{\rm d}^Dl}{(2\pi)^D}
\frac{( l \cdot \ne)^2}{d_0 d_1 d_2 d_3 d_4} \to 0,
\ee
the second choice is preferable. Indeed, since the new master integral
that we introduced to account for the need to employ dimensional
regularization does not contribute in the $D \to 4$ limit, all
four-dimensional relations between various integrals are automatically
accounted for.  
Therefore this coefficient is only needed as a subtraction term in the determination of 
lower point coefficients. Experience shows that adopting this alternative definition 
of the pentagon coefficient leads to improved numerical
stability in practical computations~\cite{Ellis:2008qc,Lazopoulos:2009zn}.

To find the other coefficients, we follow the strategy already discussed
in the context of two-dimensional computations. For example, having
determined the five-point functions, we subtract their coefficients
from the left-hand side of \Eq(\ref{x11}) and consider all the
subsets of four propagators. We focus on one subset, $d_0,...,d_3$
whose contribution is described by the coefficient
\be
\label{eq:dtilde0123}
{\tilde d}_{0123} = 
{\tilde d}_0 + {\tilde d}_1 (l \cdot n_4) + {\tilde d}_2 (l \cdot \ne)^2 
+ {\tilde d}_3 (l \cdot \ne)^2 (l \cdot n_4) + {\tilde d}_4 ( l \cdot \ne)^4.
\ee
To determine  ${\tilde d}_{0123}$, we find a momentum  
$l$ that satisfies  $d_0(l) = 0$, $d_1(l) = 0$, $d_2(l) = 0$, $d_3(l) = 0$ 
and write  it as 
\be
\begin{split} 
& l^\mu = V^\mu +  l_\perp ( \cos \phi \; n_4^\mu + \sin \phi \;  \ne^\mu ),
 \\
& V^\mu = -\frac{1}{2} \sum \limits_{i}^{3} v_i^{\mu} 
\left ( q_i^2 - m_i^2 + m_0^2 \right ),  
\end{split} 
\ee
with $l_\perp = \sqrt{l_\perp \cdot l_\perp}$. 
It is sufficient to consider
$l$ to be five-dimensional.
The length of the projection of the vector $l$ on the transverse space 
is fixed
\be
l_\perp^2 = m_0^2 - V_\mu V^\mu. 
\ee

To find the ${\tilde d}_{0},...{\tilde d}_{3}$ coefficients, we take,
for instance, $\sin \phi = 0, \cos \phi = \pm 1$, denote $l_\pm^\mu =
V^\mu \pm l_\perp n_4^\mu $, calculate the numerator for these values
of the loop momenta and find
\be
\begin{split} 
\tilde d_0 = \frac{ {\rm Num}(l_+) +{\rm Num}(l_-)}{2},
\;\;\;\;\tilde d_1 = \frac{{\rm Num}(l_+) - {\rm Num}(l_-)}{2 l_\perp}. 
\label{howtod}
\end{split}
\ee

To find $\tilde d_{2,3,4}$,  we need to do a little bit more. 
First, we take $\cos \phi = \sin \phi = \pm 1/\sqrt{2}$, 
denote the loop momentum as 
${\tilde l}_{\pm} = V \pm  l_\perp ( n_4 +  \ne)/\sqrt{2}$, 
and find
 \be
\begin{split} 
& \tilde d_2 + \tilde d_4 \frac{l_\perp^2}{2} = 
l_\perp^{-2}
\left( {\rm Num}({\tilde l}_+) + {\rm Num}({\tilde l}_-) - 2\, \tilde d_0\right), 
\\
& \tilde d_3 = \sqrt{2}\;l_\perp^{-3}
\left( {\rm Num}({\tilde l}_+) - {\rm Num}({\tilde l}_-) - \sqrt{2} \tilde d_1 l_\perp\right ). 
\end{split}
\ee
We need yet another equation to resolve the $\tilde d_2 - \tilde d_4$
degeneracy. It is convenient to take $l_\ep^\mu = V^\mu + l_\perp
\ne^\mu $; this leads to
\be
\tilde d_2 + \tilde d_4 l_\perp^2 =  \frac{{\rm Num}(l_\ep) - \tilde d_0}{l_\perp^2}. 
\ee
We find 
\be
\begin{split}
& \tilde d_2 =   \frac{1}{l_\perp^2}
\left(2 {\rm Num}({\tilde l}_+) +2 {\rm Num}({\tilde l}_-)-{\rm Num}(l_\ep) - 3 \tilde d_0\right)\,,
\\
& \tilde d_4 =   \frac{2}{l_\perp^4} 
\left( {\rm Num}(l_\ep) 
- {\rm Num}({\tilde l}_+) - {\rm Num}({\tilde l}_-)+ \tilde d_0\right). 
\end{split}
\ee

 We next discuss how to compute the coefficients of the 
 three-point functions.  As an illustration, we choose 
 a three point function with denominators $d_0,d_1,d_2$; 
 its contribution is described by a coefficient 
 \be
  \begin{split} 
 & {\tilde c}_{012} = {\tilde c}_0 + {\tilde c}_1 (l \cdot n_3) 
 + {\tilde c}_2 (l \cdot n_4) 
 + {\tilde c}_3 ( ( l  \cdot n_3)^2 -  (l \cdot n_4)^2 )
 \\
 &  
 + {\tilde c}_4 ( l  \cdot n_3) (l \cdot n_4) 
 + {\tilde c}_5  ( l  \cdot n_3)^3 
 + {\tilde c}_6  (l \cdot n_4)^3
 \\
 &  + {\tilde c}_7  (l \cdot \ne)^2 
  + {\tilde c}_8  (l \cdot \ne)^2  ( l  \cdot n_3) 
  + {\tilde c}_9  (l \cdot \ne)^2 (l \cdot n_4) .
 \label{eq70}
 \end{split} 
 \ee
 We choose the loop momentum that satisfies $d_0(l) = d_1(l) = 
 d_2(l) = 0$ and parametrize it as  
 \be
 l^\mu = V^\mu + l_\perp( x_3 n_3^\mu + x_4 n_4^\mu + x_\ep \ne^\mu) .
 \ee
We  consider momenta with $x_\ep = 0$; such a choice 
 allows us to determine the  coefficients $\tilde c_{0,.,6}$. 
 If $x_\epsilon = 0$, $x_3^2 + x_4^2 = 1$, so that we can take 
 $x_3 = \cos \phi, x_4 = \sin \phi$. It is convenient then to rewrite 
 \Eq(\ref{eq70}) as a polynomial in $t = e^{i\phi}$. 
 \Eq(\ref{eq70}) becomes 
 \be
 {\tilde c}_{012}(t)= \sum \limits_{k=-3}^{3} c_k t^k,
 \label{z10}
 \ee
 where the coefficients $c_k$ read 
 \be
 \begin{split} 
 & c_{\pm 3} = \frac{{\tilde c}_5 \pm i {\tilde c}_6}{8} l_{\perp}^3,\;\;\;
 c_{\pm 2} = \frac{2 {\tilde c}_3  \mp i {\tilde c}_4}{4}l_{\perp}^2,\\
 & c_{\pm 1} = 
 \left(\frac{1}{2}\;{\tilde c}_1\mp \frac{i}{2}\;{\tilde c}_2\right)l_\perp
 +\left(
  \frac{3}{8}\;{\tilde c}_5 \mp \frac{3i}{8} \;{\tilde c}_6  
 \right)l_\perp^3\,,
 \end{split} 
 \label{Defnofccoeficients}
 \ee 
 and $c_0 = {\tilde c}_0$. We can now use the technique of discrete
 Fourier transform, first discussed in the context of the OPP
 reduction, in Refs.~\cite{Mastrolia:2008jb,Berger:2008sj}.
 Application of the discrete Fourier transform allows us to write
 explicit expressions for the coefficients $c_k$ in a straightforward
 way. Indeed, they are given by
 \be
 c_{m} = 
 \frac{1}{7} \sum \limits_{n=0}^{6} 
 {\tilde c}_{012}(t_n)\; t_n^{-m},
 \label{z11}
 \ee
 where $t_n = e^{2\pi\; i\; n/7}$.  To prove this equation, note that 
 \be
 \sum \limits_{n=0}^{k} e^{\frac{2 \pi i n}{k+1} r} 
 = \delta_{r0} (k+1).
 \label{z12}
 \ee
 Substituting \Eq(\ref{z10}) into the right hand side of
 \Eq(\ref{z11}) and carrying out the summation over $n$ using
 \Eq(\ref{z12}), we can easily show that the right hand side of
 \Eq(\ref{z11}) is indeed equal to one of the $c$-coefficients.  
 \Eq(\ref{z11}) provides a convenient way to find the
 cut-constructible coefficients of the three-point function.  Finally,
 to determine the rational part coefficients in \Eq(\ref{eq70}), we
 take vectors $l$ that have non-vanishing projections on either $n_3$
 and $\ne$ or on $n_4$ and $\ne$. Since we already know all the
 cut-constructible coefficients, it is straightforward to find $\tilde
 c_{7,8,9}$.

 We note that the discrete Fourier transform is just one of many ways
 to solve the linear system of equations required to obtain the
 coefficients $\tilde c_0,...\tilde c_6$.  It is a convenient,
 easy-to-code-up procedure, but it is neither unique nor superior to
 other ways. In fact, it is clear that in certain cases it is better
 to avoid using the discrete Fourier transform method and to solve the
 system of equations by other means.
 
 To see why this might be the case, we discuss the computation of the
 reduction coefficients for the two-point function with two propagators
 $d_0$ and $d_1$.  Then, the physical space is one-dimensional and the
 transverse space is three-dimensional.  The momentum parametrization
 therefore reads
 \be
  l^\mu = x_1 q_1^\mu + l_\perp 
 \left ( \sum_{i=2}^{4} x_{i} n_i^\mu + x_\ep \ne^\mu \right ),\;\;\;
 q_1 \cdot n_{i \ge 2}  = 0,\;\;\; n_i \cdot n_j = \delta_{ij}.
 \label{eq20}
 \ee
 Using \Eq(\ref{eq20}), we find  that  components of the momentum
 $l$ for which  $d_{0,1} = 0$ are subject to the following constraints
 \be
 \begin{split} 
 & x_1 = \frac{( m_1^2 - m_0^2 - q_1^2)}{2q_1^2},\;\;\; 
 l_\perp^2 = m_0^2 - x_1^2 q_1^2,\;\;\;
 x_2^2 + x_3^2 + x_4^2 + x_\ep^2 = 1.
 \label{eq23}
 \end{split} 
 \ee
 The general parametrization of the ${\tilde b}$-coefficient reads 
 \beqn
 {\tilde b}_{01} &=& \tilde b_0 
 + {\tilde b}_1 (l \cdot  n_2) 
 + {\tilde b}_2 (l \cdot  n_3) 
 + {\tilde b}_3 (l \cdot  n_4) 
 \nonumber \\
 & +& {\tilde b}_4 ( (l \cdot  n_2)^2 - (l \cdot  n_4)^2) 
 +{\tilde b}_5 ( (l \cdot  n_3)^2 - (l \cdot  n_4)^2) 
 +{\tilde b}_6 (l \cdot n_2) (l \cdot  n_3) 
 \nonumber \\
 & + & {\tilde b}_7 (l \cdot n_3) (l \cdot  n_4) 
 + {\tilde b}_8 (l \cdot n_2) (l \cdot n_4) 
 + {\tilde b}_9 (l \cdot  \ne)^2.
 \label{generalbparam}
 \eeqn
 
 Similar to the case of the three-point function, there are infinitely
 many loop momenta that satisfy the constraints shown in
 \Eq(\ref{eq23}). Therefore, to find the cut-constructible
 coefficients, we can proceed as before, parameterizing
 \be
 l_\perp^\mu = l_\perp
 \left ( \sin \theta \cos \phi \; n_2^\mu
 + \sin \theta \sin \phi \; n_3^\mu
 +  \cos \theta \; n_4^\mu \right ), 
 \ee
 and then applying the technique of the discrete Fourier transform to
 determine $\tilde b_0, \dots \tilde b_8$.  Note, however, that the
 application of the discrete Fourier transform requires division by
 $l_\perp$, c.f.\ \Eq(\ref{Defnofccoeficients}),
 and this may lead to potential trouble. 
 Indeed, according
 to \Eq(\ref{eq23}), $l_\perp$ vanishes if $m_0^2 = x_1^2 q_1^2$
 which corresponds to $ q_1^2 = (m_0-m_1)^2$ or $q_1^2 = (m_0 +
 m_1)^2$.  These kinematic points are not dangerous if only massless
 virtual particles are considered.  However, the situation may become
 problematic if virtual massive particles are present in the
 calculation.  Note also that close to those exceptional values of
 $q_1^2$, $l_\perp$ can be small, so that division by $l_\perp$ may
 lead to numerical instabilities.
 
 To handle the case of small $l_\perp$ in a numerically stable way,
 the method of discrete Fourier transform is not directly applicable
 and the system of equations must be solved differently.  There are
 many ways to solve a system of linear equations avoiding division by
 $l_\perp$; one option is described below.  We begin by choosing
 $l_\perp^\pm = x_\perp n_2 \pm x_3 n_3$, $l^\pm_\perp \cdot
 l^\pm_\perp = l_\perp^2$. Recall that $l_\perp^2$ is fixed by the
 on-shell condition \Eq(\ref{eq23}) and therefore $x_3$ is expressed
 through $x_\perp$, $x_3 = \sqrt{l_\perp^2 - x_\perp^2}$.  We
 calculate $b_\pm = b(l^\pm)$ and eliminate $x_3^2$ in favor of
 $l_\perp^2$ and $x_\perp$ where possible. We obtain
 \ba
 b_{\pm } =
 \tb_0+ \tb_1 x_\perp \pm  x_3 \tb_2
 + \tb_4 x_\perp^2 
 + \tb_5 x_3^2 \pm  \tb_6 x_\perp x_3.
 \ea
 Taking the sum and the difference of $b_{\pm}$, we arrive at
 \be
 \frac{\left ( b_++ b_- \right )}{2} =
 \tb_0^{\rm eff} + \tb_1 x_\perp  + \tb_4^{\rm eff}  x_\perp^2 ,\;\;\;
 \frac{\left (b_+- b_- \right )}{2x_3}  =  \tb_2  + \tb_6 x_\perp\,,
 \label{eqbpm}
 \ee
 where 
\be
 \tb_0^{\rm eff} = \tb_0 +\tb_5 l_\perp^2,
\;\;\;\;\;\;\;\;
 \tb_4^{\rm eff} = \tb_4-\tb_5\,.
 \label{beffective}
\ee 
 %
 The right hand sides of these equations are polynomials in
 $x_\perp$. Therefore, we can apply a discrete Fourier transform with
 respect to $x_\perp$ to find coefficients $\tb_1, \tb_4^{\rm eff},
 \tb_0^{\rm eff}$ as well as $\tb_2,\tb_6$ in \Eq(\ref{eqbpm}).
 
 To determine the remaining  coefficients, we make five choices of the 
 loop-momentum, satisfying the on-shell condition.
 We choose for instance 
 \beqn
 l^{(a)}   &=& x_1 q_1  +x n_2 +  y n_4\, , \nn \\
 l^{(b)}   &=& x_1 q_1  -x n_2 +  y n_4\, , \nn \\
 l^{(c)}   &=& x_1 q_1  -x n_2  - y n_4\, , \\
 l^{(d)}   &=& x_1 q_1  +x n_4  + y n_3\, , \nn \\
 l^{(e)}   &=& x_1 q_1  +x n_2  + y \ne\, ,  \nn
 \eeqn
 where $x^2+y^2 =l_\perp^2$.  We use the notation $b_{\alpha} =
 b(l^{(\alpha)})$. With the coefficients $\tb_0^{\rm
   eff},\tb_1,\tb_2, \tb_4^{\rm eff}$ and $\tb_6$ in hand, we determine
 the other coefficients in the sequence,
 $\tb_8,\tb_3,\tb_5,\tb_7,\tb_9,\tb_0,\tb_4$. The results are
 \beqn
 \tb_8 &=&\frac{(\frac{1}{2}(b_a-b_b)-x \tb_1)}{xy} \, , \nn \\
 \tb_3 &=&\frac{\frac{1}{2}(b_a-b_c)-\tb_1 x}{y} \, , \nn \\
 \tb_5 &=&\frac{\tb_0^{\rm eff}+\tb_3 y+y x \tb_8+x \tb_1
 +(x^2-y^2) \tb_4^{\rm eff}-b_a}{3 y^2}  \, ,  \\
 \tb_7 &=&\frac{(b_d-y^2 \tb_5+\tb_5 x^2+\tb_4 x^2-\tb_3 x-y \tb_2-\tb_0)}{x y}  \, , \nn \\
 \tb_9 &=&\frac{(b_e-\tb_4 x^2-\tb_1 x-\tb_0)}{y^2} \nn \, .
 \eeqn
 The coefficients $\tb_0$ and $\tb_4$ are determined using
 \Eq(\ref{beffective}) once $\tb_5$ has been fixed.
 
 We have discussed a method to calculate the coefficients $\tb_{1,
   \dots ,9}$ in a numerically stable way for small values of
 $l_\perp$. Note that we used the fact that even for arbitrarily small
 $l_\perp^2$ we can choose large, complex values of $x,y$ with $x^2+y^2
 =l_\perp^2$.
 In a numerical program, one can switch from the discrete Fourier
 transform to the solution just described, depending on the value of
 $l_\perp$.  However, the proposed  methods can only work {\it if}
 the decomposition of the loop momentum, as in \Eq(\ref{eq20}),
 exists.  A glance at \Eq(\ref{eq23}) makes it clear that the
 decomposition fails for the {\it light-like} momentum, $q_1^2 = 0$,
 and we have to handle this case differently.  We describe a possible
 solution below.
 
First, some clarifications are in order.
Because we are interested in
one-loop calculations for infrared
safe observables, it is
reasonable to assume that
the vector $q_1$ can  be  {\it exactly} light-like but
it is impossible for that vector to be {\it nearly} light-like,
since such kinematic configurations are, typically, rejected by
cuts\footnote{External particles with small masses are
obvious exceptions but rarely do we need to know observables for, say,
massive $b$-quarks in a situation when all kinematic invariants
are large.}.
Hence,  we have to modify the above analysis to allow for an exactly 
light-like external momentum.
%
 %
 To this end, we choose a frame where the four-vector in
 \Eq(\ref{eq20}) reads $q_1 = (E,0,0,E)$. We introduce a complementary
 light-like vector $\bar q_1 = (E,0,0,-E)$.  The loop momentum is
 parametrized as $l= x_1 q_1 + x_2 \bar q_1 + l_\perp$.  We denote the
 basis vectors of the transverse space as $n_{3,4}$; they satisfy 
$n_i \cdot n_j = \delta_{ij}$, $q_{1} \cdot n_{3,4} = 0$, $ \bar q_{1} \cdot
 n_{3,4} = 0$.
 The on-shell condition for the loop momentum fixes $x_2$
 \be
 x_2 = \frac{m_1^2 - m_0^2}{s},\;\;\; s = 2 q_1 \cdot \bar q_1,
 \label{eqlperp1}
 \ee
 and a linear combination of $x_1$ and $l_\perp^2$
 \be
 l_\perp^2 + m_1^2x_1 - m_0^2 (1+x_1) = 0.
 \label{eqlperp}
 \ee
 Compared to the case when the reference vector $q_1$ is not on the
 light-cone, we write now the parametrization of the function ${\tilde
   b}$ using $n_4\cdot l$. We choose it to be
 \ba
 &&  {\tilde b}(l)  = \tb_0 + \tb_1 (\bar q_1 \cdot l)  
 + \tb_2 (n_3 \cdot l) + \tb_3 (n_4  \cdot l)
 + \tb_4 (\bar q_1 \cdot  l) (\bar q_1 \cdot  l)
 \nonumber \\
 &&\;\;\;\;\;\;\;\; 
 + \tb_5 (\bar q_1 \cdot  l) (n_3 l)
 +\tb_6 (\bar q_1 \cdot l) (n_4 \cdot l)
 + \tb_7 ( (n_3 \cdot l)^2 - (n_4 \cdot l)^2 )
 \nonumber \\
 &&\;\;\;\;\;\;\;\;  + \tb_8 (n_3 \cdot l) (n_4 \cdot l)+\tb_9 ( l \cdot \ne)^2.
 \label{eq154}
 \ea
 
 We describe a procedure to find the coefficients $\tb_0, \dots \tb_9$
 in a numerically stable way.  We begin by choosing 
$x_1 = 0.5$ and this choice is not particularly well-motivated.  This
 fixes $l_\perp^2$, and $x_2$ is fixed by the on-shell condition
 \Eq(\ref{eqlperp1}).  The freedom remains to choose the {\it
   direction} of the vector $l_\perp$ in the $(n_3,n_4)$
 plane. Consider four different vectors
 \be
 l_\perp^{(a)} =  y n_3 + x n_4,\;
 l_\perp^{(b)} = - y n_3 + x n_4,\;
 l_\perp^{(c)} = y n_3 - x n_4,\;
 l_\perp^{(d)} = -y n_3 - x n_4,
 \label{eq30}
 \ee
 where $x^2+y^2 =l_\perp^2$. We use vectors
 $l^{(\alpha)} = x_1 q_1 + x_2 \bar q_1 + l_\perp^{(\alpha )}$, $\alpha = a,b,c,d,$
 to calculate the  function $b^{(\alpha)} = \tilde b(l^{\alpha})$.
 Using $b_a,...b_d$,  we can immediately find the coefficient
 $\tilde b_8$
 \be
\tilde  b_8 = \frac{1}{4 x y}
 \left ( b^{(a)} - b^{(c)} - b^{(b)} + b^{(d)} \right ).
 \ee
 For the determination of the remaining coefficients,
 it is convenient to introduce two linear combinations
 \be
 \begin{split}
 & b_{36} =
 \frac{1}{4 x} \left ( b^{(a)} - b^{(c)} + b^{(b)} - b^{(d)} \right ),
 \\
 & 
 b_{25} = \frac{1}{2} \left ( b^{(a)} - b^{(b)} - 2 x y b_8 \right )\, .
 \end{split}
 \label{eq09}
 \ee

As the next step, we choose $x_1 = -0.5$. Note that this changes the
value of $l_\perp^2$ according to \Eq(\ref{eqlperp}).  We then repeat
the calculation described above.  Our choices of momenta in the
transverse plane $l_\perp$ are the same as in \Eq(\ref{eq30}) but, to
avoid confusion, we emphasize that $x$ and $y$ have to be calculated
with the new $l_\perp^2$. We will refer to $b$ computed with those new
vectors as $\bb^{(a)}$, $\bb^{(b)}$, etc. We calculate $\bb_{36,25}$
by substituting $b^{(\alpha)} \to \bb^{\alpha}$ in \Eq(\ref{eq09}).
It is easy to see that simple linear combinations give the desired
coefficients
\ba
\tilde b_3 = \frac{1}{2} \left ( b_{36} + \bb_{36} \right ),\;\;\;\;
\tilde b_6 = \frac{2}{s} \left ( b_{36} - \bb_{36} \right ),
\nonumber \\
\tilde b_2 = \frac{1}{2} \left ( b_{25} + \bb_{25} \right ),\;\;\;\;
\tilde b_5 = \frac{2}{s} \left ( b_{25} - \bb_{25} \right ).
\ea

Other coefficients, required for the complete parametrization of the
function $\tilde b(l)$ in \Eq(\ref{eq154}), are obtained along similar
lines; we do not discuss this further. However, we emphasize that the
procedure that we just described is important for the computation of
one-loop virtual amplitudes in a situation where both massless and
massive particles are involved.  In particular, it is heavily used in
computations of NLO QCD corrections to top quark pair production
discussed in Refs.~\cite{Melnikov:2010iu,Melnikov:2009dn}.

As a final remark, we note that there is another important difference
between reducing the two-point function to scalar integrals for a
light-like and a non-light-like vector. 
Consider only cut-constructible
terms.  Then, for $q_1^2 \neq 0$ the integration over the transverse space
can be immediately done, leading to

\be
\int \frac{{\rm d}^Dl}{(2\pi)^D} \frac{\tb (l)}{d_0 d_1}
= {\tb}_0 \int \frac{{\rm d}^Dl}{(2\pi)^D} \frac{1}{d_0 d_1}.
\ee
Hence, the only integral we need to know in $q_1^2 \neq 0$ case is the
scalar two-point function. However, in case of a light-like vector
$q_1^2=0$, {\it three} master integrals
contribute to the cut-constructible part even after averaging over the
directions of the vector $l$ in the (two-dimensional) transverse space
\be
\int \frac{{\rm d}^Dl}{(2\pi)^D} \frac{\tilde{b}(l)}{d_0 d_1}
= \int \frac{{\rm d}^Dl}{(2\pi)^D} \frac{
{\tb}_0 + {\tb}_1 (\bar q_1 \cdot  l)
+ {\tb}_4 (\bar q_1 \cdot  l)^2 
}{d_0 d_1}.
\ee
Those integrals must be included in the basis of master integrals in
the case when double cuts are considered with a light-like external vector.  
The calculation of those additional master integrals is straightforward.
For the sake of example, we give the results below for the equal mass case
$m_0 = m_1 = m$. We introduce
$d_0 = l^2 - m^2$, $d_1=(l+q_1)^2 - m^2$,
$q_1^2 = 0$, $\bar q_1 \cdot q_1 = r$,
$c_\Gamma = (4\pi)^{\epsilon-2}
\Gamma(1+\epsilon) \Gamma(1-\epsilon)^2/\Gamma(1-2\epsilon)$
and find ($D = 4-2\epsilon$)
\ba
\label{masslessbubble}
\frac{\mu^{2\epsilon}}{\mathrm{i} r_\Gamma}
\int \frac{{\rm d}^Dl }{(2\pi)^D}\;\;
\frac{l \cdot \bar q_1}{d_0 d_1}
&=&
-\frac{ r }{2}\;
\left ( \frac{1}{\epsilon} + \ln \left( \frac{\mu^2}{m^2}\right) \right ),
 \\
\frac{\mu^{2\epsilon}}{\mathrm{i}r_\Gamma}
\int \frac{{\rm d}^Dl }{(2\pi)^D}\;\;
\frac{(l \cdot \bar q_1)(l \cdot \bar q_1)}{d_0 d_1}
&=&\frac{r^2}{3}\;
\left ( \frac{1}{\epsilon} + \ln \left( \frac{\mu^2}{m^2}\right) \right ).
\ea

\subsection{Comments on the rational part}
\label{sec_rational}

The most general parameterizations of ${\tilde e},{\tilde d},{\tilde c},{\tilde
  b}$ and ${\tilde a}$-functions contain two types of terms. First,
there are terms that involve scalar products of the loop momenta with
four-dimensional vectors from various transverse spaces. Second, there
are terms that involve scalar products of the loop momentum with the
$(D-4)$-dimensional components of the vectors spanning the transverse
space.  These latter terms require going beyond the four-dimensional
loop momentum and give rise to the {\it rational part}.

The rational part is related to the ultraviolet behavior of the
theory; the naive expectation is that the better the UV behavior, the
``smaller'' the rational part. If the integral is free from the
rational part and, therefore, can be fully obtained 
by considering loop momenta confined to the four-dimensional 
space, it is said to be ``cut-constructible''.  A natural
expectation is that the rational part is absent in UV-finite
integrals.  As we explain below, this expectation turns out to be
wrong; the correct result is that a Feynman $N$-point integral is cut
constructible, provided that tensor rank, $r$, of the integral
satisfies the following condition \cite{Bern:1994cg}
\begin{equation} \label{BDKcccondition}
r < \mbox{max}\{(N-1),2\}\,.
\end{equation}

\begin{figure}[t]
\begin{center}
\includegraphics[angle=-90,scale=0.40]{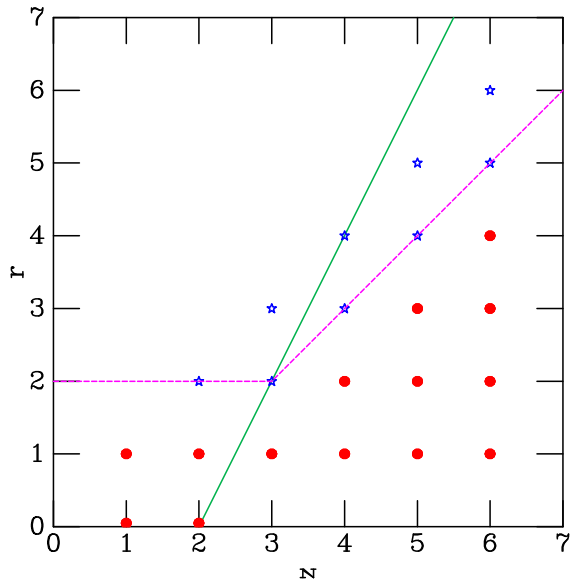}
\end{center}
\caption{Diagram showing tensor $N$-point integrals of rank $r$. Integrals
shown by bullets (red) are cut-constructible, integrals denoted by stars (blue)
contain rational terms. The UV finite integrals lie  beneath the solid (green) line, whereas
the cut constructible integrals lie beneath the dashed (purple) line.}
\label{fig:rat}
\end{figure}

The condition is illustrated in Fig.~\ref{fig:rat}. If it 
is violated the integral will contain rational parts.  Explicitly,
\Eq(\ref{BDKcccondition}) implies that the UV finite rank-two
four-point function is cut-constructible, whereas the UV-finite
rank-three four-point function is not.

To prove this assertion, we will consider all the integrals that occur 
in a renormalizable theory case by case.
To this end, consider a class of higher-rank three-point
functions that are present in renormalizable quantum field theories;
they include rank-one, rank-two and rank-three tensor three-point
functions. The rank-one three-point function is cut-constructible
since its reduction to scalar integrals can not contain terms with 
$(l \cdot \ne)$ tensor structure.
Following the discussion at the beginning of this Section, it is easy
to see that reduction of a general rank-two three-point function
contains an integral
\be
\int \frac{{\rm d}^Dl }{(2\pi)^D}
\frac{(l \cdot \ne)^2}
{
d_0 d_1 d_2
}
\neq 0~~{\rm in ~the~limit~}\ep \to 0.
\ee
Hence, the reduction of a rank-two three-point function to scalar
integrals contains the remnant of the analytic continuation of the
loop momentum to $D$-dimensions -- a sign that the rank-two
three-point function has a rational part.  Because the reduction of a
rank-three three-point function immediately leads to either rank-two
three-point functions or rank-two two-point functions, we conclude
that the reduction of a rank-three three-point function contains a
rational part as well.  Note that while the rational part is a remnant
of UV-sensitivity of a Feynman integral, it is a {\it finite}
contribution to the final answer; 
all UV-divergent contributions reside in scalar 
two- and one-point scalar integrals and are, in fact,
cut-constructible.

It is now easy to generalize the above arguments to higher-point
functions.  For example, the rank-one four-point function is clearly
cut-constructible. The rank-two four-point functions are also
cut-constructible but in a slightly more subtle manner.  Indeed, the
rank-two four-point functions are reduced to rank-one four-point
functions and rank-one three-point functions, both of which are
cut-constructible, {\it and} to a rank-two four-point integral of the
type
\be
\int \frac{{\rm d}^Dl }{(2\pi)^D}
\frac{(l \cdot \ne)^2}
{
d_0 d_1 d_2 d_3
}.
\ee
It is not difficult to see that this integral can be disregarded in
the $\epsilon \to 0$ limit and this is the reason why the rank-two
four-point function does not have a rational part.  Finally, since the
reduction of a rank-three four-point function leads to a host of
rank-two three-point functions, we conclude that the rank-three
four-point function is not cut-constructible. The results for the
rational parts of certain tensor integrals are given in \ref{app:RT}.

Finally, we comment on the computation of the rational part advocated
in Ref.~\cite{Ossola:2008xq}. The idea is to split the numerator
function of any Feynman diagram into ``four-dimensional'' terms and
$(D-4)$-dimensional terms.  The calculation of the rational part,
related to first class of terms -- ${\cal R}_1$ -- is complicated and
requires full machinery behind the OPP technology, that we discussed 
in this Section. One can think of ${\cal R}_1$ as the rational part 
that appears in ultraviolet-finite, pure four-dimensional 
integrals as a consequence  of the tensor reduction.  On the contrary, 
the rational part ${\cal R}_2$ is associated with explicit 
$(D-4)$-dimensional terms in Feynman diagrams and, for this reason, 
its calculation is easy.  Indeed, consider a Feynman diagram with a
numerator ${\rm Num}(l)$ that depends on the loop momentum
$l=(l_4,l_\epsilon)$, where $l_4$ is the four-dimensional part of the
loop momentum and $l_\epsilon$ is its $(D-4)$-dimensional part.  We
write the numerator as
\be
{\rm Num}(l) = {\rm Num}_4(l_4) + {\rm Num}_\ep(l_\ep,\ep).
\label{woo1}
\ee
The rational part ${\cal R}_2$ by definition comes from the second
term in the above equation. We now explain why it is easy to compute
${\cal R}_2$.  First note, that the exact form of ${\cal R}_2$ depends
on the regularization scheme. For example, in the four-dimensional
helicity scheme \cite{Bern:2002zk}, 
explicit $\ep$-terms in \Eq(\ref{woo1}) are not
needed.  Second, since momenta and polarization vectors of all external 
particles in \Eq(\ref{woo1}) are kept four-dimensional, ${\rm
  Num}_\ep$ in \Eq(\ref{woo1}) can only depend on $l_\ep^2$. As such,
it can only contain terms with the loop momentum squared. Third, if
a particular diagram is ultraviolet-finite, we can neglect 
${\rm Num}_\ep(l_\ep,\ep)$ in \Eq(\ref{woo1}), so that the particular 
diagram does not contribute to $R_2$. 
We conclude that  ${\cal R}_2$ {\it only comes from divergent
  graphs}, in sharp contrast to ${\cal R}_1$.  Since there is  a
small number of divergent one-loop graphs in any renormalizable
theory, the computation of the rational part ${\cal R}_2$ becomes
straightforward. Note that because the ultraviolet sensitivity of a particular 
diagram depends on the gauge-fixing condition, 
results for ${\cal R}_2$ are, in general, gauge-dependent. 

To illustrate a typical calculation of ${\cal R}_2$, we compute
it in QED.  We begin by  choosing  the Landau
gauge.  In this gauge, there is one divergent diagram in QED at the 
one-loop order -- the fermion loop contribution to photon vacuum polarization 
diagram.  The
${\cal R}_2$ for QED is then easily computed 
%
\be
\label{eq_r2}
\begin{split}
& {\cal R}_2 \left [ \Pi_{\mu \nu} \right ] 
= 
{\cal R}_2 \left [ i e^2 \int \frac{{\rm d}^D l }{(2\pi)^D}
\frac{{\rm Tr} \left [ 
\gamma_\mu \left ( \hat l + m  \right 
) \gamma_\nu ( \hat l + \hat q + m) 
\right ]
}{d_0 d_1 } \right ] 
\\
& =
{\cal R}_2 \left [ i e^2 \int \frac{{\rm d}^D l }{(2\pi)^D}
\frac{{\rm Tr} \left [ 
\gamma_\mu \hat l  \gamma_\nu \hat l 
\right ]
}{d_0 d_1 } \right ]
 =  - 4 i e^2 g_{\mu \nu} \int \frac{{\rm d}^D l }{(2\pi)^D}
\frac{( l \cdot \ne)^2 }{d_0 d_1 }
\\
&
\;\;\;\;\;\;\;\;\;\;\;\;\;
=  
- g_{\mu \nu}  \frac{\alpha}{2\pi} 
\left ( 2 m^2 - \frac{q^2}{3} \right ).
\end{split} 
\ee
If we switch to the Feynman gauge, 
the above result stays the same
but two additional divergent diagrams appear -- the one-loop
correction to the photon-electron interaction vertex and the one-loop
correction to the fermion self-energy diagram.  In the
four-dimensional helicity scheme, the fermion self-energy diagram does
not contribute to ${\cal R}_2$, since the rank of the tensor integrals
present there is not high enough. The contribution of the one-loop
photon-fermion vertex correction to ${\cal R}_2$ in the
four-dimensional helicity scheme reads
\be
{\cal R}_2 [V_\mu ]
= - i e \gamma_\mu
\int \frac{{\rm d}^D l }{(2\pi)^D}
\frac{(- 4i e^2) (l \cdot \ne)^2 }{d_0 d_1 d_2 }
= 
-i e \gamma_\mu \times \frac{\alpha}{2 \pi}.
\ee
The explicit results for the rational part ${\cal
  R}_2$ in various theories, including QCD and the Standard Model were
recently given in
Refs.~\cite{Garzelli:2010qm,Garzelli:2009is,Draggiotis:2009yb}.
Finally we note that yet another unitarity-based method for the calculation 
of the rational part has been  developed \cite{Bern:2005cq,Berger:2006ci}.
It is based on the recursion relations for on-shell matrix elements.

\subsection{Rational terms by Passarino-Veltman reduction}

In this Section we give an alternative proof of the condition that an
integral has to satisfy for being cut-constructible,
\Eq(\ref{BDKcccondition}). This proof is based on the
Passarino-Veltman reduction.  We will proceed case-by-case for the
two-, three- and four-point integrals which occur in a renormalizable
theory. The extension to higher-point integrals will be performed at
the end. We first note that the Passarino-Veltman decomposition
described in Section \ref{sec:tradoneloop} and \ref{app:PV}, yields
the coefficients of the scalar integrals $D_0,C_0,B_0,A_0$ for
arbitrary values of the number of dimensions.  Since the rational terms
are related to ultraviolet singularities they will show up at the end of the
reduction as terms of the form 
\beq \mbox{Rational terms}~\sim \epsilon B_0(p,m_i,m_j),\; \epsilon A_0(m_i), 
\eeq 
because $A_0$ and $B_0$ are the only ultraviolet-divergent
scalar integrals. Such terms can only arise if the reduction involves
the dimensional parameter $D$. This means that integrals of rank $r$
less than two will always be cut-constructible, since their reduction
coefficients are $D$-independent.  On the contrary, ultraviolet-divergent
integrals of rank two or greater (e.g.\
$D_{iiii},C_{iii},C_{ii},B_{ii}$) will  give rise to
rational parts. Thus it only remains to discuss the ultraviolet finite
integrals $D_{iii},D_{ii}$.  The integral $D_{iii}$ contains a
ultraviolet-divergent integrals of rank greater than two in its reduction
paths, $D_{iii}\to C_{ii}$, see Table~\ref{PVreductionpaths}, and hence
it will have a rational part.  This leaves the special case $D_{ii}$, a
finite integral which can contain a ultraviolet-divergent integral in its
reduction path, namely $B_0$. However since the starting integral is
ultraviolet finite, the ultraviolet poles all cancel.  Moreover the coefficients of
$B_0$ are all $\epsilon$-independent, since the only $D$ dependence enters through
$D_{00}$, which does not contain $B_{0}$ in its reduction path. Hence
the rank-two, four-point integral is cut constructible.

In a renormalizable theory the higher-point functions are not 
ultraviolet-divergent. Moreover the most 
ultraviolet-singular terms in their reduction
paths reduce both $N$ and $r$ by one unit. Therefore the reduction
paths of these ultraviolet-finite integrals can only generate a rational part
if the rank of the integral has $r \geq N-1$.  This observation extends
\Eq(\ref{BDKcccondition}) to $N$ greater than four.

\section{Managing the color}
\label{sect6}

In this Section, we describe how to connect the OPP ideas, discussed
in the previous Section in the context of individual Feynman diagrams,
with unitarity ideas.  The object that we need to calculate is the
one-loop scattering amplitude ${\cal A}(\{k\},\{\epsilon\},\{a\})$,
where external particles of definite types have on-shell momenta
$\{k\}$, polarizations $\{\epsilon\}$  and color indices $\{a\}$.
Before discussing unitarity ideas, we explain how color degrees of
freedom can be treated, in order to simplify the computation of
scattering amplitudes.

Any Feynman diagram that contributes to a scattering amplitude of
colored particles involves a color part and a space-time part.  Since
the presence of color degrees of freedom causes some additional
complexity in the evaluation of scattering amplitudes, it is important
to simplify the treatment of color as much as possible.  To deal with
color parts of the amplitude in a systematic manner, it is customary
to choose a basis in color space and express the color parts of all
diagrams as linear combinations of the basis elements
\cite{Berends:1987cv,Mangano:1988kk}.  This procedure is known as the
``color decomposition''.  A pedagogical description of the color
decomposition can be found in
Refs.~\cite{Mangano:1990by,Dixon:1996wi}.  In what follows we present
a simple but concise discussion of the color decomposition, using
basis-independent features to the extent possible.

\subsection {$n$-gluon amplitudes}
As a first step, we discuss the color decomposition of 
$n$-gluon tree amplitudes in  $SU(N_c)$ gauge theories.
The scattering amplitude 
$
{\cal A}_n^{\rm tree}
$
depends on the gluon color quantum numbers $a_i=1,\ldots,(N_c^2-1)$,
the helicities $h_i=\pm 1$ and the momenta $k_i$. We assume that the
momenta of all gluons are outgoing, so that momentum conservation
reads $k_1+k_2 \ldots + k_n=0$.  In the remainder of this Section, we
suppress helicity and momentum labels when writing scattering
amplitudes.

The generators of the Lie algebra of the $SU(N_c)$ gauge group in the
fundamental representation are determined by the following sets of
equations
\beq
\label{commT}
[T^a,T^b]=i\sqrt{2}f^{abc} T^c = -F^a_{bc}T^c\,,\quad \,\quad {\rm Tr}(T^aT^b)=\delta_{ab}.
\eeq
The color factors of the three- and four-gluon vertices are given by
the structure constants $f^{abc}$ of the $SU(N_c)$ group.  It is
well-known that structure constants can be used to define Lie algebra
generators in the adjoint representation
\beq
\label{commF}
[F^a,F^b]=-F^a_{bc}F^c\,,\quad F^a_{bc}=-i\sqrt{2}f^{abc} 
\,,\quad {\rm Tr}(F^a F^b)=2\Nc\delta_{ab}.
\eeq
The first equation in \Eq(\ref{commF}) follows from the Jacobi identity
and the commutator algebra of fundamental representation, \Eq(\ref{commT}).
Various normalization factors are
chosen for convenience.  We therefore conclude that the color part of
any contributing Feynman diagram is a product of $(n-2)$ color
matrices in the adjoint representation
\be
(F^{a_2}  F^{a_3} \ldots   F^{a_{n-2}}  F^{a_{n-1}})_{a_1a_n},
\ee
and, possibly, some other terms obtained by permutations of the color
indices of the gluons.

It follows from \Eqs(\ref{commT},\ref{commF}) that the adjoint color matrix
$F^a$ can be specified also as the trace of three color matrices of
fundamental -- and in fact any -- representation.  We obtain
\beq
\label{comm_forF}
\left(F^{a_1}\right)_{a_2a_3} =
-\frac{1}{2\Nc}{\rm Tr}\left([F^{a_1},  F^{a_2}]F^{a_3}\right)
=-{\rm Tr}\left([T^{a_1},  T^{a_2}]T^{a_3}\right).
\eeq

We can use \Eq(\ref{comm_forF}), to convert the product of color
matrices in the adjoint representation into linear combinations of
traces of strings of $F$- and $T$-matrices
\be
\label{multiplecomm}
\begin{split}
(  F^{a_2}   F^{a_3} \ldots  
&     F^{a_{(n-2)}}  F^{a_{(n-1)}}
                 )_{a_1a_n}
\\ &
=\frac{1}{2\Nc}{\rm Tr}\left(
                     \left[ [
                              \ldots [[F^{a_1} , F^{a_2}],F^{a_3}],\ldots 
,F^{a_{n-2}}\right]
[F^{a_{n-1}} , F^{a_n}]
                                     \right) 
\\&
=\phantom{\frac{1}{2\Nc}}{\rm Tr}\left(
                     \left[ [
                              \ldots [[T^{a_1} , T^{a_2}],T^{a_3}],\ldots 
,T^{a_{n-2}}\right]
[T^{a_{n-1}} , T^{a_n}]
                                     \right).
\end{split}
\ee
For example, \Eq(\ref{multiplecomm}) gives the following 
$F$-matrix identities for $n=4$ and $5$, 
\be
\begin{split} 
& (F^{a_2} F^{a_3})_{a_{1}a_{4}} 
=\frac{1}{2\Nc}{\rm Tr}\left([F^{a_1} , F^{a_2}][F^{a_{3}}, F^{a_{4}}]\right), 
\\
& (F^{a_2} F^{a_3}F^{a_4})_{a_{1}a_{5}} 
=\frac{1}{2\Nc}{\rm Tr}\left([[F^{a_1} , F^{a_2}],F^{a_3}][F^{a_{4}}, F^{a_{5}}]\right). 
\end{split}
\ee

\begin{figure}[t]
\begin{center}
\includegraphics[scale=0.4]{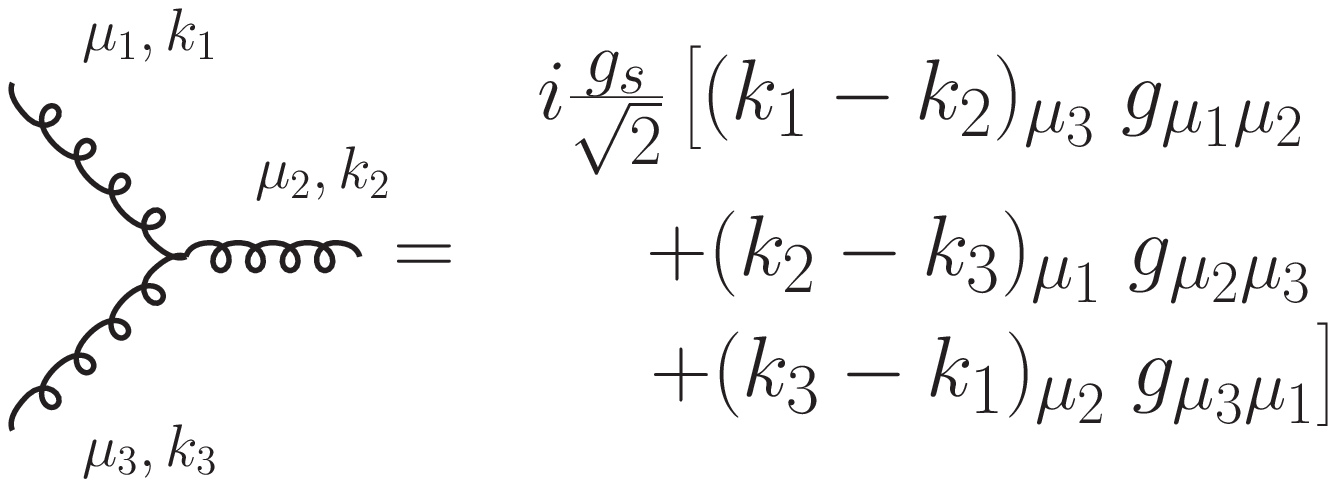}\;\;\;\;\;\;
\includegraphics[scale=0.4]{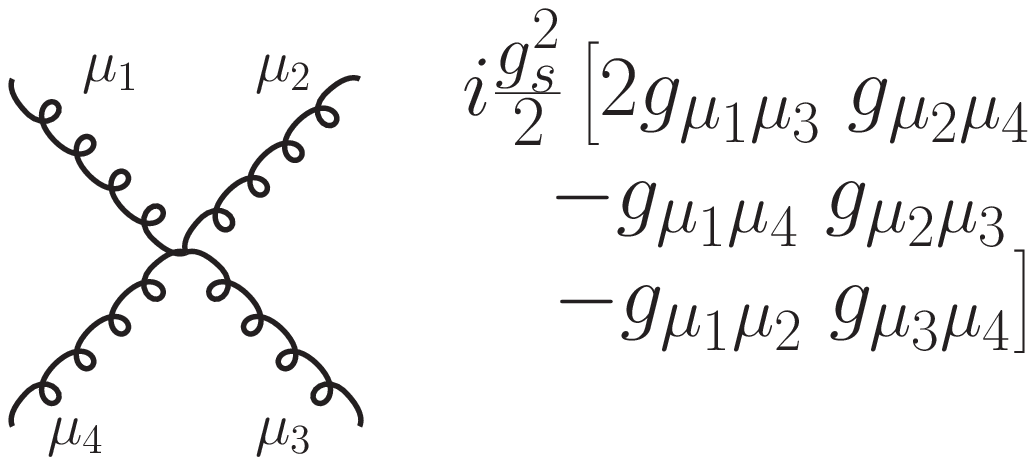}\\
\vspace*{5pt}
\includegraphics[scale=0.4]{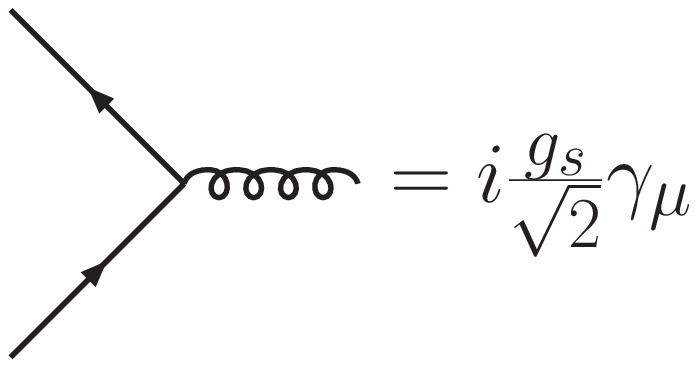}\;\;\;\;\;\;
\includegraphics[scale=0.4]{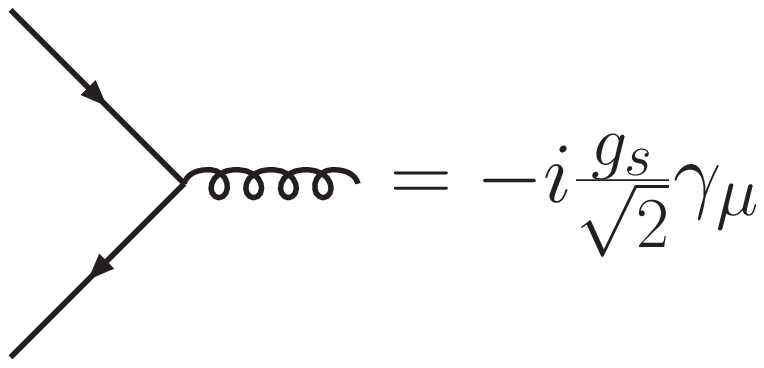}
\end{center}
\caption{Color-ordered Feynman rules. All gluon momenta are outgoing.}
\label{fig6.1}
\end{figure}

It follows from \Eq(\ref{multiplecomm}) that, apart from the
normalization factor, the identities obeyed by $F$- and $T$-matrices
are the same.  Expanding out commutators in \Eq(\ref{multiplecomm})
and collecting identical terms using the cyclic property of the trace,
we conclude that sets of traces of strings of $n$ color matrices in
any representation with all $(n-1)!$ non-cyclic permutations of the
color labels of the gluons included, give a color basis that
decomposes the tree level $n$-gluon amplitudes into colorless ordered
amplitudes
\be
\label{orderedF}
 {\cal A}^{\rm tree}_n =\frac{g_s^{n-2}}{2\Nc} \sum \limits_{\sigma \in S_n/Z_n}
{\rm Tr} \left ( 
F^{a_{\sigma(1)}} F^{a_{\sigma(2)}}
F^{a_{\sigma(3)}}\ldots F^{a_{\sigma(n)}} 
\right ) 
A ^{\rm tree}_{n,\sigma},
\ee
\be
\label{orderedT}
{\cal A}^{\rm tree}_n
=g_s^{n-2} \sum \limits_{\sigma \in S_n/Z_n}
{\rm Tr} \left ( 
T^{a_{\sigma(1)}} T^{a_{\sigma(2)}}
T^{a_{\sigma(3)}}\ldots T^{a_{\sigma(n)}} 
\right ) 
A ^{\rm tree}_{n,\sigma} \, .
\ee
In the above equations, the sum runs over all non-cyclic permutations
of $n$-gluons and $g_s$ is the QCD coupling constant.  
Because of Bose symmetry, all color-ordered
sub-amplitudes $A_{n,\sigma}^{\rm tree}$ are described by a single
function, computed for different permutations of its arguments
\be
A ^{\rm tree}_{n,\sigma} = m _n(g_{\sigma(1)},
g_{\sigma(2)},g_{\sigma(3)},\ldots, g_{\sigma(n)}), 
\label{cr5}
\ee
where $g_i$ is a generic notation for an external gluon with momentum
$k_i$ and helicity $h_i$.  Since, as follows from
\Eq(\ref{comm_forF}), the color factor of any diagram can be rewritten
through traces of $F$- or $T$-matrices by means of the same
mathematical operations, the two bases define the same colorless
ordered amplitudes.  The computation of the function $m
_n(g_1,g_2,\ldots,g_n)$ is simplified if one
notices~\cite{Berends:1987cv,Mangano:1988kk} that only diagrams where
gluons appear in the same order as their color factors in the color
trace, can contribute to a particular color-ordered tree amplitude.
In fact, these sub-amplitudes can be calculated without any reference
to color degrees of freedom, using {\it color-stripped}
Feynman rules based on ordered colorless
vertices~\cite{Bern:1994fz,Dixon:1996wi,Bern:1996je}.  We show
the color-stripped Feynman rules in Fig.~\ref{fig6.1}.  As the name implies, 
these Feynman rules yield color-ordered Feynman diagrams where the color part 
has been stripped off.~\footnote{ Note that the sign in the quark-gluon-antiquark
  vertex depends on the orientation of the diagram since
  $(T^a)_{i \jb}$ is the color matrix for a quark
  $-(T^{*\,a})_{j \ib}$ is the color matrix for an antiquark.}
They can also be used to calculate colorless ordered amplitudes using,
for instance, Berends-Giele recursion relations.

We turn to the discussion of properties of color-ordered tree
amplitudes.  Gauge invariance of the full amplitude ${\cal A}_n^{\rm
  tree}$ and the fact that the color factors in
\Eqs(\ref{orderedF},\ref{orderedT}) are linearly independent, ensure
that the sub-amplitudes $m_{n}$ are gauge invariant for each gluon
permutation separately.  In addition, they obey a number of
relations. The cyclic identity
\be
m _n(g_1,g_2,g_3,\ldots,g_n) = m_n(g_2,g_3,\ldots,g_n,g_1)\,, 
\label{cy1_id}
\ee
follows from \Eqs(\ref{commF},\ref{commT}), if we   
use the linear independence and the cyclic properties of the 
color trace.  The reflection identity 
\be
m_n(g_1,g_2,g_3,\ldots,g_{n-1},g_n) = (-1)^n m_n(g_n,g_{n-1},\ldots,g_2,g_1)\,, 
\label{cy_id}
\ee
follows from the antisymmetry of the $SU(3)$ generators in the adjoint 
representation $F^a_{bc} = -F^a_{cb}$ and the related identity for 
color traces 
${\rm Tr} \left ( F^{a_1} F^{a_2} \ldots F^{a_n} \right ) 
= (-1)^n {\rm Tr} \left (F^{a_n} \ldots F^{a_2} F^{a_1} \right ).
$

It is straightforward to verify that \Eqs(\ref{cy1_id},\ref{cy_id})
reduce the number of independent color-ordered tree amplitudes in the
pure gluon case from $n!$ to $(n-1)!/2$. However, as we now show, more
relations exist between the various color-ordered amplitudes.  In
particular, relations between different color-ordered amplitudes arise
if we consider color ordering in a theory with the gauge group given
by the direct product of the two gauge $SU(\Nc)$
groups~\cite{Berends:1988zn}.  The assumption that the two groups in
the direct product are the same is not essential, but we will use it
in what follows.  We note that gauge fields in such a theory are
charged under either one or the other gauge group, but not under both.
The Lagrangian of the theory is the sum of two terms, each containing
the square of the field strength tensor for the appropriate gluon
field. Therefore the ``gluons'' of the two gauge groups do not
interact with each other.

Given the Lagrangian of the theory, it is clear that scattering
amplitudes that involve gluons of {\it both types} should vanish.
This decoupling feature is perfectly obvious at the Lagrangian level,
but it becomes obscure once the color ordering is performed. Indeed,
the Lie algebra of the extended gauge group $SU(\Nc) \otimes SU(\Nc)$
is defined by generators ${\cal T}^a$, $a = 1,2,\ldots,2(\Nc^2-1)$.
Those generators satisfy the commutation relations $[{\cal T}^a,{\cal
  T}^b] = i g^{abc} {\cal T}^c$, where $g^{abc} = \sqrt{2} f^{abc}$
provided that $[a,b,c] \in S_1 = [1,\ldots,\Nc^2-1]$ or $[a,b,c] \in
S_2 = [\Nc^2,\ldots,2( \Nc^2 -1 ) ]$ and $g^{abc} = 0$ otherwise.  We
now consider a scattering amplitude of gluons in the theory with such
a gauge group; the interaction vertices are specified in terms of the
structure constants $g^{abc}$. The resulting color decomposition of
amplitudes is identical to that in \Eq(\ref{orderedT}) up to an
obvious replacement $T^a \to {\cal T}^a$.  Of course, the
color-stripped sub-amplitudes are, by their nature, unaffected by the
changes in the color gauge group and, therefore, remain the same. We
obtain
\be
{\cal A}^{\rm tree}_n = g_s^{n-2} \sum \limits_{\sigma \in S_n/Z_n}^{}
{\rm Tr} \left ( 
{\cal T}^{a_{\sigma(1)}} {\cal T}^{a_{\sigma(2)}} 
{\cal T}^{a_{\sigma(3)}}...{\cal T}^{a_{\sigma(n)}} 
\right ) 
m_n(g_{\sigma(1)},...,g_{\sigma(n)}). 
\label{cr41}
\ee

We now consider the scattering amplitude of $n_1$ gluons charged under
the first gauge group and $n-n_1$ gluons charged under the second
gauge group.  As we already explained, the resulting scattering
amplitude must vanish.  This feature is {\it not apparent} from
\Eq(\ref{cr41}).  In fact, as we show now, the vanishing of the full
amplitude implies non-trivial identities between different colorless
sub-amplitudes.

To see how these relations appear, we note that color factors on the
right hand side of \Eq(\ref{cr41}) can be simplified since Lie algebra
generators with indices in the two lists $S_1$ and $S_2$ commute with
each other
\be
\left [ {\cal T}^{a}, {\cal T}^{b} \right ] =0, \;\;\; a \in S_1, \;\;\;
b \in S_2. 
\ee 
Hence, the color weights in \Eq(\ref{cr41}) can be re-written by
commuting all generators ${\cal T}^{a_1}$ with $a_1 \in S_1$ to the
left of the color trace; of course, the {\it relative} ordering of
gluons with indices in the same list can not be changed.  Upon
performing this procedure several times, we arrive at the color
weights of the following form
\be
{\rm Tr} \left (
{\cal T}^{a_{\sigma(1)}}...{\cal T}^{a_{\sigma(n_1)}}
\times {\cal T}^{b_{\sigma(n_1+1)}}...{\cal T}^{b_{\sigma(n)}} 
\right ),
\ee
with $a_i \in S_1$ and $b_i \in S_2$. Since these color weights are
independent, their coefficients must vanish. These coefficients are
given by sums of color-ordered amplitudes where orderings of gluons
with indices from the lists $S_1$ and $S_2$ are kept fixed, while all
permutations between entries of different lists are allowed. For this
reason we conclude that the color-ordered amplitudes satisfy a set of
identities
\be
m_n(1,\underline{2,\ldots n_1},\overline{n_1+1,\ldots,n})\equiv
\sum \limits_{\sigma(n)}
m_n(g_1,g_{\sigma(2)},g_{\sigma(3)},\ldots,g_{\sigma(n)}) = 0.
\label{abelident}
\ee
The sum in this equation runs over all mergings of the two lists
$\{2,..,n_1\}$ and $\{n_1+1,..,n \}$, that preserve relative ordering
of elements in each list. The two lists to be merged are denoted by
under- and over-lining.  For example, in case of $n=6$, we can place
three gluons into one list and two gluons into the other list.  The
full set of permutations that we have to consider reads \be
\label{abellist}
\begin{split}
\theta({234}|{56}) = & 
[(23456),(23546),(25346),(52346), (23564),
\\
& (25364),(52364),(25634),(52634),(56234)].
\end{split}
\ee
An interesting special case occurs if we consider a list
$\theta(2|34\ldots n)$ which corresponds to the direct product of
$U(1)$ and $SU(N_c)$ gauge groups.  The gauge boson of the $U(1)$
group is referred to as a ``photon''. The ``photon'' decoupling
identities read
\be
\begin{split}
{ m}_n(1,\underline{2},\overline{3,\ldots,n}) \;\equiv \;&
m_n(g_1,g_2,g_3,\dots,g_n) + m_n(1,g_3,g_2,\dots,g_n)
\\ &
\ \ +\ldots m_n(g_1,g_3,\dots,g_n,g_2)=0.
\end{split}
\ee

It is clear that \Eq(\ref{abelident}) gives a set of relations for
scattering amplitudes; such relations are usually referred to as
{\it Abelian relations}~\cite{Berends:1988zn}.  They reduce the number
of independent color-ordered sub-amplitudes in pure gluon case from
$(n-1)!/2$ to $(n-2)!$.  Simultaneous application of the reflective
and Abelian identities leads to many new identities, such as e.g.  the
Kleiss-Kuijf relations~\cite{Kleiss:1988ne}. The Kleiss-Kuijf
relations can also be used to reduce the number of independent
color-ordered amplitudes to $(n-2)!$ and, in this sense, they are
equivalent to the Abelian identities.

The color decomposition for gluon scattering amplitudes that involves
$(n-2)!$ truly independent color structures was first derived in
Ref.~\cite{DelDuca:1999rs}.
It is remarkably simple
%
\begin{equation}
\begin{split} 
& {{\cal A}}^{\rm tree}_n (1,2,3, \ldots, n)=
g_s^{n-2} \sum_{\sigma={\cal P} 
(2,3,\ldots, n-1)} \left(F^{a_{\sigma(2)}}\ldots F^{a_{\sigma(n-1)}}\right)_{a_1a_n} 
\\
&\;\;\;\;\;\;\;\ \;\;\;\;\;\;\;\; \;\;\;\;\;\;\;\;\;\;\; \times 
m_{n}(g_1, g_{\sigma(2)},g_{\sigma(3)},\ldots,g_{\sigma(n-1)},g_n)\,.
\end{split} 
\label{TreeCol3}
\end{equation}
We sketch here a derivation of \Eq(\ref{TreeCol3}) which is based on
the color commutator algebra, a simple decomposition of the sum over
permutations and the Abelian identities~\cite{GieleColor}.  We first
introduce compact notations for products of color matrices
\be
\begin{split}
& \{  a_{i_1}a_{i_2} \ldots a_{i_n}  \} _{uv} \equiv 
\left(
F^{a_{i_1}} F^{a_{i_2}}\ldots F^{a_{i_n}} \right)_{uv}\, ,
\\
& \{  a_{i_1}a_{i_2} \ldots a_{i_n}  \} \equiv 
{\rm Tr}\left(F^{a_{i_1}} F^{a_{i_2}}\ldots F^{a_{i_n}} 
\right),
\\
& \{[a,b]\}_{cd}\equiv ([F^a,F^b])_{cd}\,,\quad\{\,\}_{ab}\equiv\delta_{ab}\, .
\label{eqcompact1}
\end{split}
\ee
In the new notation \Eq(\ref{commF}) becomes
\be
\label{commFs}
\{[a,b]\}_{cd}=-\{a\}_{by}\{y\}_{cd}\,,
\quad\{ab\}= 2\Nc \delta_{ab}.
\ee

We start with a list of identities between the products of color
matrices.  With the use of the commutator~\Eq(\ref{commFs}) it is easy
to derive
\be
\label{4gcolcomm}
\begin{split}
&\{a_2a_1\}_{uv}=\{a_1a_2\}_{uv} -\{a_2\}_{a_1y}\{y\}_{uv},
\\
&\{a_2a_3a_1\}_{uv}=\{a_1a_2a_3\}_{uv} -\{a_2\}_{a_1y}\{ya_3\}_{uv}
-\{a_3\}_{a_1y}\{ya_2\}_{uv}
\\
& \phantom{xxxxxxxxxx} + \{a_3a_2\}_{a_1y}\{y\}_{uv},
\\
&\{a_2a_3a_4a_1\}_{uv}=\{a_1a_2a_3a_4\}_{uv} 
-\{a_2\}_{a_1y}\{ya_3a_4\}_{uv}
\\ &
\phantom{xxxxxxxxxx}
-\{a_3\}_{a_1y}\{ya_2a_4\}_{uv}
 -\{a_4\}_{a_1y}\{ya_2a_3\}_{uv} 
\\ &\phantom{xxxxxxxxxx}
 + \{a_3a_2\}_{a_1y}\{ya_4\}_{uv}
+ \{a_4a_2\}_{a_1y}\{ya_3\}_{uv}
\\ &
\phantom{xxxxxxxxxx}
+ \{a_4a_3\}_{a_1y}\{ya_3\}_{uv}
- \{a_4a_3a_2\}_{ya_1}\{y\}_{uv}.
\end{split}
\ee
The generalization of these identities to products of an arbitrary
number of color matrices can be easily deduced from
\Eq(\ref{4gcolcomm}).

As a next step, we consider the four-gluon scattering amplitude and
show how to verify \Eq(\ref{TreeCol3}) in that case. This is a good
example since the algebra in the four-gluon case is easy but the
calculation generalizes to the $n$-gluon case.  We express the
four-gluon scattering amplitude through color-ordered amplitudes
fixing the position of the fourth gluon and keeping the sum over the
permutations of the second and third gluons.  We use a simplified
notation for the sum over permutations and write
\be
\label{tree4col}
\begin{split}
{\cal A}^{\rm tree}_4 =&
\frac{g_s^2}{2N_c}\sum_{{\cal P} (2,3)}\Big[
\{a_1 a_2 a_3 a_4\} 
m_4(g_1, g_2,g_3, g_4 )
\\ &
+ \{ a_2a_1 a_3a_4\} 
m_4( g_2,g_1,g_3, g_4 )
+
\{a_2 a_3a_1a_4\} 
m_4( g_2,g_3,g_1, g_4 )\Big ].
\end{split}
\ee
We employ  relations shown in \Eq(\ref{4gcolcomm}) to  
move the  color matrix  $a_1$  to the left; this generates 
seven terms on the right hand side in \Eq(\ref{tree4col}).
Collecting common color factors, we obtain 
\be
\label{tree4unitary}
\begin{split}
&{\cal A}^{\rm tree}_4 =
\frac{g_s^2}{2N_c}\sum_{{\cal P} (2,3)} \Big [
\{a_1a_2a_3a_4\} 
{ m}_4(\underline{1},\overline{ 2,3}, 4 )
\\ & 
+
\{a_2 a_3\}_{a_1y}\{ya_4\} 
m_4( g_2,g_3,g_1, g_4 )
-\{ a_2\}_{a_1y}\{y a_3,a_4\} 
{ m}_4( \underline{2,1},\overline{3}, 4 )\Big ]. 
\end{split}
\ee
The first and the last terms on the right hand side of \Eq(\ref{tree4unitary}) 
vanish because of the Abelian identities, 
$ { m}_4(\underline{1},\overline{ 2,3}, 4 ) = 0 $ and  
${ m}_4( \underline{2,1},\overline{3}, 4 ) = 0$. Finally, we use 
$\{ya_4\}=2\Nc\delta_{y a_4}$ and the cyclic and reflection identities to 
transform \Eq(\ref{tree4unitary}) to the desired result 
\be
{\cal A}^{\rm tree}_4 =  g_s^2 \sum \limits_{{\cal P}(2,3)}^{}
\{a_2 a_3\}_{a_1 a_4 }m_4(g_1,g_2,g_3,g_4).
\label{eq627}
\ee

It is easy to generalize this proof to the $n$-gluon case. We write
\be
\begin{split}
\label{ntree1}
&{\cal A}^{\rm tree}_n
=\frac{g_s^{n-2}}{2\Nc}\sum_{{\cal P}(1,\cdots ,n-1)}\{a_1a_2\cdots a_n\}\,
m_n(g_1,g_2,\ldots,g_n)
\\
&\quad\quad=\frac{g_s^{n-2}}{2\Nc}\sum_{{\cal P}(2,\cdots ,n-1)}
\Big [      \{a_1a_2\cdots a_n\}\, m_n(g_1,g_2,\ldots,g_n) 
\vphantom{\sum_a^b}
\\ 
& \;\;\;\;\;\;
+\sum_{k=2}^{n-1}   \{a_2\cdots a_{k}a_1a_{k+1}\cdots, n\}
\,m_n(g_2,\ldots,g_k,g_1,g_{k+1},\ldots,g_n)
\Big ],
\end{split} 
\ee
and, collecting the relevant terms, obtain 
\be
\begin{split} 
\label{ntree2}
&{\cal A}^{\rm tree}_n
=\frac{g_s^{n-2}}{2\Nc}\sum_{{\cal P}(2,\cdots, n-1)}
\Big [
\vphantom{ \sum_a^b } 
\{a_1a_2\cdots a_n\} { m}_n(\underline{1},\overline{2,\ldots,n-1},n)
\\
 &
\;\;\;\;\;\;+\sum_{k=2}^{n-2}(-1)^{k+1}
\{  a_{2}\cdots a_{k} \}_{a_1y}
\{ya_{k+1}\cdots a_n\}\,
\\ 
& 
\;\;\;\;\;\;
\times { m}_n( \underline{k,\ldots,2,1},\overline{k+1,\ldots,n-1},n)
\\
&\phantom{\frac{1}{\Nc}\sum_{{\cal P}(2,\cdots ,n-1)}}
+ (-1)^n\{a_2\cdots a_{n-1}\}_{a_1y}\{ya_n\} 
  m_n(g_{n-1},\ldots,g_1,g_n)
\Big ] . 
\end{split}
\ee
Similar to the four-gluon case, because of the Abelian identities,
(\Eq(\ref{abelident})), only the last term on the right hand side in
\Eq(\ref{ntree2}) does not vanish.  Application of the reflection and
cyclic identities gives
\be
\label{Ngluon}
\begin{split}
{\cal A}^{\rm tree}_n&=
\frac{g_s^{n-2}}{2\Nc}
\sum_{{\cal P}(2,\cdots ,n-1)}
\{a_2\cdots a_{n-1}\}_{a_1y}
\{ya_n\}m_n(g_1,g_2,\ldots,g_{n-1},g_n)
\\
&
=g_s^{n-2}\sum_{{\cal P}(2,\cdots ,n-1)}\{a_2\cdots a_{n-1}\}_{a_1a_n}
m_n(g_1,g_2,\ldots,g_{n-1},g_n). 
\end{split}
\ee
It follows from the derivation that we just presented that the
colorless ordered amplitudes in this representation are the same as in
the color decompositions using $F$ and $T$ bases.

An important feature of this color basis is that two of the color
indices do not appear in the permutation sum.  Hence, this color basis
is not manifestly symmetric. However, it is a unitary basis in the
sense that on each pole of the tree amplitude the color factor of a
given colorless ordered amplitude also factorizes. Indeed, poles in
tree color-ordered amplitudes appear when a linear combination of
momenta of some set of neighboring gluons becomes light-like. For
definiteness, we assume that those gluons are $g_1,..,g_m$.  The
color-ordered amplitude then factorizes into the product of two
amplitudes
\be
m_n(g_1,..,g_n) \to k_{v}^{-2}\;
m_{m+1}(g_1,\ldots,g_m,g_v) 
m_{n-m+1}(g_v,g_{m+1},
\ldots,g_n),\;
\label{cdec}
\ee
where $k_{v} = -\sum \limits_{i=1}^{m} k_{i}$ and the summation over
the helicities of the intermediate gluon is assumed.  The ``unitary''
nature of the color decomposition in \Eq(\ref{Ngluon}) can be best
seen if we associate natural color factors with the three amplitudes
in \Eq(\ref{cdec}) and sum over colors of the internal gluon. The
color factor that we associate with the right-hand side of
\Eq(\ref{cdec}) is
\be
\sum_{a_v} \{a_2 ...a_m \}_{a_1 a_v} \{a_{m+1},....a_{n-1} \}_{a_v a_n} =
\{a_2 ...a_{n-1} \}_{a_1 a_n},
\ee
which is indeed the color factor of the full amplitude 
$m_{n} (g_1,..,g_n)$.

The above argument explains why the color basis in \Eq(\ref{Ngluon})
is particularly suitable for one-loop computations, using generalized
unitarity.  Indeed, in the context of generalized unitarity, we expect
to reconstruct one-loop amplitudes from tree amplitudes.  Summation
over color indices of intermediate cut lines is performed in exactly
the same manner as in the tree level case, described above.  As a
result, we expect that the color basis in \Eq(\ref{Ngluon}) remains
valid also when one-loop amplitudes are considered.  To see this
explicitly, we write the full amplitude as a sum of terms where gluons
are ordered
\be
\label{orderedgluon}
{\cal A}^{\rm 1-loop}_n=g_s^n \cg \sum_{ \sigma \in {S_{n-1} }}
 A^{(1)}_n(g_1,g_{\sigma{(2)}},\ldots,g_{\sigma{(n)}}),
\ee
The factor $\cg$ is the standard term that appears in dimensionally regulated
one-loop calculations,
\begin{equation}
\cg = 
\frac{1}{(4\pi)^{2-\ep}}
\frac{\Gamma(1+\ep)\Gamma^2(1-\ep)}{\Gamma(1-2\ep)} = 
\frac{(4 \pi)^\ep }{16 \pi^2} \frac{1}{\Gamma(1-\ep)}\ .
\end{equation}
We now consider a double-cut of one of the terms in the sum in \Eq(\ref{orderedgluon}).
The cut splits the ordered momenta in two groups, 
say $[1,..,k]$ and $[k+1,..,n]$. We write the imaginary 
part of the amplitude in that channel as
\be
\label{cutglu}
\begin{split}
&  {\rm Im}_{(k,n)} \left [  A^{(1)}_n(g_1,g_2,\ldots,g_n) \right ] 
  =
 \{a_1a_2\cdots a_k\}_{vu}\{a_{k+1}\cdots a_n\}_{uv}
\\
&\quad \quad \quad \times
m_{k+2}(g_v, g_1,g_2,\ldots,g_k, g_u)
 m_{n-k+2}(g_u,g_{k+1},\ldots,g_n,g_v) 
\\
& \quad \quad \quad = \{a_1a_2\cdots a_n\}\ {\rm Im}_{(k,n)} \left [ 
m^{(1)}_n \left(g_1,g_2,\ldots ,g_n\right) \right ].
\end{split}
\ee
The color decomposition of the one-loop amplitude is independent of
whether we take the imaginary part  or not. We conclude that one-loop
gluon amplitudes\footnote{In this subsection we only deal with
  $SU(N_c)$ pure gauge theory and do not consider matter fields.}
obey a color decomposition of the following form~\cite{DelDuca:1999rs}
\be
\label{1loopglu}
{\cal A}^{\rm 1-loop}_n=g_s^n \cg \sum_{{{\cal P } (2,\cdots, n)}/{\cal R} }
\{F^{a_1},\ldots,F^{a_n}\}
m^{(1)}_n \left(g_1,g_2,\ldots,g_n\right).
\ee
where ${\cal R}$ is the reflection transformation, which needs 
to be factored out  to remove a symmetry factor that appears otherwise. 
The cyclic property and reflection symmetry remain valid. Hence, we
conclude that the number of independent one-loop amplitudes is
$(n-1)!/2$.

Before finishing this subsection we note that the color-ordered tree
amplitudes also satisfy the Bern-Carrasco-Johansson (BCJ)
relation~\cite{Bern:2008qj}. The BCJ relations for gluon amplitudes
read
\be
\begin{split}
s_{12}\,&m_n(g_1,g_2,g_3,\ldots,g_n)+(s_{12}+s_{23})\,
m_n(g_1,g_3,g_2,g_4,\ldots,g_n)+\cdots
 \\ & 
+(s_{12}+s_{23}+\cdots s_{2(n-1)})\,
m_n(g_1,g_3,g_4,\ldots,g_{n-1},g_2,g_n)=0,
\end{split}
\ee
where $s_{ij}= 2 k_i \cdot k_j$.  They have been proven first using
the field theory limit of monodromy relations in string
theory~\cite{BjerrumBohr:2010zs} and later derived using BCFW
recursion relations~\cite{Feng:2010my,Chen:2011jx}\footnote{In
  Ref.~\cite{Feng:2010my} the reflection identity, the Abelian
  identities and the BCJ relations are derived using only BCFW
  recursion relations.}.  When all relations between different
color-ordered gluon scattering amplitudes are combined, the number of
independent color-ordered amplitudes reduces to $(n-3)!$.

\subsection{$\qb q+(n-2)$-gluon  amplitudes}

The color parts of Feynman diagrams that contribute to tree $q\bar
q+(n-2)$-gluon amplitudes have the generic form of a product of color
matrices in the fundamental representation contracted in their adjoint
indices with products of color matrices in the adjoint representation
\be
\begin{split}
&\left(    T^{b_{1}}\ldots 
T^{b_k} \ldots \right)_{j \ib}
\times \left(   F^{a_1}\ldots F^{a_r}  \right)_{{b_1}{a_{r+1}}}
\ldots \left(   F^{a_{p}}\ldots F^{a_t-1}  \right)_{{b_k}{a_{t}}}\ldots
\end{split}
\ee 
The product of $F$-matrices can be converted into traces of
multiple commutators of the $SU(N_c)$ generators in the fundamental
representation using \Eq(\ref{multiplecomm}).
In order to convert complicated expressions to simple products of
$T$-matrices, we can use the identity
\be
\left (T_X T^a T_Y \right )_{j \ib} {\rm Tr}\left([T^a,T^b] T_Z \right)
=\left (T_X[T^b,T_Z]T_Y \right )_{j \ib}\,,
\ee
where $T_{X,Y,Z}$ denote generic products of $T$ matrices.  Therefore,
we conclude that all color factors of the $\qb q + n~{\rm gluon}$
amplitudes can be transformed into a linear combination of products of
$T$-matrices
$  
\left( T^{a_{i_1}} \ldots T^{a_{i_n}} \right)_{j \ib}.
$
All the independent terms of such type form a color basis for the
decomposition of tree amplitudes
\be
\begin{split}
\label{orderedTqq}
{\cal A}^{\rm tree}_n({\qb}_1, q_2, g_3,\ldots,g_n)
=& g_s^{n-2} \sum \limits_{\sigma \in S_{n-2}}
\left ( 
T^{a_{\sigma(3)}} T^{a_{\sigma(4)}}...T^{a_{\sigma(n)}} 
\right )_{i_2 \ib_1}\\
 &
\times m_n({\bar{q}}_1,q_2,g_{\sigma(3)},\ldots,g_{\sigma(n)}). 
\end{split}
\ee
In \Eq(\ref{orderedTqq}),
$m_n({\bar{q}}_1,q_2,g_{\sigma(3)},\ldots,g_{\sigma(n)})$ stands for
the colorless ordered tree amplitude for $\qb q + ( n-2)$-gluon
scattering.  In conjunction with two-particle unitarity cuts,
\Eq(\ref{orderedTqq}) can be used to obtain the color decomposition of
a quark-loop contribution to one-loop gluon-gluon scattering amplitude
\be
\label{nfpart}
\begin{split}
{\cal A}^{\rm 1-loop}_{n;n_f}
=&
g_s^n \cg n_f \sum_{\sigma \in S_{n-1}} 
 {\rm Tr} (T^{a_1} T^{a_{\sigma(2)}} \ldots T^{a_{\sigma{n}}})\; 
m^{(1)}_{n;n_f}\left( g_1,g_{\sigma{(2)}},\ldots,g_{\sigma{(n)}}\right),  
\end{split}
\ee
where $n_f$ is the number of quark flavors.  In QCD the full one-loop
$n$-gluon amplitude is the sum of the right hand sides of
\Eq(\ref{1loopglu}) and \Eq(\ref{nfpart}).

The color basis of \Eq(\ref{orderedTqq}) for the $\qb+q+(n-2)$-gluon
amplitudes keeps the quark and anti-quark indices fixed in the sense
that they do not participate in the permutation sum. Using the
commutator identities we will transform this color basis to another
one, where the positions of one gluon and the anti-quark are kept
fixed, while the positions of the quark and other gluons are
arbitrary.  As we will show, this new basis is an unitary color basis
and, because of that, it is well-suited for constructing the color
decomposition of one-loop amplitudes.

We begin by extending the compact notations introduced in
\Eq(\ref{eqcompact1}) for the products of $F$-matrices, to also
describe products of $T$-matrices 
\be
\begin{split}
&( a_{i_1}a_{i_2} \ldots a_{i_n}  )_{j \ib}
\equiv 
\left( T^{a_{i_1}} T^{a_{i_2}}\ldots T^{a_{i_n}} \right)_{j \ib}\,,
\\
&(  a_{i_1}a_{i_2} \ldots a_{i_n}  ) 
\equiv 
{\rm Tr} \left(T^{a_{i_1}} T^{a_{i_2}}\ldots T^{a_{i_n}} 
\right),
\\
& ([a,b])_{j \ib}=  -\{a\}_{by} (y)_{j \ib}
\,,\quad (ab ) \equiv\delta_{ab}\,.
\end{split}
\ee
The commutator identities will have the same structure as in the pure
gluon case~(\ref{4gcolcomm}) except that curly brackets will have to
be replaced by ordinary ones, when products of $T$-matrices are
involved.  We find 
\be
\label{q4gcolcomm}
\begin{split}
&(a_2a_1)_{j \ib}=(a_1a_2)_{j \ib} -\{a_2\}_{a_1y}(y)_{j \ib},
\\
&(a_2a_3a_1)_{j \ib}=(a_1a_2a_3)_{j \ib} -\{a_2\}_{a_1y}(ya_3)_{j \ib}
\\
& 
\phantom{xxxxxxxxxx}
-\{a_3\}_{a_1y}(ya_2)_{j \ib}
+ \{a_3a_2\}_{a_1y}(y)_{j \ib},
\\
&(a_2a_3a_4a_1)_{j \ib}=(a_1a_2a_3a_4)_{j \ib} 
-\{a_2\}_{a_1y}(ya_3a_4)_{j \ib}
\\ &
\phantom{xxxxxxxxxx}
-\{a_3\}_{a_1y}(ya_2a_4)_{j \ib}
 -\{a_4\}_{a_1y}(ya_2a_3)_{j \ib} 
\\ &\phantom{xxxxxxxxxx}
 + \{a_3a_2\}_{a_1y}(ya_4)_{j \ib}
+ \{a_4a_2\}_{a_1y}(ya_3)_{j \ib}
\\ &
\phantom{xxxxxxxxxx}
+ \{a_4a_3\}_{a_1y}(ya_2)_{j \ib}
- \{a_4a_3a_2\}_{ya_1}(y)_{j \ib}.
\end{split}
\ee
As a  first simple step, 
we consider the case of the  $\bar{q}q+3$~gluon amplitude.
We write permutations of gluon $g_3$ explicitly
\be
\label{qqggg1}
\begin{split}
&{\cal A}^{\rm tree}_5({\qb}_1,q_2,g_3,g_4,g_5)=
g_s^3 \sum_{{\cal P}(3,4,5)} (a_3 a_4a_5)_{i_2 \ib_1}\   
m_5({\qb}_1,q_2,g_3,g_4,g_5)
\\
&\;\;\;\;\;\;\;\;\;\;\;\;\;\;= g_s^3\sum_{{\cal P}(4,5)} 
\Big [ ( a_3a_4a_5)_{i_2 \ib_1} \, 
 m_5({\qb}_1,q_2,g_3,g_4,g_5)
\\
& 
+
(a_4a_3a_5)_{i_2 \ib_1}  
m_5({\qb}_1,q_2,g_4,g_3,g_5 )
+(a_4a_5a_3)_{i_2 \ib_1} m_5({\qb}_1,q_2, g_4,g_5,g_3) \Big ],
\end{split}
\ee
and use the color identities \Eq(\ref{q4gcolcomm}) to move the color
index $a_3$ to the first position.  Upon doing that, we find that the
first term in \Eq(\ref{qqggg1}) does not change, the second term
generates two terms and the third generates four terms.  We can
combine those terms by exploiting the fact that there is a summation
over the permutations of gluons $g_4$ and $g_5$ in \Eq(\ref{qqggg1}),
which allows us to interchange them. We obtain
\ba
\label{qqggg2}
&&{\cal A}^{\rm tree}_5({\qb}_1,q_2,g_3,g_4,g_5)=
g_s^3 \sum_{{\cal P}(4,5)} \Big [ \{ \}_{a_3y}
(ya_4a_5)_{i_2 \ib_1}\ 
m_5({\qb}_1,q_2,\underline{3},\overline{4,5})
\\ &&
-\{a_4\}_{a_3y}(ya_5)_{i_2 \ib_1}\ 
m_5({\qb}_1,q_2,\underline{4,3},\overline{5}) 
+\{a_4a_5\}_{a_3y}(y)_{i_2 \ib_1}\ 
m_5({\qb}_1,q_2,g_5,g_4,g_3)\Big ]\, .
\nonumber 
\ea
To rewrite \Eq(\ref{qqggg2}) in a way that allows further generalization, 
it is convenient to introduce the notation 
\be
\label{qqggg2b}
\begin{split}
&
{\tilde m}_5({\qb}_1,g_5,g_4,q_2,g_3)=
-m_5({\qb}_1,q_2,\underline{3},\overline{4,5}),
\\ &
{\tilde m}_5({\qb}_1,g_5,q_2,g_4,g_3)=
m_5({\qb}_1,q_2,\underline{4,3},\overline{5}),
\\ &
{\tilde m}_5({\qb}_1,q_2,g_5,g_4,g_3)=
-m_5({\qb}_1,q_2,g_5,g_4,g_3).
\end{split}
\ee
We note that the two lists of gluons, whose members can be permuted,
are separated in the argument of amplitudes ${\tilde m}_5$ by $\qb$ and
$q$ labels.  A convenient way to visualize these amplitudes is to
imagine that the quark line runs horizontally, from $\qb$ to $q$
and that, in the argument of ${\tilde m}_5$, gluons that are to the
right of $\qb$ and to the left of $q$ are drawn below the quark
line, while gluons to the right of $q$ are drawn above that line.  We
will use this way of visualizing these objects in Sect.~\ref{bgampl}
when we discuss the Berends-Giele recursion relations for colorless
tree amplitudes.

We write our final  result for the full five-point  amplitude  as
\ba
\hspace*{-0.3cm}&&{\cal A}^{\rm tree}_5({\qb}_1,q_2,g_3,g_4,g_5)=
-g_s^3  \sum_{{\cal P}(4,5)} \Big [
\{\}_{a_3 y} (ya_4a_5)_{i_2 \ib_1}\ 
{\tilde m}_5({\qb}_1,g_5,g_4,q_2,g_3)
\\ 
\hspace*{-0.3cm}&&
+\{a_4\}_{a_3 y} (ya_5)_{i_2 \ib_1}\ 
{\tilde m}_5({\qb}_1,g_5,q_2,g_4,g_3)
+\{a_4a_5\}_{a_3 y} (y)_{i_2 \ib_1}\ 
{\tilde m}_5({\qb}_1,q_2,g_5,g_4,g_3)\Big ].
\nonumber 
\ea
Note that the positions of the first gluon and the anti-quark are
fixed, while we have a permutation sum over the positions of the other
gluons and the quark.
This pattern remains valid also for $n$-gluons. We introduce a special
notation for objects that multiply color structures in $\qb q
+n$~gluon scattering amplitude
\be
\hspace*{-0.2cm} 
{\tilde m}_n({\qb}_1,g_n,.., g_{(k+1)},q_2,g_k,..,g_3)
=
(-1)^k m_n({\qb}_1,
q_2,\underline{k,...,3},\overline{(k+1),..,n}).
 \label{eq_prim}
\ee
The  color decomposition in the new basis reads
\be
\label{eq_cdec1}
\begin{split}
&{\cal A}^{\rm tree}_n({\qb}_1, q_2, g_3,\ldots,g_n)
=g_s^{n-2} (-1)^n
\sum_{k=3}^n\sum_{{\cal P}(4,\ldots,n)}
 (y\,  a_{\si(k+1)}..a_{\si(n)})_{ i_2 \ib_1}
 \\ & 
\times \{a_{\si(4)}\ldots a_{\si(k)} \}_{a_3y} \;
{\tilde m}_n({\qb}_1,g_{\si(n)},\ldots,g_{\si(k+1)},
q_2,g_{\si(k)},\ldots,g_{3}).
\end{split}
\ee
It is clear from the derivation that alternative color decompositions
are possible, e.g. by fixing positions of the quark $q$ and the gluon
$g_n$, instead of $g_1$ and $\bar{q}$.  In this case we obtain 
\ba
\label{eq_cdec2}
&&{\cal A}^{\rm tree}_n({\qb}_1, q_2, g_3,..,g_n)
=(-1)^n g_s^{n-2} 
\sum_{k=2}^{n-1}\sum_{{\cal P}(3,..,n-1)}
 (  a_{\si(3)}..a_{\si(k)}\,y  )_{ i_2 \ib_1}
 \\ 
&& 
\times \{ a_{\si(k+1)}..a_{\si(n-1)} \}_{ya_n} \;
{\tilde m}_n(\qb_1,
g_{\si(k)},..,g_{\si(3)},q_2,g_n,g_{\si(n-1)},..,g_{\si(k+1)}).
\nonumber 
\ea
We shall refer to the new fully ordered amplitudes ${\tilde m}$
defined in \Eq(\ref{eq_prim}), appearing in
\Eqs(\ref{eq_cdec1},\ref{eq_cdec2}), as tree left and right {\it primitive}
amplitudes. They are given by linear combinations of tree
color-ordered amplitudes, so that the usefulness of these objects may
not be immediately clear.  Nevertheless, the primitive amplitudes are
useful since they become basic objects in one-loop computations, as we
explain below.  Also, the primitive amplitudes can be computed in a
straightforward way using Berends-Giele recurrence relations and
color-stripped Feynman rules, see Sect.~\ref{bgampl}.

We can now explain why the color decomposition in terms of tree
primitive amplitudes appearing in \Eqs(\ref{eq_cdec1},\ref{eq_cdec2}))
is unitary. We begin by writing the full amplitude as a linear
combination of sub-amplitudes where all particles are ordered
\be
\begin{split}
\label{orderedqqgluon}
{\cal A}^{\rm 1-loop}_n&(\qb_1,q_2,g_3,\ldots,g_n)
=
\\
&
g_s^n \cg \sum_{ \sigma
  \in S_{n-2 }}\sum \limits_{k=3}^{n}
  A^{(1)}(\qb_1, g_{\si(3)},\ldots,g_{\si(k)}, q_2,g_{\si(k+1)},
\ldots,g_{\si(n)} ).
\end{split}
\ee
\begin{figure}[t]
\begin{center}
\includegraphics[scale=0.43]{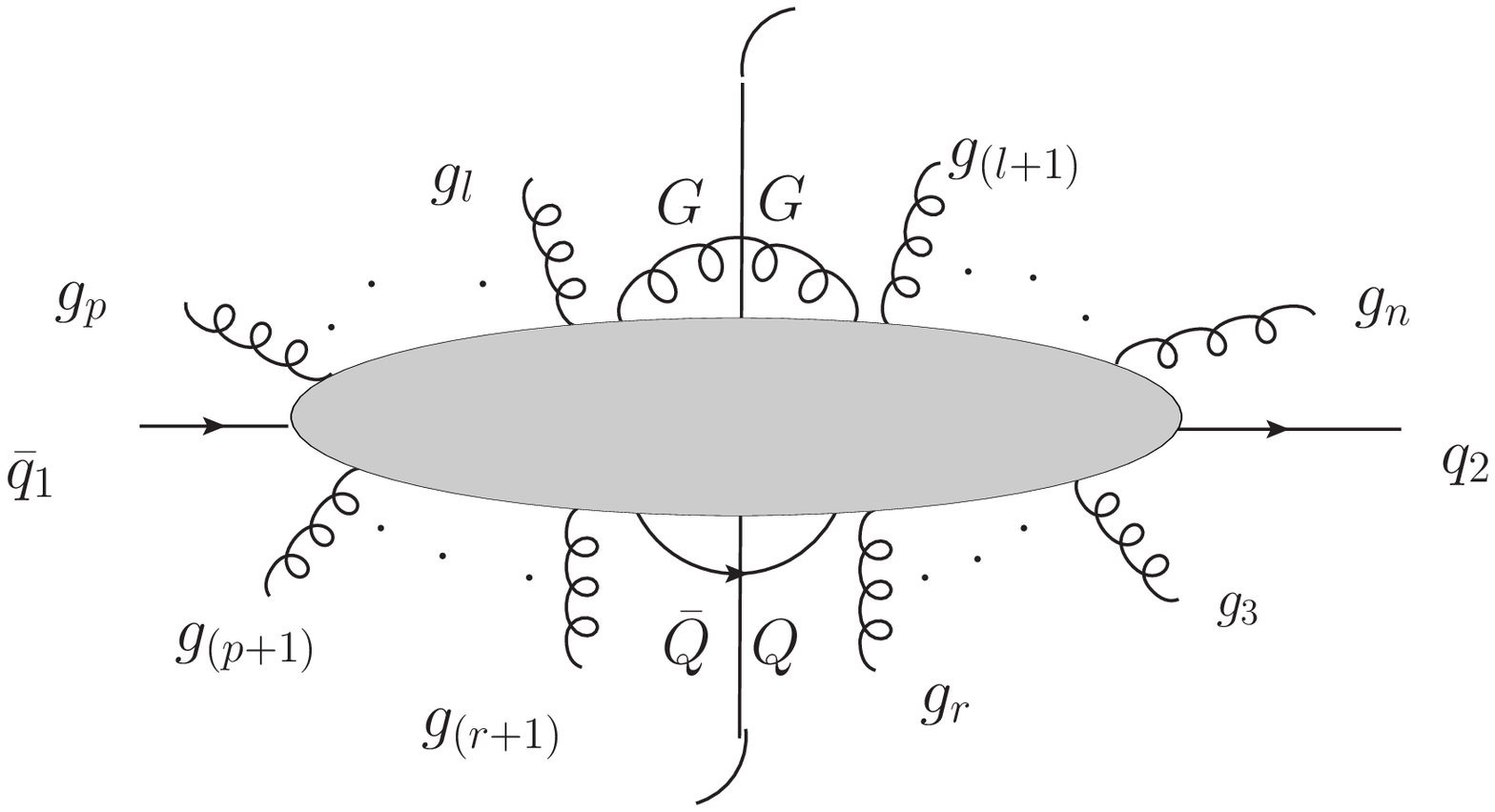}
\quad \qquad
\raisebox{-0.70cm}{
\includegraphics[scale=0.43]{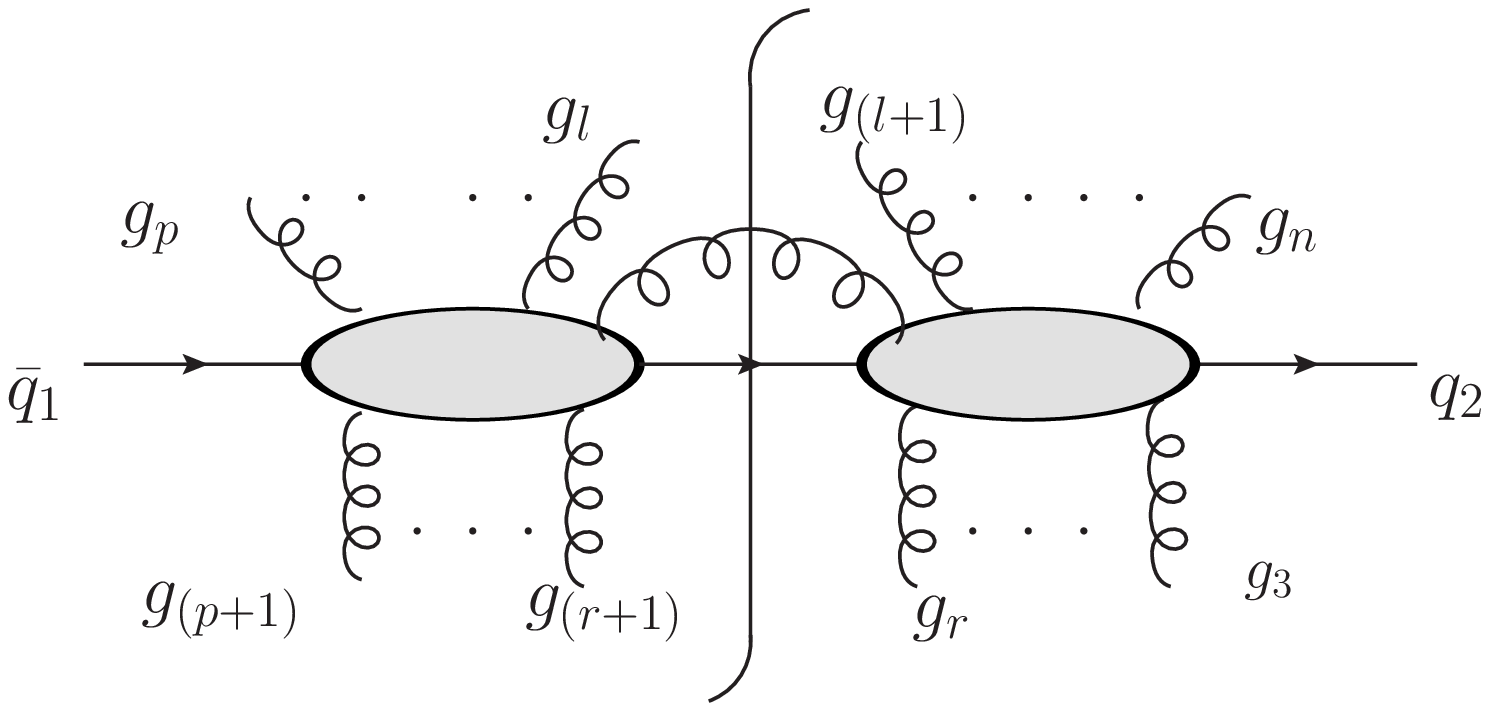}}
\quad \qquad 
\end{center}
\caption{Quark-gluon cut of the ordered colorless one-loop amplitude \Eq(\ref{cutqqglu}).}
\label{qqggcut_fig}
\end{figure}
Next, we consider the imaginary part of one of the ordered amplitudes $
A^{(1)}(\qb_1, g_p, . .  $ $. ,g_3, q_2,g_n,\ldots, g_{p+1} )$
obtained by cutting a quark propagator between outgoing gluons at
positions $r$ and $r+1$ with $r \le p$ and a gluon propagator between
gluons at positions $l$ and $l+1$ with $l\ge p+1$. The cut one-loop
ordered amplitude is given by products of tree amplitudes and we
choose the color representation where the positions of (anti-) quarks
and gluons are not subject to permutations.  Using the color bases of
\Eqs (\ref{eq_cdec2}) and (\ref{eq_cdec1}), and suppressing summation
over spin quantum numbers of the cut lines, we obtain the imaginary parts
\be
\label{cutqqglu}
\begin{split}
{\rm Im}_{(r,l)} &\left [ 
 A^{(1)}(\qb_1, g_p,..,g_{r+1}\cut{Q}{\Qb}
g_r,..,g_3, q_2,g_n,..,g_{l+1}\cut{G}{G}g_l,..,g_{p+1} )\right]
\\ & =
 \{a_{p}..a_{l}\}_{x_2a_G}
  (a_{r+1}.. a_{p+1}x_2)_{j~\ib_1}
(x_1a_3..a_{r})_{i_2~\bar j}
\{a_{l+1}.. a_{n}\}_{a_G x_1}
\\ & 
\times (-1)^{r+n-l}{\tilde m}(\Qb,g_r,\ldots,g_3,q_2,g_n,\ldots,g_{l+1}, G)
\\ &
\times (-1)^{l-r}{\tilde m}
(G,g_l,\ldots,g_{p},\qb_1,g_{p+1},\ldots,g_{r+1},Q) 
\\ &
= (-1)^n(x_1 a_{3}.. a_{p+1} x_2)_{i_2 \ib_1}\{ a_{p}.. a_n\}_{x_2x_1} 
\\ & \phantom{xxxxxxxxx}
\times {\rm Im }_{ (r,l)} \left [ 
m^{(1)}_n \left(\qb_1,g_p,..,g_3,q_2,g_n,..,g_{p+1}\right) 
\right ].
\end{split} 
\ee The cut is illustrated in Fig.~\ref{qqggcut_fig}.  Since the color
decomposition of an amplitude does not depend on whether it is cut or
not, from \Eq(\ref{cutqqglu}) we read off the following color
decomposition \cite{DelDuca:1999rs} of a one-loop $q \bar q + n$~gluon
amplitude
\be
\begin{split}
\label{qqbarng1loop}
&{\cal A}_{n}^{\rm 1-loop}(\qb_1,q_2,g_3,..,g_n)= 
g_s^n \cg \sum_{p=2}^n \sum_{\si \in S_{n-2}} (T^{x_2} T^{a_{\si_3}}..T^{a_{\si_p}} T^{x_1})_{i_2 \ib_1} 
 \\
&\times (F^{a_{\si_{p+1}}}..F^{a_{\si_n}})_{x_1 x_2} \;(-1)^n {\tilde m}^{(1)}_n(\qb_1,g_{\si(p)},..,g_{\si(3)},
q_2,g_{\si(n)},..,g_{\si(p+1)}). 
\end{split}
\ee
We note that for $p=2$ the factor $(T\cdots T)_{i_2 \ib_1}$
becomes $(T^{x_2}T^{x_1})_{i_2\ib_1}$ and for $p=n$ the factor
$(F\cdots F)_{x_1x_2}$ becomes $\delta_{x_1x_2}$.  We also note that
${\tilde m}^{(1)}_n$ is the left primitive amplitude introduced in
Ref.~\cite{Bern:1994fz}.
Although we showed that this color decomposition of a one-loop
amplitude for $\bar q q + n$~gluons is unitarity by considering only a
double cut through one quark and one gluon propagator, we find the
same color decomposition if we consider cuts of one-loop amplitudes
through two gluon lines or two quark lines.

\subsection{Amplitudes with multiple quark pairs}

As we explained in the previous subsections, it may be beneficial to
employ unitary color decomposition in one-loop computations.  Such
decomposition is known for a $n$-gluon and $\bar q q +$ $n$-gluon
scattering amplitudes, but not for scattering amplitudes that involve
multiple quark pairs. For this reason, amplitudes with multiple quark
pairs represent a special case.  Indeed, it is easy to understand how
such amplitudes can be written in terms of color-ordered amplitudes
and how primitive amplitudes can be constructed, but it is not
straightforward to connect the color-ordered and primitive amplitudes.
Below we explain how the relationship between color-ordered and
primitive amplitudes can be established considering a simple process $
0\to \qb q \Qb Q g $.

We begin by discussing the decomposition of full scattering amplitudes
into color-ordered amplitudes for $m$ non-identical quark pairs.  The
total number of quarks and anti-quarks is $n = 2m$.  Because quark
lines are connected by gluon lines, we can write a color factor
associated with a particular quark line $m_1$ as a matrix element of a
product of certain number of $T$-matrices, taken between quark and
anti-quark color states
$(T^{a_1}T^{a_2}\ldots T^{ a_r})_{i_{2m_1} \ib_{2m_1-1}}$. 
The adjoint labels $(a_1,\ldots, a_r)$ are internal indices; they are
contracted with similar indices carried by other quark lines or with
the adjoint indices that enter three- and four-gluon vertices.  All
these contractions can be turned to products of Kronecker
delta-symbols with repeated use of the identity
\be
(T^a)_{i_2 \ib_1} (T^a)_{i_4 \ib_3}=\delta_{i_2 \ib_3}
\delta_{i_4 \ib_1} 
-\frac{1}{N_c}\delta_{i_2 \ib_1} \delta_{i_4 \ib_3} \, .
\label{eqk}
\ee
It follows from \Eq(\ref{eqk}) that, to leading order in $N_c$, the
color flow of anti-quarks is flipped. The sub-leading term in $N_c$
ensures that the decomposition respects the tracelessness of $T$-matrices.
Because two quark lines can only be connected by one gluon line, we
conclude that a convenient basis for the color decomposition is given by
the set of Kronecker deltas
\be
C(\sigma,r_\sigma)=\left \{ \frac{(-1)^{r_\sigma}}{N^{r_\sigma}_c}
\delta_{i_2 \bar  i_{(2\sigma(1)-1)}}
\delta_{i_4 \ib_{(2\sigma(2)-1)}}
\ldots \delta_{i_{2m} \ib_{2\sigma(m)-1}}
\right \},
\label{quark_bas}
\ee
where $\sigma$ denotes permutations of $m$ anti-quark indices and
$r_\sigma$ is the rank of the permutation $\sigma$. The rank is
computed by counting how many times the equation $\sigma(k)=k$,
$k=1..m$, is satisfied for a given permutation $\sigma$.  If
$\{\sigma(k)\}=\{k\} $ for all $k$, $r_\sigma=m-1$.  At one loop the
basis remains the same, except for an overall factor of $N_c$.
Finally, we note that we can describe scattering amplitudes with
$m$-quark pairs and $n-2m$ gluons by replacing Kronecker delta-symbols
in \Eq(\ref{quark_bas}) with products of $T$ matrices. This leads to a
new basis set
\be
\label{eq652}
\begin{split}
\vspace*{-0.5cm} C(\sigma,r_\sigma,\{n_i\}) =
\left \{ \frac{(-1)^r}{N^{r_\sigma}_c}
( T^{a_1 :a_{n_1}})_{i_2 \bar  i_{\sigma_1}}
\ldots
( T^{a_{n_{m-1}} :a_{n_m}})_{i_{2m} \bar  i_{\sigma_m}}
\right   \},
\end{split}
\ee
where we use the notation $\sigma_k = 2\sigma(k)-1$ and $T^{a_{n_1}
  :a_{n_2}} = T^{a_{n_1}}....T^{a_{n_2}}$. In addition, the set
$\{n_i\}$, $i=1,...,m$, contains all possible partitions of $n-2m$
gluons into $m$ subsets.  We do not pursue the general
color-decomposition discussion in what follows and turn, instead, to
an example.

As clearly follows from the above discussion, the color-decomposition
of amplitudes with multiple fermion pairs can be performed in a
straightforward way.  However, within the unitarity framework we
compute primitive, rather than color-ordered, amplitudes. We can use
color-stripped Feynman rules and the fact that {\it all} particles are
ordered in a given primitive amplitude, to construct the primitive
amplitudes directly.  The non-trivial step is to connect the primitive
amplitudes, so constructed, with color-ordered or full amplitudes.  We
will explain how to do that by considering the process
$
 0\to \qb q \Qb Q  g. 
$ 
Applying the general results discussed above, 
we find  the color decomposition for tree and one-loop amplitudes
\be
\begin{split}
\B^\tree (\qb_1,q_2,\Qb_3,&Q_4,g_5) =  g_s^3 \Bigg[
(T^{a_5})_{i_4 \ib_1} \delta_{i_2 \ib_3} B^\tree_{5;1}
 +\frac{1}{N_c}
(T^{a_5})_{i_2 \ib_1} \delta_{i_4 \ib_3} 
B^\tree _{5;2}  \\
&+ (T^{a_5})_{i_2 \ib_3} \delta_{i_4 \ib_1} 
B^\tree _{5;3}
    +\frac{1}{N_c} (T^{a_5})_{i_4 \ib_3} 
\delta_{i_2 \ib_1}B^\tree _{5;4}\Bigg],
\end{split}
\ee
\be
\label{B.oneloop}
\begin{split}
\B^\oneloop (&\qb_1,q_2,\Qb_3,Q_4,g_5) =   g_s^5 \Bigg[
N_c (T^{a_5})_{i_4 \ib_1} \delta_{i_2 \ib_3} B_{5;1}
 +(T^{a_5})_{i_2 \ib_1} \delta_{i_4 \ib_3} 
B_{5;2} 
\\
& +N_c (T^{a_5})_{i_2 \ib_3} \delta_{i_4 \ib_1} 
B_{5;3}
    +(T^{a_5})_{i_4 \ib_3} 
\delta_{i_2 \ib_1}B_{5;4}\Bigg]
=\sum_{i=1}^4C_iB_{5;i}\,,
\end{split}
\ee
where we introduced the notation $C_i$ for the elements of the color
basis.  Each of these one-loop color-ordered amplitudes can be written
as a sum of two terms
\be
B_{5;i} = B^{[1]}_{5;i}+ \frac{n_f}{N_c} B^{[1/2]}_{5;i},\qquad i=1,2,3,4,\ 
\ee
to separate diagrams with a closed fermion loop from the other ones.
The amplitudes $B^{[1]}_{5;i}$ and $B^{[1/2]}_{5;i}$ can be expressed
through linear combinations of primitive amplitudes.  We will
explicitly show how to do that for amplitudes with a closed fermion
loop $B^{[1/2]}_{5;i}$.
\begin{figure}[t]
\begin{center}
\includegraphics[scale=0.43]{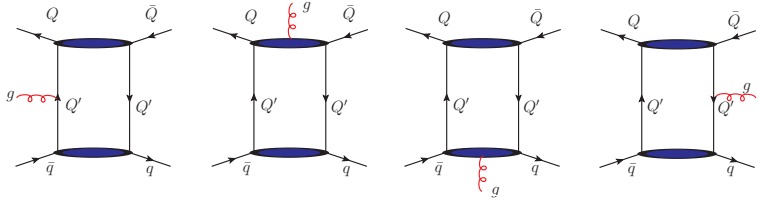}
\quad \qquad
\end{center}
\caption{Prototypes of four parent diagrams that define the primitive
  amplitudes for the $n_f$ part of the $0 \to \qb q \Qb Q g$
  amplitude.  The solid blobs denote ``dummy lines'' described in the
  text.}
\label{Fig6.3.1}
\end{figure}

Before proceeding with this discussion, we explain the general
strategy.  Since particles in each primitive amplitude are ordered,
primitive amplitudes can be characterized by ``parent diagrams''.  A
parent diagram for a primitive amplitude is the diagram with the
ordering of external particles consistent with the ordering of
external particles in the primitive amplitude and the maximal number
of propagators that depend on the loop momenta.  The primitive
amplitudes can be constructed from the parent diagrams by pinching
internal propagators and resolving illegitimate vertices by creating
propagators that do not depend on the loop momentum.  Since pinching
and pulling does not change the ordering of external particles, every
diagram obtained by this procedure contributes to the same primitive
amplitude.  Therefore, if we take a Feynman diagram, we can study its
color factor to find the color-ordered amplitudes that it contributes
to. On the other hand, since the ``pinching and pulling'' technique
connects primitive amplitudes with the color-stripped Feynman
diagrams, we can find primitive amplitudes that receive contributions
from a particular Feynman diagram.  Because color-ordered amplitudes
are given by linear combinations of primitive amplitudes with
coefficients that only depend on the number of colors $N_c$, it is
sufficient to analyze a number of Feynman diagrams to establish the
connection uniquely.

\begin{figure}[t]
\begin{center}
\includegraphics[scale=0.43]{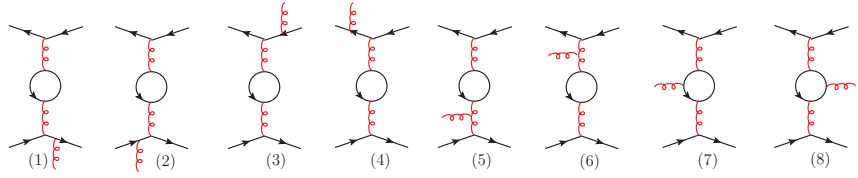}
\quad \qquad 
\end{center}
\caption{Eight diagrams that contribute 
to the  $n_f$ part of the $0 \to \qb q\Qb Q g$ amplitude.}
\label{Fig6.3.3}
\end{figure}

We now illustrate this general strategy by considering amplitudes with
a closed fermion loop $B^{[1/2]}_{5;i}$.  Since the maximal number of
propagators in the loops that contribute to $0 \to q \bar q Q \Qb
g$ is five but the fermion loop contribution has at most three
propagators, we find it to be convenient to introduce dummy lines, to
get a uniform graphical representation of the primitive amplitudes. A
dummy line represents a propagator that is not part of the loop.
Contributions obtained by cutting dummy lines vanish.  There are four
primitive amplitudes $A^{[1/2]}_{i}$, that are obtained if a gluon is
inserted in different ways between four quark lines.  They are shown
in Fig.~\ref{Fig6.3.1}. We introduce the notation \be
\begin{split}
& A^{[1/2]}_{1} = A^{[1/2]}_L(\qb_1 , g_5,Q_4,{\Qb}_3,q_2),
\;\;\;\;A^{[1/2]}_{2} = A^{[1/2]}_L(\qb_1 ,Q_4,\Qb_3,q_2,g_5), 
 \\
& A^{[1/2]}_{3} = A^{[1/2]}_L(\qb_1, Q_4,\Qb_3,g_5,q_2),
\;\;\;\;\;
A^{[1/2]}_{4} = A^{[1/2]}_L(\qb_1, Q_4,g_5,\Qb_3,q_2).
\end{split} 
\ee
Because of Furry's theorem, we have only three independent
primitive amplitudes
\be
\label{prim.decoupling}
\sum_{i=1}^{4} A^{[1/2]}_{i}=0.
\ee 
In general, the color-ordered amplitudes are given by linear combinations 
of primitive amplitudes 
\be
B^{[1/2]}_{5;i} = \sum \limits_{j=1}^{4} x_{ij} A^{[1/2]}_{j}.
\label{eq_x}
\ee
If the primitive amplitudes are represented by diagrams with dummy
lines, then the same primitive amplitude can have more than one
equivalent parent Feynman-diagram. They have the same ordering and can
be transformed into each other by pulling and pinching propagators as
explained above.  In our case we have eight Feynman diagrams shown in
Fig.~\ref{Fig6.3.3}.  Each diagram is factorized into a color factor
and a color-stripped part.  The colorless diagrams with all vertices
oriented clockwise are obtained by setting the color matrices for
quarks and gluons to $-1$.  The relations between the color-ordered
amplitudes and the primitive amplitudes are established by expanding
the color factors of the parent diagrams in the color basis of
\Eq(\ref{B.oneloop}).  Since we have only three independent primitive
amplitudes we need to choose three independent parent diagrams and
trace how they are mapped onto color-ordered amplitudes. For the sake
of definiteness, we choose the first three diagrams of
Fig.~\ref{Fig6.3.3} and write 
\be
\label{B.dia} 
\B^{\rm \oneloop}_{n_f}|
= {N_c^{-1}}\sum_{i=1}^4 C_iB^{[1/2]}_{5;i} 
  = \sum_{i=1}^3 {\rm Col}_i D_i + \ldots,
\ee
where the ellipsis stands for contributions of diagrams four to eight. 
The decomposition of   the color factors  of these diagrams 
into the color basis reads 
\be
\begin{split}
\label{ColiCj}
{\rm Col}_1&=(T^aT^x)_{i_2 \ib_1}(T^x)_{i_4 \ib_3}=N_c^{-1}(C_3-C_2),\\
{\rm Col}_2&=(T^xT^a)_{i_2 \ib_1}(T^x)_{i_4 \ib_3}=N_c^{-1}(C_1-C_2),\\
{\rm Col}_3&=(T^xT^a)_{i_4 \ib_3}(T^x)_{i_2 \ib_1}=N_c^{-1}(C_3-C_4).
\end{split}
\ee
The primitive amplitudes are given by linear combinations of
colorless ordered parent diagrams.  Using color-stripped Feynman 
rules, we find 
\be
A^{[1/2]}_{1}  = - D_2+ \ldots,\;\;\;\
A^{[1/2]}_{2} = D_1 + D_2\ldots,\;\;\;
A^{[1/2]}_{3} = -D_1 - D_3\ldots,
\ee
where ellipsis stand for the contributions of diagrams four to eight.
We derive the inverse relations 
\be
\label{Di.parent}
D_2=-A^{[1/2]}_{1} ,\quad
D_1= A^{[1/2]}_{1}+A^{[1/2]}_{2},\quad
D_3=-A^{[1/2]}_{1}-A^{[1/2]}_{2} - A^{[1/2]}_{3},\\
\ee
and 
use \Eqs~(\ref{ColiCj}), (\ref{Di.parent}) and
(\ref{prim.decoupling}) to get 
\be
\label{B.primitive} 
\B^{\rm \oneloop}_{n_f}
=N_c^{-1}\sum_{i=1}^4 C_iB^{[1/2]}_{5;i}
=-N_c^{-1}\sum_{i=1}^4 C_i A^{[1/2]}_{i}.
\ee
Comparing \Eq(\ref{B.primitive}) 
with \Eq(\ref{eq_x}), 
we conclude that $x_{ij}=-\delta_{ij}$.
Explicitly, the relation between 
color-ordered and primitive amplitudes reads
\be
\begin{split}
& B^{[1/2]}_{5;1} = -A^{[1/2]}_L(\qb_1 , g_5,Q_4,{\Qb}_3,q_2), \\
& B^{[1/2]}_{5;2} = -A^{[1/2]}_L(\qb_1 ,Q_4,\Qb_3,q_2,g_5), \\
& B^{[1/2]}_{5;3} = -A^{[1/2]}_L(\qb_1, Q_4,\Qb_3,g_5,q_2), \\
& B^{[1/2]}_{5;4} = -A^{[1/2]}_L(\qb_1, Q_4,g_5,\Qb_3,q_2). 
\end{split}
\ee 
We note that our derivation relies on a particular choice of
parent diagrams for primitive amplitudes but it is straightforward to
check that the final result is independent of this choice.

\begin{figure}[t]
\begin{center}
\includegraphics[scale=0.43]{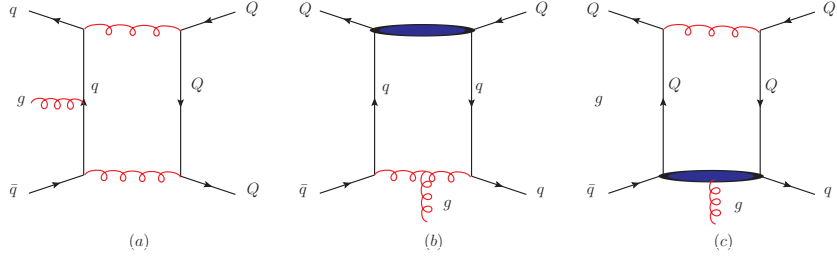}
\quad \qquad 
\end{center}
\caption{Prototype of parent diagrams for primitive amplitudes with
  four quarks and one gluon. The gluons can be inserted in four
  possible ways into the prototype diagrams, leading to four primitive
  amplitudes. The solid blobs denote dummy lines described this
  section.}
\label{Fig6.3.2}
\end{figure}

The $n_f$-independent part of the amplitudes $B^{[1]}_{5;i}$ have
contributions of three classes of primitive amplitudes as indicated in
Fig.~\ref{Fig6.3.2}.
Following the same procedure as in the case of the $n_f$-dependent
part, we can work out relations between the color-ordered amplitudes
for each class and the corresponding primitive amplitudes. Such
relations can be found in Ref.~\cite{Ellis:2008qc} and we do not repeat
them here.

Finally, we note that the decomposition of color-ordered amplitudes in
terms of primitive amplitudes becomes more and more cumbersome as the
number of quark pairs increases.  Partially motivated by this
difficulty a different approach was suggested in
Ref.~\cite{Giele:2009ui}. The starting point is the observation that,
to implement unitarity, complete ordering is not necessary.  We only
need to consider all partitions of the external particles into subsets
of particles that can be connected to cut propagators in the loop. The
particles within a given subset are not ordered.  In addition, the
color degrees of freedom are treated at the same footing as other
quantum numbers.  This treatment is facilitated by the use of the
color-flow representation and the color-dressed Berends-Giele
recursion relations \cite{Gleisberg:2008fv}.  By considering gluon
scattering amplitudes it was shown in Ref.~\cite{Giele:2009ui} that
scaling of the computer time and the stability of the numerical
evaluation is competitive with the implementation based on ordered
approach.  To understand merits of this implementation, further work
is required.

\subsection{Singular behavior of one-loop amplitudes}

\label{sec:singular}
In this Section we discuss the singular behavior of one-loop amplitudes  
\cite{Kunszt:1994mc,Catani:1998bh}.
Although this topic is interesting in its own right, the reason we present it 
here  is a particular representation of color algebra employed in Ref.~\cite{Catani:1998bh}
to present singular limits of one-loop amplitudes in a simple way.
This representation of the color
algebra has the advantage that it treats the color degrees of freedom of 
quarks and gluons on the same footing.
Below we explain how to connect this representation 
with more conventional treatment of color degrees of freedom presented above.


The infrared singular behavior of one-loop amplitudes is completely
known, specifically in the case when the singularities are regularized
by dimensional regularization.  Although the singular terms do not in
the end contribute to physical answers, they provide an important
check on the correctness of intermediate results. This check can be performed
both when setting up an analytic formula, and also when performing
numerical evaluation of the answer, to control the numerical
stability, see Section~\ref{sec:checks}.

The results for the infrared-singular behaviour of on-shell QCD
amplitudes at one- and two-loop orders with massless particles have been given
in Ref.~\cite{Catani:1998bh}.  The generalization to the case when
massive particles are present is given at one-loop order in
Ref.~\cite{Catani:2000ef}. In this Section we shall consider only the
case where there are no massive partons and refer the reader to
Ref.~\cite{Catani:2000ef} for more the complicated case.  The results
of Ref.~\cite{Catani:1998bh} are presented using a color space
notation, which has the advantage that it is very compact and 
can deal with both quarks
and gluons in a seamless way. However, this notation  is  
not particularly well-known and, in what  follows,  we give several
examples to elucidate the notation and connect it to the discussion 
of color given earlier in this Section. 

We begin by writing the perturbative expansion 
of  scattering amplitudes following 
Ref.~\cite{Catani:1998bh}
\beq \label{Amplitudeexpansion}
| \cm\ra =
g_s^{r} \left ( 
| \cm^{(0)}(\{p,\lambda\})\ra
+g_s^2 \cg 
| \cm^{(1)}(\mu^2;\{p,\lambda\})\ra+...\right  ),
\eeq
where $r$ is the overall power of the (unrenormalized) 
strong coupling constant $g_s$ and $\{p,\lambda\}$ is 
a set of momenta and helicities of the external particles.  
The amplitude is written  using 
a bra and ket notation, to indicate that it is a vector in color space.
The one-loop subamplitude $| \cm^{(1)}(\mu^2; \{p\}) \ra $ has both double
and single poles in $1/\ep$. Catani's result \cite{Catani:1998bh}
is that  these singularities can be obtained from the tree amplitudes by 
operating with the color-space operator ${\bom I}^{(1)}$,
\beq
\label{ff1loop}
| \cm^{(1)}(\mu^2;\{p,\lambda\}) \ra
= {\bom I}^{(1)}(\ep,\mu^2;\{p\}) 
\; | \cm^{(0)}(\{p,\lambda\}) \ra
+ O(\epsilon^0)\,.
\eeq
The operator ${\bom I}^{(1)}$ has the following explicit expression 
in terms of color charges of the external particles 
\beq
\label{Catani}
{\bom I}^{(1)}(\ep,\mu^2;\{p\}) =  
\frac{r}{2} \frac{\beta_0}{\ep} + 
\sum_i \;\frac{1}{{\bom T}_{i}^2} \;{\cal
V}_i^{{\rm sing}}(\ep)
\; \sum_{j \neq i} {\bom T}_i \cdot {\bom T}_j
\; \left( \frac{\mu^2}{-s_{ij}-i0} \right)^{\ep} \;\;,
\eeq
where $s_{ij}=2 p_i \cdot p_j$, 
$\beta_0 = 11/3 C_A - 2/3 T_R N_f$ is the first coefficient in the expansion 
of the QCD $\beta$-function.
The singular 
function ${\cal V}_i^{{\rm sing}}(\ep)$ depends only on the 
parton type $i$ and is given by
\beq
\label{calvexp}
{\cal V}_i^{{\rm sing}}(\ep) =  \frac{{\bom T}_{i}^2}{\ep^2} 
+  \;\frac{\gamma_i}{\ep}.
\eeq
Squares of the color-charge operators ${\bom T}_{i}^2$
and coefficients $\gamma_i$ for quarks and gluons  are
\beqn
{\bom T}_q^2 = {\bom T}_{\qb}^2 = C_F, 
\;\;\;\;\;\;&&{\bom T}_g^2 = C_A, \nonumber \\
\gamma_q = \gamma_{\qb} = \frac{3}{2} \,C_F, 
\;\;\;\;\;\;&&\gamma_g = \frac{\beta_0}{2}.
\eeqn

We note that in \Eq(\ref{ff1loop}) the double poles in $\ep$ are 
factorized completely. If we expand \Eq(\ref{Catani}) in powers of $\ep$ 
and then use the color
conservation relation, $\sum_{j \neq i} {\bom T}_j = - {\bom T}_i$, we 
obtain the result
\beq
\label{ieeexp}
{\bom I}^{(1)}(\ep,\mu^2;\{p\}) =  
\sum_i \frac{1}{\ep^2} \; \sum_{j \neq i} {\bom T}_i \cdot {\bom T}_j
+ {\cal O}\left(\frac{1}{\ep}\right) = - \frac{1}{\ep^2} \sum_i {\bom T}_i^2
+ {\cal O}\left(\frac{1}{\ep}\right),
\eeq
that explicitly shows the absence of color correlations at ${\cal O}(1/\ep^2)$.
Note however that the single $1/\ep$ poles have color correlations. 
We will explicitly show how to compute those by considering a few examples 
below.

\subsubsection{Divergence structure of the one-loop $\qb q g g$ amplitude}
As a first example we compute the singular behavior of the one-loop $\qb
q gg $ amplitude, which illustrates how color operators act on both
quark and gluon fields. Here we shall present the singularity structure
of primitive amplitudes, since primitive amplitudes are the building
blocks in one-loop calculations.  To arrive at the divergent structure
of the primitive amplitudes, we will write the divergences of the
one-loop $\qb q g g$ amplitude using the color basis shown in
\Eq(\ref{qqbarng1loop}).  We begin by acting with the color vector on
the tree amplitude. We recover the regular color-ordered amplitudes for a
particular choice of quark and antiquark color indices, 
$i_2,\ib_1$ and gluon indices, $a_3,a_4$ \be
\begin{split}
 | \cm^{(0)} \rangle & = | \cm^{(0)}_{34} \rangle
+ |\cm^{(0)}_{43} \rangle ,
\\
  \la \ib_1 i_2 a_3 a_4 | \cm^{(0)}_{34} \ra  & =
(T^{a_3} T^{a_4})_{i_2 {\ib}_1} m_4(\qb_1,q_2,g_3,g_4),
\end{split}
\eeq
where $m_4(\qb_1,q_2,g_3,g_4)$ is the tree color-stripped amplitude.
The tree-graph amplitude for $\qb q g g$ scattering is proportional to
the second power of the strong coupling constant, so that 
$r$ as defined in \Eq(\ref{Amplitudeexpansion}) is equal to two.
As a result the contributions of $\beta_0$ and $\gamma_g$ in
\Eq(\ref{Catani}) cancel. This leads to the following structure for
the infrared poles 
\be
\begin{split}
\la \ib_1 i_2 a_3 a_4 | \cm^{(1)}\ra = &
\sum_k \sum_{n \neq k} 
\left(\frac{1}{\eps^2}+\frac{1}{\eps}\left(
\frac{\gamma_q}{T_q^2} \left( \delta_{k1} +\delta_{k2} \right )
+ L_{kn}\right)\right) 
\\
&
\times \la \ib_1 i_2 a_3 a_4 |
{\bom T}_k \cdot {\bom T}_n
| \cm^{(0)} \rangle\,, 
\end{split}
\label{eq:cataniexp}
\ee
where we have introduced the notation 
$L_{kn} = \ln \left( \mu^2 /(-s_{kn}-i0)\right)$. 
To calculate the matrix element of the product of two color charge 
operators between a state of definite color 
and the amplitude $| \cm^{(0)} \rangle$, we insert a complete 
set of color states 
\be
\begin{split}
& \la \ib_1 i_2 a_3 a_4 |{\bom T}_k \cdot {\bom T}_n | \cm^{(0)} \rangle =
\sum \limits_{{\cal P}(3,4)}^{}
\la \ib_1 i_2 a_3 a_4 |{\bom T}_k \cdot {\bom T}_n | 
\cm^{(0)}_{\sigma(3),\sigma(4)} \rangle, 
\\
& 
\la \ib_1 i_2 a_3 a_4 |{\bom T}_k \cdot {\bom T}_n | \cm^{(0)}_{34} \rangle  =
\sum \limits_{\kappa }^{}
\la \ib_1 i_2 a_3 a_4 |{\bom T}_k \cdot {\bom T}_n | \kappa \ra
\la  \kappa | \cm^{(0)}_{34} \rangle, 
\end{split}
\ee
where we use the notation $|\kappa \ra = |\jb_1 j_2 b_3 b_4\ra$.
We can evaluate the matrix elements of the color charge operators 
using the following matrix elements 
$\la i_k |{\bom T}_k^x|j_k \ra = 
T^x_{i_k j_k}/\sqrt{2}$  if $k$ is a quark, 
$\la {\ib}_k |{\bom T}_k^x| {\jb}_k \ra = 
-T^x_{{\jb}_k {\ib}_k}/\sqrt{2}$  if $k$ is an-antiquark,
and  $\la a_k |{\bom T}_k^x|b_k \ra = 
F^x_{a_k b_k}/\sqrt{2}$  if $k$ is a gluon. Also, we note that 
the color-charge 
operator ${\bom T}_k$  only acts on the color index of a parton $k$.

We now write down the color correlation matrix elements 
for $q \qb$, $gg$, $ q g$ and $\qb g$ and 
bring those color structures 
to a form consistent with \Eq(\ref{qqbarng1loop}). For 
the $\qb q gg$ case, this implies that the 
result should be  expressed through the following color 
structures 
\be
\begin{split}
&\left ( T^{x_1} T^{x_2} \right )_{i_2 {\ib}_1} (F^{a_4} F^{a_1})_{x_1 x_2},
\;\;\;
\left ( T^{x_1}T^{a_3}T^{x_2} \right )_{i_2 {\ib}_1} 
\left ( F^{a_4} \right )_{x_1 x_2}, 
\\
& ~~~~~~~~~~~~~~~~~~~~~~~~~~~~~~~
\left ( T^{x_1} T^{a_3} T^{a_4} T^{x_2}\right )_{i_2 \ib_1} \delta_{x_1 x_2},
\end{split} 
\label{color_basis}
\ee
where the summation over repeated indices is assumed and 
$3 \leftrightarrow 4$ permutations are not shown.

We begin with the $q g_3$ case. 
The matrix element reads 
\beq
\la \ib_1 i_2 a_3 a_4 |{\bom T}_2 \cdot {\bom T}_3 | \jb_1 j_2 b_3 b_4\ra = 
\frac{1}{2} (T^x)_{i_2 j_2} (F^x)_{a_3 b_3} 
\delta_{a_4 b_4} \delta_{\ib_1 \jb_1}. 
\eeq
This implies 
\be
 \la \ib_1 i_2 a_3 a_4 |{\bom T}_2 \cdot {\bom T}_3 | \cm^{(0)}_{34} \ra
= \frac{1}{2} (F^x)_{a_3 b_3}  
\left ( T^x T^{b_3} T^{a_4}  \right )_{i_2 {\ib}_1 } m_4(\qb_1,q_2,g_3,g_4).
\ee
The color factor in the above equation is not yet in the right form
(c.f.  \Eq(\ref{color_basis})). We therefore commute 
$T^{b_3}$ to the right using $[T^{a},T^{b}]=-F^a_{by}T^y$ and obtain 
\be
F^x_{a_3 b_3}  
\left ( T^x T^{b_3} T^{a_4}  \right )_{i_2 {\ib}_1 }
= 
-F^{a_3}_{x b_3} (T^x T^{a_4} T^{b_3})_{i_2 {\ib}_1} 
   - (F^{a_3} F^{a_4})_{x y}   (T^x T^y)_{i_2 {\ib}_1}.
\ee
This leads to 
\be
\begin{split}
&  \la \ib_1 i_2 a_3 a_4 |{\bom T}_2 \cdot {\bom T}_3 | \cm^{(0)}_{34}\rangle = 
-\frac{1}{2}
\Big [(T^{x}T^{a_4}T^y)_{i_2 {\ib}_1} (F^{a_3})_{xy} 
\\
& + (T^{x}T^y)_{i_2 {\ib}_1} (F^{a_3} F^{a_4})_{xy} \Big ]
m_4(\qb_1,q_2,g_3,g_4). 
\end{split}
\ee
On the other hand, 
considering the color correlation $q g_4$
we find 
\beq
\la \ib_1 i_2 a_3 a_4 |{\bom T}_2 \cdot {\bom T}_4 | \jb_1 j_2 b_3 b_4\ra = 
\frac{1}{2} (T^x)_{i_2 j_2} (F^x)_{a_4 b_4} 
\delta_{a_3 b_3} \delta_{\ib_1 \jb_1},
\eeq
so that color factors are  already in the right form
\be
 \la \ib_1 i_2 a_3 a_4 |{\bom T}_2 \cdot {\bom T}_4 | \cm^{(0)}_{34} \ra
= -\frac{1}{2} (F^{a_4})_{x b_4}  
\left ( T^x T^{a_3} T^{b_4}  \right )_{i_2 {\ib}_1 } m_4(\qb_1,q_2,g_3,g_4).
\ee
Similarly for the case ${\bom T}_2 \cdot {\bom T}_1 $ we have 
\beq
\la \ib_1 i_2 a_3 a_4 |{\bom T}_2 \cdot {\bom T}_1 | \jb_1 j_2 b_3 b_4\ra = 
-\frac{1}{2} (T^x)_{i_2 j_2} 
\delta_{a_3 b_3} \delta_{a_4 b_4}  (T^x)_{\jb_1 \ib_1}, 
\eeq
so that 
\be
  \la \ib_1 i_2 a_3 a_4 |{\bom T}_2 \cdot {\bom T}_1 | 
\cm^{(0)}_{34} \ra
= -\frac{1}{2} 
\left ( T^x T^{a_3} T^{a_4}T^{x} \right )_{i_2 {\ib}_1} m_4(\qb_1,q_2,g_3,g_4). 
\ee
Evaluating the remaining matrix elements we obtain 
\be 
\begin{split}
&  \la \ib_1 i_2 a_3 a_4 |{\bom T}_3 \cdot {\bom T}_4 | \cm^{(0)}_{34}\rangle = 
-\frac{1}{2} \left ( T^{x}T^y \right )_{i_2 {\ib}_1} \left (
F^{a_3} F^{a_4} \right )_{xy}  m_4(\qb_1,q_2,g_3,g_4),  \\ 
&  \la \ib_1 i_2 a_3 a_4 |{\bom T}_3 \cdot {\bom T}_1 | \cm^{(0)}_{34}\rangle = 
-\frac{1}{2}
\left ( T^{x}T^{a_4}T^{y} \right )_{i_2 {\ib}_1} \left ( F^{a_3} \right )_{xy}  
m_4(\qb_1,q_2,g_3,g_4),  \\ 
&  \la \ib_1 i_2 a_3 a_4 |{\bom T}_4 \cdot {\bom T}_1 | \cm^{(0)}_{34}\rangle = 
-\frac{1}{2} \Big [(T^{x}T^{a_3}T^{y})_{i_2 {\ib}_1} 
(F^{a_4})_{xy} 
\\
& 
\;\;\;\;\;\;\;\;\;\;\;\;\;\;
+ (T^{x}T^y)_{i_2 {\ib}_1} (F^{a_3} F^{a_4})_{xy} \Big ] 
m_4(\qb_1,q_2,g_3,g_4).
\label{eqs:titjqqgg}
\end{split}
\ee
Finally, putting everything together we find 
\be
\label{eq_qqgg_div}
\begin{split}
&  \la i_2 a_3a_4\ib_1 | \cm^{(1)} \rangle = - \sum_{P(3,4)} 
m_4(\qb_1,q_2,g_3,g_4)\;\bigg[ \\
&
 \left(T^{x_{1}} T^{x_{2}}\right)_{i_2 {\ib}_1} 
\left(F^{a_3} F^{a_4} \right)_{x_{1} x_{2}}
\left(\frac{3}{\eps^2} +
\frac{1}{\eps}
\left(
\frac{3}{2}
+\sum_{i=2,4} L_{i,i+1}
\right)
\right)
 \\  
& + \left(T^{x_{1}} T^{a_4} T^{x_{2}}\right)_{i_2 {\ib}_1} 
\left(F^{a_3}\right)_{x_{1} x_{2}}
\left(\frac{2}{\eps^2} +
\frac{1}{\eps}
\left(
\frac{3}{2}+L_{23}+L_{13}
\right)
\right)
\\  
& + \left(T^{x_{1}} T^{a_3} T^{x_{2}}\right)_{i_2 {\ib}_1} 
\left(F^{a_4}\right)_{x_{1} x_{2}}
\left(\frac{2}{\eps^2} +
\frac{1}{\eps}
\left(
\frac{3}{2}+L_{13}+L_{14}
\right)
\right)
 \\  
& + \left(T^{x_{1}} T^{a_3} T^{a_4} T^{x_{1}}\right)_{i_2 {\ib}_1} 
\left(\frac{1}{\eps^2} +
\frac{1}{\eps}
\left(\frac{3}{2}+L_{12}
\right)
\right) \bigg]
+  {\cal O}(\eps^0)\,.  
\end{split}
\ee 
It follows from \Eq(\ref{qqbarng1loop}) that each of the color
structures in \Eq(\ref{eq_qqgg_div}) is uniquely associated with
the primitive amplitudes for the $ \qb q g g$ process. In particular,
the structure of the poles of primitive amplitudes that appears in
\Eq(\ref{eq_qqgg_div}) is remarkably simple: the coefficient of the
double pole is equal to the number of gluon propagators in the parent
diagram of the given primitive. The single pole contains, beyond 
the terms $\gamma_q/C_F$ associated with the quark line, only logarithms
of scalar products of pairs of parton momenta, where the relevant pairs of momenta
are connected by gluon propagators in the parent diagram. The generalization 
of this result to the case of primitive amplitudes for the 
$\qb q$ $+(n-2)$-gluon case reads 
\begin{equation}
\begin{split}
\hspace*{-0.2cm}
& {\tilde m}_n^{(1)}({\qb}_n,g_{k+1},...,g_{n-1},
q_2,g_3,...g_k)  ={\tilde m}_n({\qb}_n,g_{k+1},...,g_{n-1},
q_2,g_3,...g_k) 
\\
&\;\;\;\;\;\;\;\;\;\;\;\;\;\;\;  \times \left [ 
 -\frac{k}{\ep^2} - \frac{1}{\ep} \left ( \frac{3}{2} 
+  \sum \limits_{i=1}^{k-1}L_{i \, i+1} + L_{kn}
\right ) \right ] +  {\cal O}(\eps^0)\,, 
\end{split}
\end{equation}
where the tree primitive amplitudes ${\tilde m}_n$ 
are defined  in \Eq(\ref{eq_prim}).
In the following subsection we derive the corresponding 
result for pure gluon amplitudes for all values of $n$.

\subsubsection{Divergence structure of n-gluon amplitudes} 

As our second example we use \Eq(\ref{Catani})
to derive the divergences of the primitive amplitudes for an
$n$-gluon scattering, neglecting loops of virtual fermions, 
$N_f = 0$. For an $n$-gluon scattering, 
$q$ equals $n-2$. Expanding \Eq(\ref{Catani}) in $\ep$, 
we find 
\be
| \cm^{(1)} \rangle =
-\frac{11}{3 \ep} C_A | \cm^{(0)} \rangle 
+ \sum \limits_{i}^{} \sum_{j \neq i}^{} \left [ 
\frac{1}{\ep^2}  + \frac{1}{\ep} L_{ij} \right ] {\bom T}_i \cdot {\bom T}_j 
 | \cm^{(0)} \rangle ,
\label{cat_form}
\ee
where $L_{ij} = \ln [\mu^2/(-s_{ij}+i0)]$ and we suppressed helicity 
and momentum labels of the amplitudes.  We can use 
\be
\frac{11}{3} C_A = \frac{11}{3} 
{\bom T}_g^2 = - 
\sum \limits_{i}^{n} 
\sum \limits_{j \neq i} \frac{11}{3n} \; {\bom T}_i \cdot {\bom T}_j,
\ee
to cast \Eq(\ref{cat_form}) into the following form 
\be
| \cm^{(1)} \rangle =
\sum \limits_{i}^{} \sum_{j \neq i}^{} \left [ 
\frac{1}{\ep^2}  + \frac{1}{\ep} \left ( \frac{11}{3n} + 
L_{ij} \right )  \right ] {\bom T}_i \cdot {\bom T}_j 
 | \cm^{(0)} \rangle .
\label{cat_form2}
\ee

If we take the matrix element of $|\cm^{(1,0)} \rangle$ with the state of 
$n$-gluons with particular color indices $|a_{1:n}\rangle \equiv
|a_1,...a_n \rangle$ we recover 
the normal color-ordered amplitude, see \Eq(\ref{TreeCol3}).
For example, at tree-level we find 
\be
 \langle a_{1:n}| \cm^{(0)} \rangle 
= \sum _{{\cal P}(2,..n-1)} \left ( F^{a_2}...F^{a_{n-1}} \right )_{a_1 a_n}
m_n(g_1,g_2,...g_n),
\ee
while the one-loop decomposition reads 
\be
g_s^4 \cg \langle a_{1:n}| \cm^{(1)} \rangle 
= \sum _{{\cal P}(2,..n)/{\cal R}} {\rm Tr} 
\left ( F^{a_1} F^{a_2}...F^{a_{n-1}} F^{a_n} \right ) m_n^{(1)}(g_1,g_2,...g_n),
\label{eq_1l}
\ee
where ${\cal R}$ is the reflection transformation. 

Calculating a similar matrix element in \Eq(\ref{cat_form2}), we 
find  
\be
\begin{split}
\langle a_{1:n}| \cm_1 \rangle 
= 
 \sum \limits_{i}^{} \sum_{j \neq i}^{} \left [ 
\frac{1}{\ep^2}  
+ \frac{1}{\ep} \left ( \frac{11}{3n} + L_{ij} \right ) \right ]
\langle a_{1:n}| {\bom T}_i \cdot {\bom T}_j | \cm^{(0)} \rangle.
\label{eq676}
\end{split} 
\ee
We now explain how to compute the matrix elements between 
the state of definite  color $ | a_{1:n} \rangle $, 
the color charges ${\bom T}_i$ and  the amplitudes $|\cm^{(0)} \rangle $.
To do so, we insert the complete set of color states 
\be
\langle a_{1:n}| {\bom T}_i \cdot {\bom T}_j | \cm^{(0)} \rangle
= \sum \limits_{b}^{} 
\langle a_{1:n}| {\bom T}_i \cdot {\bom T}_j |b_{1:n} \rangle 
\langle b_{1:n} | \cm^{(0)} \rangle,
\ee
and use 
\be
\langle a_{1:n}| {\bom T}_i \cdot {\bom T}_j |b_1,..,b_n \rangle 
= \delta_{a_1 b_1} ..if_{a_i c b_i}... if_{a_j c b_j}..  \delta_{a_n b_n},
\ee
which is true in our case since both partons $i$ and $j$ are gluons.
Because $f_{acb} \equiv i F^{a}_{cb}/\sqrt{2}$, we write 
\begin{eqnarray}
&&  \langle a_{1:n}| {\bom T}_i \cdot {\bom T}_j | \cm^{(0)} \rangle
= - \frac{1}{2}
\sum \limits_{{\cal P} [ \theta_{ij}]}^{} 
\left (F^{a_1}..F^{a_n} \right )_{b_i b_j} F^{a_i}_{c b_i} F^{a_j}_{b_j c} 
m_n(g_i,g_1,..,g_n,g_j),
\nonumber \\
&&= 
- \frac{1}{2} \sum \limits_{{\cal P} [ \theta_{ij}]}^{} 
{\rm Tr} \left  
( F^{a_i} F^{a_1}..F^{a_n} F^{a_j} \right )
m_n(g_i,g_1,..,g_n,g_j),
\end{eqnarray}
where $\theta_{ij} = (1,..i-1,i+1,.,j-1,j+1,..n)$.
Inserting this equation back into \Eq(\ref{eq676}),
we find 
\be
\begin{split} 
 \langle a_{1:n}| \cm^{(1)} \rangle = &
-\frac{1}{2} \sum \limits_{i}^{} \sum_{j \neq  i} 
\left [ 
\frac{1}{\ep^2}  + \frac{1}{\ep} \left ( \frac{11}{3n} + 
L_{ij} \right )  \right ]
\\
& \times \sum \limits_{{\cal P} [ \theta_{ij}]}^{} 
{\rm Tr} \left  
( F^{a_i} F^{a_1}..F^{a_n} F^{a_j} \right )
m_n(g_i,g_1,...g_n,g_j).
\end{split} 
\label{eq_1la}
\ee
To read off the divergences of the primitive amplitudes 
from \Eq(\ref{eq_1la}), we need to identify color structures 
in the right hand side of that equation with 
color structures shown in \Eq(\ref{eq_1l}). This can be achieved 
if we re-arrange the summation order in \Eq(\ref{eq_1la}). Upon reflection, 
it is easy to realize that the summation over $i$ and $j$ in \Eq(\ref{eq_1la}) 
reduces to the summation over indices of adjacent particles (cf. positions 
of $i$ and $j$ in the primitive amplitudes $m_n$ in \Eq(\ref{eq_1la})),
while the summation over all permutations in \Eq(\ref{eq_1la})
gives, upon 
using the reflection identities,
twice the sum over permutations in \Eq(\ref{eq_1l}). 
As the result, we arrive at an   
extremely simple formula for divergences 
of $n$-gluon primitive amplitudes 
\be
m_n^{(1)}(g_1,g_2,...g_n)
= - 
\sum_{i=1}^{n} \left [ \frac{1}{\ep^2} 
+ \frac{1}{\ep} \left (\frac{11}{3 n} + L_{i,i+1}
\right ) \right ] m_n^{(0)}(g_1,g_2,...g_n),
\ee
where $L_{n,n+1} = L_{n,1}$.

\section{Colorless amplitudes}
\label{sec:7new}

\subsection{Colorless loop integrand in $D$-dimensions}

We are now in a position to discuss the color-stripped cyclic-ordered
$N$-particle one-loop scattering amplitudes.  In general, these
amplitudes are not primitive amplitudes, although in the case of
gluons color-ordered and primitive amplitudes coincide.

Since permutations of external particles are not allowed,\footnote{We
  discuss here the case where all particles have non-zero color
  charges.} when a primitive amplitude is constructed, we know all
the propagators that can appear in such an amplitude.  We imagine that
we start from a parent diagram, so that all propagators are uniquely
defined. We then find all the diagrams that contribute to the
primitive amplitude by pinching propagators in the parent diagram.
Hence, we write
\beq\label{ScatterAmp}
{\cal A_{(D)}}(\{p_i\},\{J_i\})=\int \frac{d^D\, l}{i \pi^{D/2}}
\frac{{\cal N}(\{p_i\},\{J_i\};l)}{d_0d_1\cdots d_{N-1}}\ ,
\eeq
where $\{p_i\}$ and $\{J_i\}$ are the two 
sets that represent momenta and sources (polarization vectors, 
spinors, etc.) of the external particles.
The numerator structure
${\cal N}(\{p_i\},\{J_i\};l)$ depends on the particle content
of the theory.
The denominator is a product of inverse propagators
\beq
d_i=d_i(l)=(l+q_i)^2-m_i^2\,,
\eeq
where the four-vector $q_0 = 0$, by convention. 

In principle, the one-loop amplitude is just a collection of Feynman
diagrams; therefore, it can be investigated by means of the OPP
method, discussed in the previous Section.  However, we 
point out that a key element of the OPP method is the observation
that, to calculate ${\cal A_{(D)}}$, we need to know ${\cal
  N}(\{p_i\},\{J_i\};l)$ for values of the loop momentum $l$ where
some subset of inverse propagators $(d_0,...,d_{N-1})$ vanishes.  When
an inverse propagator vanishes, the corresponding particle goes on the
mass shell and the flow of the loop momentum terminates. In fact, if
we put one propagator on the mass shell, the one-loop integrand
becomes just a tree amplitude; if we put two propagators on the mass
shell, the integrand becomes a product of two tree amplitudes, etc.
Hence, we see right away that the OPP procedure is related to
unitarity.

However, we need to do calculations in $D$-dimensions to regularize
the ultraviolet and infrared divergences. The simplest option
\cite{Giele:2008ve}, explained in Sect.~\ref{sect5}, is to extend the
OPP procedure so that the unitarity cuts are performed in
$D$-dimensions.  This implies two things.  First, as we already
discussed in the context of the OPP reduction in the previous Section,
the cut momentum is not four-dimensional and the parameterization of
the residues requires introducing tensors that contain the
$(D-4)$-dimensional components of the loop momentum. Second, the
Lorentz indices of the internal particles must also be treated in
$D$-dimensions. Ultimately, this second feature is related to the fact
that not only loop momenta but also polarization vectors of various
particles must be continued to higher-dimensional spacetimes, for a
consistent regularization procedure.  As a result, the number of spin
eigenstates changes and becomes $D$-dependent.

We will be concerned with the cases where there is massless gauge
boson or (massless or massive) quark in the loop.  Then, a massless
spin-one particle in $D_s$ dimensions has $D_s-2$ spin eigenstates
while Dirac spinor in $D_s$ dimensions has $2^{(D_s-2)/2}$ spin
eigenstates.  In the latter case, $D_s$ should be even. The spin
density matrix for a massless spin-one particle with momentum $l$ and
polarization vectors $e_{\mu}^{(i)}$ is given by
\be
\sum \limits_{i=1}^{D_s-2} e_{\mu}^{(i)}(l)e_{\nu}^{(i)}(l)=
-g_{\mu\nu}^{(D_s)}+\frac{l_{\mu}b_{\nu}+b_{\mu}l_{\nu}}{l\cdot b},
\ee
where $b_{\mu}$ is an auxiliary light-cone vector required to fix the
gauge.  The spin density matrix for a fermion with momentum $l$ and
mass $m$ is given by
\be
\sum_{i=1}^{2^{(D_s-2)/2}} u^{(i)}(l)\overline{u}^{(i)}(l)
=\sum_{\mu=1}^{D} l_{\mu}\Gamma^{\mu}+m\ ,
\ee
where $\Gamma_\mu$ is the $D$-dimensional generalization of the 
Dirac matrix. 

In the preceding discussion, we introduced a special notation for the
dimensionality of the internal spin space $D_s$, to distinguish it
from the dimensionality of the space-time $D$ where the loop momentum
lives.  For consistency, we must choose $D \le
D_s$~\cite{Bern:2002zk}.  We now generalize the notion of dimensional
dependence of the one-loop scattering amplitude in
\Eq(\ref{ScatterAmp}) by extending the sources of all unobserved
particles to a $D_s$-dimensional space-time
\be
{\cal A}_{(D,D_s)}(\{p_i\},\{J_i\}) =\int \frac{{ d}^D\, l}{i\pi^{D/2}}
\frac{{\cal N}^{(D_s)}(\{p_i \},\{J_i\};l )}{d_0d_1\cdots d_{N-1}}.
\label{eq1}
\ee
The numerator function ${\cal N}^{(D_s)}(\{p_i\},\{J_i \}; l)$ depends
explicitly on $D_s$ through the number of spin eigenstates of virtual
particles.

A simple, but important observation is that the numerator function
${\cal N}^{(D_s)}$ (without closed fermion loops) depends 
{\it linearly} on $D_s$ provided that the external momenta and polarization
vectors are four-dimensional.  To generate an explicit dependence on
$D_s$ we need to have a closed loop of contracted Lorentz-vector
indices coming from vertices and propagators.  At one loop we can
generate at most one trace of the metric tensor.  Therefore, the
numerator functions of the integrand of one-loop amplitudes without
closed fermion loops can always be parametrized as
\be
{\cal N}^{(D_s)} (l) = {\cal N}_0(l) + (D_s - 4)\, {\cal N}_1(l).
\label{eq2}
\ee
We emphasize that there is no explicit dependence on either $D_s$ or
$D$ in the functions ${\cal N}_{0,1}$.  In the case of amplitudes with
one closed fermion loop an additional $D_s$ dependence comes from the
normalization of the trace of the $D_s$-dimensional Dirac matrices
\be 
\label{eq6.73}
{\rm Tr} \left ( \Gamma^\mu \Gamma^\nu \right ) 
= t_{D_s} \;4 g_{D_s}^{\mu \nu},\;\;\;\;
t_{D_s}=\frac{1}{4} {\rm Tr(1)}=2^{(D_s-4)/2}\,.
\ee
This is an overall normalization factor.  In traditional calculations
with Feynman-diagrams its expansion in $\eps$ can be postponed until
after the cancellation of the singularities has been carried out.
Then the $\eps$-dependent part in \Eq(\ref{eq6.73}) becomes
irrelevant.  In the $D$-dimensional unitarity cut method this
additional $D_s$ dependence has to be taken into account and we obtain
the following result for the fermion loops
\be
{\cal N}_f^{(D_s)} (l) = t_{D_s}\left({\cal N}_{0,f}(l) + 
 (D_s-4) {\cal N}_{1,f}(l)\right)\,,
\label{eq2nf}
\ee
At one loop we can have either a closed loop of Lorentz tensors or a
closed loop of Dirac matrices. Mixed cases can appear only in the case
of two or more loops. We conclude that ${\cal N}_{1,f}(l)=0$ and that
the numerator function for closed fermion loops can be calculated in
four dimensions.  The $D_s$ dependence in that case is entirely due to
the trace of the identity matrix.

For amplitudes without closed fermion loops, in a numerical
implementation we need to separate the two functions ${\cal
  N}_{0,1}$. To do so, we compute the left hand side of \Eq(\ref{eq2})
for $D_s = D_1$ and $D_s =D_2 $ and, after taking appropriate linear
combinations, we obtain
\be
\label{eq3}
\begin{split} 
& {\cal N}_0(l) = \frac{(D_2-4){\cal N}^{(D_1)}(l) - (D_1-4){\cal 
N}^{(D_2)}(l)}{D_2-D_1}, \\
& {\cal N}_1(l) =  
\frac{{\cal N}^{(D_1)}(l) - {\cal N}^{(D_2)}(l)}{D_2-D_1}\,.
\end{split} 
\ee
In the case of amplitudes with only bosons in the loop the numerator
functions are well defined numerically for any integers $D_{1,2}>4$.
If fermions are also present in the loop, then we have to choose {\it
  even} values for $D_{1,2}$, subject to the constraints $D_{1,2} \geq
4$.

Having established the $D_s$-dependence of the amplitude, we discuss
the analytic continuation for sources of unobserved particles.  We can
interpolate $D_s$ either to $D_s \to 4-2\ep$ or to $D_s\to 4$. The
first case is known as the 't~Hooft-Veltman (HV)
scheme~\cite{tHooft:1972fi}, the second case -- as the
four-dimensional helicity (FDH) scheme~\cite{Bern:2002zk}.  The latter
scheme is of particular interest in supersymmetric (SUSY) calculations
since all SUSY Ward identities are preserved. We see from
\Eq(\ref{eq2}) that the difference between the two schemes is simply
$-2\ep{\cal N}_1$.

We now substitute \Eq(\ref{eq3}) into \Eq(\ref{eq1}) and obtain
explicit expressions for one-loop amplitudes in the HV and the FDH
schemes. We derive
\ba
{\cal A}^{\rm FDH} &=&
\left(\frac{D_2-4}{D_2-D_1}\right){\cal A}_{(D, D_s =D_1)}
-\left(\frac{D_1-4}{D_2-D_1}\right){\cal A}_{(D, D_s =D_2)}, \nonumber \\
{\cal A}^{\rm HV} &=& {\cal A}^{\rm FDH}
-\left(\frac{2\ep}{D_2-D_1}\right)\left({\cal A}_{(D, D_s =D_1)}
-{\cal A}_{(D, D_s =D_2)}\right).
\label{eq2.4}
\ea
We emphasize that $D_s=D_{1,2}$ amplitudes on the r.h.s. of
\Eq(\ref{eq2.4}) are conventional one-loop scattering amplitudes whose
numerator functions are computed in higher-dimensional space-time,
i.e. all internal metric tensors and Dirac gamma matrices are in
integer $D_s = D_{1,2}$ dimensions.  The loop momentum is in $D\leq
D_s$ dimensions.  It is important that no explicit dependence on the
regularization parameter $\ep = (4-D)/2$ is generated by spin density
matrices in these amplitudes.  For this reason, \Eq(\ref{eq2.4})
allows numerical implementation.  In particular, calculation of
unitarity cuts in $D_s$ dimensions is now straightforward, as cut
internal lines possess well-defined spin density matrices.  Below we
discuss the construction of the spin states in $D$-dimensional
space-time.

\subsection{Polarization states in $D$-dimensions}
\label{sec:polD}

We begin by reminding the reader about the four-dimensional case.  In
four dimensions, gluons and massless quarks have two polarization
states.  One can use helicity states of fermions to find a useful
representation of the gluon or photon polarization vectors that leads
to significant simplifications of the analytic computations.  Such
methods are reviewed in e.g. Ref.~\cite{Dixon:1996wi}
and in Appendix~\ref{App:SH}.

However, as we have seen in the previous Section, 
$D$-dimensional unitarity requires an analytic continuation of spin
degrees of freedom to higher-dimensional space-times.  In the
spinor-helicity formalism, polarizations of particles of integer spin
are expressed in terms of spinor solutions.  It is not trivial to do
this in higher-dimensional space-times, although some work in this
direction has been done
recently~\cite{Cheung:2009dc,CaronHuot:2010rj}.  We will not discuss
these issues in this Section, focusing instead on the construction of
polarization states for both quarks and gluons for which the
continuation to higher-dimensional spaces is straightforward.

In $D=4$ the two polarization vectors for massless gauge bosons with
momentum $p$ satisfy the constraint $p \cdot \epsilon_{\lambda} = 0$
($\lambda= \pm 1$).  We consider a general parametrization of the
massless momentum
$
p = E(1,\sin \theta \cos \phi, \sin \theta \sin \phi, \cos \theta),
$
and find the two polarization vectors 
\be
\begin{split} 
& \epsilon_{\lambda}(p) = \frac{1}{\sqrt{2}} 
\Big ( 0 , c_\theta c_\phi - {\rm sgn}(E) \lambda i s_\phi,
c_\theta s_\phi + {\rm sgn}(E) 
\lambda i c_\phi, -s_\theta \Big  ),
\end{split} 
\label{w1}
\ee
where $c_{\theta} = \cos \theta, s_\theta = \sin \theta$, 
$c_{\phi} = \cos \phi$ and $s_{\phi} = \sin \phi$.

To describe the explicit solution of the massless Dirac equation
$\slsh{p} u = 0$, we need to fix the representation of the Dirac
matrices. For massless fermions, it is convenient to choose the Weyl
representation where the four-dimensional $\gamma$-matrices are given
by
\begin{equation} \label{eq:gammamatrices}
\gamma^0=\ \left(
\begin{matrix}{
\bf0}&{\bf1}\cr{\bf1}&{\bf0}\cr
\end{matrix}
\right)\ ,\,\,
\gamma^i\ =\ \left(
\begin{matrix}
{\bf0}&-{\bf\sigma}^i\cr
                           {\bf\sigma}^i&{\bf0}\cr\end{matrix}
\right)\ ,\,\, 
\gamma^5\ 
\ =\ \left(
\begin{matrix}
{\bf1}&{\bf0}\cr{\bf0}&-{\bf1}
\cr\end{matrix}
\right).
\end{equation}
In this equation the boldface symbols indicate $2\times 2$ submatrices
and ${\bf \sigma}^i$ are the Pauli matrices.  Consider a massless
fermion with momentum $p = (E,p_x,p_y,p_z)$ and let $p_+ =
E+p_z$. Solving the Dirac equation for massless quarks, we find
\be \label{eq:Diracsolutions}
u_{\lambda = 1}(p) =
\left (
\begin{array}{c}
\sqrt{p_+} \\
(p_x + i p_y)/\sqrt{p_+} \\
0 \\
0 \\
\end{array} 
\right ),\;\;
u_{\lambda = -1}(p) =
\left (
\begin{array}{c}
0\\
0\\
(p_x - i p_y)/\sqrt{p_+} \\
-\sqrt{p_+} \\
\end{array} 
\right ),
\ee
where $\lambda=\pm 1$ refers to the helicity of an incoming fermion.
One can easily  check  that if $u(p)$ satisfies the massless Dirac
equation then
\beq
\frac{\vec{p}\cdot \vec{\Sigma}}{\abs{\vec{p}}} \, u_\lambda(p)=\frac{1}{2}\gamma_5 \, u_\lambda(p)
=\frac{1}{2}\lambda \, u_\lambda(p)~\mbox{where}~\vec{\Sigma}^i=
\frac{i}{4}\epsilon^{ijk}\gamma^j\gamma^k\, . 
\eeq
For anti-fermions the sign of the helicity  is flipped
so that $v_{\lambda}(p)=u_{-\lambda}(p)$.
Because $p_+$ vanishes for $E = -p_z$, the solution for the fermion moving in the
$-z$ direction requires care. Taking the limit, we arrive at
\be
u_{\lambda = 1}(p) =
\left (
\begin{array}{c}
0 \\
\sqrt{2 E}\\
0 \\
0 \\
\end{array} 
\right ),\;\;\;\;
u_{\lambda = -1}(p) =
\left (
\begin{array}{c}
0\\
0\\
\sqrt{2E} \\
0 \\
\end{array} 
\right ).
\ee

We are now in position to generalize the description of the
polarization states to the $D$-dimensional case.  We begin with the
discussion of a massless gauge boson.  From previous considerations we
know that we must consider two dimensionalities of the loop momentum
vector space, $D=4,5$ and two dimensionalities of the space where
spins are embedded $D_s \ge D$.  For example, if no quarks are present
in the loop, we may take $D=4,5$ and $D_s = 5,6$. In practice, we
always compute the cut-constructible coefficients, which require only
four-dimensional information, using $D=4,D_s=4$ to minimize the amount
of algebra.

We first take $D_s=5$ and denote the polarization vector by $\ep =
({\boldsymbol \epsilon}_4, \ep_5)$, where ${\boldsymbol \epsilon}_4$
are the four components of the polarization vector that are embedded
in the four-dimensional space.  Then, if the momentum of a gluon is
four-dimensional, $p = ({\bf p}_4,0)$, there are three obvious choices
for the polarization vectors: $\ep_{\pm} = ({\boldsymbol
  \ep}_4^{\pm},0)$ and $\ep_0 = ({\bf 0}_4,1)$.  However, if the
momentum is five-dimensional $p = ({\bf p}_4, p_5) $, the situation is
different.  Note that, from a four-dimensional point of view, the
massless boson with five-dimensional momentum corresponds to a massive
boson in four dimensions, with the mass $p_5^2$.  Hence, we can choose
four-dimensional polarization vectors of a massive gauge boson, to
describe the required polarization states.  Explicitly, if we write
the five-dimensional momentum as
$$
p = (E,p_4\sin \theta \cos \phi, p_4\sin \theta \sin \phi, p_4\cos \theta,
p_5),\;\;\;\;E^2 - p_4^2 = p_5^2,\;\;\;\; E> 0, 
$$
a convenient choice of the polarization vectors is
\be
\begin{split} 
& \epsilon^{\pm} = \frac{1}{\sqrt{2}}
\left ( 
0, \cos \theta \cos \phi \mp i \sin \phi, \cos \theta \sin \phi \pm i \cos \phi,
-\sin \theta, 0 \right ),
\\
& \epsilon^{0} = p_5^{-1} \left ( p_4,E
\sin \theta \cos \phi, E\sin \theta \sin \phi, 
E\cos \theta, 0 \right ).
\end{split} 
\ee

It is straightforward to discuss the $D_s = 6$ case. Indeed, since the
gluon momentum does not have a six-dimensional component and since we
need only one additional polarization vector for $D_s=6$, compared to
$D_s = 5$, we can choose this additional vector to be $\epsilon_L =
({\bf 0}_5,1)$ since such a choice clearly satisfies the
transversality constraint.  This construction generalizes to
higher-dimensional space-times, provided that the dimensionality of
the loop momentum vector is restricted to five or less.

We now turn to the discussion of fermions.  In $D_s$-dimensions
fermions have $2^{D_s/2-1}$ polarization states, given by
$2^{D_s/2}$-component Dirac spinors $u_j^{(s)}(p)$, $s=1,\ldots
,2^{D_s/2-1}$, $j=0,\ldots ,2^{D_S/2}-1$.  These spinors are
solutions of the massive or massless Dirac equation in $D_s$
dimensions
\be
\sum_{\mu=0}^{D-1} \left [ p_{\mu}\Gamma^{\mu}_{(D_s)} 
- m \right ] \  u^{(s)}(p)\ = \ 0,\;\;\;
\quad \mu\ =\ 0,\ldots, D-1 <  D_s, 
\label{ww2}
\ee
where $m$ denotes the mass of the fermion. We note that in
\Eq(\ref{ww2}) we treat the loop momentum as a $D$-dimensional vector,
$D \le D_s$, in accord with our previous discussion.

To find solutions to \Eq(\ref{ww2}), we need to explicitly construct
Dirac matrices in higher dimensions.\footnote{
With a slight abuse of notation, we shall talk 
about $D$-dimensional Dirac algebra, rather than $D_s$-dimensional
Dirac algebra throughout the text until \Eq(7.30).} 
To this end, we consider even
space-time dimensionalities and follow the recursive definition given
in Ref.~\cite{collins}.  The Dirac matrices have to satisfy the
anti-commutation relation
\be
\Gamma_{(D)}^{\mu} \Gamma_{(D)}^{\nu} + \Gamma_{(D)}^{\nu} \Gamma_{(D)}^{\mu} 
= 2 g^{\mu \nu},\;\;\;\;
\mu,\nu=0,\ldots,D-1.
\label{eq2222}
\ee

We assume that the Dirac matrices for $D = D_0$ are known and we need
to construct Dirac matrices for $D = D_0 +2$.  When we move from $D =
D_0$ to $D = D_0+2$, the dimensionality of the matrices increases by a
factor of two.  \Eq(\ref{eq2222}) can be satisfied by choosing the
first $D_0$ of the $D_0+2$ matrices to be given by
$2^{(D_0+2)/2}\times 2^{(D_0+2)/2}$ block-diagonal matrices defined as
follows
\be
\Gamma^0_{(D_0+2)} = 
\left (
\begin{array}{cc}
\Gamma^0_{(D_0)} & 0 \\
0 & \Gamma^0_{(D_0)} 
\end{array}
\right ),\;\;\;
\Gamma^{i=1,\dots,D_0-1}_{(D_0+2)} =
\left (
\begin{array}{cc}
\Gamma^i_{(D_0)} & 0 \\
0 & \Gamma^i_{(D_0)} 
\end{array}
\right ).
\label{firstD0matrices}
\ee
The remaining two matrices need to anticommute with the matrices
defined in \Eq(\ref{firstD0matrices}).  They can be obtained by
constructing an analog of the four-dimensional $\gamma_5$-matrix in
$D_0$-dimensions; this is achieved by taking the product
\be
\hat{\Gamma}_{(D_0)} = i^{{D_0}/2-1} 
\Gamma^{0}_{(D_0)} \Gamma^{1}_{(D_0)} ... \Gamma^{D_0-1}_{(D_0)}. 
\ee
The matrix $\hat{\Gamma}_{(D_0)}$ satisfies 
\be
\hat{\Gamma}_{(D_0)} \Gamma_{(D_0)}^\mu + 
\Gamma_{(D_0)}^\mu \hat{\Gamma}_{(D_0)} = 0, \;\;\; \mu = 0,...,D_0-1.
\label{Gamma_5_in_D0_dimensions}
\ee 
Using $\hat{\Gamma}_{(D_0)}$, we can construct the two missing matrices, 
to complete the algebra of Dirac matrices in $D_0+2$ dimensions 
\be
\Gamma_{(D_0+2)}^{D_0} 
= 
\left (
\begin{array}{cc}
0  & \hat{\Gamma}_{(D_0)} \\
-\hat{\Gamma}_{(D_0)} & 0
\end{array}
\right ),\;\;\; 
\Gamma^{D_0+1}_{(D_0+2)} =
\left (
\begin{array}{cc}
0  & i \hat{\Gamma}_{(D_0)}\nn \\
i \hat{\Gamma}_{(D_0)} & 0
\end{array}
\right )\ .
\ee

The recursive construction requires an initial condition. In this
case, the appropriate condition is clear since it is given by the
familiar algebra of Dirac matrices in four-dimensional space-time.  We
note that the above construction is independent of the
four-dimensional representation of the Dirac matrices, except for the
initial condition. This is a welcome feature since it allows us to use
the same formalism for both Weyl and Dirac representations.  In the
following we suppress the dimensionality index $D$, so that
$\Gamma^{\mu}$ denotes a $D$-dimensional gamma-matrix.
We also note that within this framework, the {\it four-dimensional}
$\gamma_5$ matrix is continued to $D$-dimensions following the
prescription by 't~Hooft and Veltman \cite{tHooft:1972fi}.  Indeed,
they defined the continuation of $\gamma_5$ as a matrix that commutes
with $\gamma_\mu$ if $\mu > 3$ and anticommutes otherwise.  It is easy
to see that the appropriate continuation of $\gamma_5$ to $D_0+2$
dimensions is given by a block-diagonal matrix with four-dimensional
$\gamma_5$ along the diagonal.  As an example, in six dimensions we
have
\beq
\Gamma^{\gamma_5}=
\left(
\begin{matrix}
{\bf \gamma}_5&{\bf 0}\cr{\bf0}&{\bf\gamma}_5\cr
\end{matrix}
\right).
\eeq 

The recursive structure of the $D$-dimensional gamma-matrices allows
us to solve the Dirac equation recursively. We first discuss massless
solutions.  Consider a massless fermion with momentum $p$, in
$D$-dimensions.  To construct explicit polarization states for such a
fermion, we employ an auxiliary light-like vector $n$ such that $n
\cdot p \neq 0$. We write
\be
u^{(s)}(p,n) = \frac{\hat p}{\sqrt{2 p\cdot n}} \chi^{(D)}_s(n),
\;\;\;
\bar u^{(s)}(p,n) = \bar \chi^{(D)}_s(n) \frac{\hat p}{\sqrt{2 p\cdot n}}.
\ee
Here $\hat p = p_\mu \Gamma^\mu$, where the index $\mu$ runs over
$0,1,...D-1$ components.  The index $s$ specifies the spinor
polarization states.  Note that when the Dirac conjugate spinor is
constructed, the spinor  momentum $p$ is {\it not} complex conjugated.
This is an irrelevant detail if the on-shell momentum is real, but it
becomes important for consistent applications within generalized
unitarity where calculations with on-shell complex momenta are
required.

We choose $D$-dimensional, $p$-independent spinors $\chi^{(D)}_s(n)$
in such a way that their direct product reads 
\be
\sum_{s=1}^{2^{(D/2-1)}}  \chi^{(D)}_s(n) \otimes \bar \chi^{(D)}_s(n) = \hat n.
\ee
Then, it is easy to see that the $u_j(p,n)$ spinors satisfy both the
Dirac equation for massless fermions and the completeness relation
\be
\sum_{s=1}^{2^{(D/2-1)}} u^{(s)}(p,n) \otimes \bar u^{(s)}(p,n) = 
\frac{\hat p \hat n \hat p}{2p\cdot n} = \hat p. 
\label{comp1}
\ee
We conclude that $u^{(s)}(p,n)$ is a valid choice for the on-shell 
fermion wave functions.

The above construction involves an auxiliary vector $n$ and, for this
reason it is quite flexible. Having such a flexibility turns out to be
important, especially since we have to construct on-shell spinors for
complex momenta.  We give a few examples below.  We consider a
$D$-dimensional vector $n = (n_0,n_x,n_y,n_z,\{ n_{i \in (D-4)} \})$,
choose $n_0 = 1/2, n_z = 1/2$ and set all other components to zero.
We need to find the spinors $\chi$ such that
\be
\sum_{s=1}^{2^{(D/2-1)}}  \chi^{(D)}_s(n) \otimes \bar \chi^{(D)}_s(n) = \hat n
= \frac{1}{2} \left ( \Gamma^0 - \Gamma^z \right ).
\ee
Since $\Gamma^{0,x,y,z}$ are all block-diagonal \cite{collins}, with
``blocks'' being $4 \times 4$ matrices, a $D$-dimensional spinor is
constructed by simple iteration of the four-dimensional spinor.  We
assume that the four-dimensional Weyl representation is extended to
$D$-dimensions.  The four-dimensional spinors are given by
\be
\chi^{(4)}_1 = \left (
\begin{array}{c}
1\\
0\\
0\\
0
\end{array}
\right )
,\;\;\;\
\chi^{(4)}_2 = \left ( 
\begin{array}{c}
0\\
0\\
0\\
-1
\end{array}
\right ).
\ee
In six dimensions the eight-component  
spinors are chosen to be
\be
\chi^{(6)}_{1,2} = \left (
\begin{array}{c}
\chi^{(4)}_{1,2}\\
0\\
\end{array}
\right )
,\;\;
\chi^{(6)}_{3,4} = \left ( 
\begin{array}{c}
0\\
\chi^{(4)}_{1,2}
\end{array}
\right ).
\label{eqeta6}
\ee
The case $D=8$ is obtained again by  iteration. We find
\be
\begin{split}
& \chi^{(8)}_{1,2} = \left (
\begin{array}{c}
\chi^{(6)}_{1,2}\\
0\\
\end{array}
\right )
,\;\;
\chi^{(8)}_{3,4} = \left ( 
\begin{array}{c}
\chi^{(6)}_{3,4}\\
0
\end{array}
\right ),
\;\;\;
\\
& \chi^{(8)}_{5,6} = \left (
\begin{array}{c}
0\\
\chi^{(6)}_{1,2}
\end{array}
\right )
,\;\;
\;\;\;
\chi^{(8)}_{7,8} = \left ( 
\begin{array}{c}
0\\
\chi^{(6)}_{3,4}
\end{array}
\right ).
\label{eqeta8}
\end{split}
\ee

We now present two alternative procedures to define fermionic spinors
which we employ when a particular choice of the vector $n$ leads to
numerical instabilities. This occurs for the on-shell momentum $p =
(p_0, 0, 0, p_0)$ since $(p \cdot n)= 0$. To handle this case, we
change the vector $n$ to $(1/2,0,0,-1/2,0_{D-4})$ in the above
formulas. However, even this can be insufficient. Indeed, note that a
complex momentum $p = (0,p_x,p_y,0)$ can be light-like and, therefore,
be a valid on-shell momentum for a massless fermion. Clearly $p \cdot
n = 0$ in this case, for the two choices of the vector $n$ that we
discussed.  To deal with this case, we need to choose yet another $n$.
We can take $n = (1,1,0,0,0_{D-4})$ and choose the following
four-dimensional spinors
\be
\chi^{(4)}_1 = \left (
\begin{array}{c}
1\\
0\\
0\\
1
\end{array}
\right )
,\;\;\;
\chi^{(4)}_2 = \left ( 
\begin{array}{c}
0\\
1\\
1\\
0
\end{array}
\right ).
\ee
The higher-dimensional spinors are obtained from these
four-dimensional solutions along the lines discussed above.

This procedure can be easily extended to the case of massive fermions.
We will discuss such an extension assuming that the four-dimensional
gamma-matrices are in the Dirac representation.  We need to explicitly
construct the $2^{D_s/2-1}$ spin polarization states $u_j^{(s)}(p,m)$
that satisfy the Dirac equation
\Eq(\ref{ww2}) and the completeness relation 
\be
\sum \limits_{s = 1}^{2^{(D_s/2-1)}} u_i^{(s)}(p,m) \bar u_j^{(s)}(p,m)
= \sum_{\mu=0}^{D-1} p_\mu \Gamma_{ij}^\mu + m\times\delta_{ij}.
\label{eq.comp}
\ee
The on-shell condition for a fermion with the mass $m$ and momentum
$p$ is taken to be $p^2=m^2$.  To construct a set of $2^{D_s/2-1}$
spinors satisfying the Dirac equation we generalize the procedures
used in the massless case. We define the spinors
\be
u^{(s)}(p,m) = \frac{(p_\mu \Gamma^\mu + m)}{\sqrt{p_0 + m}}\eta_{D_s}^{(s)},
\;\;\;\; s=1,\ldots,2^{D_s/2-1}\ .
\ee
For $D_s=4$ we choose
\be
\eta_1^{(4)}=\left (\begin{array}{c} 1 \\ 0 \\ 0 \\ 0 \\ \end{array}\right),\;\;\;\
\eta_2^{(4)}=\left (\begin{array}{c} 0 \\ 1 \\ 0 \\ 0 \\ \end{array}\right)\ ,
\ee
and construct recursively the $D_s=6$ eight-component basis spinors
\be
\eta_{1,2}^{(6)} = 
\left ( \begin{array}{c} \eta_{1,2}^{(4)} \\ 0 \end{array} \right ),\;\;\;\;
\eta_{3,4}^{(5)} = \left ( \begin{array}{c} 0 \\ \eta_{1,2}^{(4)} 
\end{array} \right ).
\ee
The eight spinors for $D_s=8$ are obtained using the obvious
generalization.  It is easy to see that the spinors constructed in
this way do indeed satisfy the Dirac equation.  To check the
completeness relation, we define the Dirac conjugate spinor to be
\be
\bar u^{(s)}(p,m) = \bar \eta_{D_s}^{(s)} 
\frac{(p_\mu \Gamma^\mu + m)}{\sqrt{p_0 + m}}.
\label{eq.ncc}
\ee
Then it is easy to see that  \Eq(\ref{eq.comp}) is satisfied.

\subsection{Berends-Giele recursion relations} 
\label{bgampl}

Any calculation based on unitarity cuts requires on-shell scattering
amplitudes for complex on-shell momenta.  For $D$-dimensional
generalized unitarity, these amplitudes must be computed in
$D$-dimensional space-time which implies that  momenta of some
particles in the scattering amplitude and their polarization states
are continued to $D$-dimensions.  How do we  obtain such
scattering amplitudes? It turns out that the most robust method that
allows a fast and efficient computation of the on-shell scattering
amplitudes is based on the recursion relations by Berends and Giele
\cite{BG}.  Here we summarize the main idea of this method and give
some examples. 

\begin{figure}[t]
\begin{center}
\hspace*{-0.1cm}
\includegraphics[scale=0.4]{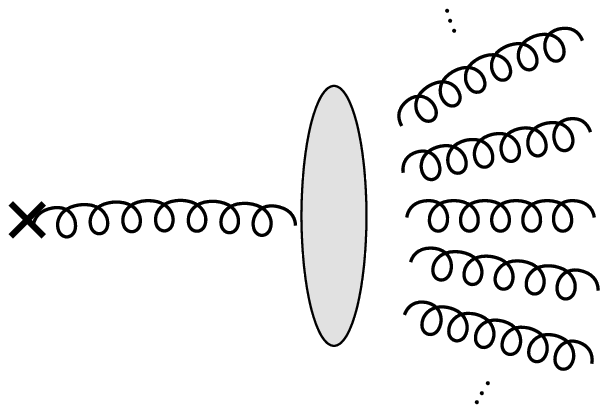}
\quad \qquad
\raisebox{-0.2cm}{
\includegraphics[scale=0.3]{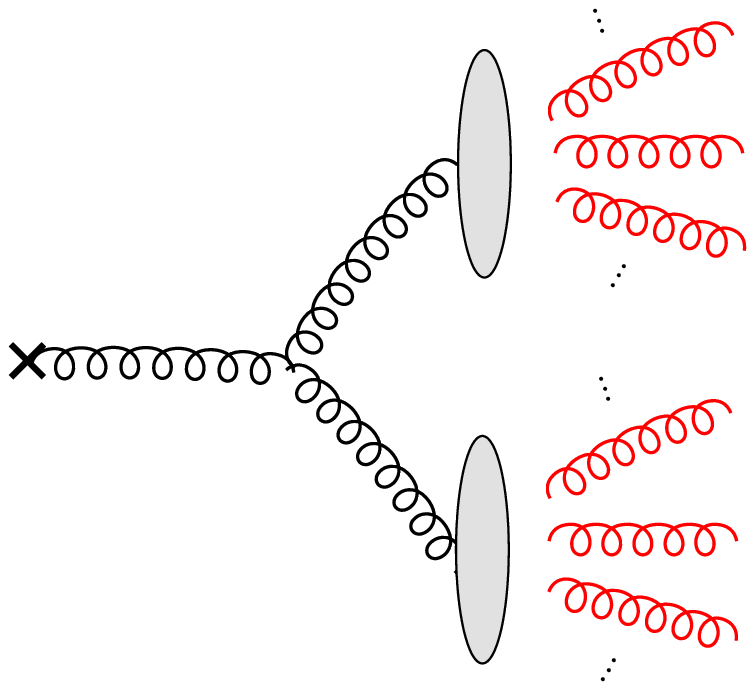}
\quad \qquad \qquad \,\,\, 
\includegraphics[scale=0.3]{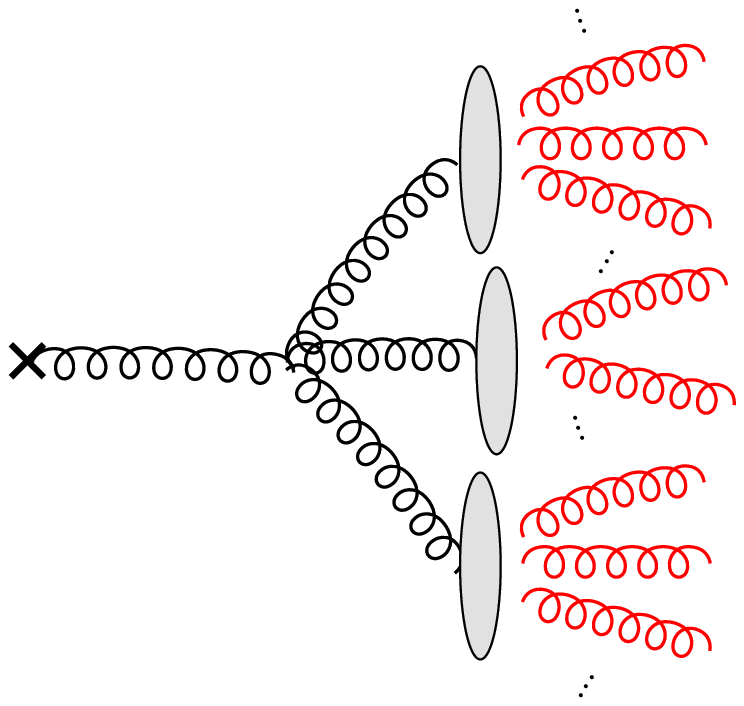}}
\end{center}
\caption{Recursion relation for the $n$-gluon current, \Eq(\ref{bg1}),
  the integers denote the number of gluons involved in the current.}
\begin{picture}(0,0)
\put(80,100){$ \displaystyle n$}
\put(197,95){$ \displaystyle {\color{red}m}$}
\put(197,45){$ \displaystyle {\color{red}n-m}$}
\put(323,94){$ \displaystyle {\color{red} m}$}
\put(323,70){$ \displaystyle {\color{red}k-m}$}
\put(323,50){$\displaystyle {\color{red}n-k}$}
\put(95,70){$ \displaystyle = \sum_{m=1}^{n-1}$}
\put(200,70){$ \displaystyle + \sum_{m=1}^{n-1} \sum_{k=m+1}^{n-1}$}
\end{picture}
\label{fig:ngluon}
\end{figure}

Consider a color-ordered $(n+1)$-gluon scattering amplitude at tree
level.  We take one of the gluons off the mass-shell, and remove its
polarization vector.  The resulting object is called {\it the gluon
  current}~\footnote{Note that in Ref.~\cite{BG} the currents also
  include the gluon propagator of the off-shell gluon. We choose to
  define currents without that propagator, to be as close as possible
  to a numerical implementation of Berends-Giele recursion
  relations.}; it is denoted by $G^{\mu}(g_1,g_2,..,g_n)$. The
arguments of the current refer to ordered gluons with outgoing momenta
and polarization vectors, i.\ e.\ $g_i = (k_i,\epsilon_i)$. The
outgoing momentum of the off-shell line is $-k_1 - k_2-.... -k_n$, due
to the momentum conservation.  A simple and remarkably  robust
observation is that there exists a recursion relation that the gluon
current satisfies.  One can understand this equation by tracing back
the ``off-shell'' gluon and realizing that it can split into two or
three off-shell currents, thanks to the presence of three- and
four-gluon vertices in QCD.  Because such a splitting can not violate
the ordering of the external gluons, the number of terms that
contribute is limited.  The recurrence relation reads
\ba
&& G^\mu(g_1,g_2,...g_n)  = 
\sum \limits_{m=1}^{n-1}  
V_3^{\mu \nu \rho}(-k_{1;n},k_{1;m},k_{m+1;n}) 
S_{\nu \nu'}^G(k_{1;m}) S_{\rho\rho'}^G(k_{m+1;n}) \nonumber 
\\
&&
\;\;\;\;\;\;\;\; \times G^{\nu'}(g_1,g_2,\dots,g_m) G^{\rho'}(g_{m+1},\dots,g_n) 
\label{bg1}
\\
&& +  \sum \limits_{m=1}^{n-1} 
\sum \limits_{k = m+1}^{n-1} 
 V_4^{\mu \nu \rho \sigma}(-k_{1;n},k_{1;m},k_{m+1;k},k_{k+1;n})  
S_{\nu\nu'}^G(k_{1;m}) S_{\rho\rho'}^G(k_{m+1;k}) 
\nonumber \\
&&  \times S_{\sigma\sigma'}^G(k_{k+1;n}) 
G^{\nu'}(g_{1},\dots,g_{m}) G^{\rho'}(g_{m+1},\dots,g_{k})
G^{\sigma'}(g_{k+1},\dots,g_{n}). 
\nonumber 
\ea
In \Eq(\ref{bg1}), $V_{3,4}$ are the three- and four-gluon
color-stripped vertices, see Fig.~\ref{fig:ngluon}, $S_{\mu
  \nu}^{G}(p)$ is the gluon propagator and $k_{i;j} = \sum
\limits_{m=i}^{j} k_{m}$.  The on-shell color-ordered physical
amplitude for the $(n+1)$-gluon scattering is obtained from the
on-shell limit $k_{1;n}^2 \to 0$ of the $n$-gluon current
$G^\mu(g_1,g_2,...g_n)$, contracted with the polarization vector of
the gluon with momentum $k_{1;n}$.  The Berends-Giele procedure is
recursive since it expresses currents of higher multiplicities through
currents of lower multiplicities.  The initial condition for the
recursion is $G^\mu(g_1) = \epsilon_1^\mu$, so that the one-gluon
current is just the polarization vector of that gluon.  It is an
important feature of the Berends-Giele construction that currents can
be easily computed for both, complex momenta and in space-times of
higher dimensionality. This makes Berends-Giele recursion an ideal
tool to use with generalized $D$-dimensional unitarity.

\begin{figure}[t]
\begin{center}
\includegraphics[scale=0.57]{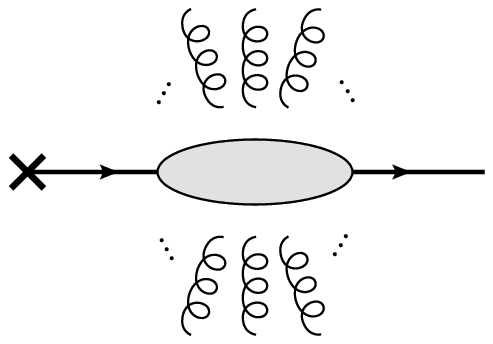}
\quad \qquad
\raisebox{-0.70cm}{
\includegraphics[scale=0.57]{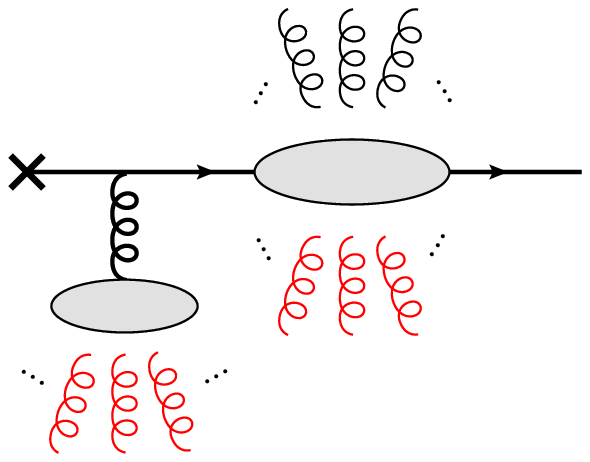}}
\quad \qquad 
\includegraphics[scale=0.57]{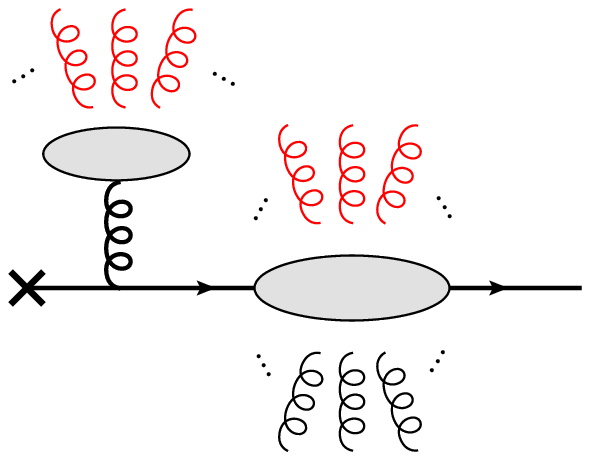}
\end{center}
\caption{Recursion 
relation for the two-quark $n$-gluon current, \Eq(\ref{bg2}).}
\begin{picture}(0,0)
\put(38,120){$ \displaystyle n_1$}
\put(38,45){$ \displaystyle n_2$}
\put(168,120){$ \displaystyle n_1$}
\put(124,30){$ \displaystyle {\color{red}n_2-m}$}
\put(168,50){$ \displaystyle {\color{red}m}$}
\put(260,135){$ \displaystyle {\color{red}m}$}
\put(288,120){$ \displaystyle {\color{red}n_1-m}$}
\put(298,50){$ \displaystyle n_2$}
\put(88,83){$ \displaystyle = \sum_{m=0}^{n_{2}-1}$}
\put(217,83){$ \displaystyle + \sum_{m=1}^{n_{1}}$}
\end{picture}
\label{qqBG_fig}
\end{figure}

The Berends-Giele recursion can be easily generalized to cases with
fermions, gluons and electroweak vector bosons.  Recall that when
fermions are involved, we compute primitive, rather than
color-ordered, amplitudes using color-stripped Feynman rules,
Fig.~\ref{fig6.1}.  An interesting consequence of color ordering is
that the sign of the color-stripped quark-gluon vertex depends on
whether, when progressing along the fermion line towards the vertex
facing in the direction of the arrow, the gluon appears on the
right-hand side or on the left-hand side.  The simplest current that
one can define in this case is the current that gives the primitive
amplitude for two-quark and $n$-gluon scattering.  We define the
current by taking the incoming fermion line off shell; we denote the
corresponding current by $\bar
Q(g_1,g_2,...,g_{n_1};q;g_{n_1+1},g_{n+2}...g_{n_1+n_2})$, with $q$
being the outgoing fermion.  The recursion relation for $\bar Q$ reads
\be
\begin{split}
\bar Q = & 
\frac{i}{\sqrt{2}}  
\sum \limits_{m=0}^{n_2-1} 
G^\nu(g_{n_1+m+1},..,g_{n_1+n_2})
\bar Q (g_1,..g_{n_1};q;g_{n_1+1},..g_{n_1+m})
\, \\ &
\times 
\hat S^F(p_{q}+k_{1;n_1+m})\hat \Gamma^\mu   S^G_{\mu\nu}(k_{n_1+m+1;n_1+n_2}) 
\\
 - &
\frac{i}{\sqrt{2} } 
\sum \limits_{m=1}^{n_1} 
G^\nu(g_{1},..,g_{m})
\bar Q (g_{m+1},..g_{n_1};q;g_{n_1+1},..g_{n_2})
\\&
\times \hat S^F(p_{q}+k_{m+1;n_1+n_2}) \hat \Gamma^\mu 
S^G_{\mu\nu}(k_{1;m})
\,,
\end{split} 
\label{bg2}
\ee
where $p_{q}$ is the momentum carried by the outgoing fermion and $
\hat S^F(p) = i/(\hat p - m)$ is a fermion propagator.  We have used
the notation $\hat \Gamma_\mu$ in \Eq(\ref{bg2}) to denote the
$D$-dimensional Dirac matrix and to emphasize that the recursion
relation can be continued to higher-dimensional space-times in a
straightforward way.  This relation is illustrated in
Fig.~\ref{qqBG_fig}.  The initial condition for the current $\bar Q$
is given by the massless or massive Dirac-conjugate spinor $\bar u $.

\begin{figure}[t]
\begin{center}
\hspace{-0.5cm}
\includegraphics[scale=0.57]{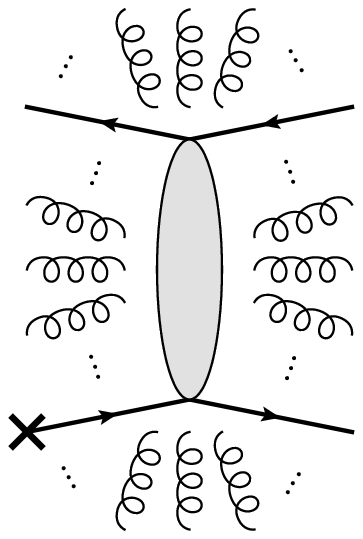}
\hspace{0.5cm}
\quad \qquad
\raisebox{-0.00cm}{
\includegraphics[scale=0.57]{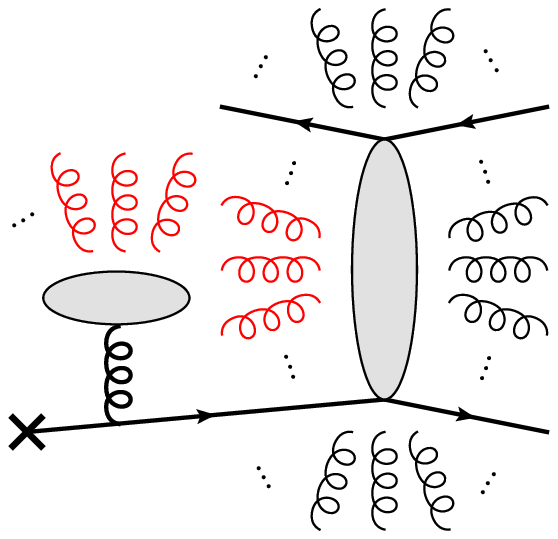}}
\quad \qquad 
\includegraphics[scale=0.57]{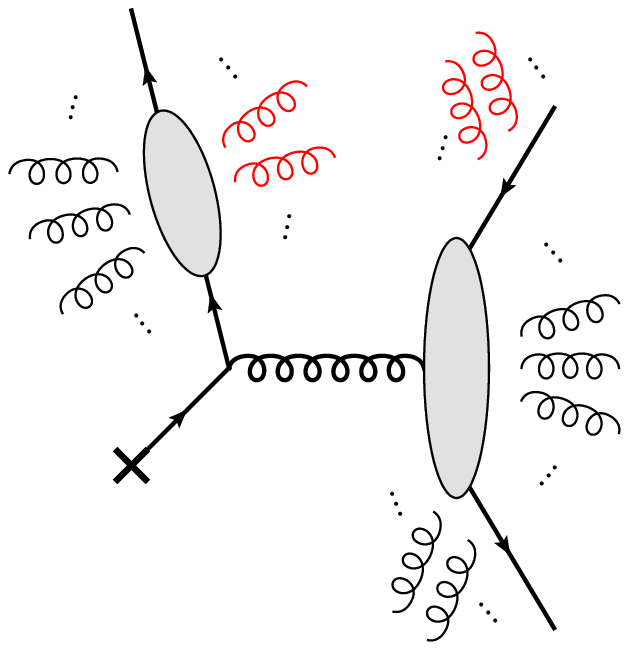}

\vspace{0.5cm}

\hspace{2cm}\quad \qquad 
\raisebox{-0.0cm}{
\includegraphics[scale=0.57]{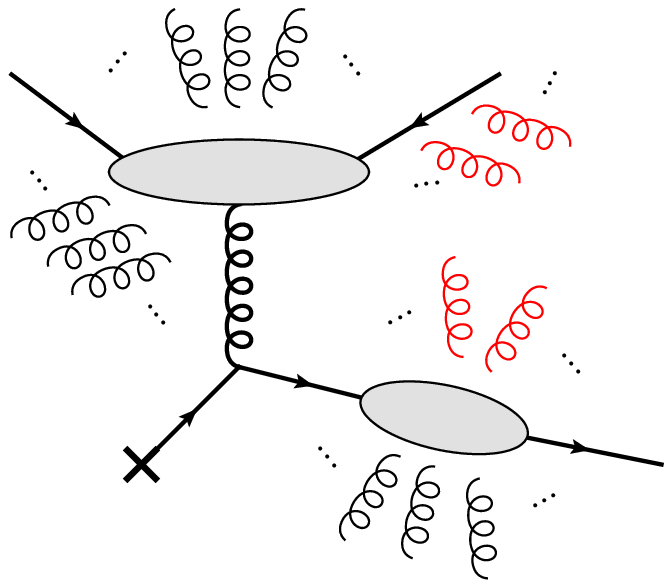}}
\quad \qquad 
\includegraphics[scale=0.57]{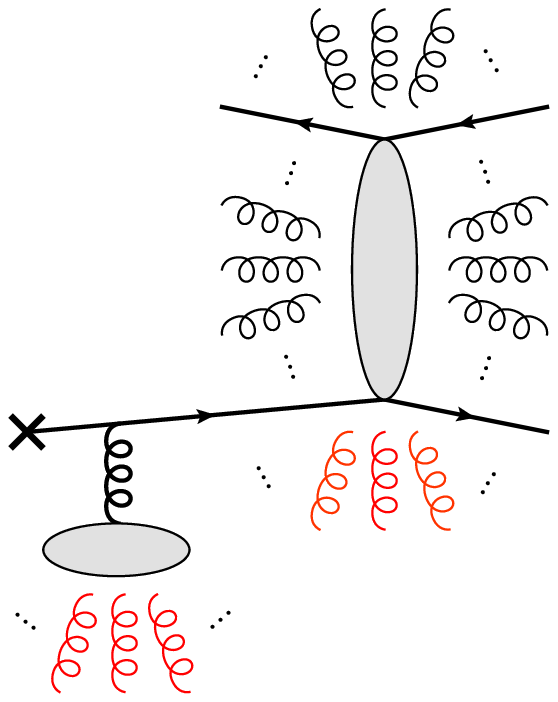}

\end{center}
\caption{Recursion relation for the four-quarks and n-gluon current, \Eq(\ref{bg3}).}
\begin{picture}(0,0)
\put(-15,215){$ \displaystyle n_1$}
\put(22,265){$ \displaystyle n_2$}
\put(60,215){$ \displaystyle n_3$}
\put(22,160){$ \displaystyle n_4$}
\put(123,240){$ \displaystyle {\color{red}m}$}
\put(136,195){$ \displaystyle {\color{red}n_1-m}$}
\put(283,238){$ \displaystyle {\color{red}m}$}
\put(286,277){$ \displaystyle {\color{red}n_2-m}$}
\put(197,115){$ \displaystyle {\color{red}m}$}
\put(170,98){$ \displaystyle {\color{red}n_3-m}$}
\put(250,33){$ \displaystyle {\color{red}n_4-m}$}
\put(300,58){$ \displaystyle {\color{red}m}$}
\put(72,210){$ \displaystyle = \sum_{m=1}^{n_{1}}$}
\put(205,210){$ \displaystyle + \sum_{m=0}^{n_{2}}$}
\put(72,83){$ \displaystyle + \sum_{m=0}^{n_{3}}$}
\put(205,83){$ \displaystyle + \sum_{m=0}^{n_{4}-1}$}
\end{picture}
\label{fig:4Qrec}
\end{figure}

As a further example, we present the recursion relation for the
current that defines the primitive amplitude that involves two
quark-pairs of the same flavour and an arbitrary number of gluons. To
define the current, we take one of the incoming fermions off the mass
shell. All partons are ordered; the current is written as
$\bar Q_{Q \bar Q Q}(g_{1...n_1}; q_1;g_{n_1+1, ... , n_{12}};
\bar q_2;g_{n_{12}+1, ... ,n_{123}}; q_3;g_{n_{123}+1, ... ,n_{1234}})$.
Here, $q_1, \bar q_2$ and $q_3$ are the  outgoing quark,
anti-quark and quark respectively, $n_{ijk\dots}=n_i+n_j+n_k+\dots$,
and $g_{i_1+1, ... i_2}$ denote the  $i_2-i_1$ gluons
$g_{i_1+1} \dots g_{i_2}$.  The recursion relation for 
$\bar Q_{Q \bar Q Q}$
reads 
\ba
&&\bar Q_{Q \bar Q Q}(g_{1...n_1}; q_1;g_{n_1+1, ... , n_{12}};
\bar q_2;g_{n_{12}+1, ... ,n_{123}}; q_3;g_{n_{123}+1, ... ,n_{1234}})= 
\nonumber 
\\
&& 
\frac{-i}{\sqrt{2}}
\sum_{m=1}^{n_{1}} 
\bar Q_{Q \bar Q Q} \left ( g_{m+1,..,n_1}; q_1;g_{n_1+1,..,n_{12}};
\bar q_2;g_{n_{12}+1,.. ,n_{123}};
q_3;g_{n_{123}+1,..,n_{1234}} \right )
\nonumber  \\
&& 
\;\;\;\;\;\;\;\;\; \times \hat S^F(q_{123}+g_{m+1;n_{1234}}) 
\hat \Gamma^\mu 
G^{\nu}(g_{1\dots m})
S^G_{\mu\nu}(g_{1;m}) 
\nonumber  \\
&&+
\frac{i}{\sqrt{2}}
\sum \limits_{m=0}^{n_2} 
\bar Q (g_1,..g_{n_1};q_1;g_{n_1+1},..g_{n_1+m})
\hat S^F(q_{1}+g_{1;n_1+m}) \hat \Gamma^\mu 
\label{bg3}
 \\
&& 
\times  S^G_{\mu\nu}(q_{23}+g_{n_1+m+1;n_{1234}}) 
G_{\bar QQ}^{\nu}(g_{n_1+m+1;n_{12}};\bar q_2;g_{n_{12}+1, ... ,n_{123}}; q_3;g_{n_{123}+1, ... ,n_{1234}})
\nonumber \\
&&- 
\frac{i}{\sqrt{2}}
\sum \limits_{m=0}^{n_3} 
\bar Q(g_{n_{12}+m+1, ... ,n_{123}}; q_3;g_{n_{123}+1, ... ,n_{1234}})
\hat S^F(q_{3}+g_{n_{12}+m+1,n_{1234}}) \hat \Gamma^\mu 
\nonumber \\
&&
\times G_{Q \bar Q}^{\nu}(g_{1...n_1}; q_1;g_{n_1+1, ... , n_{12}};
\bar q_2;g_{n_{12}+1, ... ,n_{12}+m})
S^G_{\mu\nu}(q_{12}+g_{1;n_{12}+m}) 
\nonumber \\ 
&&+ 
\frac{i}{\sqrt{2}}
\sum_{m=0}^{n_{4}-1} 
\bar Q_{Q\bar Q Q}(g_{1,...,n_1}; q_1;g_{n_1+1, ... , n_{12}};
\bar q_2;g_{n_{12}+1, ... ,n_{123}}; q_3;g_{n_{123}+1, ... ,n_{123}+m}) 
\nonumber \\
&& \times \hat S^F(q_{123}+g_{1;n_{123}+m}) \hat \Gamma^\mu 
G^{\nu}(g_{n_{123}+m+1\dots n_{1234}})
 S^G_{\mu\nu}(g_{n_{123}+m+1;n_{1234}})\,. 
\nonumber 
\ea
In \Eq(\ref{bg3}) $G_{Q\bar Q}^\mu$ and $G_{\bar Q Q}^\mu$ denote
currents with an off-shell gluon, a quark-antiquark pair and an
arbitrary number of gluons in the final state. We do not give the
explicit recursion relations for these currents, since they are very
similar to the recursion relation shown in \Eq(\ref{bg3}).
\Eq(\ref{bg3}) holds also for four-quark amplitudes with two
distinct quark flavours as long as the gluon currents $G_{Q \bar Q}$
and $G_{\bar Q Q}$ are set to zero.  It is clear that the right hand
side of \Eq(\ref{bg3}) involves simpler lower-multiplicity currents.
This relation is illustrated in Fig.~\ref{fig:4Qrec}.  The initial
condition for the current can be obtained by setting $n_i=0$,
$i=1,2,3,4$, in \Eq(\ref{bg3}). One obtains 
\be
\begin{split}
{\bar Q}_{Q \bar Q  Q}(q_1;\bar q_2;q_3)&= 
\frac{i}{\sqrt{2}}
\bar Q (q_1)
\hat S^F(q_{1}) \hat 
\Gamma^\mu G^\nu_{\bar Q Q}(\bar q_2;q_3)\hat S^G_{\mu\nu}(q_{23}) 
\\
&-\frac{i}{\sqrt{2}}
\bar Q(q_3) \hat S^F(q_{3}) \hat \Gamma^\mu   G^\nu_{Q \bar Q}(q_1;\bar q_2)
\hat S^G_{\mu\nu}(q_{12})\,. 
\\ 
\end{split} 
\label{bg3init}
\ee

The above construction generalizes to even more complicated currents,
with e.g. larger number of fermion pairs.
Currents that satisfy recurrence relations can also be constructed to
compute scattering amplitudes with color-neutral particles, that are
required to describe the production of the electroweak gauge bosons
and the Higgs boson.  Such currents have been heavily used in recent
one-loop computations, see e.g.
Refs.~\cite{Melia:2010bm,KeithEllis:2009bu,Frederix:2010ne,
  Melnikov:2011ta,Melia:2011dw}.

\subsection{Britto-Cachazo-Feng-Witten relations for on-shell 
amplitudes}
\label{bcfw}

Berends-Giele recursion relations constitute an important tool for
{\it numerical} computations of one-loop scattering amplitudes. There
are however also a number of cases when those relations can be used to
derive properties of scattering amplitudes analytically. Interesting
examples can be found in a review~\cite{Dixon:1996wi} and we do not
repeat them here. In this Section, we describe a different example of
the application of the Berends-Giele recursion -- a derivation of
relations between tree-level on-shell scattering amplitudes involving
a different number of particles, first discovered by Britto, Cachazo,
Feng and Witten (BCFW) \cite{Britto:2005fq}.

We begin by considering the $n$-gluon color-ordered tree-level
scattering amplitude ${\cal
  M}(g_{\lambda_1},g_{\lambda_2},..,g_{\lambda_n})$, where all gluon
momenta are incoming. We assume that $n > 4$.  It is well-known that,
in order to be non-zero, the amplitude must contain at least two
helicities (say, minus one) that differ from all the other helicities
(say, plus one).  We fix the helicity of the first gluon to be
negative; then, it is always possible to find a gluon of positive
helicity in a position that is not adjacent to the gluon $g_1$.  We
will therefore study the amplitude ${\cal M}(1^-, \ldots, j^+, \ldots
,n)$, $j \neq 2,n$.

Following \cite{ArkaniHamed:2008yf}, it is convenient to choose a
reference frame where the momentum of the gluon $g_1$ is $p_1 = \mu
(1,{\bf 0}_\perp,1)$ and the momentum of the gluon $g_j$ is $p_j = \mu
(1,{\bf 0}_\perp,-1)$, where $\mu$ is a constant with the mass
dimension one.  We choose the polarization vectors to be $\epsilon_1^+
= \epsilon_j^- = q/\sqrt{2}$, where $q = (0,1,i,0)$.  We can now
deform the momenta of the gluons $g_{1,j}$ by shifting them in the
$q$-direction
\be
p_1 \to p_1(z) = p_1 +  z\;\mu q,\;\;\;
p_j \to p_j(z) = p_j -  z\; \mu q,
\label{eq685}
\ee
where $z$ is a complex parameter.  It follows from \Eq(\ref{eq685})
that $p_1(z) + p_j(z) = p_1 + p_j$ and, since vectors $p_{1,j}$ and
$q$ are orthogonal, $p_1^2(z) = p_j^2(z) = 0$.  Finally, $p_1(z) \cdot
\epsilon_1 = 0$ and $p_j(z) \cdot \epsilon_j = 0$, which implies that
both $\epsilon_1$ and $\epsilon_j$ are valid polarization vectors for
gluons with shifted momenta.

If the amplitude ${\cal
  M}(g_{\lambda_1,z},...g_{\lambda_j,z},...g_{\lambda_n})$ is
calculated with the momenta $p_{1,j}(z)$, it becomes a rational
function of $z$ since the only place such a dependence enters is
through the momenta of gluons $g_{1}$ and $g_{j}$.  If we assume that
the amplitude ${\cal M}(z)$ vanishes at $z = \infty$, so does the
integral of ${\cal M}(z)$ over an infinitely remote contour
\be
\oint \limits_{|z|=\infty}^{} \frac{{\rm d} z}{z} {\cal M}(z) = 0.
\label{eqww13}
\ee
The integration in \Eq(\ref{eqww13}) can be performed using 
Cauchy's theorem. We obtain 
\be
{\cal M}(0) + \sum \limits_{z = z_\alpha \neq 0}^{}
{\rm Res} \left[  \frac{{\cal M}(z)}{z} \right ] = 0,
\label{eqbasic}
\ee
where the sum goes over all poles of the amplitude 
${\cal M}(z)$ in the complex $z$-plane.

To understand where these poles come from, we 
consider all (cyclic) 
ordered partitions of the set $ \pi = \{1,2,..j,j+1, \ldots n \}$ 
into two sets $\pi = \left \{ \pi_1^{\alpha} \cup \pi_j^{\alpha} \right \}$,
$\alpha =1, \ldots \alpha_{\rm max}$  such 
that $\pi_1^{\alpha}$ contains $g_1$ and $\pi_j^{\alpha}$ contains 
$g_j$.  We consider now the value of the parameter 
$z$ close to $z_\alpha$, 
defined by the following equation 
\be
z_{\alpha} = -\frac{P_{\pi_1^{\alpha}}^2}{2 \mu P_{\pi_1^{\alpha}} \cdot q},
\;\;\;P_{\pi_1^{\alpha}} = \sum \limits_{i \; \subset \; \pi_1^{\alpha}} p_i, 
\ee
and observe that the dominant contribution to the scattering amplitude 
comes from the resonant term 
\be
\begin{split} 
\label{eqbasic1}
\lim \limits_{z \to z^\alpha}  
  \frac{{\cal M}(z)}{z} &  \approx 
  \frac{-i {\cal M}^\mu (\pi_1^{\alpha}) 
g_{\mu \nu} {\cal M}^\nu (\pi_j^{\alpha})   
}{2 \mu \left ( P_{\pi_1^\alpha}  \cdot q 
\right ) \left ( z - z_\alpha \right ) z_\alpha}
\\
& = \frac{i}{\left ( z - z_\alpha \right ) P_{\pi_1^{\alpha}}^2} 
\times \sum \limits_{\lambda=\pm} 
 {\cal M} (\pi_1^{\alpha},\lambda) 
{\cal M} (\pi_j^{\alpha},\lambda).
\end{split}
\ee
In the second step in \Eq(\ref{eqbasic1}), we used gauge invariance of
on-shell amplitudes to replace the metric tensor $g_{\mu \nu}$ with
the sum over polarizations of the nearly on-shell, intermediate gluon
line.  We use \Eq(\ref{eqbasic1}) to compute the residue in
\Eq(\ref{eqbasic}) and obtain the final expression for the on-shell
scattering amplitude
\be
{\cal M}(0)  = - i
\sum \limits_{\alpha=1}^{\rm \alpha_{\rm max}} 
\sum \limits_{\lambda=\pm} 
 \frac{{\cal M}(\pi_1^{\alpha},\lambda) 
 {\cal M} (\pi_j^{\alpha},\lambda)}{P_{\pi_1^{\alpha}}^2}.
\label{eqbasic2}
\ee
The striking feature of \Eq(\ref{eqbasic2}) is that it expresses the
desired on-shell scattering amplitude through on-shell scattering
amplitudes of {\it lower multiplicities}, evaluated at complex
on-shell momenta.  This relation between scattering amplitudes was
derived in Ref.~\cite{Britto:2005fq}; it is known as the BCFW
relation.

The derivation of the BCFW relation, that we just described, relies on
the fact that the amplitude ${\cal M}(z)$ vanishes at $z \to
\infty$. To prove this feature of QCD amplitudes, we will use
Berends-Giele recursion relations.  We begin by deriving constraints
imposed by gauge invariance.  If we write the scattering amplitude as
${\cal M}(z) = \epsilon_{1,\mu}(z) A^{\mu}(z)$, gauge invariance
requires $p_1^\mu(z) A_{\mu}(z) = 0$. Using the explicit form of
$p_1(z)$ and the fact that $\epsilon_{1,\mu} = q_\mu/\sqrt{2}$, we
find
\be
\epsilon_{1,\mu} A^{\mu}(z) = -\frac{p_1^\mu A_\mu(z)}{\sqrt{2}\mu z}.
\label{bg6}
\ee
To compute $A_\mu(z)$, we can use Berends-Giele recursion relation for
the gluon current. The recursion relation for the gluon current
involves two or three lower-multiplicity currents. One of those
currents includes gluon $g_j$ that, in the limit of large $z$, carries
large momentum; we will refer to such a current as ``hard'' .  We
denote the hard (soft) current by $H_\mu$ ($S_\mu$) and the (outgoing)
momentum that it carries by $p_H$ ($p_S$). In contrast to the previous
Section, it is convenient to include propagators associated with
off-shell legs into the definition of the currents.  Contracting the
hard and soft currents with three- and four-gluon vertices, and using
the transversality of the currents, we obtain
\be
\begin{split}
H_\mu  \sim & (p_H+p_S)^{-2} 
\Big (
2 (p_H \cdot S) H_\mu + (p_S - p_H)_\mu (H \cdot S)
\\
 & - 2 (p_S \cdot H) S_\mu 
+ V_{4,\mu}(S, H, S) 
\Big ). 
\label{eq_rr}
\end{split}
\ee
The large-$z$ limit corresponds to $p_H = - z \mu q + {\cal O}(1) $,
$p_S \sim O(1)$. We will make the assumption that the hard current
scales with $z$ as $H \sim {\cal O}(1)$ and we will show that the
recursion relation is consistent with this scaling.  To this end note
that, under the scaling assumption, the four-gluon vertex contribution
in \Eq(\ref{eq_rr}) is suppressed as $z^{-1}$ and can be disregarded.
We obtain
\be
\begin{split} 
H_\mu  & \sim z^{-1}  
\left ( 2z (q \cdot S) H_\mu - z q_\mu (H \cdot S) 
-2 (p_s \cdot H) S_\mu \right ) 
\\
& \sim  (2 q \cdot S) H_\mu - q_\mu (H \cdot S) + {\cal O}(z^{-1}),
\end{split}
\label{bg4}
\ee
so that indeed $H_\mu \sim {\cal O}(1)$ in the large-$z$ limit.
We note that at large $z$ there 
are two terms in the recursion relation -- one that is proportional 
to $H_\mu$ and the other proportional to $q_\mu$. Since the initial condition 
for the recursion is $H_\mu = \eps_{j,\mu} \sim q_\mu$, we conclude that 
the solution of the recursion relation is 
\be
H^\mu \sim J q^\mu + {\cal O}(z^{-1}),
\ee 
where the constant $J$ is $z$-independent. 
To obtain the amplitude from the current, it needs to be multiplied by 
the off-shell  propagator $(p_H+p_S)^2$ that  scales as $z$. 
Therefore, we find 
\be
A^\mu(z) \sim z H^\mu \sim J z q^{\mu} + {\cal O}(1).
\label{eqfirst_a}
\ee
We use \Eq(\ref{eqfirst_a}) in  \Eq(\ref{bg6}) and 
derive the scattering amplitude
\be
{\cal M}(z) = \epsilon_{1,\mu} A^{\mu}(z) = -\frac{p_1^\mu A^\mu(z)
}{\sqrt{2} \mu z} 
= 
- \frac{J p_1 \cdot q}{\sqrt{2}\mu} + {\cal O}(z^{-1}) = {\cal O}(z^{-1}),
\label{eqfin}
\ee
where the last step follows from the orthogonality of $p_1$ and $q$.
\Eq(\ref{eqfin}) proves that the on-shell scattering amplitude
vanishes in the limit $z \to \infty$.  Finally, we note that, since
the Berends-Giele recursion relation can be easily continued to
higher-dimensional space-times, the above derivation shows that the
BCFW relation \Eq(\ref{eqbasic2}), originally derived in four
dimensions, remains valid in a space-time of {\it arbitrary
  dimensionality}.

\subsection{Computations with massive particles}

Unitarity-based computations are often discussed in the context of
massless particles. Apart from well-known technical simplifications
that are possible in massless theories, but are harder to achieve when
massive particles are involved, there are important conceptual issues
that must be understood before incorporating massive particle into the
unitarity-based framework for loop computations.  Potential
complications can already be seen from the fact that when massive
particles are involved, the full basis of master integrals includes
tadpole integrals that do not have a discontinuity in any variable
related to the external kinematics.  However, when the generalized
unitarity technique for one-loop computations is viewed as the
consequence of the OPP procedure, it becomes perfectly clear that
massive particles are straightforwardly incorporated into the
unitarity framework.  Nevertheless, peculiar features appear when
unitarity and massive particles are combined; the goal of this Section
is to mention them. In our discussion we closely follow
Ref.~\cite{Ellis:2008ir}; we also note that all aspects of the
following discussion have been checked in realistic computations of
NLO QCD corrections to top quark pair production reported in
Refs.~\cite{Melnikov:2010iu,Melnikov:2009dn,Melnikov:2011ta}.
Analytic computations of one-loop helicity amplitudes for $t \bar t$
production within generalized unitarity framework can be found in
\cite{Badger:2011yu}.

An important difference between computations with massless and massive
particles is that in the former case there is no renormalization
beyond the coupling constant. This happens because all one-particle
reducible diagrams, usually associated with mass- and wave-function
renormalization of external particles vanish in the massless case,
provided that dimensional regularization is used.  However, those
quantities do not vanish in the massive case. If we define the
renormalization constants for the quark mass $m$ and for the quark
wave function $\psi$ in the on-shell scheme as
\be
m_0 = Z_m m,\;\;\;\;\psi_0 = \sqrt{Z_2} \psi,
\ee
in {\it any} covariant gauge we obtain 
\be
Z_m = Z_2 = 1 - 
C_F g_s^2 c_\Gamma \left ( \frac{\mu^2}{m^2} \right )^\ep 
\left ( \frac{3}{\epsilon} + 5 - \eta \right ). 
\label{eqwf}
\ee
We use parameter $\eta$ to distinguish between the four-dimensional
helicity scheme ($ \eta = 0) $ and the 't~Hooft-Veltman scheme ($\eta
= 1$).  We point out that the equality at one-loop between the two
renormalization constants is fortuitous, since $Z_m$ has only
ultraviolet divergences while $Z_2$ contains both ultraviolet and
infrared divergences.

The need to apply wave-function renormalization to remove ultraviolet
divergences from a scattering amplitude is related to the fact that
one-particle reducible self-energy corrections to external lines are
non-vanishing.  When one-loop scattering amplitudes are computed in a
conventional diagrammatic framework, one-particle reducible diagrams
are simply discarded, and their effect is accounted for by the
wave-function renormalization constants.  Disregarding one-particle
reducible contributions is straightforward in the diagrammatic
approach but it becomes more subtle, if the scattering amplitude is
computed from unitarity cuts.

\begin{figure}[t]
\begin{center}
\includegraphics[scale=0.4,angle=-90]{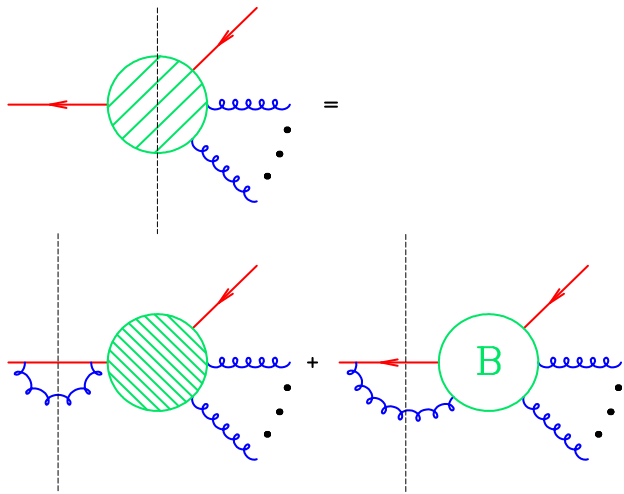}
\end{center}
\caption{A particular cut of a general amplitude with massive quarks
  and gluons that splits it into a self-energy contribution and a
  higher-point function. The different shadings of the blobs represent
  different content. }
\label{figtop}
\end{figure}

To illustrate this, we consider one-loop corrections to a scattering
amplitude of a pair of massive quarks and any number of gluons.  We
focus on a particular double-cut of that scattering amplitude, where
the heavy quark is on one side of the cut and every other particle is
on the other side, see Fig.~\ref{figtop}. The double cut of the
one-loop amplitude is given by the product of two tree amplitudes,
\be
{\rm Res}[{\cal M}(q,\{q,g\})] =
\sum \limits_{\rm states} {\cal M}_0 (q,q^*, g^*) {\cal M}(q^*,g^*,\{q,g \}).
\label{t4}
\ee
The amplitude ${\cal M}_0$ describes the splitting of a massive fermion 
into an on-shell fermion and a gluon; in turn, the amplitude 
${\cal M}(q^*,g^*,\{q,g\})$ describes scattering of the quark $q^*$ 
and the gluon $g^*$ into the final state particles. On general grounds, 
this  amplitude can be written as a sum of two terms
\be
{\cal M}(q^*,g^*,\{q,g\}) = 
\frac{R(p_{q^*},p_{g^*},\{q,g\})\}}{(p_{q^*} + p_{g^*})^2 - m^2}
+ B(p_{q^*},p_{g^*},\{q,g\}),
\label{t5}
\ee
where the $B$-amplitude is non-singular in the limit $(p_{q^*} +
p_{g^*})^2 \to m^2$. In the case of interest, momentum conservation
forces the sum of the two momenta $p_{q^*},p_{g^*}$ to be equal to
$p_q$, so that $(p_{q^*} + p_{g^*})^2 = p_q^2 = m^2$ and the first
term in \Eq(\ref{t5}) becomes infinite.  By reconstructing the diagram
whose double cut corresponds to the product of ${\cal M}_0(q,q^*,g^*)$
and the singular term in \Eq(\ref{t5}), it is easy to understand that
it corresponds to a {\it one-particle reducible} diagram -- the
self-energy insertion on an external massive fermion line.

In principle, when the one-loop amplitude is calculated recursively,
it is easy to disregard the singular term by truncating the recursion
steps.  This corresponds to setting
\be
M(q^*,g^*,\{q,g\}) \to  B(p_{q^*},p_{g^*},\{q,g\})
\ee
in \Eq(\ref{t5}).  Such a replacement makes the matrix element finite
but it introduces a problem since $B(p_{q^*},p_{g^*},\{q,g\})$ is not
gauge-invariant.  Gauge invariance is restored when the $B$-amplitude
is combined with the on-shell wave-function renormalization constant
shown in \Eq(\ref{eqwf}).\footnote{ We note that the on-shell
  wave-function renormalization constant $Z_2$, computed in
  dimensional renormalization, is gauge-parameter independent in
  $R_\xi$-covariant gauges through two loops in QCD
  \cite{Broadhurst:1991fy}.  However, it is gauge-dependent since
  $Z_2$ in e.g. the light-cone gauge and $Z_2$ in the covariant gauges
  differ.} This requires that the $B$-amplitude and the wave function
renormalization constant are computed in the same gauge.  It is
easiest to compute the wave-function renormalization constant in one
of the covariant gauges; the result is given in
\Eq(\ref{eqwf}). Hence, if we want to use $Z_2$ shown in
\Eq(\ref{eqwf}) in an actual computation, we need to compute the
$B$-amplitude also in a covariant gauge and we choose the Feynman
gauge for that purpose.  We implement the Feynman gauge in our
calculation by introducing two unphysical polarizations for the
``cut'' gluon line with momentum $g^*$, so that the sum over
polarizations reads
\be
\sum \limits_{\lambda=1}^{4} 
\epsilon^\mu(\lambda) \epsilon^{\nu,*}(\lambda) = 
- g^{\mu \nu}.
\ee
We emphasize that this replacement should only be done for the
computation of the $B$-amplitude, for a very special type of double
cut. For other cuts, it is sufficient to sum over physical
polarization states, thanks to gauge invariance of tree amplitudes.

Another important feature of calculations with massive {\it internal}
particles is the existence of a non-vanishing double-cut contribution
for light-like momenta.  Because the square of the light-like momentum
vanishes, the construction of the traditional van Neerven - Vermaseren
basis becomes impossible. We have explained in the final paragraph of
Sect.~\ref{sec5.2} how to proceed in this case; but we mention this
feature here for completeness.  Finally, we point out that when
massive internal particles are present, single-particle cuts appear.
While the calculation of those cuts is straightforward, it is
important to keep in mind that also in that case one-particle
reducible contributions to amplitudes cause trouble. The problematic
contributions correspond to massive tadpoles diagrams and, just as is
done in a diagrammatic computation, these contributions should be
discarded. As we already pointed out in the context of the double-cut
computation, within the unitarity framework this is accomplished by
truncating steps in Berends-Giele recursion.

\section{Analytic computations} 
\label{analytic}

In this Section we discuss some recent ideas related to the
possibility to compute the reduction coefficients directly.  We
describe approaches suggested by Forde \cite{Forde:2007mi} and
Mastrolia \cite{Mastrolia:2009dr} to the computation of
cut-constructible parts of one-loop amplitudes and by Badger
\cite{Badger:2008cm} to the computation of the rational part.  Our
discussion is motivated by
Refs.~\cite{Forde:2007mi,Mastrolia:2009dr,Badger:2008cm}, but some
details are different. In particular, in contrast to original
references we do not use the spinor-helicity decomposition of the loop
momentum. Furthermore, we attempt to provide an intuitive geometric
picture behind the sophisticated choices of integration variables in
Refs.~{\cite{Forde:2007mi,Mastrolia:2009dr,Badger:2008cm}}.  
In this respect, our
discussion partially overlaps with the treatment of one-loop
amplitudes in general quantum field theories, given in
Ref.~\cite{ArkaniHamed:2008gz}.  For simplicity, we consider only
massless cases in this Section. The extension of the method of
Ref.~\cite{Forde:2007mi} to massive theories is given in
Ref.~\cite{Kilgore:2007qr}.

\subsection{Direct computation of the cut-constructible coefficients}
\label{direct}

To motivate the discussion that follows, we remind the reader that the
cut-constructible part of any one-loop amplitude can be written as a
linear combination of box, triangle, bubble and tadpole one-loop
integrals.  There is just one coefficient per master integral in such
a linear combination but, within the OPP framework, we calculate {\it
  two} coefficients per box integral, {\it seven} coefficients per
triangle, {\it nine} coefficients per bubble and {\it five}
coefficients per tadpole.  All but one of the coefficients for each master
integral vanishes after integration over the loop momentum.  Therefore,
the very fact that those terms are  computed, seems to imply that
the OPP procedure is not as efficient as, perhaps, it is possible to
make it.  Of course, as should be clear from the discussion in
preceding Sections, the large number of coefficients is computed on
purpose since we want to subtract the {\it full} contributions of
higher-point integrands to lower-point residues.  Therefore, any
inefficiency that may be present is not so easy to get rid of, but it is
interesting to investigate if the computation of certain terms can be avoided.
In this Section we explain, following Ref.~\cite{Forde:2007mi}, how
to obtain the required minimal set  of the reduction 
coefficients by simple  algebraic manipulations 
with  cuts of one-loop scattering amplitudes.

We begin with the easiest case --  the reduction coefficients for  
four-point functions.  As we explained above, there are two of them, 
but one vanishes upon integration over the loop momentum. Hence,  
we would like to avoid computing that coefficient.

For definiteness, we consider the four-point
master integral that contains inverse Feynman propagators
$d_0,d_1,d_2,d_3$.  The corresponding reduction coefficient is
calculated from the quadruple cut \cite{Britto:2004nc}
\be
{\tilde d}^{(0)}_{0123} =
\frac{1}{2} \sum \limits_{i= \pm}^{} 
A_1 (l_i) A_2 (l_i)  A_3 (l_i)  A_4 (l_i)\,, 
\label{quadd}
\ee
where $A_{1, \dots ,4}$ are the tree on-shell scattering amplitudes 
and 
$l_\pm$ are the two cut momenta that satisfy $d_j(l^\pm) = 0$,
for all $j=0,1,2,3$.  The result shown in \Eq(\ref{quadd}) immediately follows
from the discussion in the previous Section and, in particular, from
\Eq(\ref{howtod}).  We stress that $\tilde {d}_{0123}^{(0)}$ is the
only reduction coefficient that we need, to determine the contribution
of a particular four-point function to a one-loop scattering
amplitude.

Now, consider the triple-cut, that corresponds to the vanishing of the
inverse propagators $d_0,d_1,d_2$. The triple cut is parametrized by 
seven reduction coefficients; only one of them does not vanish 
upon integration over the loop momentum. We now discuss the algebraic 
procedure  that allows us to obtain the non-vanishing reduction 
coefficient directly. 

As explained in Sect.~\ref{sec5.2},
we can write the loop momentum on the cut as
\be
l^\mu = V^\mu + l_\perp (\cos \varphi \; n_3^\mu  + \sin \varphi \; n_4^\mu).
\label{eq546}
\ee
We remind the reader that in this Section we only address the
cut-constructible part. Therefore,  the vector $l$ is purely
four-dimensional and there is no dependence on $\ne$ in
\Eq(\ref{eq546}).  In addition, we reiterate that we are treating the
massless case, so the internal masses in the propagators are equal to
zero.  In \Eq(\ref{eq546}), $V^\mu$ is a constant vector, given by a
linear combination of vectors in the physical space (i.e.\ orthogonal
to $n_3$ and $n_4$ that span the four-dimensional transverse space)
and from $l^2=0$ it follows that $l_\perp = \sqrt{-V^2}$.  Introducing
$t_+ = e^{i\varphi}$, we can rewrite \Eq(\ref{eq546}) as
\be
l_+^\mu = V^\mu + l_{\perp} \left(t_+  n_-^\mu + t_+^{-1} n_+^\mu\right),
\label{eq547}
\ee
where $n_{\mp} = (n_3 \mp i n_4)/2$, so that $n_-^2 = n_+^2 = 0$ and
$2n_- n_+ = 1$.  The triple-cut of the one-loop amplitude, evaluated
at the momentum $l = l_+$, reads
\be
A_1(t_+) A_2 (t_+)  A_3(t_+) 
= {\tilde c}_{012}(l_+)\,
+ \sum \limits_{i=3}^{N}  
\frac{{\tilde d}^{(0)}_{012i} + {\tilde d}^{(1)}_{012i} 
({\tilde n}_i \cdot l_+)}{
d_i(l_+)}, 
\label{AAA}
\ee
where $\tilde n_i$ is the four-vector orthogonal to $q_1$, $q_2$ and
$q_i$. When writing \Eq(\ref{AAA}) we used the fact the loop momentum 
in \Eq(\ref{eq547}) depends on a single parameter $t_+$, therefore the
tree amplitudes depend on the loop momentum only through $t_+$. 
The inverse propagator $d_{i}(l_+)$,  $i \in \{3,..,N\}$, reads 
\be
d_{i}(l_+) = (q_i+l_+)^2 
= \Delta_i + 2 l_\perp \left ( q_i \cdot n_- \right )  t_+ + 
2 l_\perp \left ( q_i \cdot n_+ \right ) t_+^{-1}, 
\label{eq:denom1}
\ee
where $\Delta_i = q_i^2 + 2 q_i \cdot V$.  Also, since $V^{\mu}$ is a
linear combination of the propagator offset momenta $q_1^{\mu}$ and
$q_2^\mu$ in $d_1$ and $d_2$ respectively and since ${\tilde n}_i^\mu$
is orthogonal to both of these momenta, we find
\be
{\tilde d}^{(0)}_{012i} 
+{\tilde  d}^{(1)}_{012i} ({\tilde n}_i \cdot l_+)
= 
{\tilde d}_{012i}^{(0)} + 
l_\perp {\tilde  d}^{(1)}_{012i}  \left [
( {\tilde n}_i \cdot n_-) t_+ 
+ ( {\tilde n}_i \cdot n_+ ) t_+^{-1} \right].
\label{eq:num1}
\ee
We use \Eqs(\ref{eq:denom1},\ref{eq:num1}) to perform partial
fractioning with respect to the variable $t_+$.  We obtain
\be
\begin{split}
\frac{{\tilde d}^{(0)}_{012i} 
+{\tilde  d}^{(1)}_{012i} ({\tilde n}_i \cdot l_+)}{d_i(l_+)}
= {\tilde d}^{(1)}_{012i} 
\frac{( {\tilde n}_i \cdot n_- ) }{\left (2 q_i \cdot n_- \right  )} 
+ \frac{r_{i,1}}{t_+-t_{i}^{(1)}} + \frac{r_{i,2}}{t_+-t_i^{(2)}}\,,
\end{split}
\label{d0012i}
\ee
where $t_i^{(1)}, t_i^{(2)}$ are values of $t_+$ for which $d_i(l_+)$
vanishes and $r_{i,1}, r_{i,2}$ are independent of $t^+$.
 
The most general parametrization of the triangle residue ${\tilde
  c}_{012}(l_+)$ is given in \Eq(\ref{eq_c_coeff}). We can set
$n_\epsilon \to 0$ there since in this Section we are interested in
the cut-constructible part only.  If we write ${\tilde c}_{012}$ in
terms of the variable $t_+$, we find
\be
\label{eq7.8}
{\tilde c}_{012}(l_+) = {\tilde c}_{012}^{(0)} 
+ \sum \limits_{k=-3,k\neq 0}^{3} {c}^{(k)}_{012} t_+^k,
\ee
where ${\tilde c}^{(0)}_{012}$ is the reduction coefficient of the
three-point function.  Putting box and triangle coefficients together,
using \Eqs(\ref{AAA},\ref{d0012i},\ref{eq7.8}), 
we derive the following decomposition of the triple cut of the
integrand
\be
\begin{split}
A_1(t_+)A_2(t_+)A_3(t_+) & = 
 \sum \limits_{i=3}^{N} 
\left(
\sum \limits_{j=1}^{2} \frac{r_{i,j}}{t_+-t_i^{(j)}}
+
{\tilde d}^{(1)}_{012i} 
\frac{( {\tilde n}_i \cdot n_- ) }{\left (2 q_i \cdot n_- \right  )} 
\right) \\  
&+ {\tilde c}^{(0)}_{012}
+ \sum \limits_{k=-3, k \neq 0}^{k=3} {c}^{(k)}_{012} t_+^k.
\label{eq255}
\end{split}
\ee
The question we address in the following is how to identify and
extract the ${\tilde c}_{012}^{(0)}$ coefficient if the left hand side
of \Eq(\ref{eq255}) is {\it analytically} known as a function of $t$.
To this end, we define an operator ${\cal L}_{t,m}$ such that when it
acts on a rational function $F(t)$, it picks up the coefficient of the
${\cal O}(t^m)$ term in the Laurent expansion of $F(t)$ at $t =
\infty$.\footnote{This operator is similar to the operator ${\rm Inf}$
  introduced in Ref.~\cite{Forde:2007mi}.}  Applying this operator to
both sides of \Eq(\ref{eq255}), we obtain
\be
{\cal L}_{t_+,0}[A_1(t_+) A_2(t_+) A_3(t_+) ] 
= {\tilde c}_{012}^{(0)} 
+ \sum \limits_{i=3}^{N} {\tilde d}^{(1)}_{012i} 
\frac{( {\tilde n}_i \cdot n_- ) }{\left (2 q_i \cdot n_- \right  )}. 
\label{eq458}
\ee
\Eq(\ref{eq458}) shows that we almost succeeded in computing the
reduction coefficient of the triple-cut directly, except that in
\Eq(\ref{eq458}) there are still contributions from the {\it
  evanescent coefficients} of the quadruple residue, ${\tilde
  d}^{(1)}_{012i}$, that need to be removed.  To accomplish this, we
simply repeat the whole procedure taking the cut momentum in
\Eq(\ref{eq546}) to be
\be
l_-^\mu = V^\mu + l_{\perp} \left(t_-  n_-^\mu + t_-^{-1} n_+^\mu\right),
\label{eq549}
\ee
with $t_{-} = 1/t_{+}$.  Following the discussion that leads to
\Eq(\ref{eq546}), with the obvious change $t_+ \to t_-$ where
appropriate, we find
\be
{\cal L}_{t_-,0}[A_1(t_-) A_2(t_-) A_3(t_-) ] 
= {\tilde c}^{(0)}_{012} + \sum \limits_{i=3}^{N} 
{\tilde d}^{(1)}_{012i} 
\frac{( {\tilde n}_i \cdot n_+ ) }{\left (2 q_i \cdot n_+ \right  )}. 
\label{eq459}
\ee
Taking the average of \Eqs(\ref{eq458},\ref{eq459}), we obtain the
final result for the triple cut coefficient
\be
{\tilde c}^{(0)}_{012} = \frac{1}{2} \sum \limits_{i=\pm}^{} 
{\cal L}_{t_i,0}[A_1(t_i) A_2(t_i) A_3(t_i) ].
\label{eq460}
\ee
In deriving \Eq(\ref{eq460}), we used the identity
\be
\label{eq461}
\begin{split}
\frac{( {\tilde n}_i \cdot n_- ) }{\left ( q_i \cdot n_- \right  )} 
+ \frac{( {\tilde n}_i \cdot n_+ ) }{\left ( q_i \cdot n_+ \right  )}
& = 
\frac{
( 
{\tilde n}_i \cdot n_-) \left ( q_i \cdot n_+ \right  )
+ 
( {\tilde n}_i \cdot n_+ ) \left ( q_i \cdot n_- \right  )
}{
\left ( q_i \cdot n_- \right  ) 
\left ( q_i \cdot n_+ \right  )
} 
\\
& = 
\frac{ {\tilde n}_i^\mu \omega_{\mu \nu}(q_1,q_2)  q_i^\nu 
}
{
2\left ( q_i \cdot n_- \right  ) 
\left ( q_i \cdot n_+ \right  )
} = 
\frac{ {\tilde n}_i \cdot q_i 
}
{
2\left ( q_i \cdot n_- \right  ) 
\left ( q_i \cdot n_+ \right  )
} = 
0,
\end{split}
\ee
where $\omega_{\mu\nu}(q_1,q_2)$ is the metric tensor of the
transverse space orthogonal to vectors $q_1$ and $q_2$. Also, we note
that in the last step in \Eq(\ref{eq461}), we used the orthogonality
of vectors ${\tilde n}_i$ and $q_1,q_2,q_{i \in [3,..N]}$.
\Eq(\ref{eq460}) provides us with the local momentum-space relation
between products of on-shell amplitudes on the triple cut and the
reduction coefficient of the three-point function.

Next, we extend this discussion to double cuts.  For definiteness, we
consider a double cut specified by vanishing of two inverse Feynman
propagators. In the massless case that we study here,  these
are given by $d_0= l^2$ and $d_1= (l+q_1)^2$.  It follows from the
discussion in the previous Section, that the general expression for
the double cut is
\be
\left [ A_1 A_2 \right ](l_d) 
= {\tilde b}_{01}(l_d) 
+ \sum \limits_{2 \leq i < j}^{N} 
\frac{{\tilde d}_{01ij}(l_d)}{d_i(l_d) d_j(l_d)}
+ \sum \limits_{i=2}^{N} \frac{{\tilde c}_{01i}(l_d)}{d_i(l_d)}\,, 
\label{eq999}
\ee
where $l_d$ is the momentum that satisfies the double-cut constraints
$d_0(l_d) = 0$, $d_1(l_d) = 0$.  Similar to the discussion of a
triple-cut case, the goal is to find a procedure that allows the
computation of the two-point function reduction coefficient ${\tilde
  b}_{01}^{(0)}$ without having to deal with many other terms present
in \Eq(\ref{eq999}).  To motivate the loop momentum parametrization
that we use for the double cut, we first discuss the genuine bubble
coefficient $\tilde b_{01}(l_d)$ and ignore quadruple- and triple-cut
remnants in \Eq(\ref{eq999}).

As we explained in the previous Section, a generic parametrization of
the loop momentum on the double-cut is
\be
l_d^\mu = -\frac{q_1^\mu }{2}
+ l_\perp  ( n_2^\mu \cos \theta 
+ n_3^\mu \sin \theta \cos \varphi + n_4^\mu  \sin \theta \sin \varphi ),
\label{eq_double}
\ee
where $l_\perp = \sqrt{-q_1^2}/2$.  It is straightforward to rewrite
this expression in terms of the variable $t= e^{i\varphi}$, that we
introduced earlier. We obtain
\be
l_d^\mu = -\frac{q_1^\mu}{2}  + l_\perp \left ( n_2^\mu \cos \theta 
+ t \sin \theta n_-^\mu   +  t^{-1} \sin \theta n_+^\mu   \right ).
\label{param0}
\ee
To motivate the parametrization of $\cos \theta$ and $\sin \theta$
that we are about to introduce, we note that the full tensor structure of
the bubble coefficient shown in \Eq(\ref{eq987}) is a linear
combination of the constant term, that we need, and spherical
harmonics $Y_l^m(\theta,\varphi)\propto t^m$ with $l \le 2$. We would like to
find a way to identify and project out the spherical harmonics by
working with rational functions.  It is easy to realize that spherical
harmonics with non-vanishing $m$ can be isolated by performing the
Laurent expansion in the variable $t$ at $t = \infty$, and by picking
up $t$-independent terms.  Spherical harmonics that are not removed by
this procedure correspond to $l=0,m=0$, $l =1$, $m=0$ and $l=2,
m=0$. Hence, by performing Laurent expansion in $t$ and by picking up
${\cal O}(t^0)$ terms, we isolate\footnote{To avoid confusion, we
  reiterate that we are are not discussing remnants of four- and
  three-point functions at the moment.} the following structure on the
double cut
\be
{\tilde b}_{01}(l_d) \Rightarrow  b_{01}(\theta) = {\tilde b}_{01}^{(0)} 
+ {\tilde b}_{01}^{(1)} \cos \theta 
+ {\tilde b}_{01}^{(2)} ( 3 \cos^2 \theta - 1).
\label{eq_a_1}
\ee 
We need a rational parametrization of $\cos \theta$ in terms of some
variable, for which another Laurent expansion can be formulated.
Typically, such parametrization will not lead to a rational
parametrization of the $\sin \theta$ function in \Eq(\ref{param0}).
For this reason, it is convenient to change variables $t \to z$, where
$ t= z/\sin \theta$, and use \cite{Forde:2007mi} $\cos \theta = 1-2 y$
and $\sin^2 \theta = 4 y(1-y)$. As a result, the momentum
parametrization in \Eq(\ref{param0}) becomes
\be
l^\mu = -\frac{q_1^\mu}{2}  + l_\perp \left ( n_2^\mu (1-2y) 
+ z  n_-^\mu   +  \frac{4 y(1-y)}{z}  n_+^\mu   \right ).
\label{param1}
\ee

The simplest way to extract the coefficient ${\tilde b}_{01}^{(0)}$
from the function $b_{01}(\theta)$ in \Eq(\ref{eq_a_1}) is to
integrate over $\cos \theta$
\be
\frac{1}{2} \int \limits_{-1}^{1} 
{\rm d} \cos \theta \; b_{01}(\theta) = \tilde b_{01}^{(0)}.
\ee
We would like to implement this integration as an algebraic procedure.
To this end, note that because $-1 < \cos \theta < 1$, the integration
region for $y$ is $ 0 < y < 1$, and the following integration rule is
valid
\be
\int \limits_{0}^{1} {\rm d} y \, y^m = 
\frac{1}{m+1} \int \limits_{0}^{1}{\rm d} y.
\label{eq_ysubs}
\ee
Therefore, a substitution $y^m \to f_m = 1/(m+1)$, suggested in
Ref.~\cite{Forde:2007mi}, is {\it equivalent} to integration over
$\cos \theta$, which removes ${\tilde b}_{01}^{(1)} \cos \theta$ and
${\tilde b}^{(2)}_{01} ( 3 \cos^2 \theta - 1)$ in
\Eq(\ref{eq_a_1}). Hence, we find that if we use the parametrization
\Eq(\ref{param1}) for the cut loop momentum, we arrive at the simple
formula that extracts the reduction coefficient of the double-cut
integral
\be
{\tilde b}_{01}^{(0)} = 
\left[ {\cal L}_{z,0} \left [ 
{\tilde b}_{01}(l_d) \right ] \right]^{y^m \to f_m}.
\label{eq1000}
\ee
In the above equation, the Laurent expansion in $z$ removes all the
spherical harmonics with non-trivial dependence on the azimuthal angle
$\varphi$, while the substitutions $y^m \to f_m$ help us to integrate
over $\cos \theta$, removing contributions of $Y^{0}_{l=1,2}(\cos
\theta)$ and leaving the constant term ${\tilde b}_{01}^{(0)}$ only.
Note that one can accomplish the same goal in a slightly different way, by
assuming the following parametrization for $\cos \theta$
\be
\cos \theta = \frac{1}{2} \left ( w + \frac{2}{3 w} \right). 
\ee
This parametrization has the property that $w$-independent term drops
out both in $\cos \theta$ and in $ 3 \cos^2 \theta -1 $. This implies
that the term in the Laurent expansion that scales like ${\cal O}(w^0
z^0)$ is the reduction coefficient of a two-point function ${\tilde
  b}_{01}^{(0)}$; no integration over any parameter is required.

According to Ref.~\cite{Forde:2007mi}, the full double-cut reduction
coefficient can be calculated with the help of the following equation
\be
\begin{split}
b_{01}^{(0)} 
& 
= \left [ 
{\cal L}_{z,0} \left [ 
{\cal L}_{y,\ge 0} \left[ A_1 A_2 \right ]^{y^m \to f_m}  \right ]  
\right ]
\\
& 
- \frac{1}{2} \sum \limits_{i,\alpha= \pm} 
\left [ 
{\cal L}_{z,\ge 0}  [ A_1 A_2 A_3]^{(i)}(z,y_{\alpha}^{(i)})
\right ]^{z^n \to Z(n)},
\label{eq_forde}
\end{split}
\ee
where the sum runs over all propagators $d_i$ that, when taken
together with $d_0$ and $d_1$, make a valid triple cut of the one-loop
amplitude. Also, $f_m = 1/(m+1)$, and $Z(n)$, $n=0,1,2,3$ are functions
of external momenta and the integer $n$; they can be found in
Ref.~\cite{Forde:2007mi} and we also derive them in what follows.  The
Laurent expansion operator ${\cal L}_{{x},\ge 0}$ implies that after
performing the Laurent expansion in the variable $x$ at infinity, only
non-negative powers of that variable must be kept.  It is assumed that
the momentum parametrization of \Eq(\ref{param1}) is used in
\Eq(\ref{eq_forde}) and $y_\pm^{(i)}$ are the two values of the
variable $y$ for which $d_i(l_d(z,y^{(i)}_{\pm}))=0$.

We would like to prove the validity of \Eq(\ref{eq_forde}).  Since the
most general parametrization of the double and triple cuts is provided
by the OPP parametrization of the residues, we can check
\Eq(\ref{eq_forde}) directly.  We first explain why 
remnants of the quadruple cuts 
(the terms $\tilde d_{01ij}(l_d)/d_i(l_d)/d_j(l_d)$) do not contribute to \Eq(\ref{eq_forde}).  
It happens for two different reasons. One such contribution 
is removed by an operator ${\cal L}_{y,\ge 0}$
from the products of two on-shell
amplitudes $A_1 A_2$  in \Eq(\ref{eq_forde}), because 
by simple power counting remnants of the quadruple cuts 
vanish in the large $y$-limit.  Interestingly, in the case of 
products of three on-shell amplitudes $A_1 A_2 A_3$, remnants 
of quadruple cuts  produce finite $z$-independent terms, besides terms
that vanish in the large-$z$ limit.  However, as we explain below, we have to set 
$Z(0) = 0$, for consistency; this completely removes contributions of
quadruple cuts from the right hand side \Eq(\ref{eq_forde}).  

We turn
to the discussion of double and triple cuts contributions 
to \Eq(\ref{eq_forde}).
The relevant expression for the double cut is given in
\Eq(\ref{eq999}).  Since we explained that the quadruple cut
contributions are immaterial, the product of three on-shell amplitudes
is given by the $c(l)$-functions, evaluated with the double-cut loop
momentum. Since 
${\tilde b}_{01}(l_d)$ is a polynomial in $y$, we can use the method
explained in the previous Section to extract ${\tilde
  b}_{01}^{(0)}$. We find
\be
{\cal L}_{z,0} \left [ {\cal L}_{y,\ge 0} 
\left [ {\tilde b}_{01}(l_d) \right ] ^{y^m \to f_m} \right ] 
= {\tilde b}_{01}^{(0)}.
\ee
We apply ${\cal L}_{z,0} [\dots]^{y^m\to f_m}$ to $A_1 A_2$ in
  \Eq(\ref{eq999}). As discussed quadruple cuts do not contribute,
  therefore \Eq(\ref{eq_forde}) can only be valid if the
triple-cut functions satisfy the equation
\be
\begin{split}
&{\cal L}_{z,0} \left [ {\cal L}_{y,\ge 0} 
\left [ \frac{{\tilde c}_{01i}(l_d(z,y))}{d_i(l_d)} 
\right ]^{y^m \to f_m} \right ]
= \phantom{XXXXXXXXXX}
\\ 
&\phantom{XXXXXX}\frac{1}{2} 
\left [ {\cal L}_{\rm z, \ge 0} \left [{\tilde c}_{01i}(l_d(z,y_+^{i})) 
+ {\tilde c}_{01i}(l_d(z,y_-^{i})) \right ] \right ]^{z^n \to Z(n)}, 
\label{eq_check}
\end{split}
\ee
for each value of $i$. In general, ${\tilde c}_{01i}(l)$ is the
rank-three tensor but, for the sake of simplicity, we will begin by
considering it to be rank-two. We comment on the rank-three
contributions to ${\tilde c}_{01i}(l)$ at the end of this Section. We
write
\be
\label{eq_c_param}
{\tilde c}_{01i}(l) 
= {\tilde c}_{01i}^{(0)} + s_{\mu} l^\mu + t_{\mu \nu} l^\mu l^\nu, 
\ee
where $s^\mu$ is a vector and $t^{\mu \nu}$ is a traceless tensor that
are both transverse to $q_1$ and $q_i$.

As a first step, we compute the left hand side in
\Eq(\ref{eq_check}).  We perform a Laurent expansion in $y$ at
infinity, keeping non-negative powers in the expansion.  We integrate
over $y$ from zero to one, which is equivalent to making the
substitutions $y^m \to 1/(m+1)$.  Then, we perform a further Laurent
expansion in $z$ at $z = \infty$, and pick up the $z$-independent term
${\cal O}(z^0)$.  Finally, we obtain a simple expression
\be
{\cal L}_{z,0} \left [ {\cal L}_{y, \ge 0} \left [ 
\frac{c_i(l_d(z,y))}{d_i(l_d)} 
\right ]^{y^m \to f_m} \right ]
= \frac{1}{2 u_+} \; s \cdot n_{+}
- 
\frac{ \Delta_i }{4u_+^2} \;t_{\mu \nu} n_{+}^{\mu} n_+^{\nu},
\label{eq_simp1}
\ee
where $\Delta_i = q_i^2 - q_i \cdot q_1$ and $u_+ = q_i \cdot n_+$.  We now
check if a similar expression can be obtained by computing the right
hand side of \Eq(\ref{eq_check}).  To do that, we need the double-cut
momenta evaluated at two particular values of the variable $y$, namely
$y = y_\pm^{(i)}$.  It is convenient to introduce two auxiliary
vectors $L = (l(z,y_+) + l(z,y_-))/2$ and $K = (l(z,y_+) -
l(z,y_-))/2$.  After some algebra, we find
\be
L^\mu = -\frac{q_1^\mu}{2}
+ l_\perp \left ( - \frac{\Delta_i}{2 l_\perp u_+} n_+^\mu
+ \frac{z}{2u_+} \; \omega^{\mu \nu}(q_1) q_{i,\nu}
- z \frac{q_i^\alpha \omega_{\alpha \beta}(q_1) q_i^\beta}{2 u_+^2} n_+^\mu
\right ),
\label{eq_L}
\ee
where $\omega_{\mu \nu}(q_1)$ is the metric tensor of the
three-dimensional space which is transverse to the vector $q_1$.  In
deriving \Eq(\ref{eq_L}), it is important to use the completeness
relations for vectors $n_2,n_\pm$
\be
2 n_+^\mu n_-^\nu + 2 n_-^\mu n_+^\nu 
+ n_2^\mu n_2^\nu   =\omega^{\mu \nu}(q_1) = g^{\mu \nu}  - \frac{q_1^\mu q_1^\nu}{q_1^2}.
\ee
Because we will only use the vector $L^\mu$ in formulas where it is
contracted with vectors or tensors that are transverse to $q_1$ and
$q_i$, we can drop $q_1^\mu$ and $\omega^{\mu \nu}(q_1)q_{i,\nu}$ in
\Eq(\ref{eq_L}). Upon doing that, we get a very simple expression
\be
L^\mu \to L^\mu = - \frac{n_+^\mu  }{2u_+}  
\left (\Delta_i 
+ z \frac{q_{i,\perp}^2 l_\perp }{u_+}
\right ),
\label{eq_L_fin}
\ee
where $q_{i,\perp}^2 = q_i^\alpha \omega_{\alpha \beta}(q_1) q_i^\beta$.
The vector  $K_\mu$ is obtained using similar arguments. We find 
\be
K^\mu = l_\perp (y_- - y_+) \kappa^\mu,\;\;\;
\kappa^\mu = n_2^\mu - \frac{u_2}{u_+} n_+^\mu,
\ee
where $u_2 = q_i \cdot n_2$.  It is easy to establish that only an
{\it even} number of vectors $K_\mu$ can enter the computation (as can
be seen explicitly from \Eq(\ref{eq_sum}); as a
result, we need
\be
(y_- - y_+)^2 = 1 + \frac{q_{i,\perp}^2}{4 u_+^2}z^2 
+ \frac{\Delta_i z}{2 l_\perp u_+}.
\label{eq_diff_sq}
\ee

We are now in position to compute 
${\bar c}_i = {\tilde c}_{01i}(z,y_+)/2+ {\tilde c}_{01i}(z,y_-)/2$. We obtain 
\be
\begin{split}
& \bar c
= {\tilde c}_{01i}^{(0)} 
+ s_\mu L^\mu + t_{\mu \nu} L^\mu L^\nu + t_{\mu \nu} K^\mu K^\nu
\\
& = {\tilde c}_{01i}^{(0)}  
- \frac{s_\mu n_+^\mu  }{2u_+}  
\left (\Delta_i 
+ z \frac{q_{i,\perp}^2 l_\perp }{u_+}
\right )
+ t_{\mu \nu} \frac{n_+^\mu n_+^\nu}{4 u_+^2} 
 \left (\Delta_i 
+ z \frac{q_{i,\perp}^2 l_\perp }{u_+}
\right )^2 
\\
& + l_\perp^2 \left(   1 + \frac{q_{i,\perp}^2}{4 u_+^2}z^2 
+ \frac{\Delta_i z}{2 l_\perp u_+} \right ) 
t_{\mu \nu} \kappa^\mu \kappa^\nu,
\label{eq_sum}
\end{split} 
\ee
which implies that $\bar c$ is a polynomial in $z$.  Forde 
\cite{Forde:2007mi} suggests to
match \Eq(\ref{eq_sum}) and \Eq(\ref{eq_simp1}) by {\it defining}
mappings of powers of $z$ on to some functions $z^n \to Z(n)$. Since 
linearly independent tensor structures must satisfy such
mappings separately, the possibility to do that is not 
obvious.  We now show that it is possible.

To construct such a mapping, we note that \Eq(\ref{eq_simp1}) does not
contain ${\tilde c}_{01i}^{(0)}$, while \Eq(\ref{eq_sum}) does.  This
suggests that all $z$-independent terms in \Eq(\ref{eq_sum}) must be
set to zero, $z^0 \to Z(0) = 0$; \Eq(\ref{eq_sum}) becomes
\be
\begin{split}
\bar c  = &
- s_\mu n_+^\mu\;\frac{z q_{i,\perp}^2 l_\perp }{2 u_+^2}
+ t_{\mu \nu} \frac{n_+^\mu n_+^\nu}{4 u_+^2} 
 \left (z \frac{2 \Delta_i q_{i,\perp}^2 l_\perp }{u_+}
+ \frac{ z^2 q_{i,\perp}^4 l_\perp^2 }{u_+^2}
\right ) 
\\
& + l_\perp^2 \left( \frac{q_{i,\perp}^2}{4 u_+^2}z^2 
+ \frac{\Delta_i z}{2 l_\perp u_+} \right ) 
t_{\mu \nu} \kappa^\mu \kappa^\nu.
\label{eq_sum_1}
\end{split} 
\ee

Comparing \Eq(\ref{eq_sum_1}) and \Eq(\ref{eq_simp1}), we see that
expressions for the rank-one tensors match, provided that we make the
substitution
\be
z \to Z(1) = -\frac{u_+}{l_\perp q_{i,\perp}^2}.
\ee 
As a result, \Eq(\ref{eq_sum_1}) becomes 
\be
\begin{split}
\bar c & =
 \frac{s_\mu n_+^\mu}{2 u_+} 
+ 
 \left (
 \frac{ z^2 q_{i,\perp}^4 l_\perp^2 }{u_+^2}
-2 \Delta_i
\right ) \; t_{\mu \nu} 
\left ( 
\frac{n_+^\mu n_+^\nu}{4 u_+^2} 
+\frac{\kappa^\mu \kappa^\nu}{4q_{i,\perp}^2}
\right ).
\label{eq_sum_2}
\end{split} 
\ee

When we compare tensor structures in \Eq(\ref{eq_sum_2}) and
\Eq(\ref{eq_simp1}), there appears to be a problem because in
\Eq(\ref{eq_simp1}) the tensor structure involves $n_+^\mu n_+^\nu$
but in \Eq(\ref{eq_sum_2}) $\kappa^\mu \kappa^\nu$ appears in
addition. The two equations get reconciled if we use properties of the
tensor $t_{\mu \nu}$ and the vectors $n_+^\mu$ and $\kappa^\mu$.

First, we note that $t_{\mu \nu}$ is symmetric, traceless rank-two
tensor in the space that is transverse to $q_1$ and $q_i$. Because
vector $\kappa$ satisfies $\kappa \cdot q_i = \kappa \cdot q_1 = 0$,
$\kappa^2 =1 $, we can use it as one of the basis vectors of the
required transverse space.  We call the other basis vector $\sigma$,
$\sigma^2=1$.  Then we write the most general parametrization for the
tensor $t_{\mu \nu}$
\be
t_{\mu \nu} = t_1 \left ( \sigma_\mu \sigma_\nu - \kappa_\mu \kappa_\nu \right ) 
+ \frac{t_2}{2} \left ( \sigma_\mu \kappa_\nu + \kappa_\nu \sigma_\mu \right).
\ee
When contracting $t_{\mu \nu}$ with $n_+^\mu n_+^\nu$ and $\kappa^\mu
\kappa^\nu$, we use the fact that $\kappa \cdot n_+ = 0$, $\kappa
\cdot \sigma = 0$. Also, we need $(\sigma \cdot n_+ )^2$ and we
compute it using the completeness identity
\be
\omega^{\mu \nu}(q_1,q_i) = \kappa^\mu \kappa^\nu + \sigma^\mu \sigma^\nu.
\ee
Contracting  $\omega^{\mu \nu}$  with $n_+^\mu n_+^\nu$, we obtain 
\be
(\sigma \cdot n_+ )^2 = n_+^\mu \omega_{\mu \nu}(q_1,q_i) n_+^\nu = 
- \frac{u_+^2}{q_{i,\perp}^2}.
\ee
Putting everything together, we arrive at 
\be
t_{\mu \nu} 
\left ( \frac{n_+^\mu n_+^\nu}{4 u_+^2}
+ \frac{\kappa^\mu \kappa^\nu}{4 q_{i,\perp}^2} 
\right )
= t_1 \left ( \frac{(\sigma \cdot n_+)^2}{4 u_+^2} - \frac{1}{4 q_{i,\perp}^2} 
\right ) = -\frac{t_1}{2 q_{i,\perp}^2}.
\ee
We then find 
\be
\begin{split}
\bar c & =
 \frac{s_\mu n_+^\mu}{2 u_+} 
- 
 \left (
 \frac{ z^2 q_{i,\perp}^4 l_\perp^2 }{u_+^2}
-2 \Delta_i
\right ) 
\frac{t_1}{2 q_{i,\perp}^2}. 
\label{eq_sum_3}
\end{split} 
\ee
Requiring that \Eq(\ref{eq_sum_3}) matches  \Eq(\ref{eq_simp1}), 
we find the substitution rule 
\be
\begin{split}
z^2 \to Z(2) = \frac{3 \Delta_i u_+^2}{2 q_{i,\perp}^4 l_\perp^2}.
\label{eq_sum_4}
\end{split} 
\ee

Finally, we discuss an extension of these results to the rank-three 
case. Consider the additional term in \Eq(\ref{eq_c_param}) 
\be
{\tilde c}_{01i}(l) \to {\tilde c}_{01i}(l) + 
t^{\mu \nu \alpha} l_\mu l_\nu l_\alpha. 
\ee
When $t^{\mu \nu \alpha}$ is expressed through basis vectors $\kappa$
and $\sigma$, four terms appear
\be
t^{\mu \nu \alpha} = 
a_1 \omega^{\{\mu \nu} \kappa^{\alpha \}} 
+a_2 \omega^{\{\mu \nu} \sigma^{\alpha \}} 
+a_3 \kappa^\mu \kappa^\nu \kappa^\alpha
+a_4 \sigma^\mu \sigma^\nu \sigma^\alpha. 
\label{four_terms}
\ee
In \Eq(\ref{four_terms}), we introduce the metric tensor of the vector
space transverse to $q_1$ and $q_i$, $\omega_{\mu \nu} = \omega_{\mu
  \nu}(q_1,q_i)$.  Also, indices in curly brackets need to be
symmetrized.  Computations in this case are straightforward, although
more tedious than for the rank-one and rank-two. We skip all the
details and only present the result.  First, the
Laurent expansion, {\it restricted to the rank three terms}, gives
\be
\begin{split}
{\cal L}_{z,0} \left [ {\cal L}_{y, \ge 0} \left [ 
\frac{c_i(l_d)}{d_i(l_d)} \right]^{y^m \to f_m} \right ]
& =  a_2 (\sigma \cdot n_+)
\left ( - \frac{3 \Delta_i^2}{8 q_{i,\perp}^2 u_+}
 + \frac{3 l_\perp^2}{2 u_+} \right )
\\
& + 
a_4 (\sigma \cdot n_+)
\left ( - \frac{\Delta_i^2}{8 q_{i,\perp}^2 u_+}
 + \frac{ l_\perp^2}{3 u_+} \right ).
\label{eq_a}
\end{split} 
\ee
Second, calculating the function ${\tilde c}_{01i}$ for
$l_d(z,y_{\pm})$, and using the substitution rules for $z^{n} \to
Z(n)$, $n = 0,1,2$, that we derived in this Section, we obtain
\be
\begin{split} 
\frac{1}{2} ( c(z,y_+) + c(z,y_-)) 
& = 
a_2 (\sigma \cdot n_+)
\left ( - \frac{3 \Delta_i^2}{8 q_{i,\perp}^2 u_+}
 + \frac{3 l_\perp^2}{2 u_+} \right )
\\
& + 
a_4 (\sigma \cdot n_+)
\left (  \frac{3\Delta_i^2}{16 q_{i,\perp}^2 u_+}
 + \frac{ z^3l_\perp^3 q_{\perp, i}^4 }{8 u_+^4} \right ).
\label{eq_b}
\end{split} 
\ee
Comparing \Eq(\ref{eq_a}) and \Eq(\ref{eq_b}), we find 
that they can be matched 
by the substitution 
\be 
z^3 \to Z(3) =  \frac{8 u_+^3}{3 l_\perp q_{i,\perp}^4}
 -\frac{5 \Delta_i^2 u_+^3}{2 q_{i,\perp}^6 l_\perp^3}.
\label{eq_r_3}
\ee
Because the highest rank of a tensor in the function $c(l)$ is three,
\Eq(\ref{eq_r_3}) completes the list of substitutions that are
required to prove the validity of \Eqs(\ref{eq_forde}) and (\ref{eq_check}).

\subsection{An alternative formula for bubble reduction coefficients}
\label{alternative}

In this Section we describe another approach to the calculation of
bubble reduction coefficients, suggested by Mastrolia in
Ref.~\cite{Mastrolia:2009dr}.  We consider a double-cut of the
one-loop scattering amplitude that is defined by the condition that
two inverse propagators, $d_0$ and $d_1$, vanish.  We use the
parametrization of the loop momentum given in \Eq(\ref{eq_double}),
take the double-cut of the amplitude given in \Eq(\ref{eq999}), and
integrate over $\theta$ and $\varphi$. We use the expression for the
coefficients $\tilde d(l), \tilde c(l), \tilde b(l)$ in \Eqs(\ref{x0123},\ref{eq_c_coeff},\ref{eq987}), neglect
terms proportional to $\ne$, since here we are interested in the
cut-constructible part only, and drop terms that vanish after the
integration over the solid angle. 
We obtain \be
\label{eq5000}
\begin{split} 
\int \frac{{\rm d} \Omega}{4\pi} 
[A_1 A_2](l_d) & = {\tilde b}_{01}^{(0)} 
+ \sum \limits_{i} \int \frac{{\rm d} \Omega}{4\pi} 
\frac{{\tilde c}_{01i}^{(0)}}{d_i(l_d)} 
+ \sum \limits_{ij} \int \frac{{\rm d} \Omega}{4\pi}  
\frac{{\tilde d}_{01ij}^{(0)}}{d_i(l_d) d_j(l_d)}, 
\end{split} 
\ee
where $ {\rm d}\Omega = {\rm d}\cos \theta \; {\rm d} \varphi $ is an
element of the solid angle.  We stress that remnants of three- and
four-point functions in the right hand side of \Eq(\ref{eq5000}) are
multiplied by the corresponding reduction coefficients ${\tilde
  c}_{01i}^{(0)}$ and $d_{01ij}^{(0)}$ and not by the full
$l$-dependent functions $c(l)$ and $d(l)$.  We now rewrite the
integration over the solid angle by performing the standard
change of variables 
\beq
\rho = \tan \frac{\theta}{2},
\eeq
to express $\cos \theta,\sin \theta$ as rational functions of $\rho$
\be
\cos \theta = \frac{1-\rho^2}{1+\rho^2},\;\;\;
\sin \theta = \frac{2 \rho}{1+\rho^2},\;\;\;
\tan \theta = \frac{2 \rho}{1-\rho^2}.
\ee
We further introduce a complex variable $z = \rho e^{i \varphi}$,
denote $\bar z = \rho e^{-i \varphi}$, and 
write 
\be
\cos \theta = \frac{1 - z \bar z}{1+ z \bar z},\;\;\;
\tan \theta e^{i\varphi} = \frac{2 z }{1 - z \bar z},\;\;\;
\tan \theta e^{-i\varphi} = \frac{2 \bar z }{1 - z \bar z}.
\ee
The final parametrization of the loop momentum reads 
\be
l_d^\mu = -\frac{q_1^\mu}{2}
+\frac{\sqrt{-q_1^2}}{2} \frac{1-z \bar z}{1+ z \bar z}
\left ( n_2^\mu
+\frac{2z}{1-z \bar z} n_-^\mu 
+ \frac{2\bar z}{1-z \bar z} n_+^\mu 
\right ).
\label{paramz}
\ee

Given the mapping between $\theta, \varphi$ and $z, \bar z$ variables, 
it is easy to find the relation of the integration measures
\be
 {\rm d} \Omega = \frac{2 {\rm d} \bar z \wedge {\rm d} z}{i (1+z \bar z)^2}.
\label{eq_mes}
\ee
Integration over $z$ and $\bar z$ extends through  the entire 
complex plane $D_{\infty}$. We therefore rewrite \Eq(\ref{eq5000}) as 
\be
\label{eq5001}
{\tilde b}_{01}^{(0)} = 
\frac{1}{2\pi i} 
 \int \limits_{D_\infty}
\frac{{\rm d} {\bar z} \wedge {\rm d} z }{(1+z \bar z)^2}
f(z,\bar z),
\ee
where 
\be
f(z,\bar z) = [A_1 A_2](l_d) 
- \sum \limits_{i} 
\frac{{\tilde c}_{01i}^{(0)}}{d_i(l_d)} - 
\sum_{ij} \frac{{\tilde d}_{01ij}^{(0)}}{d_i(l_d) d_j(l_d)}\,,  
\label{eq5002}
\ee
and the cut momentum $l_d$ is given by \Eq(\ref{paramz}).

As recognized in Ref.~\cite{Mastrolia:2009dr}, \Eq(\ref{eq5001}) has a
structure that can be integrated using the generalized Cauchy (or
Cauchy-Pompeiu) theorem.  The theorem states that for a rational
function $F(z,\bar z)$ defined in a domain $D_c$ of a complex plane
that is bounded by a contour $L_c$, the following identity is valid
\be
\begin{split}
&\frac{1}{2 \pi i} \oint_{L_c} dz \; F(z,\bar z) 
- \frac{1}{2\pi i}  \int_{D_c}
\; \frac{\partial F(z,\bar z)}{\partial \bar z} \; 
{\rm d} {\bar z} \wedge {\rm d}  z 
\\
& \;\;\;\;\;\;\;\;\;\;= \sum {\rm Res} (F(z,\bar z)).
\label{pompeiu}
\end{split}
\ee
The sum on the right-hand side of \Eq(\ref{pompeiu}) runs over all the
$z$-poles in the domain $D_c$.  To apply this theorem to the
calculation of the double-cut coefficient, we identify
\be
\frac{\partial F(z, \bar z)}{\partial {\bar z} } 
= \frac{f(z, \bar z)}{(1+z \bar z)^2}.
\label{der}
\ee
We also identify $D_c$ with $D_\infty$; as a consequence, the
contour $L_c$ in \Eq(\ref{pompeiu}) runs at infinity.  Because the limit 
$z, \bar z \to \infty$ 
corresponds to {\it finite} cut-momenta $l_d$ (see \Eq(\ref{paramz})), 
the function $f(z, \bar
z)$ remains finite in that limit.  Then, as a consequence of
\Eq(\ref{der}), we can choose the function $F(z,\bar z)$ such that it
vanishes at the complex infinity as $1/(z ^2 \bar z)$.  
Hence the first term in the
Cauchy-Pompeiu formula \Eq(\ref{pompeiu}) can be dropped and we find
\be
\sum \limits_{z~{\rm poles~}\in D_\infty}^{} {\rm Res} \left [ 
F(z,\bar z) \right ] = 
  - \frac{1}{2\pi i} \int_{D_c}
\; \frac{\partial F(z,\bar z)}{\partial \bar z} \; 
{\rm d} {\bar z} \wedge {\rm d}  z 
= - {\tilde b}_{01}^{(0)}, 
\label{fun_eq}
\ee
where in the last equation we simply inserted \Eq(\ref{eq5001}). 

It follows from \Eq(\ref{fun_eq}) that, to calculate the coefficient
${\tilde b}_{01}^{(0)}$, we need to find the anti-derivative function
$F(z,\bar z)$ and compute its residues in the entire complex plane.
According to \Eqs(\ref{eq5002},\ref{der}), the function $F(z,\bar z)$
can be written as a sum of three terms
\be
F(z, \bar z) = F^{(A)}(z,\bar z)
- \sum \limits_{i}^{} F^{(c_i)}(z,\bar z) 
- \sum \limits_{ij}^{} F^{(d_{ij})}(z, \bar z),
\ee
where 
\be
\frac{\partial F^{(A)}(z, \bar z)}{\partial {\bar z} } 
= (1+z \bar z)^{-2}\;[ A_1 A_2](z, \bar z),
\label{eq_FA}
\ee
and $F^{(c_i),(d_{ij})}$ are anti-derivatives 
due to contributions of three- and four-point 
functions in \Eq(\ref{eq5002}). It is easy to see that, 
$F^{(c_i),(d_{ij})}$ have the following form
\be
{\rm Res} \left [ F^{(c,d)}(z,\bar z) \right ] \sim 
R(z,\bar z) \ln( Q(z,\bar z)),
\ee
where $R(z,\bar z)$ and $Q(z, \bar z)$ are some rational functions.
Those terms subtract similar contributions to function $F^{(A)}(z,\bar
z)$; as a result they do not affect the evaluation of the double-cut
reduction coefficient ${\tilde b}_{01}^{(0)}$.  This observation was
used in Ref.~\cite{Mastrolia:2009dr} where it was suggested that one
can drop $F^{(c,d)}(z,\bar z)$ {\it and} all the logarithmic terms in
$F^{(A)}(z, \bar z)$ to obtain a simpler formula for the reduction
coefficient
\be
 {\tilde b}_{01}^{(0)} = 
- \sum \limits_{z~{\rm poles~}\in D_\infty}^{} {\rm Res} 
\left [ F^{(A),\rm rat}(z,\bar z) \right ].
\label{eq_simp}
\ee
In \Eq(\ref{eq_simp}) $F^{(A),\rm rat}(z,\bar z)$ is the
anti-derivative as in \Eq(\ref{eq_FA}), from where all the logarithmic
terms are omitted.

There is a subtlety in proving \Eq(\ref{eq_simp}),
which is not mentioned in the literature.  Indeed, consider a
representation of the amplitudes $A_1 A_2$ on a double cut in terms of
the OPP reduction coefficients
\be
[A_1 A_2](l_d) = {\tilde b}_{01}(l_d) + \sum \frac{{\tilde
    c}_{0i1}(l_d)}{d_i(l_d)} + \ldots 
\ee
Using the explicit parametrization of the cut momentum $l_d$ as in
\Eq(\ref{paramz}), it is straightforward to prove that ${\tilde
  b}_{01}(l_d)$ only contributes to $F^{(A),\rm rat}(z, \bar z)$.  The
term ${\tilde c}_{01i}^{(0)}/d_i$ contributes to the logarithmic part of
$F^{(A)}(z,\bar z)$ and hence it is discarded when $F^{(A),\rm rat}$ is
constructed.

A more complicated situation occurs however with the spurious contributions to
${\tilde c}_{01i}(l_d)$.  Consider the rank-one tensor ${\tilde
  c}_{01i}^{(1)}\; (s \cdot l_d) / d_i(l_d)$ as an example.  Using the
parametrization for $d_i(l_d)$ in terms of $z, \bar z$, we write
\be
d_i(z, \bar z) = \frac{P_{12}(z) + \bar z P_{34}(z)}{1 + z \bar z}\,, 
\label{eq45}
\ee
where $P_{12}(z) = p_1 + z p_2$, $P_{34}(z) = p_3 + z p_4$ 
and 
\be 
\begin{split} 
& p_1 = q_i^2 - q_i \cdot q_1 + 2 l_\perp n_2 \cdot q_i, 
\;\;\;
 p_2 = 4 l_\perp n_- \cdot q_i,\;\;\;\;\;\;
\\
& p_3 = 4 l_\perp n_+ \cdot q_i,\;\;\;\;\;\;
 p_4 = q_i^2 - q_i \cdot q_1 - 2 l_\perp n_2 \cdot q_i.
\label{eq46}
\end{split} 
\ee
Computing the contribution of those terms to $F^{(A)}(z,\bar z)$ 
we find 
\be
F^{(A)}_{c_i^{(1)}}(z,\bar z) = 
\frac{N(z,\bar z) \ln \left ( d_i(z , \bar z)  \right )}
{ \left ( P_{12}(z) z - P_{34}(z)  \right )^2}  
 + R(z,\bar z).
\label{eq_spur}
\ee
The second power of the rational function in the denominator of the
logarithmic term in the above equation is a direct consequence of the
fact that we are dealing with the  rank-one tensor.  In general, a
rank-$n$ tensor integral leads to the appearance of terms $\ln \left [
  d_i(z,\bar z) \right ] /(P_{12}(z) z - P_{34}(z))^{n+1}$ in the
anti-derivative.

Because our original expression is a spurious term, it should
integrate to zero. This implies that the following equation holds
\be
\sum \limits_{{z~\rm poles~}\in D_\infty}^{} {\rm Res} 
\left[ F^{(A)}_{c_i^{(1)}}(z,\bar z)  \right ] = 0.
\ee
Writing $P_{12}(z) z - P_{34}(z) = p_2 ( z - z_+) (z - z_-)$ and 
computing residues at, say, $z = z_+$, we find 
\be
\begin{split} 
& {\rm Res} 
\left[ F^{(A)}_{c_i^{(1)}}(z,\bar z)  \right ]_{z =z_+} 
= \frac{Y_+}{p_2^{2} (z_- - z_+)^3 } \ln (d_i(z_+, \bar z_+) ) 
\\
& +  \frac{N(z_+,\bar z_+) }{p_2^2(z_+-z_-)^2}
\frac{\partial \ln d_i(z,\bar z)}{\partial z} |_{z=z_+}
+ {\rm Res}_{z=z_+} 
\left [ R(z,\bar z) \right  ],
\label{eq_res_+}
\end{split}
\ee 
where 
\be
Y_+ =  2 N(z_+,\bar z_+)  
+ (z_--z_+) \frac{\partial N(z,\bar z)}{\partial z} |_{z = z_+}.
\ee 

It is clear that {\it after} calculating residues of
$F^{(A)}_{c_i^{(1)}}(z,\bar z)$, logarithmic and rational functions of
$z$ should vanish separately.  Therefore, \Eq(\ref{eq_res_+}) implies
that $Y_+=0$ and 
\be
\begin{split}
\sum \limits_{z = z_\pm} \left [ 
\frac{N(z,\bar z) }{p_2^2(z_+-z_-)^2}
\frac{\partial \ln d_i(z,\bar z)}{\partial z}
+ {\rm Res} 
\left [ R(z,\bar z) \right  ] \right ]
= 0.
\label{eq_drop}
\end{split}
\ee
\Eq(\ref{eq_drop})
is striking since it implies that {\it neglecting
  logarithmic terms before computing the residues} could, potentially,
be problematic since logarithmic and rational functions mix. This is a
consequence of the fact that higher-order poles in the $z$-complex
plane appear in anti-derivatives of spurious terms. Mixing of rational
and logarithmic functions is controlled by the $z$-derivatives of
propagator $d_i$ evaluated at the position of the poles $z_{\pm}$.
Using \Eqs(\ref{eq45},\ref{eq46}) we find that
\be
\frac{ \partial^n d_i(z,\bar z)}{\partial z^n}|_{z = z_\pm, \bar z = 
\bar z_{\pm}} = 0, 
\ee
for $n \ge 1$. Hence, it follows from \Eq(\ref{eq_drop}) that mixing
of logarithmic and rational terms is, indeed, absent.  This justifies
\Eq(\ref{eq_simp}) as a valid way to compute the double-cut reduction
coefficient.

Before closing this Section, we point out that the method of computing
the double-cut reduction coefficient \cite{Mastrolia:2009dr}, that is
described in this Section, is applicable in a more  general case,
when arbitrary masses are allowed. Indeed, the original
parametrization of the double-cut loop momentum in terms of polar and
azimuthal angles \Eq(\ref{eq_double}) is valid independently of
masses. The mass dependence appears in the absolute value of the
transverse part of the double-cut momentum $l_\perp$, since the
on-shell conditions change $ l^2 = m_0^2$, $(q_1+l)^2 = m_1^2$.  Also,
we note that when arbitrary masses are allowed, the coefficients $p_{1,4}$
in \Eq(\ref{eq46}) receive equal shifts of their $l_\perp$-independent
parts.  However, because the derivation of \Eq(\ref{eq_simp}) only
depends on the combination $p_1 - p_4$ and is valid for arbitrary
$l_\perp$, we conclude that \Eq(\ref{eq_simp}) allows the computation
of the double-cut cut-constructible reduction coefficient in the most
general case.  We will comment on the applicability of this method to
the computation of the rational part in the next Section.

\subsection{Direct computation of the rational part}
\label{ratpart_bad}

In Section~\ref{direct} we described how to compute the
cut-constructible reduction coefficients directly using Forde's
method \cite{Forde:2007mi}.  Badger pointed out in
Ref.~\cite{Badger:2008cm} that this method can be further extended to
compute the rational part. Below we explain how this can be done.  We
focus on the contribution of a massless scalar field to one-loop gluon
scattering amplitudes.  As follows from the supersymmetric
decomposition of one-loop scattering amplitudes discussed in
Ref.~\cite{Bern:1994cg}, the rational part of gluon amplitudes can be
extracted if such contribution is known.  We work in the
four-dimensional helicity scheme.  In this Section, we will use the
notation $\mu^2 = (l \cdot \ne)^2$.

We explained the parametrization of all unitarity cuts in
Section~\ref{sect5}, including their dependence on $\mu^2$.  In what
follows we start the discussion with the pentuple cut and then move to
quadruple, triple and bubble cuts.  We make maximal use of the
discussion in Section~\ref{direct} since, as we will see, we need only
minimal modifications to obtain the rational part. We always assume
that we cut a subset of $d_0,d_1,..,d_4$ inverse propagators, as
appropriate for a particular cut.

As discussed in Section~\ref{sec5.2}, the parametrization of the
pentuple cut is
\be
{\tilde e}_{01234}(l) = {\tilde e}_{01234}^{(0)} \mu^2.
\ee
The power counting implies that the five-point master integral
vanishes in the limit $D \to 4$,
\be
\int \frac{{\rm d}^D l}{(2\pi)^D} \frac{\mu^2}{d_0 d_1 d_2 d_3 d_4}  \to 0,
\ee
so that we do not need to compute the reduction coefficients ${\tilde
  e}_{01234}^{(0)}$.

Next, we consider the quadruple cut and choose the momentum
parametrization to be
\be
l_{q,\pm}^\nu = V_4^\nu \pm  l_\perp n_4^\nu + \mu n_\ep^\nu. 
\ee
As explained in Sect.~\ref{sect5}, $V_4$ is a constant vector
orthogonal to $n_4$ and $n_\ep$ and $l_\perp^2 = -V_4^2 -
\mu^2$. Taking the quadruple cut we find
\be
\begin{split}
\left [ A_1 A_2 A_3 A_4 \right ](l_{q}) 
& = 
{\tilde d}_{0123}^{(0)} 
+ {\tilde d}_{0123}^{(1)} (l_{q} \cdot n_4) 
+ {\tilde d}_{0123}^{(2)} \mu^2
\\
& 
+ {\tilde d}_{0123}^{(3)} \mu^2 (l_{q} \cdot n_4) 
+ {\tilde d}_{0123}^{(4)} \mu^4
+ \sum \limits_{i}^{} \frac{\tilde e_{0123i}^{(0)} \mu^2}{d_i(l_{q})}.
\end{split}
\ee
By power counting it is easy to understand that only the term
$\tilde d^{(4)}_{0123} \; \mu^4$ contributes to the rational part.
Since the inverse Feynman propagator $d_i$ scales as $d_i \sim l_\perp
\sim \mu$ in the limit of very large value of $\mu$, we conclude that
performing Laurent expansion at $\mu = \infty$ and picking up the
$\mu^4$ term gives the coefficient of the only master integral related
to the quadruple cut contribution to the rational part. We find
\be
\tilde d_{0123}^{(4)} = 
{\cal L}_{\mu^2,4} \left [ A_1 A_2 A_3 A_4 \right ](l_{q,\pm}).
\label{qcutrat}
\ee 

As a next step we discuss the triple cut.  For definiteness, we assume
that it corresponds to zeros of inverse propagators
$d_{0},d_{1},d_{2}$.  The momentum parametrization reads
\be
l^\nu_{t,\pm} = V_3^\nu + l_\perp \left ( 
t n_\mp^\nu  + t^{-1} n_\pm^\nu \right ) 
+ \mu \ne^\nu.
\label{eq_ltpm}
\ee 
Similar to the case of the quadruple cut that we have already
discussed, $V_{3}^\mu$ is a constant vector and $l_\perp^2 = -V_3^2
-\mu^2$. Taking the triple cut of the amplitude, we find
\be
\left [ A_1 A_2 A_3 \right ](l_t) = 
{\tilde c}_{012}(l_t) + \sum \limits_{i} \frac{{\tilde d}_{012i}(l_t)}{d_i(l_t)}
+ \sum \limits_{ij}^{} \frac{{\tilde e}_{012ij}^{(0)} \mu^2}{d_i(l_t) d_j(l_t)},
\ee
where ${\tilde c}_{012}(l_t)$ is given in \Eq(\ref{eq_c_coeff}).  The
rational part of the function ${\tilde c}_{012}$ reads 
\be
{\tilde c}_{012}^{\rm rat}(l_t) =  {\tilde c}_{012}^{(7)}  \mu^2
 + {\tilde c}_{012}^{(8)}   ( l_t  \cdot n_3) \mu^2 
 + {\tilde c}_{012}^{(9)}   (l_t \cdot n_4) \mu^2,
\ee
and ${\tilde c}_{012}^{(7)}$ is the relevant, non-evanescent reduction
coefficient.  To project onto ${\tilde c}_{012}^{(7)}$, we begin by
performing Laurent expansion at $t = \infty$, as we did for the
cut-constructible part.  We find
\be
{\cal L}_{t,0} \left [ {\tilde c}_{012}(l_{t,+}) \right ] 
= {\tilde c}_{012}^{(0)} + {\tilde c}_{012}^{(7)} \mu^2,\;\;\;
{\cal L}_{t,0} 
\left [ \frac{{\tilde e}_{012ij}^{(0)} \mu^2}
{d_i(l_{t,+}) d_j(l_{t,+})} \right ]
= 0,
\ee
and 
\be
{\cal L}_{t,0} \left [ \frac{{\tilde d}_{012i}(l_{t,+})}{d_i(l_{t,+})} \right ] 
= \left ( {\tilde d}_{012i}^{(1)} + \mu^2 {\tilde d}_{012i}^{(3)} 
\right ) \frac{{\tilde n}_4^{(i)} \cdot n_+ }{2 q_i \cdot n_+}.
\ee
As we explained in Section~\ref{direct}, the contribution of the
quadruple-cut coefficient is removed by taking the sum over two
solutions $l_{t,\pm}$ \Eq(\ref{eq_ltpm}) which, as follows from the
parametrization, corresponds to taking the large-$t$ limit along both
$n_+$ and $n_-$ directions.  Hence, we find
\be
\frac{1}{2} \sum \limits_{i=\pm} 
{\cal L}_{t,0}[A_1 A_2 A_3 ](l_i)
= {\tilde c}_{012}^{(0)} + {\tilde c}_{012}^{(7)} \mu^2.
\ee
The coefficient ${\tilde c}_{012}^{(7)}$ is then obtained by the
application of an additional Laurent expansion operator, designed to
pick up the ${\cal O}(\mu^2)$ term, to the triple cut
\be
{\tilde c}_{012}^{(7)} = 
\frac{1}{2} \sum \limits_{i=\pm} 
{\cal L}_{\mu^2,2} \left [ {\cal L}_{t,0}[A_1 A_2 A_3 (l_i) ] \right ].
\label{tripcutrat}
\ee

Finally, it is straightforward to extend this analysis to the double
cut, following the discussion of the cut-constructible part. The
parametrization of the momentum shown in \Eq(\ref{param1}) is extended
in a straightforward way. We write
\be
l^\mu = -\frac{q_1^\mu}{2}  + l_\perp \left ( n_2^\mu (1-2y) 
+ z  n_-^\mu   +  \frac{4 y(1-y)}{z}  n_+^\mu   \right )
+ \mu \ne^\mu,
\label{param1_ep}
\ee
where $l_\perp^2 = -q_1^2/4 - \mu^2$. The function ${\tilde
  b}_{01}(l)$ gets an additional contribution ${\tilde b}_{01}^{(9)}
\mu^2$ that leads to a rational part.  Given the argument described in
connection with the cut-constructible part, about the relation between
$y$ and $\cos \theta$ integrations, it is clear that
\be
{\cal L}_{z,0} \left[ 
{\cal L}_{y, \ge 0} \left [ {\tilde b}_{01}(z,y) \right]^{y^m \to f_m}  
\right] = {\tilde b}_{01}^{(0)} + \mu^2 {\tilde b}_{01}^{(9)}.
\ee
The final ingredients we need to compute the
contribution of a double cut to the rational part, are the terms that
come from the remnants of the triple- and other higher-multiplicity
cuts. Similar to the cut-constructible part, we only look at the
remnants of the triple cuts.

In line with what we did in the computation of the cut-constructible
part, we apply the Laurent expansion procedure to the remnant of the
triple cut.  We can easily extend \Eq(\ref{eq_forde}) to include
$\mu^2$ by promoting all vectors and constants to $\mu^2$-dependent
quantities.  Then, because in the analysis of the cut-constructible
part we never used the explicit form of $l_\perp$ and this is the {\it
  only} $\mu$-dependent quantity, that analysis is applicable also for
the rational part.  In particular, substitutions $z^n \to Z(n)$ that
we derived in Sect.~\ref{direct}, remain unchanged.  As a result, we
find that \Eq(\ref{eq_forde}) is valid, provided that the left-hand
side is changed to $b_{01}^{(0)} \to b_{01}^{(0)} + b_{01}^{(9)}
\mu^2$.  Therefore, to find the reduction coefficient for the rational
part of the two-point function, we only need to apply another Laurent
expansion operator at $\mu =\infty$, to \Eq(\ref{eq_forde}).  We find
\be
\begin{split} 
{\tilde b}_{01}^{(9)} 
= & {\cal L}_{\mu,2} \left [ 
{\cal L}_{z,0} \left [ {\cal L}_{y,\ge 0} \left[ A_1 A_2 \right ]^{y^m \to f_m}  
\right ] \right] 
\\
- & \frac{1}{2} \sum_{i} \sum \limits_{\alpha= \pm} 
{\cal L}_{\mu,2} 
\left [ {\cal L}_{z,\ge 0}[A_1 A_2 A_3]^{(i)}(z,y_{\alpha}^{(i)})
\right ]^{z^n \to Z(n)},
\label{eq_forde_2}
\end{split}
\ee 
which is the formula derived in 
Ref.~\cite{Badger:2008cm}.

As our final comment, we point out that the procedure for calculating
the double-cut cut-constructible coefficient suggested in
Ref.~\cite{Mastrolia:2009dr} and explained in
Section~\ref{alternative} can be easily generalized to deal with the
rational part.  Indeed, the double-cut loop momentum parametrization
in terms of polar and azimuthal angles \Eq(\ref{eq_double}) is valid
even if a $(D-4)$-dependent component of $l$ is added.  Clearly, the
on-shell condition gets modified and reads $ l_\perp^2 + \mu^2 +
q_1^2/4 = 0$.  While it does change $l_\perp$, it is easy to realize
that the exact form of $l_\perp$ is irrelevant for the argument in
Section~\ref{alternative}. Hence, we conclude that if the
$(D-4)$-dimensional part of the loop momentum is kept when the double
cut of a one-loop scattering amplitude is computed, \Eq(\ref{eq_simp})
generalizes in a sense that the left hand side becomes a rank-two
polynomial of $\mu$
\be
 {\tilde b}_{01}^{(0)} +\mu^2 {\tilde b}_{01}^{(9)}= 
-\sum \limits_{z~{\rm poles~}\in D_\infty}^{} {\rm Res} 
\left [ F^{(A),\rm rat}(z,\bar z) \right ].
\label{eq_simp_1}
\ee
The rational part of a double cut is then obtained by picking up the
$\mu^2$-dependent term after computing the right hand side of
\Eq(\ref{eq_simp_1}).

\subsection{Using the helicity formalism}
\label{helicity-formalism}
The method of generalized unitarity, as described above,
provides an algorithm for computation of 
one-loop multi-leg amplitudes
by extracting reduction  coefficients of one-loop
integrals. Since the algorithm is numerical,
it can be easily  implemented using the conventional 
relativistic formalism of Dirac spinors and polarization vectors.
Tree amplitudes, required for the determination of the 
reduction coefficients, are computed numerically using 
recursion relations. This enables us to avoid 
dealing  with analytic expressions for scattering amplitudes, making 
the method robust. 

On the other hand, this procedure is amenable to significant simplifications 
if scattering amplitudes of massless particles are considered. 
In this case, 
analytic expressions for scattering amplitudes are often available, 
leading to deeper insights into the structure of gauge field 
theories.   This is particularly true for
amplitudes in  non-abelian gauge field 
theories,  where color-ordered 
$n$-gluon maximally
helicity violating (MHV) tree amplitudes
are  given by a remarkably simple expression~\cite{Parke:1986gb}
\be 
m_n(1^+,2^+,\ldots,i^-,\ldots, j^-, \ldots, n^+)=
i\frac{\la ij \ra^4}{\la  12\ra\la 23\ra\ldots \la n1\ra}.
\ee
This  result was  conjectured as a generalization of analytic results
obtained with  helicity methods
for  four-, five- and six-gluon scattering amplitudes and later  proven 
for an arbitrary  number of gluons \cite{Berends:1987cv}.

The spinor-helicity methods  are also  useful in loop computations 
\cite{Kunszt:1993sd}.  However,  because of the four-dimensional 
nature of spinor-helicity variables,  these methods 
can not be used directly for computing the rational parts.
With the advent  of the unitarity methods
simple  analytic expressions   
could be derived 
for tree and loop amplitudes in QCD, 
in $N=4$ Super-Yang-Mills
theory and  in  $N=8$ supergravity.  Since a number of  excellent  
reviews of spinor-helicity formalism are 
available in the literature 
~\cite{Mangano:1990by,Dixon:1996wi,Peskin:2011in,Drummond:2010ep}, 
we do not discuss it here. 
Nevertheless, 
since the helicity method is an important part of the 
toolkit of analytic computations in gauge field theories, 
we  summarize the basics of the  method in Appendix D. We also 
demonstrate  in Sec.~9.6 how the spinor-helicity method can be used 
in conjunction with generalized unitarity
 by deriving compact results   for the one-loop amplitude of  
the $q\bar q gg$ subprocess.

\section{Examples}
\label{sec:examples}

The goal of this Section is to discuss examples where the computation
of the full answer, or some well-defined parts of it, can be
performed with relative ease.  We are particularly interested in cases
where the peculiar nature of the rational part, and its relation to
anomalous behavior of quantum field theories, becomes explicit.  This
feature is best illustrated by computing one-loop quantities that are
finite but whose computation in four-dimensions -- or better to say in
unregularized quantum field theory -- would have been impossible.  In
the literature, some calculations of the rational part of multi-gluon
amplitudes are reported \cite{NigelGlover:2008ur,Badger:2008cm}; our
examples below refer to a slightly different physics.

\subsection{Higgs decay to two photons through massless scalars}

\begin{figure}[t!]
\begin{center}
\includegraphics[angle=0,scale=0.70]{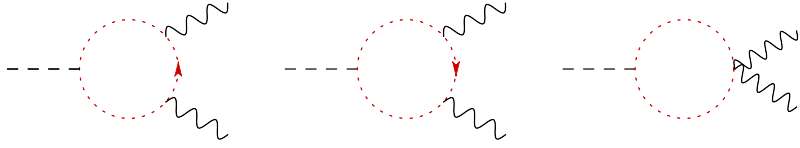}
\caption{The three diagrams for the $H \to \gamma \gamma$ decay
  mediated by a loop of charged scalars.}
\end{center}
\label{Hgamgam} 
\end{figure}

We begin by considering the decay of a scalar particle -- we will call
it the Higgs boson --  to two massless gauge bosons -- the photons --
through a loop of massless charged scalars.  We label those scalars as
$\varphi^\pm$.  If physical justification is needed, the massless
scalars can be thought of as longitudinal modes of $W$-bosons;
treating longitudinal $W$-bosons as massless scalar particles in Higgs
boson decays is justified by the Equivalence Theorem
\cite{Lee:1977eg,Chanowitz:1985hj} in the limit $m_H \gg m_W$.  We
will denote the coupling of the massless scalars to the Higgs boson as
$g_H$, so that the $H \varphi^2$ vertex is $-ig_H$.  We consider the
decay of the Higgs boson with the mass $m_H$ in its rest frame. The
photon momenta are taken to be $k_{1,2} = (m_H/2,0,0,\pm m_H/2)$. For
the decay of the Higgs boson to occur, the photons should have equal
helicities. Their polarization vectors are chosen to be
\be
e_{1,2} = \frac{1}{\sqrt{2}} (0,1,\pm i,0).
\ee
There are three different diagrams that contribute to the 
decay of the Higgs boson to two photons shown in Fig.~\ref{Hgamgam}. 
Given that virtual particles, as well as photons, are massless, the
unitarity calculation includes two triple cuts and a double cut.  The
double cuts that have a single photon on one side of the cut can be
disregarded because the corresponding master integrals are scaleless
and therefore vanish.

We begin the computation with one of the triple cuts. The loop momentum 
must satisfy  the cut constraints given by the following equations
\be
l^2 =0 ,\;\;\;\;(l+k_1)^2 = 0,\;\;\; (l-k_2)^2 = 0. 
\ee
These equations imply that $l$ must be orthogonal to $k_1$ and $k_2$, 
$(l\cdot k_1) =(l\cdot k_2) =0$,
therefore the solution of these equations 
is simply given by  $l^\mu  = l_\perp^\mu + (l \cdot \ne) \ne $, where 
$l_\perp \cdot k_1 = l_\perp \cdot k_2 = 0$ and 
$l_\perp^2 +(l \cdot \ne)^2 = 0$. 
The numerator of the triple 
cut is given by the product of two $\varphi \varphi \gamma$ vertices and the 
$H \varphi^2$ vertex.  
Because $k_{1,2} \cdot  e_{1,2} = 0$, the numerator is proportional to 
$(l_\perp \cdot e_1) (l_\perp \cdot e_2)$.  We can simplify 
this expression by using the specific form of the polarization vectors $e_{1,2}$.
We write
\be
\begin{split}
& 4 (l_\perp \cdot e_1 ) (l_\perp \cdot e_2) 
= 2 l_{\perp, \mu} l_{\perp, \nu}
\left ( e_1^{\mu} e_2^{\nu} + e_1^{\nu} e_2^{\mu} \right) 
\\
& \;\;\;\;=  - 2 l_\perp^{\mu} l_{\perp}^{\nu} \omega_{\mu \nu} (k_1,k_2) 
= 2 l_\perp^2 (e_1 \cdot e_2)  = - 2(l \cdot \ne)^2 (e_1 \cdot e_2),
\label{eq:lperpsimp}
\end{split}
\ee
where $\omega_{\mu \nu}$ is the projector on the full transverse space, 
c.f.\ \Eq(\ref{def:w}). In \Eq(\ref{eq:lperpsimp}) we used the fact
that only its four-dimensional part 
$\left ( e_1^{\mu} e_2^{\nu} + e_1^{\nu} e_2^{\mu} \right)$ 
contributes when contracted with $l_{\perp, \mu} l_{\perp, \nu}$. 
Hence, the numerator factor of the triple cut reads
\be
{\cal I}^{(3,a)} = -2 g_H e^2 (l \cdot \ne)^2 (e_1 \cdot e_2).
\ee
Computing also the second triple cut, we obtain
the contribution of the three-point function to the amplitude 
\be
{\cal A}_{t} = 
-2 g_H e^2  \int \frac{{\rm d}^Dl }{(2\pi)^D} \left (
\frac{(l\cdot \ne)^2 (e_1 \cdot e_2)
}{l^2(l+k_1)^2(l-k_2)^2}
+  \frac{(l\cdot \ne)^2 (e_1 \cdot e_2)}{l^2(l+k_2)^2(l-k_1)^2}
\right ).
\label{eq543}
\ee
It is easy to compute the corresponding integrals since they do not
depend on kinematics, c.f.\ \Eq(\ref{Remarkablysimple}),
\be\label{triangleintegral}
\int \frac{{\rm d}^Dl }{(2\pi)^D} 
\frac{(l \cdot \ne)^2}{d_1 d_2 d_3 }
= -\frac{i}{32 \pi^2} +O(\epsilon).
\ee

As the next step, we need to compute the double cut, with the Higgs
boson on one side of the cut and both photons on the other.  The
momentum on the cut satisfies the constraint
\be
l^2 = 0,\;\;\;\; (l-K)^2 = 0,\;\;\;\; K = k_1 + k_2. 
\ee
The momentum $l$ that satisfies these constraints is written as 
\be
l^\mu = \frac{1}{2} K^\mu + l_\perp^\mu + (l \cdot \ne) \ne^\mu,
\label{mom_param_bub}
\ee
where $l_\perp \cdot K = l_\perp \cdot \ne  = 0$ and
\be
l_\perp^2 + (l \cdot \ne)^2 = -\frac{K^2}{4} = -\frac{m_H^2}{4}.
\ee

The residue of the one-loop amplitude on the double cut is
proportional to $ \varphi(K-l) + \varphi(l) \to \gamma(k_1) +
\gamma(k_2) $ on-shell scattering amplitude, 
\be 
A^{\rm tree}_{\varphi(K-l) + \varphi(l) \to \gamma(k_1) +\gamma(k_2)} = 
-i e^2\left[
\frac{(2l\cdot e_1)(2l \cdot e_2)}{(l+k_1)^2}+
\frac{(2l \cdot e_2)(2l\cdot e_1)}{(l+k_2)^2}+
2 e_1 e_2
\right]\,, 
\ee
where we used the fact that $k_{1,2}\cdot e_{1}= k_{1,2}\cdot
e_{2}=0$. Therefore the residue reads
\be
   {\cal I}^{(2)} =  -g_H e^2 
\left [
4 (l \cdot e_1)(l \cdot e_2)
\left (  
\frac{ 1 }{2 l \cdot k_1}
+ \frac{1 }{2 l \cdot k_2} \right )
+ 2  e_1 \cdot e_2 
\right ].
\ee
The first two terms can be simplified since 
\be
2 l \cdot k_{1,2} = k_1 \cdot k_2  + 2 l_\perp \cdot k_{1,2}
=  \frac{m_H^2}{2} \pm m_H (l \cdot n_3),
\ee
where $n_1=(0,1,0,0),n_2=(0,0,1,0),n_3=(0,0,0,1)$.
Moreover, using  
\beq 2 (l \cdot e_1) (l \cdot e_2) = 
-e_1 \cdot e_2 \left[(l \cdot n_1)^2 + (l \cdot n_2)^2\right]
=e_1 \cdot e_2 \left [ l_\perp^2 +   (l \cdot n_3)^2 \right],
\eeq
we find   
\beq
{\cal I}^{(2)} = 
8 g_H e^2 \frac{( l \cdot \ne)^2 \; e_1 \cdot e_2 
}{m_H^2 -  4 (n_3 \cdot l )^2}.
\label{eq_348}
\ee

We note that ${\cal I}^{(2)}$ in \Eq(\ref{eq_348}) is not yet a
result for the double cut that can be easily translated into the
reduction coefficient of the two-point function. For this last step,
we need to subtract from \Eq(\ref{eq_348}) the contributions of all
possible higher-point coefficients to the double cut. This procedure
is explained in detail in Sect.\ref{sect5}. In the present case we
just need to subtract from \Eq(\ref{eq_348}) the contributions of the
three-point functions in \Eq(\ref{eq543}) to the double cut. Given the
difference in the momentum parametrization in triple and double cuts,
the term that needs to be subtracted reads
\be
{\cal I}_{\rm subtr} = 
- \frac{2 g_H e^2 \; e_1 \cdot e_2 (l \cdot \ne)^2}{l^2} \Bigg |_{l \to l_1 }
- \frac{2 g_H e^2 \; e_1 \cdot e_2(l \cdot \ne)^2}{l^2} \Bigg |_{l \to l_2},
\ee
where $l_{1,2} = l - k_{1,2}$ and 
the momentum $l$ is given in  \Eq(\ref{mom_param_bub}).
A simple computation yields
\be
{\cal I}_{\rm subtr} = 
8 g_H e^2 \frac{( l \cdot \ne)^2 \; e_1 \cdot e_2 
}{m_H^2 -  4 (n_3 \cdot l )^2}.
\label{eq_349}
\ee

Comparing \Eq(\ref{eq_348}) and \Eq(\ref{eq_349}), we see that
complete double-cut contribution of the one-loop amplitude is
contained in the triple cut which implies that the two-point function
reduction coefficients vanish.  Therefore, the one-loop amplitude for
the Higgs boson decay to two photons through a massless scalar loop is
given by \Eq(\ref{eq543}) and is entirely due to the rational part of
a three-point function.  We obtain
\be
{\cal A} =  i g_H \frac{\alpha}{2\pi} (e_1 \cdot e_2 ).
\ee
Finally, as we already mentioned, this calculation describes the $m_H
\gg m_W$ limit of the $W$-boson loop contribution to $H \to \gamma
\gamma$ scattering amplitude. The phenomenology of the Higgs boson
decay into two photons was first discussed in
Ref.~\cite{Ellis:1975ap}.

\begin{figure}[t!]
\begin{center}
\includegraphics[angle=0,scale=0.60]{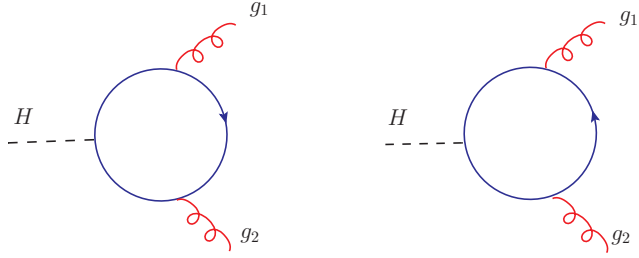}
\caption{Diagrams that contribute to Higgs boson decay to two gluons
  through a massive quark loop. }
\label{gghCut} 
\end{center}
\end{figure}

\subsection{Higgs decay to two gluons through massive quarks loop}

In this Section, we discuss the decay of the Higgs boson to two
gluons, through the triangle loop of massive quarks, see
Fig.~\ref{gghCut}. Our goal here is to illustrate two different ways --
the Passarino-Veltman reduction and the OPP procedure -- to describe
this process.  We define the amplitude for Higgs boson decay to two
gluons $H(k_3) \to g(k_1) + g(k_2)$ as
\be
{\cal M}_{h_1, h_2} 
= \delta_{a_1,a_2} \; \frac{g_H \alpha_s}{8 \pi} \;A, \quad\quad  { A}=
\int \frac{d^D l}{i\pi^{D/2}}{\cal A}_{h_1,h_2}(k_1,k_2,l),
\ee
where $h_{1,2}$ and $a_{1,2}$ are the helicity and color labels of the
two gluons, $g_H$ is the Yukawa coupling of the Higgs boson to massive
fermions and $\alpha_s$ is the strong coupling constant. Because the
Higgs boson is a spinless particle, the two gluons in the final state
must have identical helicities. Thanks to parity conservation, $M_{++}
= M_{--}$, and we only need to consider the case $h_1 = h_2 = +1$.  We
work in the rest frame of the Higgs-boson, and choose the momenta and
helicity vectors as
\be
\begin{split}
&k_{1,2}=\frac{m_H}{2}\left(  1,0,0,\pm 1\right)
\,,\quad e_{1,2}^{+}=\frac{1}{\sqrt{2}} 
\left( 0,1,\pm i,0 \right)\,,\\ &
k_{i} \cdot e^{+}_{1,2}=0 \,,\quad  e^{+}_{1}\cdot e^+_{2}=-1.
\end{split}
\ee
The integrand function can be written as 
\be
\label{hggIntegrand}
{\cal A}=\frac{{\rm Tr} 
 [(\hat{l}+m)\hat{e}_1(\hat{l}+\hat{k_1}+m)\hat{e}_2
(\hat{l}+\hat{k}_3+m)]}{d_0d_1d_3} 
 + \Big[ 1 \leftrightarrow 2 \Big],
\ee
where
\be
d_0=l^2 -m^2 ,\;d_1=(l+k_1)^2-m^2\,,\;\
d_3=(l+k_3)^2-m^2.
\ee
Evaluating the trace and setting ${\rm Tr}[1] = 2^{D/2} \to 4$, we obtain 
\be
\label{Ahgg}
{\cal A}= 4 m\left\{ 
\Big[ \frac{4(l \cdot e_1)(l \cdot e_2)+\frac{1}{2}m_H^2}{d_0d_1d_3} +\frac{1}{d_0d_3}\Big]
   +\Big[1 \leftrightarrow 2\Big]
\right\}.
\ee

We consider the OPP reduction of the amplitude in \Eq(\ref{Ahgg}).
The OPP parametrization of the integrand reads
\be
\label{AhggOPP}
{\cal A}=\frac{{\tilde c}_1(l)}{d_0d_1d_3}+\frac{{\tilde c}_2(l)}{d_0d_2d_3}
+ \frac{{\tilde b}_{03}(l)}{d_0d_3} + \ldots,
\ee
where the points of ellipsis indicate contributions to the integrand that, when
integrated over the loop momentum, can only depend on a single
parameter -- the quark mass.  We begin by calculating the triple cut,
specified by the condition $d_0=d_1=d_3=0$.  The loop momentum on this
triple cut is parametrized similar to \Eq(\ref{eq547})
\be 
\label{xitriple}
l_1^{\mu}= -k_1^{\mu} + i\frac{l_{\perp}}{\sqrt{2}}
\left(z e_1^{\mu}+\frac{1}{z}e_2^{\mu}\right)
+ \mu \ne^{\mu}\,,\quad 
 l_{\perp}^2= m^2-\mu^2,
\ee
where $\mu =  l \cdot \ne$.
Using the cut momentum \Eq(\ref{xitriple}) in  \Eq(\ref{hggIntegrand}) 
to compute the relevant residue, we obtain 
\be
\label{c_res}
{\tilde c}_1(z,\mu^2) = \lim_{d_{0,1,3} \to 0} d_0 d_1 d_3 {\cal A}
=4 m\left[
\frac{1}{2}m_H^2-2m^2+2\mu^2\right]\,.
\ee
The generic parametrization of the triple cut is given in \Eq(\ref{eq7.8});
it reads
\be
\label{c_res_param}
{\tilde c}_1(z,\mu^2)=\sum _{j=-3}^3{\tilde c}_1^{(j)}z^j 
+{\tilde c}^{(\epsilon)}_1(z) \mu^2.
\ee
Comparing \Eqs(\ref{c_res}) and (\ref{c_res_param}), we find 
\be
\label{cval}
{\tilde c}_1^{(0)}= 4m\left[
\frac{1}{2}m_H^2-2m^2\right]\,, \quad {\tilde c}_1^{(\epsilon)}(z)= 8m
\,,\quad  {\tilde c}_1^{(j)}=0,j\neq 0\,.
\ee
The coefficient of the second triple cut is obtained by the exchange
of the indices $1\leftrightarrow 2$; we conclude that $c_2=c_1$.

Next, we need to evaluate contributions from double cuts. We begin by
considering the cut $d_0(l) = 0$, $d_3(l) = 0$. The loop momentum is
parametrized as
\be
\begin{split}
&l_{03}^\mu=-\frac{k_3^\mu}{2}+il_{\perp}\left(
  \frac{k_{12}^\mu}{m_H}(1-2y)
+z\frac{e_1^\mu}{\sqrt{2}} 
+ 4y(1-y) \frac{e_2^\mu}{z\sqrt{2}}
    \right)
+ \mu \ne^\mu,
\end{split}
\ee
where $k_{12} = k_1 - k_2$ and $l_{\perp}^2=m^2-m_H^2/4 - \mu^2$.

We can compute the double-cut using this momentum parametrization in
the trace, \Eq(\ref{Ahgg}) and subtracting from it the double-cut
remnant of a triple cut contribution \Eq(\ref{c_res})
\be
\begin{split}
\label{b03l}
& {\tilde b}_{03}(l) = \lim_{d_{0,3} \to 0} d_0 d_3 
\left [ {\cal A} -  \left ( \frac{{\tilde c}_1(l)}{d_0 d_1 d_3} 
+ \frac{{\tilde c}_2(l)}{d_0 d_2 d_3} \right ) 
\right ]
\\
& = \Big[ \left(   4( l_{03} \cdot e_1 )(l_{03} \cdot e_2)+
2m^2 - 2 \mu^2 \right)\left(\frac{1}{d_1}+\frac{1}{d_2}\right) +2\Big].
\end{split}
\ee

It is easy to see that the following relations are true
\be
\begin{split}
& 4( l_{03} \cdot e_1 )(l_{03} \cdot e_2)=-2l_{\perp}^2\left[1-(1-2y)^2\right],\\
&d_1d_2=m_H^2\left[( m^2 - \mu^2) (1-2y)^2+\frac{m_H^2}{4}[1-(1-2y)^2]\right],
\end{split}
\ee
and $d_1+d_2=-m_H^2$. Inserting these  expressions  into  \Eq(\ref{b03l}) we 
find that  ${\tilde b}_{03}(l)=0$.    

We can now argue that the remaining double- and single-cut
contributions should vanish as well. Indeed, it is easy to see that
all these contributions can be written through a single {\it
  divergent} vacuum bubble master integral $I_1(m^2)$,
cf. \Eq(\ref{Scalarintegrals}). Since this integral is
ultraviolet-divergent, and the amplitude $H \to gg$ is
ultraviolet-finite, the reduction coefficient of this reduction
integral must vanish.  We conclude that the $H \to gg$ amplitude is
completely determined by the triple cut contribution.  To arrive at
the final answer, we need to use the triple cut reduction coefficient,
in conjunction with the values of master integrals. We find
\be
A=2{\tilde c}^{(0)}_1 I_3[1] + 2{\tilde c}^{(\epsilon)}_1I_3[\mu^2]
\,, \quad
I_3[x^2] = 
 \int \frac{{\rm d}^Dl }{i\pi^{D/2}} 
\frac{x^2} {d_0d_1d_3}\; ,
\ee
where $I_3[1]$ denotes the standard scalar triangle integral
and $I_3[\mu^2]$ can be obtained from \Eq(\ref{Remarkablysimple}). We find 
\be
I_3[1]=I_3(0,0,m_H^2,m^2,m^2,m^2)=\frac{2}{m_H^2} f(\tau)\,,\quad 
I_3[\mu^2]=-\frac{1}{2}\,,
\ee
where
\begin{equation} \label{ftau}
f(\tau) = \left\{ \begin{array}{ll}
  -\arcsin^2\frac{1}{\sqrt{\tau}} &, \quad \textrm{if} \ \tau > 1 \ , \\
 +\frac{1}{4} \left(\ln \frac{1+\sqrt{1-\tau}}{1-\sqrt{1-\tau}} 
- i\pi\right)^2 &, \quad \textrm{if} \ \tau \leq 1 \ .
\end{array} \right.
\end{equation}
Putting everything together, 
 we obtain the well-known result (see e.g. Ref.~\cite{Ellis:1991qj}),
\be
\label{reshgg}
A=8 m
\left[(1-\tau)f(\tau)-1\right]\,,\quad \tau=\frac{4m^2}{m_H^2}\,.
\ee

We now briefly discuss how to derive  the same result using the
Passarino-Veltman reduction.  We use the results for the
Passarino-Veltman reduction given in Section~\ref{sec:tradoneloop}.
The integral of the rank two tensor part of the first term in
\Eq(\ref{Ahgg}) is given by the $C_{00}$ function of \Eq
(\ref{Crank2text}).  We note that
\be
(l \cdot e_1)(l \cdot e_2)=-\frac{1}{2}w^{\mu\nu}l_{\mu}l_{\nu}
\,,\quad w^{\mu\nu}=-e_1^{\mu}e_2^{\nu}-e_1^{\mu}e_2^{\nu}
\,,\quad w^{\mu}_{\mu}=2,
\ee
and, therefore,
\be
\int \frac{d^D l}{i{\pi}^{D/2}}\frac{4 (l \cdot e_1)(l \cdot e_2)}{d_0d_1d_3}=
-4\; C_{00}(1,2,3)\,.
\ee
Using the reduction equations
(\ref{ldotp1}, \ref{ldotp2}, \ref{Cformfactoreq1}, \ref{C00reduction})
we obtain
\be
\label{C00ahgg}
\begin{split}
-4 \; C_{00}(1,2,3)= & -\frac{2}{D-2}(2m^2C_0(1,2,3)+B_0(1,3)) =
\\
& -2m^2 I_3[1] -I_2(m_H^2,m^2,m^2)-1.
\end{split} 
\ee 
By inserting this value of the tensor integral into the expression of
the amplitude ${\cal A}$ of \Eq(\ref{Ahgg}), the scalar bubble
integral function $ I_2(m_H^2,m^2,m^2)$ is cancelled by the
contribution from the two-denominator term in \Eq(\ref{Ahgg}) and we reproduce
the previous result (\ref{reshgg}).  As already has been pointed out
in Section~\ref{sec:tradoneloop} in the Passarino-Veltman reduction
the rational part is obtained by a cancellation of the $1/\epsilon$
singularity of the two-point scalar integral with the
$\epsilon$-dependent coefficient $2/(D-2)$, that appears in the
reduction formula, \Eq(\ref{C00ahgg}).

\subsection{The rational part of the four-photon scattering amplitudes}

Our next example concerns the photon-photon scattering through a loop
of massive fermions. This process is described by an
ultraviolet-finite scattering amplitude that, however, can not be
computed in four dimensions. As a consequence, the four-photon
scattering amplitude has a non-trivial rational part whose computation
we now describe.

We denote the incoming momentum of the $i$-th photon by $q_i$ and its
polarization vector by $e_i$, $i=1, \ldots, 4$.  There are six diagrams that
contribute to the scattering amplitude, but only three of them are
independent. This allows us to write
\be
{\cal M}_{\rm tot} = -2 i \alpha^2 {\cal M},\;\;\;
{\cal M} = \sum _{i=a,b,c} {\cal M}_i.
\ee

\begin{figure}[t!]
\begin{center}
\includegraphics[angle=0,scale=0.70]{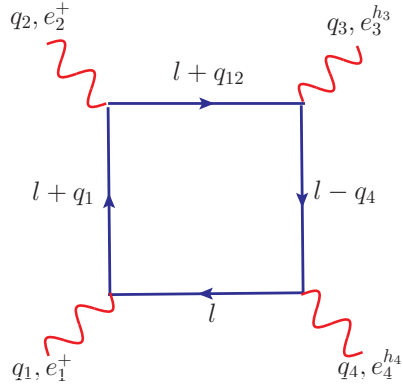}
\caption{One of the diagrams that contributes to photon-photon
  scattering.}
\end{center}
\label{figa} 
\end{figure}

Contribution of the diagram $a$, shown in Fig.~\ref{figa},  reads 
\be
{\cal M}_a = 
-i \int \frac{{\rm d}^Dl}{\pi^{D/2}}
\frac{ {\rm Num}_a(l,m,\{q_i \},\{e_i\})}{d_0 d_1 d_2 d_3},
\label{eq83_1}
\ee
where 
$d_0 = (l^2-m^2)$, $d_1 = (l+q_1)^2-m^2$, 
$ d_{2} = (l+q_{12})^2-m^2$, $d_3 = (l-q_4)^2-m^2$,  
$q_{12} = q_1 + q_2$ and 
the numerator function  is 
\be
{\rm Num}_a = 
{\rm Tr}(\hat e_1 ( \hat l + m) \hat e_4 ( \hat l - \hat q_4 + m) 
\hat e_3 ( \hat l + \hat q_{12} + m) \hat e_2 (\hat l + q_1 + m) ).
\label{eq83_2}
\ee
Contributions of diagrams $b,c$ are obtained from \Eq(\ref{eq83_1}) by
simple substitutions, ${\cal M}_b = {\cal M}_a( 3 \leftrightarrow 4) $
and ${\cal M}_c = {\cal M}_a( 2 \leftrightarrow 3)$.

The polarization vector $e$ of a massless gauge boson can be chosen in
such a way that it satisfies a constraint $ e \cdot n_{\rm aux}$,
where $n_{\rm aux}$ is an auxiliary light-like vector. We use this
freedom and choose $q_2$ to be an auxiliary vector for $e_1$, $q_1$
for $e_2$, $q_3$ for $e_4$ and $q_4$ for $e_3$.  We shall refer to
these choices as gauge fixing conditions.  As an example, we consider
positive helicities of the photons $1$ and $2$. Then we can write
\be
e_1^{+\mu} e_2^{+\nu} = \frac{1}{2} (e_1^+ \cdot e_2^+) 
\left ( \omega^{\mu \nu}(q_1 ,q_2)
- 
i (q_1 \cdot q_2)^{-1}
\epsilon^{\mu \nu \alpha \beta} q_{1,\alpha} q_{2,\beta} 
\right ),
\label{w33}
\ee
where $\omega^{\mu \nu} (q_1,q_2)$ is the metric tensor of the linear vector 
space that is transverse to the two light-like 
vectors  $q_1$ and $q_2$
\be
\omega^{\mu \nu} (q_1,q_2) = g_4^{\mu \nu} 
- \frac{q_1^\mu q_2^\nu + q_1^\nu q_2^\mu}{q_1 \cdot q_2}.
\ee
We note that $g_4^{\mu \nu}$ is the metric tensor of the four-dimensional 
space.

As follows from \Eqs(\ref{eq83_1}, \ref{eq83_2}), the photon-photon
scattering is described by the four-point tensor integrals up to rank
four.  As we explained in Sect.~\ref{sec_rational}, only rank-four and
rank-three four-point tensor integrals contain a rational part; for
this reason we focus on those integrals in what follows.  We begin by
considering the rational part of the photon scattering amplitude that
originates from the rank-four tensor integrals. We shall use the
notation ${\cal R}[{\cal O}]$, to denote the rational part of the
amplitude ${\cal O}$.  For diagram $a$ the corresponding expression
reads
\be
{\cal R} [ {\cal M}_a^{(4)} ]  = 
{\rm Tr}[ \hat e_1 \gamma^\mu \hat e_4 \gamma^\nu \hat e_3 \gamma^\alpha 
\hat e_2 \gamma^\beta ] {\cal R}^{4,a}_{\mu \nu \alpha \beta}, 
\label{r1}
\ee 
where
\be
\begin{split} 
{\cal R}^{4,a}_{\mu \nu \alpha \beta} =
-i\; {\cal R} \Bigg [\int \frac{{\rm d}^D l}{\pi^{D/2}} \times 
\frac{l_\mu l_\nu l_\alpha l_\beta}{d_0 d_1 d_2 d_3}
 \Bigg ].
\end{split}
\ee
We can calculate trace in \Eq(\ref{r1}) and simplify it by noticing
that the rank-four tensor integral is fully-symmetric and that, when
all of its indices are contracted, the rational part vanishes.  This
allows us to write
\be
{\cal R}[{\cal M}_a^{(4)}]  = \left ( 
32 e_1^\mu e_2^\nu e_3^\alpha e_4^\beta 
- 8 f_{1,4;2,3}^{\mu \nu} g^{\alpha \beta} 
- 8 f_{1,2;3,4}^{\mu \nu} g^{\alpha \beta} 
\right ) R^{4,a}_{\mu \nu \alpha \beta}, 
\ee
where $f_{ij;km}^{\mu \nu} = e_i^\mu e_j^\nu (e_k \cdot e_m) + e_k^\mu
e_m^\nu (e_i \cdot e_j) $.  A computation of a similar contribution
for other diagrams amounts to a simple permutation of momenta and
polarization vectors.  The complete result for the rational part that
originates from the rank-four tensor integrals reads
\be
\begin{split} 
& {\cal R}[{\cal M}^{(4)}] =
32 e_1^\mu e_2^\nu e_3^\alpha e_4^\beta  R^4_{\mu \nu \alpha \beta} 
- 8 f_{1,4;2,3}^{\mu \nu} g^{\alpha \beta} \left ( R^{4,a}_{\mu \nu \alpha \beta} 
+R^{4,c}_{\mu \nu \alpha \beta} \right ) \\
& 
- 8 f_{1,2;3,4}^{\mu \nu} g^{\alpha \beta} 
\left ( R^{4,a}_{\mu \nu \alpha \beta} + R^{4,b}_{\mu \nu \alpha \beta} 
\right ) 
- 8 f_{1,3;2,4}^{\mu \nu} g^{\alpha \beta} 
\left ( R^{4,b}_{\mu \nu \alpha \beta} 
+ R^{4,c}_{\mu \nu \alpha \beta} \right ),
\end{split}
\label{ff25}
\ee
where $R^4_{\mu \nu \alpha \beta} = \sum \limits_{i=a,b,c}^{} R^{4,i}_{\mu \nu \alpha \beta}$.

All but one term in \Eq(\ref{ff25}) include a contraction of the metric
tensor $g^{\alpha \beta}$ with the rational part of various rank-four
tensors. Those terms are easy to compute because such a contraction
gives a rank-two four-point function which does not have a rational
part and a rank-two three-point function whose rational part is
simple, cf.~\ref{app:RT}. We obtain \be
\begin{split} 
& g^{\alpha \beta} R^{4,a}_{\mu \nu \alpha \beta}
= g^{\alpha \beta} R^{4,c}_{\mu \nu \alpha \beta}
= \frac{1}{4} \omega_{\mu \nu}(q_2,q_3),
\\
& g_{\alpha \beta} R^{4,b}_{\mu \nu \alpha \beta}
= \frac{1}{4} \omega_{\mu \nu}(q_2,q_4).
\end{split}
\ee
Hence, the rational part of the amplitude related to the 
rank-four tensor contribution is  
\be
\begin{split} 
& {\cal R}[{\cal M}^{(4)}] =
32 e_1^\mu e_2^\nu e_3^\alpha e_4^\beta  R^4_{\mu \nu \alpha \beta} 
- 4 f_{1,4;2,3}^{\mu \nu} \omega_{\mu \nu}(q_2,q_3) 
\\
& - 2 \left ( f_{1,2;3,4}^{\mu \nu}  + f_{1,3;2,4}^{\mu \nu}  \right )
\left ( \omega_{\mu \nu}(q_2,q_3) 
+ \omega_{\mu \nu}(q_2,q_4) \right ).
\end{split}
\ee

As the next step, we consider  the rank-three part of the amplitude. 
There are more terms that contribute in this case but a  
simplification comes from the observation that if any pair of 
indices of a rank-three four-point function is contracted with 
the metric tensor, the resulting integrals have no rational part. 
We compute  traces, use the gauge fixing conditions for the 
polarization vectors and disregard all terms  where two indices of the 
rank three tensor are contracted. For the diagram $a$ we find  
\be
\begin{split}
{\cal R}[{\cal M}_a^{(3)}] & = 
-16 \Big ( 
(e_1 \cdot e_2) e_4^\mu e_3^\nu q_1^\alpha 
+ (e_2 \cdot e_3) e_4^\mu e_1^\nu q_{12}^\alpha 
\\
& - (e_4 \cdot e_3) e_1^\mu e_2^\nu q_4^\alpha 
\Big ) R^{3,a}_{\mu \nu \alpha}(q_1,q_{12},-q_4),
\label{w2}
\end{split} 
\ee
where the momenta of three propagators that enter the rank-three 
box integral are shown explicitly.  We can simplify this expression 
using the symmetry of $R_{3,a}^{\mu \nu \alpha}$ with respect to its 
Lorentz indices and  the fact that in \Eq(\ref{w2}) 
$R_{3,a}^{\mu \nu \alpha}$ is always contracted with one of the propagator 
momenta, projecting the general expression for the rational 
part of the rank-three four-point functions, that is found in~\ref{app:RT}, 
on one of the vectors in the van Neerven-Vermaseren basis. Moreover, 
for simplification purposes,  it is useful to  
write the rational part  of 
the rank-three four-point integral as 
\be
\begin{split} 
& R_{3,a}^{\mu \nu \alpha}(q_1,q_{12},-q_4) 
  = \frac{1}{8} 
\Big ( 
v_1^\alpha ( \omega^{\mu \nu}(q_3,q_4) - \omega^{\mu \nu}(q_2, q_3) ) 
 \\
& + v_2^\alpha ( \omega^{\mu \nu}(q_1,q_4) - \omega^{\mu \nu}(q_2, q_3) ) 
+ v_3^\alpha ( \omega^{\mu \nu}(q_1,q_2) - \omega^{\mu \nu}(q_2, q_3) ) 
\Big ),
\label{eq_355}
\end{split} 
\ee
where the vectors $v_{1,2,3}$ are such that the only non-vanishing 
scalar products are $v_1 \cdot q_1 = 1$, $v_2 \cdot q_{12} = 1$
and $v_3 \cdot q_4 = -1$. 
We use \Eq(\ref{eq_355}) in \Eq(\ref{w2}),  
obtain contributions of other diagrams by permutations 
of momenta and polarization vectors, and 
arrive at the following result for the contribution 
of rank-three  tensors to the rational part 
\be
\begin{split} 
& {\cal R}[{\cal M}^{(3)}] 
= -4 \sum \limits_{i \neq j \neq k \neq m} 
\left (     
e_i \wedge e_j\right)_{ij}  (e_k \cdot e_m) 
+ 4 f_{1,4;2,3}^{\mu \nu} \omega_{\mu \nu}(q_2,q_3)
\\ 
& + 2 ( f_{1,2;3,4}^{\mu \nu} + f_{1,3;2,4}^{\mu \nu}  ) 
\left ( \omega^{\mu \nu}(q_2,q_3) 
+ \omega^{\mu \nu}(q_2,q_4) 
\right )
\\
& + 32 ( e_{1}^\mu e_2^\nu e_4^\alpha (e_3 \cdot q_1)
+ e_{1}^\mu e_3^\nu 
e_4^\alpha (e_2 \cdot q_3) )R^{\mu \nu \alpha}_{3,c}(q_1,q_{13},-q_4). 
\label{e3c}
\end{split} 
\ee
In \Eq(\ref{e3c}), we introduced 
the  notation 
\be
\left ( e_i \wedge e_j \right )_{ij} 
= e_i^\mu \omega_{\mu \nu}(q_i,q_j) e_j^\nu.
\ee
The term with $R_{3,c}$ in \Eq(\ref{e3c})  does not contribute. 
We can see this using the explicit expression 
\be
\begin{split} 
& R^{\mu \nu \alpha}_{3,c}(q_1,q_{13},-q_4)  
  = \frac{1}{8} 
\Big ( 
v_1^\alpha ( \omega^{\mu \nu}(q_2,q_4) - \omega^{\mu \nu}(q_2, q_3) ) 
 \\
& + v_2^\alpha ( \omega^{\mu \nu}(q_1,q_4) - \omega^{\mu \nu}(q_2, q_3) ) 
+ v_3^\alpha ( \omega^{\mu \nu}(q_1,q_3) - \omega^{\mu \nu}(q_2, q_3) ) 
\Big ),
\end{split}
\ee 
so that 
\be
e_1^\nu e_2^\alpha R_{3,c}^{\mu \nu \alpha} = 
e_3^\nu e_4^\alpha R_{3,c}^{\mu \nu \alpha} = 0,
\ee
thanks to our gauge fixing conditions.  Finally, if we combine the
rank-four and the rank-three parts of the amplitude, terms with
$f_{ij;km}^{\mu \nu}$ cancel out and we obtain the simple result for
the full rational part of the four-photon scattering amplitude
\be
{\cal R}[{\cal M}] = 
32 e_1^\mu e_2^\nu e_3^\alpha e_4^\beta  R^4_{\mu \nu \alpha \beta} 
-4 \sum \limits_{i \neq j \neq k \neq m} 
\left ( e_i \wedge e_j\right)_{ij} (e_k \cdot e_m) .
\label{Rank4bit}
\ee

We are left with the question of how to compute the first term in
\Eq(\ref{Rank4bit}), i.e.\ the rational part of the rank-four box
contracted with the polarization vectors.  To do this, we use
\Eq(\ref{w33}) where the tensor product of two photon polarization
vectors is written as a linear combination of the metric tensor and
the Levi-Civita tensor and observe that the Levi-Civita tensor does
not contribute because of the symmetry of $R_4^{\mu \nu \alpha
  \beta}$.  We obtain
\be
\begin{split} 
& 32 e_{1,\mu}^{+} e_{2,\nu}^{+} R_4^{\mu \nu \alpha \beta} 
= 16 (e_1^+ \cdot e_2^+) \omega^{\mu \nu}(q_1,q_2) R_4^{\mu \nu \alpha \beta} 
\\
& 
= 16 (e_1^+ \cdot e_2^+) 
\left ( g_4^{\mu \nu} 
- 4\frac{q_1^\mu q_2^\nu}{q_{12}^2}
\right ) R^4_{\mu \nu \alpha \beta}
\\
& = 
16 (e_1^+ \cdot e_2^+) 
\left ( g^{\mu \nu} - g_{-2\ep}^{\mu \nu} 
- 4\frac{q_1^\mu q_2^\nu}{q_{12}^2}
\right ) R^4_{\mu \nu \alpha \beta}.
\label{w3}
\end{split} 
\ee 
Note that in the last step in \Eq(\ref{w3}), we wrote $g_4^{\mu \nu}$
-- the metric tensor of the four-dimensional space -- as a difference
of the full metric tensor $g^{\mu \nu}$ and the metric tensor of the
$(D-4)$-dimensional space. The contraction of $R_4^{\mu \nu \alpha
  \beta}$ with the tensor $g_{\mu \nu}$ is straightforward to compute
since the resulting integral is just the rank-two three point function
whose rational part is very simple.  To calculate $q_1^{\mu} q_2^{\nu}
R^4_{\mu \nu \alpha \beta}$, we contract the momentum $q_1$ with the
loop momentum $l$, rewrite the scalar product through the inverse
propagators and shift the loop momentum, as appropriate, to ensure
cancellations of (parts) of diagrams. We obtain
\be
2 q_1^{\mu} R^4_{\mu \nu \alpha \beta} 
= \frac{1}{4} 
\left ( 
q_{34}^{\{\nu } \omega^{\alpha \beta \} }(q_3,q_4) 
+q_{14}^{\{\nu } \omega^{\alpha \beta \} }(q_2,q_3) 
-q_{3}^{\{\nu } \omega^{\alpha \beta \} }(q_2,q_4) 
\right ),
\label{w55}
\ee
where $q_{14} = q_1 - q_4$. \Eq(\ref{w55}) -- a remnant of the
gauge-invariance -- allows us to get rid of all rank four-boxes and
rank-three triangles whose rational parts are complicated.  The last
thing we need is the contraction of $R_4^{\mu \nu \alpha \beta}$ with
$g_{-2\ep}^{\mu \nu}$. We obtain
\be
g_{-2\ep}^{\mu \nu} R_{\mu \nu \alpha \beta} 
= \frac{1}{24 \epsilon}
\left ( - 2\ep g_{\alpha \beta} + 2 g^{-2\ep}_{{\alpha \beta}} \right).
\ee
Putting the three terms together, we find 
\be
32 e_1^{+,\mu} e_2^{+,\nu} e_3^\alpha e_4^\beta  R^4_{\mu \nu \alpha \beta} 
= 12 (e_1^+ \cdot e_2^+) (e_3 \cdot e_4). 
\ee
Hence, the rational part of the full 
photon-photon scattering amplitude 
reads 
\be
{\cal R}[{\cal M}] = 
12 (e_1^+ \cdot e_2^+) (e_3 \cdot e_4) 
-4 \sum \limits_{i \neq j \neq k \neq m} 
\left ( e_i \wedge e_j\right) (e_k \cdot e_m) .
\ee
We can further simplify the sum in the above equation using 
the gauge fixing condition.  We find 
\be
\begin{split} 
& {\cal R} [{\cal M}^{++\lambda_3 \lambda_4}] =
8 i \alpha^2 (e_1^+ e_2^+ ) 
\Bigg [ 1 
+ \left ( 1 - \bar \lambda_{34} \right ) 
\frac{(q_1 \cdot q_2) (q_1 \cdot e_3) (q_1 \cdot e_4) }{q_1 \cdot q_3 \; q_1 \cdot q_4 }
\Bigg ],
\end{split} 
\label{w5}
\ee 
where $\bar \lambda_{34} = (\lambda_3 + \lambda_4)/2$ is the
average helicity of the photons 3 and 4.

We note that the result for the rational part of the photon-photon
scattering amplitude \Eq(\ref{w5}) agrees with the explicit
computation in Ref.~\cite{Binoth:2006hk} which was performed for {\it
  massless} virtual fermions.  Our derivation shows that the rational
part of the four-photon scattering amplitude is, in fact, {\it
  mass-independent}. In particular, the rational part is given by
\Eq(\ref{w5}) even if the energy of the photon-photon scattering is
{\it much smaller} than the fermion mass.  We emphasize that this
feature is striking because the full amplitude in this limit {\it
  does} depend on the mass of the virtual fermion. In fact, it is well
known that the full amplitude for photon-photon scattering, given by
the sum of the cut-constructible and the rational parts, scales as
\be
{\cal M} \sim e^4 \frac{s^2}{m^4},\;\;\; m \gg \sqrt{s},\;\;
\label{w6666}
\ee 
where $\sqrt{s}$ is the energy of the photon-photon collision.  It
follows that in order to make \Eqs(\ref{w5}, \ref{w6666}) consistent,
we must assume that the rational part \Eq(\ref{w5}) cancels the
mass-independent term in the cut-constructible part, making the full
amplitude consistent with the required mass scaling. However, such
cancellations signal potential sources of numerical instabilities at
low energies, $\sqrt{s} \ll m$, which may be hard to deal with.

\subsection{Rational parts of $n$-photon scattering amplitudes, $n > 4$}

We comment on the rational part of
higher-multiplicity photon scattering amplitudes.  For six-photon
scattering, the rational part was computed in
Ref.~\cite{Binoth:2006hk} and was found to be zero
\be
{\cal R} [ {\cal M}_{6\gamma}] = 0.
\ee
Later, it was argued in Ref.~\cite{Badger:2008rn} that rational parts
of $n$-photon scattering amplitudes vanish for $n \ge 6$.  In what
follows, we summarize arguments given in Ref.~\cite{Badger:2008rn}.
In the remainder of this Section, we consider photon scattering
mediated by {\it massless} fermions.

To simplify the problem, it is useful to invoke the supersymmetric
decomposition of scattering amplitudes \cite{Bern:1994cg}. This
decomposition allows us to argue that rational parts of a photon
scattering amplitude mediated by fermions are identical to the
rational parts of a photon scattering amplitude mediated by loops of
charged scalars $\varphi$.  Cuts of those one-loop amplitudes lead to
tree amplitudes that describe $0 \to 2 \varphi + n \gamma$ scattering.
We first discuss properties of tree scattering amplitudes.  Once these
amplitudes are understood, it is straightforward to present arguments
that explain the vanishing of the rational parts for multi-photon
scattering amplitudes.

We take $p_{a,b}$ as momenta of the two scalars and denote the momenta
of the photons as $k_i,\; i=1,..,n$.  In the spirit of BCFW momentum
deformation \cite{Britto:2005fq}, we shift momenta $p_{a,b}$ by the
light-like momentum $q$ that is orthogonal to $p_{a,b}$. We then write
\be
p_{a,b} \to p_{a,b}(z) = p_{a,b} \pm z q.
\ee
We are interested in the behavior of $0 \to 2 \varphi + n \gamma$
scattering amplitudes in the limit $z \to \infty$.  To simplify
studies of the large-$z$ limit, we choose vector $q$ as auxiliary
quantization vector for {\it all} photons, so that $e_i \cdot q = 0$,
$i=1,..n$.  Since the $\varphi ^2 \gamma$ vertex is proportional to
the difference\footnote{All the momenta are considered to be
  outgoing.} of the momenta of the two scalars, choosing $e_i \cdot q
= 0$ ensures that the triple vertex becomes $z$-independent. The
$\varphi^2 \gamma \gamma$ vertex, on the other hand, is
$z$-independent in any gauge and all the off-shell propagators in the
scattering amplitudes scale as ${\cal O}(z^{-1})$ in the large-$z$
limit.  Hence, the large $z$-behavior of $0 \to 2 \varphi + n \gamma$
scattering amplitudes is determined by diagrams with the {\it
  smallest} number of propagators. This requirement selects diagrams
with the {\it largest} number of quartic $\varphi^2 \gamma \gamma$
vertices.  It is easy to understand that, for $n$-external photons,
the minimal number of propagators is $n/2-1$ if $n$ is even and
$(n-1)/2$, if $n$ is odd.  Such scaling implies that
\be
\lim_{z \to \infty} \; {\cal A}_n \sim 
\left \{
\begin{array}{cc}
z^{1-n/2}, & n~{\rm even},\\
z^{(1-n)/2}, & n~{\rm odd}.
\end{array}
\right.
\label{eq_371}
\ee
However, in reality the amplitudes {\it decrease faster} in the large
$z$-limit, than indicated by \Eq(\ref{eq_371}) \cite{Badger:2008rn}.
To see why this happens, consider the $n = 4$ amplitude. Among six
diagrams with two $\varphi^2 \gamma \gamma$ vertices, there are three
groups which correspond to definite combinations of scalar products of
the polarization vectors. We write
\be
{\cal A}_4 = \sum \limits_{[ij],[lm]} 
(e_i \cdot e_j) ( e_l \cdot e_m) A_{[ij],[lm]}
+{\cal O}(z^{-2}),
\ee
where $[ij] \neq [lm] \in \{[12],[13],[14],[23],[24],[34] \}$.
Amplitude $A_{[ij],[lm]}$ is determined by the sum of two diagrams
with inverse Feynman propagators that are $ zq \cdot (k_i+k_j)$ and $z
q \cdot (k_l + k_m)$, in the large $z$-limits.  Hence, taking the sum
of the two diagrams that contribute to each of these groups, we find
\begin{equation}
\begin{split} 
& \lim \limits_{z \to \infty} 
A_{[ij],[lm]} \approx
\frac{1}{z q \cdot (k_i+k_j)}+ \frac{1}{z q \cdot (k_l+k_m)}
+ {\cal O}(z^{-2}) 
\\
& = \frac{q \cdot ( k_i + k_j + k_l + k_m)}{z q \cdot (k_i+k_j)\;
  q \cdot (k_l + k_m) }
+ {\cal O}(z^{-2}) 
\\
& = -\frac{q \cdot (p_a(z) + p_b(z))}
{z q \cdot (k_i+k_j)\;  q \cdot (k_l + k_m) } + {\cal O}(z^{-2}) 
= {\cal O}(z^{-2}).
\label{wre}
\end{split} 
\end{equation}
The very last step follows from the fact that $q$ is transverse to the
momenta of the two scalars, $q \cdot p_a = q \cdot p_b = 0$. Hence, we
conclude that for $n=4$, the amplitude scales as ${\cal A}_4 \sim
z^{-2}$ in the large-$z$ limit.  It is straightforward to see how this
cancellation mechanism generalizes to higher multiplicities.  Consider
the $n=5$ case and take all diagrams where photons $e_{1,2}$ and
$e_{3,4}$ form the $\varphi^2 \gamma \gamma$ vertices, and the photon
$e_5$ contributes to $\varphi^2 \gamma$ vertex.  These diagrams have
two propagators and, therefore, naively should scale as
$z^{-2}$. However, collecting all such diagrams -- that differ by the
permutation of $(e_1 \cdot e_2)$, $(e_3 \cdot e_4)$ and $e_5$, we find
that they are proportional to
\be
q \cdot \sum \limits_{i=1}^{5} k_i
= - q \cdot (p_a + p_b) = 0, 
\ee
which implies that ${\cal A}_{5} \sim z^{-3}$.  The mechanism
described here 
is generic and can be shown to be valid for arbitrary multiplicities
\cite{Badger:2008rn}. 
We conclude that the tree $0 \to 2\varphi + n \gamma$ amplitudes scale as
\be
 \lim_{z \to \infty} {\cal A}_n 
\sim z^{-n+2}.
\label{scale_right}
\ee
We can use very similar arguments to understand properties of the
rational part of one-loop $n$-photon scattering amplitudes, mediated
by scalars.  We shall compute the rational part directly following
Ref.~\cite{Badger:2008cm}. We explained this method in
Section~\ref{ratpart_bad}.  We need to consider quadruple, triple and
double cuts. The respective rational contribution due to the
quadruple cut is calculated following \Eq(\ref{qcutrat}), where the
parametrization of the cut loop momentum reads
$
l_{q,\pm}^\nu = V_4^\nu \pm  l_\perp n_4^\nu + \mu n_\ep^\nu.
$
We remind the reader that $l_{q,\pm}^2 = 0$ and that $l_\perp \sim
\mu$, in the limit of large $\mu$.  To connect the calculation of the
rational part with the large-$z$ scaling discussed previously, we note
that we can take $n_4$ as the auxiliary quantization vector for the
photons. Then, $\mu$ plays the role of $z$ and $n_4$ plays the role of
$q$ in the preceding discussion of the tree amplitudes.  Therefore,
taking the large-$\mu$ limit and using \Eq(\ref{scale_right}), we find
\be
\lim_{\mu \to \infty} \left [ A_1 A_2 A_3 A_4 \right ] 
\sim \mu^{-n + 8}.
\ee
Since, as follows from \Eq(\ref{qcutrat}), only the ${\cal O}(\mu^4)$
term in the product of four on-shell amplitudes can lead to a
quadruple cut contribution to the rational part, we find that if the
number of photons exceeds four, the quadruple-cut contribution to the
rational part vanishes.

As the next step, we consider a triple cut. We parametrize the loop
momentum as
$
l^\nu_{t,\pm} = V_3^\nu + l_\perp \left ( 
t n_\mp^\nu  + t^{-1} n_\pm^\nu \right ) 
+ \mu \ne
$.
According to \Eq(\ref{tripcutrat}), we need to consider the Laurent
expansion in the variable $t$ at infinity, followed by the Laurent
expansion in the variable $\mu$ at infinity.  Similar to the quadruple
cut discussed above, $l_\perp \sim \mu$, in the large-$\mu$ limit.  We
can take $n_\pm$ as auxiliary quantization vectors for photons, in
respective amplitudes.  Using \Eq(\ref{scale_right}), we find that the
product of three amplitudes scales as
\be
\lim_{\mu \to \infty} 
\lim _{t \to \infty} [A_1 A_2 A_3 ] \sim \mu^{6-n} t^{6 - n}.
\label{eq382}
\ee
According to \Eq(\ref{tripcutrat}), the rational part requires picking
up the ${\cal O}(t^0 \mu^2)$ contribution from the product of the
three amplitudes and, as follows from \Eq(\ref{eq382}) such terms do
not exist.  Finally, one can address the double-cut contribution to
the rational part using similar arguments. We do not discuss this
issue here. The interested reader may consult
Ref.~\cite{Badger:2008rn} where it was shown that double-cut
contributions also vanish for $n \ge 6$.

\subsection{The axial anomaly}

One of the manifestations of the Adler-Bell-Jackiw axial anomaly in
QED \cite{Adler:1969gk,Bell:1969ts} is the peculiar property of the
matrix element of the divergence of the axial current $J^5_\mu = \bar
\psi \gamma_\mu \gamma_5 \psi$, where $\psi$ is the ``electron''
field, taken between the vacuum and the two-photon states.  For
massless electrons, such a matrix element reads
\be
\label{ABJ}
\begin{split}
{\cal M}_{ABJ}\,=\,&
\la \gamma{(k_1,\lambda_1)} \gamma{(k_2,\lambda_2)} \mid \partial^\mu
J^5_\mu(0)\mid0\ra
 \,  = \,  
\frac{e^2}{2\pi^2}
\varepsilon^{{\mu}{\nu}{ \lambda}{\rho}}
e^*_{1\mu} k_{1\nu} e^*_{2\lambda} k_{2\rho},
\end{split} 
\ee
where $k_{1,2}$ and $e_{1,2}$ are momenta and polarization vectors of
the outgoing photons with helicities $h_{1,2}$.  The matrix element
${\cal M}_{ABJ}$ is purely rational.  Below we derive ${\cal M}_{ABJ}$
using the algorithm of $D$-dimensional unitarity.

The amplitude $M_{ABJ}$ is given by the sum of two triangle Feynman
diagrams with the electron loop.  The matrix element is written as
%
\def\khat{\hat{k}}
\def\lhat{\hat{l}}
\def\eshat{\hat{e}^*}
%
\be
\label{axialloop}
\begin{split}
{\cal M}_{ABJ} & = \frac{i e^2}{(4\pi)^{(D/2)}}
\int\frac{d^Dl}{i\pi^{(D/2)}}
\\
& 
\times {\rm Tr}
\left\{
\khat_{12}\Gamma_{\gamma_5}
\left[
\frac{\lhat\eshat_1(\lhat+\khat_1)\eshat_2(\lhat+\khat_{12})}
{l^2(l+k_1)^2(l+k_{12})^2}
\right] + (1\leftrightarrow 2)
\right\},
\end{split}
\ee
where $k_{12}=k_1+k_2$.  The external momentum and polarization
vectors are four-dimensional, whereas the loop momentum and the Dirac
matrices $\Gamma^\mu$ and the matrix $\Gamma_{\gamma_5}$ are continued
to $D$-dimensions, following the discussion in Sect.~\ref{sec:7new}.
We note that $\Gamma_{\gamma_5}$ in \Eq(\ref{axialloop}) denotes the
$D$-dimensional continuation of the matrix $\gamma_5$.  We perform
such a continuation following the t'Hooft and Veltman prescription
\cite{tHooft:1972fi}. It is defined by the set of commutation
relations
\be
\begin{split}
&\left\{\Gamma^{\mu},\Gamma_{\gamma_5}\right\}=0, \ \  {\rm for}\  \mu=0,1,2,3,
\\
&
\left[\Gamma^{\mu},\Gamma_{\gamma_5}\right]=0, \ \  {\rm for}\ \mu=4,\ldots D-1.
\end{split}
\ee

\Eq(\ref{axialloop}) defines the integrand function of the loop
momentum integral but it does not define it uniquely.  This is not a
problem since the integral is regularized dimensionally and shifts of
the loop momenta are allowed. We will exploit such shifts to simplify
the computation.  To this end, we split the integrand in
\Eq(\ref{axialloop}) using the identity
\be
\khat_{12}\Gamma_{\gamma_5}\, = \,  ( \lhat + \khat_{12}) \Gamma_{\gamma_5}  
 + \Gamma_{\gamma_5} \lhat -2 \Gamma_{\gamma_5} \lhat_{\ep},
\ee
where $l_{\ep}^\mu = (l \cdot \ne) \ne^\mu $ is the $(D-4)$-dimensional 
part of the loop momentum. 
After the split, 
the trace in \Eq(\ref{axialloop}) gets additional 
terms
\be
\begin{split}
& 
{\rm Tr}
\left\{
\khat_{12}\Gamma_{\gamma_5}
\left[
\frac{\lhat\eshat_1(\lhat+\khat_1)\eshat_2(\lhat+\khat_{12})}
{l^2(l+k_1)^2(l+k_{12})^2}
\right] + (1\leftrightarrow 2)
\right\}
= {\rm Tr}_1 + {\rm Tr}_2,
\\
& {\rm Tr}_1 =  -2 {\rm Tr} \left\{
\Gamma_{\gamma_5}
\lhat_{\ep}
\frac{\lhat\eshat_1(\lhat+\khat_1)\eshat_2(\lhat+\khat_{12})}
{l^2(l+k_1)^2(l+k_{12})^2}
+ (1\leftrightarrow 2)
\right\},
\\
&
{\rm Tr}_2 = 
{\rm Tr}
\left\{
\Gamma_{\gamma_5}
\left[
\frac{\lhat\eshat_1(\lhat+\khat_1)\eshat_2}
{l^2(l+k_1)^2} + 
\frac{\eshat_1(\lhat+\khat_1)\eshat_2(\lhat + \khat_{12})}
{(l+k_1)^2(l+k_{12})^2}
\right] + (1\leftrightarrow 2)
\right\}.
\end{split}
\ee
Fortunately, it is easy to perform shifts of the loop momenta of the
type $l \to l - k_{1,2}$ to show that the contribution due to ${\rm
  Tr}_2$ vanishes. This allows us to re-write \Eq(\ref{axialloop}) in
a simplified form
\be
\label{eqaxial1}
\begin{split} 
& {\cal M}_{ABJ} = 
\int\frac{d^Dl}{i(\pi)^{(D/2)}} 
{\cal I}_{ABJ}(k_1,k_2,e_1,e_2,l),
\\
&  {\cal I}_{ABJ}= \frac{-2 ie^2}{(4\pi)^{(D/ 2)}}
{\rm Tr}\left\{
\Gamma_{\gamma_5}
\lhat_{\ep} 
\frac{\lhat\eshat_1(\lhat+\khat_1)\eshat_2(\lhat+\khat_{12})}
{l^2(l+k_1)^2(l+k_{12})^2}
+ (1\leftrightarrow 2)
\right\}.
\end{split}
\ee
Since the integrand ${\cal I}_{ABJ}$ is proportional to $\lhat_{\ep}$,
its cut-constructible part vanishes and its OPP parametrization
becomes simple
\be
\label{ABJparam}
\begin{split}
{\cal I}_{ABJ}(l)  \,=\, &
\frac{c_1
  l_{\ep}^2}{d_0 d_1 d_{12}}
+\frac{c_2l_{\ep}^2}{d_0 d_2 d_{12}}
+\frac{b_1 l_{\ep}^2}{d_0 d_1}
+\frac{b_2 l_{\ep}^2}{d_0 d_2}
\\
& +\frac{b_3 l_{\ep}^2}{d_1 d_{12}}
+\frac{b_4 l_{\ep}^2}{d_2 d_{12}}
+\frac{b_5 l_{\ep}^2}{d_0 d_{12}}.
\end{split}
\ee
We use $ d_0 = l^2$, $d_1 = (l+k_1)^2$, $d_2 = (l+k_2)^2$ and $d_{12}
= (l+k_{12})^2$ in \Eq(\ref{ABJparam}).  The general OPP
parameterization \Eq(\ref{eq_c_coeff}) for non-cut constructible terms
would contain extra terms trilinear in the loop momentum that are not
present in \Eq(\ref{ABJparam}). The fact that these terms are absent
will be justified {\it a posteriori}.  Although we work under the
assumption that electrons are massless, we note that all the
manipulations we did up to now remain valid also for massive
electrons.\footnote{ In the massive electron case, the divergence of
  the axial current involves a canonical term $ \partial_\mu J^\mu_5 =
  2m \bar \psi \gamma_5 \psi $, which should be treated separately.}
In the massless electron case some of the bubble integrals are
scaleless and therefore vanish, but the residues of the corresponding
integrands do not vanish and are, in fact, mass-independent.  We will
show below that $c_1=c_2$ and $b_{i=1,..5}=0$.

In the case of closed fermion loops the Dirac algebra has to be
performed in six dimensions with five-dimensional loop momentum
$l=l_{(4)}+l_{\ep}$ , where using the notation
$(l_0,l_1,l_2,l_3,l_4,l_5)$ we have that $l_{\ep}=(0,0,0,0,\mu,0)$ and
$l_{\ep}^2=-\mu^2$.
As explained in Sect.~\ref{sec:polD}, for six-dimensional Dirac matrices we use
the simple representation
\be
\begin{split}
&\Gamma^0 = 
\left (\begin{array}{cc}
\gamma^0 & {\bf 0} \\
{\bf 0} & \gamma^0 
\end{array}
\right ),\;\;\;
\Gamma^{i=1,2,3} =
\left (
\begin{array}{cc}
\gamma^i & {\bf 0} \\
{\bf 0} & \gamma^i 
\end{array}
\right ),\;\;\; 
\\
&
\Gamma^{4} =
\left (
\begin{array}{cc}
{\bf 0}  & \gamma_5 \\
-\gamma_5 & {\bf 0}
\end{array}
\right ),\;\;\; 
\Gamma^{5} =
\left (
\begin{array}{cc}
{\bf 0}  & i \gamma_5 \\
i \gamma_5 & {\bf 0}
\end{array}
\right ),\;\;\;
\Gamma_{\gamma_5} =
\left (
\begin{array}{cc}
\gamma_5 & {\bf 0} \\
{\bf 0} & \gamma_5 
\end{array}
\right )\ .
\end{split}
\label{eqd1}
\ee
Note that for our choice of $l$, $\Gamma^5$ never appears in 
\Eq(\ref{eqaxial1}).
Finally, we 
choose a special reference  frame where 
\be
\begin{split}
k_{12}&=(m,0,0,0,0,0)\,,\quad k_{1,2}= \left (
\frac{m}{2},\pm \frac{m}{2},0,0,0,0 \right )\,,
\\
&e^*_{1,2}=\frac{1}{\sqrt{2}}(0,0,1,\pm i,0,0),\quad
l_{\perp}=\alpha e^*_1 + \beta e^*_2.
\end{split}
\ee
With this choice of the polarization vectors, it is clear that $e^*_i
k_j=0\,,\ (i=1,2)$.  The coefficients $c_1,c_2,b_1,\ldots b_5$ can be
obtained by evaluating triple cuts and double cuts on both sides of
\Eq(\ref{ABJparam}).

We begin by considering the triple cut specified by the condition
$d_0=d_1=d_{12}=0$.  Decomposing the loop momentum on the cut as
\be
l_{c_1}^\mu=x_1k_1^\mu+x_2 k_2^\mu +\tilde{l}^\mu,\quad  
\tilde{l}^\mu=l_{\perp}^\mu +l_{\ep}^\mu,
\ee
we find that $x_1 = -1$, $x_2 = 0$ and 
$
\tilde{l}^2=l_{\perp}^2+l_{\ep}^2=0
$.
Taking the  $d_0, d_1, d_{12}$ residue 
of the left hand side of \Eq(\ref{ABJparam}) we obtain 
\be
\label{anomaly89}
\begin{split}
&{\rm Res}\left({\cal I}_{ABJ}\right)
\left |_{d_0=d_1=d_{12}=0} \right. \;
=
\\ &
\frac{-2i e^2}{(4\pi)^{(D/2)}} 
{\rm Tr} 
\left\{
\Gamma_{\gamma_5} \lhat_{\ep}
(\lhat_{\perp}+\lhat_{\ep}-\khat_1)
\eshat_1(\lhat_{\perp}+\lhat_{\ep}) 
\eshat_2(\lhat_{\perp}+\lhat_{\ep}
+\khat_2) 
\right \}.
\end{split}
\ee

It follows from \Eq(\ref{anomaly89}) that the triple cut residue is
the fourth-order polynomial in $l_\ep$.  However, it is easy to argue
that only limited number of terms can contribute to the trace. Indeed,
for our choice of the loop momentum, $\hat l_\epsilon$ is proportional
to $\Gamma^4$ in \Eq(\ref{eqd1}) while all other terms in
\Eq(\ref{anomaly89}) are linear combinations of
$\Gamma^{0,1,2,3}$. Since the former is block off-diagonal while the
latter are block-diagonal, terms with odd number of $l_\epsilon$'s do
not contribute to the trace.  Moreover, for the trace in
\Eq(\ref{anomaly89}) to be non-zero, at least four $\Gamma$ matrices
are needed in addition to $\Gamma_{\gamma_5}$.  Since $\hat l_{\ep}$
anticommutes with all other matrices of the trace, the term with four
$\hat l_\ep$ vanishes.  We conclude that the only term that, perhaps,
contributes to the trace is quadratic in $\hat l_\ep$.  This justifies
the form of the OPP parameterization in \Eq(\ref{ABJparam}). Finally,
because $l_{\perp}$ can be written as a linear combination of $e^*_1$
and $e^*_2$, only terms that contain two $\hat l_\ep$ and no $l_\perp$
terms give non-vanishing contributions.  Taking all this into account,
we arrive at a simple expression for the trace and the residue
\be
{\rm Res} \left({\cal I}_{ABJ}\right) \left |_{d_0=d_1=d_{12}=0} \right. =
 \frac{2i e^2}{(4\pi)^{(D/2)}} 
  {\rm Tr} 
\left\{
\Gamma_{\gamma_5} \lhat_{\ep}\khat_1
\eshat_1 \lhat_{\ep} \eshat_2\khat_2 \right \}.
\ee
Since  the residue  of the right hand side in 
\Eq(\ref{ABJparam}) 
is $c_1l^2_{\ep}$ 
we derive the value of the $c_1$ coefficient 
\be
c_1= -\frac{2^{D/2+1}e^2}{(4\pi)^{\frac{D}{2}}}
\varepsilon^{{\mu}{\nu}{ \lambda}{\rho}}
e^*_{1\mu} k_{1\nu} e^*_{2\lambda} k_{2\rho}.
\label{c1res}
\ee
Finally, because of the ${1\leftrightarrow 2}$
symmetry, we find  $c_2=c_1$.

We next proceed to the double cuts. Apart from obvious changes in the
physical and transverse spaces, the only new feature is that on the
right hand side of \Eq(\ref{ABJparam}) we get the double cut
contribution also from the triple pole terms.  We illustrate the
calculation of the double-pole terms taking $d_0=d_{1}=0$, as an
example.  Although this is a double cut, the reference momentum is
light-like $k_1^2 = 0$, so the parametrization is subtle. We
parametrize the loop momentum on the double-cut as
\be
l_{b_1}^\mu =x_1k_1^\mu + x_2 k_2^\mu +  \tilde{l}^\mu, 
\ee
and use $l^2 = 0, l \cdot k_1 = 0$ to find $x_2 = 0$ and $\tilde
{l}^2= 0$, while $x_1$ is unconstrained.  We compute the $d_0,d_1$
residue of the left-hand side of \Eq(\ref{ABJparam}) using the
expression in \Eq(\ref{eqaxial1}).  We obtain
\be
\begin{split}
&{\rm Res} \left( {\cal I}_{ABJ}\right) \left |_{d_0=d_1=0} \right.
=\frac{-2i e^2}{(4\pi)^{\frac{D}{2}}} \frac{1}{(l+k_{12})^2}
 {\rm Tr} 
\Big \{
\Gamma_{\gamma_5} \lhat_{\ep}
(\lhat_{\perp}+\lhat_{\ep}+x_1\khat_1)
\\
& \times 
\eshat_1(\lhat_{\perp}+(1+x_1)\khat_1
+ \lhat_{\ep}) 
\eshat_2(\lhat_{\perp}+\lhat_{\ep}+x_1\khat_1
+\khat_{12})
\Big  \}. 
\end{split}
\ee
Similar to the triple-cut case considered earlier, only terms
quadratic in $l_{\ep}$ contribute. The non-vanishing terms are
proportional to $k_2$ and, after some algebra, we find that all terms
proportional to $x_1$ cancel and the result is simply expressed
through the $c_1$ coefficient in \Eq(\ref{c1res})
\be
{\rm Res} \left({\cal I}_{ABJ}\right) \left |_{(d_0=d_1=0)} \right. 
= c_1\frac{l_{\ep}^2}{(l+k_{12})^2}. 
\ee
Since the $(d_0,d_1)$-residue in \Eq(\ref{ABJparam}) is
$c_1l_{\ep}^2/d_{12} + b_1$, we find $b_1 = 0$.  A similar calculation
proves that other bubble coefficients also vanish.  To obtain the
final result, we need the value of the triangle integral, (see
\Eq(\ref{Remarkablysimple})),
\be
\int \frac{{\rm d}^Dl}{\pi^{D/2}}\frac{l_\ep^2}{d_0 d_1 d_2}
= -\frac{i}{2} + {\cal O}(D-4), 
\ee
and the value of the coefficient $c_1$ from \Eq(\ref{c1res}).  By
adding the two triangle contributions and setting $D=4$, we obtain the
anomalous amplitude ${\cal M}_{ABJ}$ shown in \Eq(\ref{ABJ}).

\subsection{Calculation of one-loop corrections to $\qb qgg$ }
\label{Sec:Oneloopexample}
We will illustrate the analytic methods of Section~{\ref{analytic}} 
by considering a specific case for a simple $2 \to 2$ process. 
As our example we will describe the calculation 
of a particular one-loop primitive amplitude for $ \qb q gg$ scattering.

From \Eq(\ref{orderedTqq}) we have the 
color decomposition for the tree graph amplitude for this process,
\be
\begin{split}
\label{orderedTqqbis}
{\cal A}^{\rm tree}_4(\qb_1, q_2, g_3, g_4)
= g_s^2 \Bigg[&\left ( T^{a_{3}} T^{a_{4}} \right )_{i_2 \ib_1}  m_4({\bar{q}}_1,q_2,g_3,g_4
)\\ 
 +&\left ( T^{a_{4}} T^{a_{3}} \right )_{i_2 \ib_1}  m_4({\bar{q}}_1,q_2,g_4,g_3 )\Bigg]. 
\end{split}
\ee
Further, \Eq(\ref{qqbarng1loop}) gives the 
decomposition of the full one-loop amplitude in terms of primitive amplitudes
\be
\label{qqbarng1loopbis}
\begin{split}
&{\cal A}_{4}^{\rm 1-loop}(\qb_1,q_2,g_3,g_4)
= g_s^4 \cg 
\\
& \times  \sum_{P(3,4)} \bigg[ 
 \left(T^{x_{2}} T^{x_{1}}\right)_{i_2 \ib_1} 
\left(F^{a_3} F^{a_4} \right)_{x_{1} x_{2}}
{\tilde m}^{(1)}_4(\qb_1,q_2,g_4,g_3)
 \\  
 &~~~~~~~~~~~~~ + \left(T^{x_{2}} T^{a_3} T^{x_{1}}\right)_{i_2 \ib_1} 
\left(F^{a_4}\right)_{x_{1} x_{2}}
{\tilde m}^{(1)}_4(\qb_1,g_3,q_2,g_4)
\\
&~~~~~~~~~~~~~+  \left(T^{x_{1}} T^{a_3} T^{a_4} T^{x_{1}}\right)_{i_2 \ib_1} 
 {\tilde m}^{(1)}_4(\qb_1,g_4,g_3,q_2)\bigg].
\end{split}
\ee 
For simplicity, we neglect diagrams containing fermion loops in this
discussion.
By using the identities \Eq(\ref{commT}) and the crossing relation, \Eq(\ref{eqk}),
we can write the one-loop amplitude, \Eq(\ref{qqbarng1loopbis}), 
in the form in which it would appear after full reduction of the color matrices,
\be
\begin{split}
 \A_{4}
(\qb_1,q_2,g_3,g_4)\ =\ g_s^4 \cg  \biggl[
    &N_c\,(T^{a_3}T^{a_4})_{i_2 \ib_1} \ A_{4;1}(\qb_1,q_2,g_3,g_4) \\
  + &N_c\,(T^{a_4}T^{a_3})_{i_2 \ib_1} \ A_{4;1}(\qb_1,q_2,g_4,g_3) \\
  + &{\rm Tr}(T^{a_2}T^{a_3}) \, \delta_{i_2 \ib_1} \ A_{4;3}(\qb_1,q_2,g_3,g_4)\biggr] \ .
\end{split} 
\ee
In terms of the primitive amplitudes, the amplitudes $A_{4;j}$ are given as,
\beqn
\label{eq:a41}
A_{4;1}(\qb_1,q_2,g_3,g_4) &=& {\tilde m}^{(1)}_4(\qb_1,q_{2},g_3,g_4) 
- {1\over N_c^2} {\tilde m}^{(1)}_4(\qb_1,g_4,g_3,q_{2})\,, \\
A_{4;3}(\qb_1,q_2,g_3,g_4)  
&=&{\tilde m}^{(1)}_4(\qb_1,q_2,g_4,g_3)
+{\tilde m}^{(1)}_4(\qb_1,g_4,g_3,q_2) \nonumber \\
&+&{\tilde m}^{(1)}_4(\qb_1,g_4,q_{2},g_3) 
+{\tilde m}^{(1)}_4(\qb_1,q_2,g_3,g_4)  \\
&+&{\tilde m}^{(1)}_4(\qb_1,g_3,g_4,q_2)
+{\tilde m}^{(1)}_4(\qb_1,g_3,q_{2},g_4)\,. \nonumber 
\eeqn

In this Section we shall perform the calculation of ${\tilde m}^{(1)}_4(\qb_1,g_4,g_3,q_{2})$, the color-suppressed
primitive amplitude in \Eq(\ref{eq:a41}).
As a first step we note that the singular behavior
of this primitive amplitude can be read off from \Eq(\ref{eq_qqgg_div}),
\beq
\label{eq_div}
{\tilde m}^{(1)}_4(\qb_1,g_4,g_3,q_{2}) = - m_4(\qb_1,q_2,g_3,g_4)
\;\bigg[ \left(\frac{1}{\eps^2} +\frac{1}{\eps}\left(\frac{3}{2}+L_{12}\right)\right) \bigg]
+  {\cal O}(\eps^0)\,,
\eeq
where $L_{12} = \ln \left( \mu^2 /(-s_{12}-i0)\right)$, $s_{12} = 2 p_1\cdot p_2$.

The amplitude $\tilde{m}_4^{(1)}$ can be expanded in 
terms of scalar integrals defined in \Eq(\ref{Scalarintegrals}).
In the notation of Passarino and Veltman, the scalar integrals that can contribute are
the box integral, $D_0(p_1,p_2,p_3,0,0,0,0)$, and any scalar integral 
that can be obtained from it by removing propagators.
The relevant integrals are 
\beqn
I_4(0,0,0,0;\sud,\sdt;0,0,0,0)  &\sim &  D_0(p_1,p_2,p_3,0,0,0,0)\ ,
\nn \\
I_3(0,0,\sud;0,0,0)  &\sim&  C_0(p_1,p_2,0,0,0) = C_0(p_{12},p_{3},0,0,0) \ ,  \nn \\
I_3(0,0,\sdt;0,0,0) &\sim& C_0(p_2,p_3,0,0,0) = C_0(p_1,p_{23},0,0,0)     \ , \nn \\
I_2(\sud,0,0) &\sim& B_0(p_{12},0,0) \ , \nn \\
I_2(\sdt,0,0) &\sim&  B_0(p_{23},0,0) \, .
\eeqn
The scaleless integrals $B_0(p_i,0,0),A_0(0)$ are equal to zero in dimensional regularization 
and we do not consider them.
The general expansion of the amplitude in terms of master integrals can thus be written as,
\be
\label{eq:expansioninscalars}
\begin{split}
& 
-i \ \tilde{m}_4^{(1)}(\qb_1,g_4,g_3,q_{2})=
\tilde{d}_0 \ I_4(0,0,0,0;\sud,\sdt;0,0,0,0) \\
& \;\;\;\;\;+\tilde{c}_0^{(12)} \ I_3(0,0,\sud;0,0,0)
+\tilde{c}_0^{(23)} \ I_3(0,0,\sdt;0,0,0) \\
&\;\;\;\;\; + \tilde{b}_0^{(12)} \ I_2(\sud;0,0) 
+\tilde{b}_0^{(23)} \ I_2(\sdt;0,0) + {\cal R} \; ,
\end{split} 
\ee
where ${\cal R}$ stands for the rational term. In the following subsections we shall illustrate
the calculation of the coefficients of the master integrals. 
Scalar integrals that appear in \Eq(\ref{eq:expansioninscalars})
are tabulated in Ref.~\cite{Ellis:2007qk} and are 
given in~\ref{App:int} for completeness. 

Since the divergences of the amplitude $\tilde{m}_4^{(1)}(\qb_1,g_4,g_3,q_{2})$ and of all 
 the scalar integrals in \Eq(\ref{eq:expansioninscalars}) are known  and 
since all reduction coefficients are $\ep$-independent, 
we can find relations between various reduction coefficients 
by requiring that coefficients of $1/\e^2,1/\e,L_{12}/\e$ and $L_{23}/\e$ on both sides 
in \Eq(\ref{eq:expansioninscalars}) coincide. These relations are 
\be 
\label{eq:IRrelns}
\begin{split} 
& \tilde{b}_0^{(12)} + \tilde{b}_0^{(23)} -\frac{3}{2} i \ m_4(\qb_1,q_2,g_3,g_4)=0\, , \\
& 2 \frac{\tilde{d}_0}{\sud \sdt} +\frac{\tilde{c}_0^{(12)}}{\sud}-i \ m_4(\qb_1,q_2,g_3,g_4)=0 \, , \\
& 2 \frac{\tilde{d}_0}{\sud \sdt} +\frac{\tilde{c}_0^{(23)}}{\sdt} = 0 \, .
\end{split} 
\ee

As follows from \Eq(\ref{eq:expansioninscalars}), there are six unknowns 
and three equations, cf. \Eq(\ref{eq:IRrelns}), that constrain them. 
It is possible to use these relations to express the two triangle
coefficients in terms of box and bubble coefficients. 
Therefore the full calculation reduces to the calculation of the box reduction 
coefficient $\tilde{d}_0$, only {\it one} of the two  bubble  coefficients,
$\tilde{b}_0^{(12)}$ and the rational part ${\cal R}$. 
Substituting the relations  \Eq(\ref{eq:IRrelns}) 
into \Eq(\ref{eq:expansioninscalars}) we obtain,
\beqn \label{eq:finalexpansioninscalars}
&& 
-i \ \tilde{m}_4^{(1)}(\qb_1,g_4,g_3,q_{2})= \tilde{d}_0 \Big[\ I_4(0,0,0,0;\sud,\sdt;0,0,0,0)\nn \\
 &-&\frac{2}{\sdt} I_3(0,0,\sud;0,0,0)-\frac{2}{\sud} I_3(0,0,\sdt;0,0,0)\Big]
\nn \\
&+&i \ m_4(\qb_1,q_2,g_3,g_4)\ \Big[ \sud \ I_3(0,0,\sud;0,0,0) +\frac{3}{2} I_2(\sud;0,0) \Big]
\nn \\
&+&\tilde{b}_0^{(23)} \Big[I_2(\sdt;0,0)-I_2(\sud;0,0)\Big] + {\cal R} \; .
\eeqn
\subsubsection{The calculation of box coefficients}
The box coefficient can be calculated by applying \Eq(\ref{quadd}) and using
analytic results for the amplitudes at the corners of the boxes to compute ${\tilde d}_0$.
A simple example will illustrate the power of the method.
\begin{figure}[t]
\begin{center}
\includegraphics[scale=0.35,angle=270]{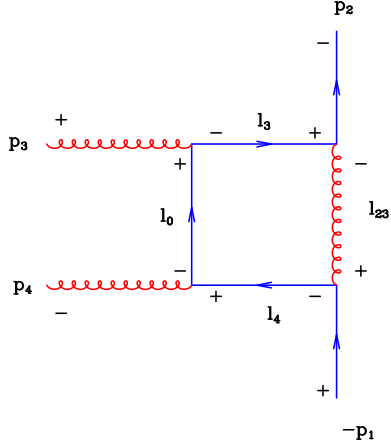}
\end{center}
\caption{A box contribution to the amplitude $\tilde{m}^{(1)}(\qb_{1}^+,g_4^-,g_3^+,q_2^-)$}
\label{qggqv}
\end{figure}
Consider the diagram shown in Fig.~\ref{qggqv}. Momentum assignments
for the internal lines are
$l_3=l_0-p_3,l_{23}=l_0-p_2-p_3,l_4=l_0+p_4$ and all external momenta are taken to be outgoing.

To parametrize the loop momentum in such a way that  four on-shell conditions
are easily solved, we choose to expand the vector $l_0$, (which is flanked on either 
side by two massless momenta $p_3$ and $p_4$), in terms of $p_3$ and $p_4$ as well
as two complex momenta formed from spinors of $p_3$ and $p_4$. We write 
\beq \label{eq:lexpansion}
l_0^\mu
 = \alpha p_3^\mu + \beta p_4^\mu + \gamma \epsilon^\mu_{34}+ \delta \epsilon^\mu_{43}\,,
\eeq
where\footnote{Our notations for the spinor-helicity variables are summarized in~\ref{App:SH}.}
\beq
\epsilon_{ij} = \frac{1}{2} \langle i- | \gamma^\mu |j - \rangle
= \frac{1}{2}\langle i | \gamma^\mu | j ].
\eeq
These vectors are light-like and transverse to the momenta $p_{i,j}$. In particular,  
\be
\epsilon_{ij} \cdot \epsilon_{ij} =0,\;\;\;\;
\epsilon_{ij} \cdot \epsilon_{ji} =-\frac{s_{ij}}{2},
\ee
where $s_{ij}=2 p_i \cdot p_j$.
Using  the mass shell conditions  for three internal lines 
$l_3^2=l_4^2=l_0^2$, we find $\alpha = 0$, $\beta = 0$ and $\gamma \delta = 0$.
We take $\delta=0$ and justify this choice 
in the next Section.
The last on-shell condition $ l_{23}^2 = 0 $ fixes the value of $\gamma$,
\beq \label{gammafix}
-\gamma \; \langle 3 | \slsh{p_2}+\slsh{p_3} |4 ] + 2 p_2 .p_3  = 
-\gamma \; \spa3.2 \spb2.4 +\spa3.2\spb2.3 =0,\;
\Rightarrow \gamma = \frac{\spb2.3}{\spb2.4}.
\eeq

\subsubsection{Box coefficients for $qgg\qb$}

The box coefficients are determined by the quadruple cuts according to
\Eq(\ref{quadd}). For the case $qgg\qb$ the on-shell tree amplitudes 
at the four corners of the diagram are three-parton vertices. 
The analytic expressions for the relevant three parton vertices 
are given in spinor notation in \ref{App:qqg}. These three-point vertices,
calculated for three complex momenta, $p_i,p_j,p_k$, are subject to 
the mass shell constraint that {\it either} $\spa i.j= \spa i.k= \spa j.k=0$
{\it or} $\spb i.j= \spb i.k= \spb j.k=0$.

We consider first the case where the external gluons have the same helicity.
This helicity amplitude vanishes because we would have three-point vertices with helicity 
assignments  (++-)  both at the vertex where $p_3$ flows out 
and at the vertex where $p_4$ flows out. This implies that the two vertices
are 
\beq
\frac{{\spb 3. {l_0}}^2}{\spb {l_3}.{l_0}} \times
\frac{{\spb 4. {l_4}}^2}{\spb {l_0}.{l_4}}\,, 
\eeq  
or consequently that both $\spa {l_0}.3 = 0$ and $\spa {l_0}.4 =0$ which cannot
simultaneously be satisfied for generic external momenta $p_3$ and $p_4$. 
This helicity amplitude thus has a vanishing
contribution to the coefficient of this box integral.

Next consider the case where the gluons have opposite helicities as shown in Fig.~\ref{qggqv}. 
The box coefficient is constructed from the product of four three-point amplitudes. We find 
\beq  \label{productofamps}
\tilde{d}_0(\qb_{1}^+,g_4^-,g_3^+,q_2^-)=
\frac{1}{2} \frac{{\spa l_{23} . 2 }^2}{\spa 2.{l_3}} \times
\frac{{\spb 3. {l_0}}^2}{\spb {l_3}.{l_0}} \times
\frac{{\spa l_{0}.4}^2}{\spa {l_0}.{l_4}} \times
\frac{{\spb 1.{l_{23}}}^2}{\spb 1.{l_{4}}}.
\eeq
To simplify this expression, we use e.g.
\be
{\spb 3. {l_0}} {\spa l_0.4} = [ 3 | \hat l_0 | 4 \rangle.
\ee
In order for this quantity to be non-vanishing, we must have $\gamma \neq 0$
and hence $\delta=0$.
This justifies setting $\delta = 0$ in our parametrization of the momentum $l_0$. 
Collecting terms 
we may write \Eq(\ref{productofamps}) as
\be
\begin{split}
\tilde{d}_0(\qb_{1}^+,g_4^-,g_3^+,q_2^-) &=
\frac{\langle 2|\slsh{l}_{23} | 1]^2
\langle 4|\slsh{l}_0 | 3]^2}
{\langle 2|\slsh{l}_3\slsh{l}_0 \slsh{l}_4 | 1]} 
=
\frac{\langle 2|\slsh{l}_{0}-\slsh{3} | 1]^2
\langle 4|\slsh{l}_0 | 3]^2}
{\spa2.3 [3|\slsh{l}_0 | 4 \rangle [41]} 
 \\
&=
\frac{(\gamma \spa2.3 \spb4.1 - \spa2.3 \spb3.1 )^2
\gamma \spa4.3 \spb4.3 }
{\spa2.3 \spb4.1 }\,,
\end{split} 
\ee
where we have used mass shell constraints and made the substitution $\slsh{l}_0= \gamma |3 \rangle[4|$.
Finally, using $\gamma =\spb2.3 /\spb2.4$ from \Eq(\ref{gammafix}) we obtain 
\be
\begin{split} 
 \tilde{d}_0(\qb_{1}^+,g_4^-,g_3^+,q_2^-) &=
\frac{1}{2} \frac{\spa2.3^2 (\spb2.3 \spb4.1 - \spb2.4 \spb3.1 )^2
\spb2.3 \spa4.3 \spb4.3 }
{\spa2.3 \spb2.4^3 \spb4.1 }  \\
& =\frac{1}{2} \frac{\spa2.3 {\spb2.1}^2 \spb2.3 \spa4.3 \spb4.3^3 }
{\spb2.4^3 \spb4.1 } 
\\
& 
= 
\frac{1}{2} \frac{\spa2.3 \spa 2.1 {\spb2.1}^2 \spb2.3 \spa4.3 \spb4.3^3 }
{\spb2.4^3 \spb4.1 \spa 2.1}  
 \\
&=\frac{1}{2} \frac{\spa 2.1 {\spb2.1}^2 \spb2.3 \spa4.3 \spb4.3^2 }
{\spb2.4^3}   \\
&
=
\frac{1}{2} \frac{s_{12}^2 s_{23}  \spb2.1 \spb4.3 }
{\spa2.3 \spb2.4^3} 
=\frac{1}{2} \frac{s_{12} s_{23}  \spb2.1^2 \spb3.4^2 }
{\spb2.4^3 \spb4.1 } \ . 
\end{split}
\ee

\subsubsection{General methods for bubble coefficients}

  In Section~\ref{alternative} we discussed the technique to obtain
     analytic results for bubble reduction coefficients. Here
	we will apply this technique to compute the reduction
coefficient ${\tilde b}_{0}^{12}$ in \Eq(\ref{eq:expansioninscalars}).
It turns out that for computations that employ spinor-helicity variables,
    it is convenient to write the momentum flowing into the bubble
	  as a sum of two light-like vectors.  Therefore, we
will use the momentum parametrization shown in \Eq(\ref{paramz}), but we identify
\beqn
q_1 = -p_3 -p_4,\;\;
n_2 = \frac{(p_3-p_4)}{\sqrt{-q_1^2}},\;\;
n_+ = \frac{\varepsilon_{34}}{\sqrt{-q_1^2}},\;\;
n_- = \frac{\varepsilon_{43}}{\sqrt{-q_1^2}}\ .
\eeqn
With this parametrization, the momentum on the double cut reads
\beq
l^\mu = \frac{1}{1+z \zb} ( p_3^\mu + z \zb p_4^\mu 
+ \varepsilon_{43}^\mu z
+ \varepsilon_{34}^\mu \zb)\,.
\eeq

According to Mastrolia's method, explained in Section~\ref{alternative}, 
 we need to compute the double cut 
residue $f(z,\zb)$, see \Eq(\ref{eq5002}), 
find the anti-derivative of the function 
$f(z,\zb)/(1+z \zb)^2$ with respect to $\zb$ and compute 
the $z$-residues of this anti-derivative in the entire complex $z$-plane. 

To get used to this parametrization, we consider how various  
integrands $f(z,\zb)/(1+z \zb)^2$
appear  when expressed in terms of 
$z$ and $\zb$. 
We write schematically 
\beqn
\frac{1}{l^2 (l-p_3-p_4)^2} & \to & \frac{1}{(1+z \zb)^2}\,, \nn \\
\frac{1}{l^2 (l-p_3)^2 (l-p_3-p_4)^2} & \to & \frac{-1}{2 p_3\cdot p_4} \frac{1}{(1+z \zb) z \zb}\,, \nn \\
\frac{l^\mu}{l^2 (l-p_3)^2 (l-p_3-p_4)^2} & \to & \frac{-1}{2 p_3\cdot p_4} 
\frac{p_3^\mu+z \zb p_4^\mu+z \varepsilon_{43}+\zb \varepsilon_{34}}{(1+z \zb)^2 z \zb}\,.
\eeqn
Obtaining the primitives with respect to $\zb$ we get
\beqn
\frac{1}{l^2 (l-p_3-p_4)^2} & \to & - \frac{1}{z (1+z \zb)}\,, \nn \\
\frac{1}{l^2 (l-p_3)^2 (l-p_3-p_4)^2} & \to & -\frac{1}{2 p_3\cdot p_4} 
\frac{\ln (\zb) - \ln( z \zb+1)}{z}\,, \nn \\
\frac{l^\mu}{l^2 (l-p_3)^2 (l-p_3-p_4)^2} & \to & \frac{1}{2 p_3\cdot p_4} 
\frac{(p_4^\mu-p_3^\mu)- z \varepsilon^\mu_{43}+\frac{1}{z} \varepsilon^\mu_{34}}
{(1+z \zb) z }\,.\nn \\ 
&+& \mbox{logarithmic~terms}
\eeqn
Discarding the logarithmic terms and calculating 
residues with respect to  $z$, 
 we find the expected contributions to the bubble coefficients
coming from the scalar bubble and the rank-one triangle.

\subsubsection{Application to $\tilde{b}_0^{(12)}(\qb_{1}^+,g_4^-,g_3^+,q_2^-)$.}
\begin{figure}
\begin{center}
\includegraphics[angle=270,scale=0.35]{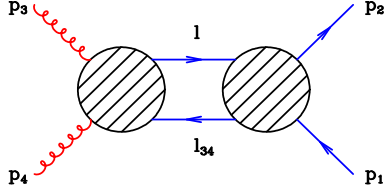}
\caption{The double cut of the amplitude $M^{(1)}(1_\qb^+,2_q^-, 3_g^+,4_g^-) $}
\label{bub}
\end{center}
\end{figure}

Now we turn to the concrete physical example, shown in Fig.~\ref{bub}.
First we should write down amplitudes for the tree level processes 
on either side of the two particle cut. The computation of these amplitudes 
is discussed in~\ref{App:SH}.  Specifically, 
the amplitude on the left hand side of the cut, $M_L$,
is the amplitude for opposite helicity gluon quark 
scattering, \Eq(\ref{qqggppmm}), which can also be written as 
\begin{eqnarray}
i M_L=  m_4({\qb}_1^-,q_2^+,g_3^+,g_4^-) &=& 
i \frac{\spa{4}.{2}^3} { \spa{2}.{3} \spa{3}.{4} \spa{1}.{2} }.
\label{Mqggqb0}
\end{eqnarray}
In addition, we shall need 
four-quark scattering, derived in \Eq(\ref{qbQqQb}),
\beq
i M_R=  m_4(\qb_1^+,q_3^-,\Qb_4^+,Q_2^-) = i \frac{\spb1.4^2}{\spb2.4 \spb3.1}.
\label{MqbQqQb0}
\eeq
Now we form the combination as indicated in Fig.~\ref{bub}.
This corresponds to inserting 
\Eq(\ref{Mqggqb0}) with $1 \to -l_{34}, 2  \to l$,
and inserting 
\Eq(\ref{MqbQqQb0}) with $3\to l_{34},4\to -l$ where $l_{34}=l+p_3+p_4$.
We find
\be
i M_L(-l_{34},l,p_3,p_4) \times i M_R(p_1,p_2,l_{34},-l) =
- \frac{\spa{4}.{l}^3}{\spa{l}.{3}\spa{3}.{4}\spa{l_{34}}.{l}} 
\times \frac{\spb{1}.{l}^2}{\spb{2}.{l} \spb{1}.{l_{34}}}. 
\ee
We eliminate $l_{34}$ using the momentum conservation relation
\begin{equation}
\spa{l}.{l_{34}}\spb{l_{34}}.{1} = \spab{l}.{\slsh{3}+\slsh{4}}.{1}
=-{\spa{l}.2} {\spb2.1},
\end{equation}
and  obtain
\begin{eqnarray}
i M_L \times i M_R &=& 
\frac{1} {\spa{3}.{4} \spb{2}.{1} }
\times \frac{ \spa{4}.{l}^3  \spb{1}.{l}^2} {\spa{2}.{l}\spb{2}.{l}\spa{3}.{l}}.
\end{eqnarray}
Putting in the integration measure and rescaling  the loop momentum 
as 
\be
\label{rescaling}
l^\mu = \frac{1}{1+z \zb} \lambda^\mu,\;\;
\Rightarrow 
\lambda^\mu = p_3^\mu+ z \zb p_4^\mu 
+\frac{z}{2} \spab {4}.{\gamma^\mu}.{3}
+\frac{\zb}{2} \spab {3}.{\gamma^\mu}.{4},
\ee
we derive an integral representation for the ${\tilde b}_0^{(12)}$ coefficient
\beqn
&& \tilde{b}_0^{(12)}
= \int d^4 l \; \delta^+(l^2) \; \delta^+((l-P)^2) \; i M_L \times i M_R  \nn \\
&=& - \int \frac{dz d \zb}{(1+z \zb)^2}
\frac{1} {\spa{3}.{4} \spb{2}.{1}}
\times \frac{ \spa{4}.{\lambda}^3  \spb{1}.{\lambda}^2} {(1+z \zb) \spb{2}.{\lambda} \spa{2}.{\lambda} \spa{3}.{\lambda}}
\label{allofit}
\eeqn

To proceed further, we shall first find the primitive with respect to $z$. To accomplish this, we rewrite \Eq(\ref{allofit}) as 
\beq 
\tilde{b}_0^{(12)}
= - \frac{s_{12}^3} {\spa{3}.{4} \spb{2}.{1} } \times 
\oint_C d \zb \int dz 
\frac{ \spa{4}.{\lambda}^3  \spb{1}.{\lambda}^2} {\spab{\lambda}.{P}.{\lambda}^3 \spb{2}.{\lambda} \spa{2}.{\lambda} \spa{3}.{\lambda}},
\label{allofit2}
\eeq
where momentum $P = p_3 + p_4$ and 
\be
\spab{\lambda}.{P}.{\lambda} = P^2 (1 + z \zb).
\ee

The $z$-dependence in \Eq(\ref{allofit2}) is now hidden in the $\lambda$-spinors, 
so we examine all $\lambda$-dependent terms in \Eq(\ref{allofit2}). 
We write them as 
\beq
A=\frac{\spa{4}.{\lambda}^3} {\spab{\lambda}.{P}.{\lambda}^3 \spa{2}.{\lambda}\spa{3}.{\lambda}}.
\label{part}
\eeq
To simplify \Eq(\ref{part}), we note that 
\be
\label{Mastrolia}
\begin{split}
\ket{\lambda} = \ket{3}+ \zb  \ket{4},\;\;\;\bra{\lambda} = \bra{3}
+ z  \bra{4}.
\end{split}
\ee
Since $\langle  \lambda | \sim z$ and $| \lambda ] \sim \zb$, we should 
examine the dependence on $\langle \lambda |$, to find anti-derivative w.r.t.
$z$.

Using \Eq(\ref{Mastrolia}) we make the following simplifications of the 
various terms in $A$, \Eq(\ref{part}),
\be
\begin{split}
& {\spa4.\lambda} =-{\spa3.4},\;\;\;\;
{\spa2.\lambda} ={\spa2.3}+z {\spa2.4},   \\
& {\spa3.\lambda} = z {\spa3.4}, \;\;\;\;
\spab{\lambda}.{P}.{\lambda} =
 {\spa{3}.{4}} ( {\spb{4}.{\lambda}}-z {\spb{3}.{\lambda}} ).  
\end{split}
\ee

Thus \Eq(\ref{part}) becomes
\beq
A= -\frac{1}{z {\spa3.4}} \frac{1}{(z {\spb3.\lambda} - {\spb4.\lambda})^3 (z {\spa2.4} + {\spa2.3})}.
\eeq
By partial fractioning, obtaining the primitive with respect to $z$, and dropping the logarithmic terms we get
\beq
        -\frac{\spb3.\lambda (2 {\spab{2}.{4}.{\lambda}} + {\spab{2}.{3}.{\lambda}}) }{\spa3.4 {\spb4.\lambda}^2 {\spab{2}.{3+4}.{\lambda}}^2 ({\spb3.\lambda} z-{\spb4.\lambda})}
        +\frac{\spb3.\lambda}{2 \spa3.4 \spb4.\lambda  {\spab{2}.{3+4}.{\lambda}} (\spb3.\lambda z-\spb4.\lambda)^2}.
\eeq
We now multiply by the missing factors from \Eq(\ref{allofit2})
$
- s_{12}^3/( \spa{3}.{4} \spb{2}.{1})  \times  
\spb{1}.{\lambda}^2/\spb{2}.{\lambda}
$
and make substitutions from \Eq(\ref{Mastrolia})
\be
\begin{split}
& {\spb2.\lambda} ={\spb2.3}+\zb {\spb2.4},\;\;\;\;\;\;   
{\spb3.\lambda} =\zb {\spb3.4},
\;\;\;\;\; {\spb4.\lambda} =-{\spb3.4}, \\
& 
{\spb1.\lambda} =-{\spb3.1}-\zb {\spb4.1} 
= -{\spb3.1}+\zb \frac{\spa2.3 \spb3.1}{\spa2.4}.  
\end{split} 
\ee
Further, using momentum conservation we have that $s_{12}=-\spa3.4 \spb3.4$.
Taking the residue at $\zb=-\spb2.3/\spb2.4$ which is the only pole that gives a contribution 
and setting $z= -\spa2.3/\spa2.4$ we get
\beq
\frac{\spa3.4 \spb2.3 (5 \spa2.4 \spb2.4  + 2 \spa2.3 \spb2.3) \spb3.1^2}
{2 \spa2.4 \spb2.4^2 (\spa2.4 \spb2.4+\spa2.3 \spb2.3) \spb1.2}.
\eeq
Simplifying using momentum conservation, 
we obtain for the coefficient of the $B_0(p_{12},0,0)$ bubble 
contribution to the amplitude,
\beqn
\tilde{b}_0^{(12)}(\qb_{1}^+,g_4^-,g_3^+,q_2^-)
&=&-\frac{3}{2} \frac{\spb2.3 \spb3.1^2}{\spb1.2 \spb3.4 \spb2.4 }+\frac{\spb2.3 \spb3.1}{\spb2.4^2}  \\
&\equiv & \frac{3}{2} i m_4({\qb}_1^+,q_2^-,g_3^+,g_4^-) 
-\frac{3}{2} \frac{\spb3.1^2}{\spb4.1\spb2.4}
+\frac{\spb2.3 \spb3.1}{\spb2.4^2}. \nn
\eeqn
Therefore using \Eq(\ref{eq:IRrelns}) the result for $\tilde{b}_0^{(23)}$ is
\beq
\tilde{b}_0^{(23)}(\qb_{1}^+,g_4^-,g_3^+,q_2^-)= 
\frac{3}{2} \frac{\spb3.1^2}{\spb4.1\spb2.4}-\frac{\spb2.3 \spb3.1}{\spb2.4^2} \, .
\eeq

\subsubsection{The rational part}
The rational part can be obtained with a number of methods
\cite{Badger:2008cm,
Bern:2005cq,Pittau:2011qp}
For this elementary example, 
the simplest method is to retain the dimension-dependent terms
in a Passarino-Veltman decomposition. The result is
\beq
{\cal R} =\frac{1}{2} \frac{{\spb2.3} {\spb3.1}^2} {{\spb2.4} {\spb3.4} {\spb2.1}} \; .
\eeq
\subsubsection{Assembling it all: Inserting the integrals}

The result for the lowest order amplitude  is, c.f.\ \Eq(\ref{qqggppmm}),
\begin{eqnarray} 
-i \ m_4({\qb}_1^+,q_2^-,g_3^+,g_4^-)&=& 
\frac{{\spb3.1}^3}{{\spb3.4} {\spb4.1} {\spb1.2}}. 
\end{eqnarray} 
The full answer for the color-suppressed primitive amplitude is
\beqn \label{ansqcd1}
&& 
-i \ \tilde{m}_4^{(1)}(\qb_1^+,g_4^-,g_3^+,q_{2}^-)= \frac{1}{2} \frac{\spb2.1^2 \spb3.4^2 }{\spb2.4^3 \spb4.1 }
\Big[\sud\sdt  I_4(0,0,0,0;\sud,\sdt;0,0,0,0)\nn \\
 &-&2 \sud I_3(0,0,\sud;0,0,0)-2\sdt I_3(0,0,\sdt;0,0,0)\Big]
\nn \\
&+&i \ m_4(\qb_1,q_2,g_3,g_4)\ \Big[ \sud \ I_3(0,0,\sud;0,0,0) +\frac{3}{2} I_2(\sud;0,0) \Big]
\nn \\
&+&
\Big(\frac{3}{2} \frac{\spb3.1^2}{\spb4.1\spb2.4}-\frac{\spb2.3 \spb3.1}{\spb2.4^2} \Big)
\Big[I_2(\sdt;0,0)-I_2(\sud;0,0)\Big] \nn \\
      &+&\frac{1}{2} \frac{{\spb2.3} {\spb3.1}^2} {{\spb2.4} {\spb3.4} {\spb2.1}} \, .
\eeqn
The final result for the color suppressed one-loop $q gg \qb $ amplitude, 
can be obtained by substituting the integrals, 
\Eqs(\ref{eq:I4},\ref{eq:I2}) in \Eq(\ref{ansqcd1}). 
The first and third square brackets in \Eq(\ref{ansqcd1}) are finite in the limit $D \to 4$.
The result assumes a remarkably simple form
\beqn
&& \tilde{m}(\qb_{1}^+,g_4^-,g_3^+,q_2^-)=m_4(\qb_1^+,q_2^-, g_3^+,g_4^-) 
\Bigg[ -\frac{1}{\epsilon^2} \Big(\frac{\mu^2}{ - s_{12}}\Big)^{\epsilon}
-\frac{3}{2 \epsilon} \Big(\frac{\mu^2}{ - s_{12}}\Big)^{\epsilon}-\frac{7}{2}\nn \\
 & - & \frac{1}{2} \frac{s_{12}}{s_{13}} \Big[\Big(1-\frac{s_{12}}{s_{13}} \ln\Big(\frac{-s_{23}}{-s_{12}}\Big) \Big)^2
 +\ln\Big(\frac{-s_{23}}{-s_{12}}\Big)+\pi^2 \frac{s_{12}^2}{s_{13}^2}\Big]\Bigg] \; .
\eeqn
After taking account of the different regularization scheme 
(four-dimensional helicity vs 't Hooft-Veltman) that we are using, 
this result is in agreement with Ref.~\cite{Kunszt:1993sd}.

\section{Numerical implementation} 
\label{sec:num}

\subsection{Comments on the numerical implementation}

In this Section we describe how $D$-dimensional generalized unitarity
can be implemented in a computer code.  We focus here only on the
implementation of the one-loop contribution to NLO cross-sections. In
the spirit of Refs.~\cite{Binoth:2010xt,Hirschi:2011pa} it is possible
to interface the one-loop calculation with a tool that provides the
real emission corrections, the subtraction terms, and can carry out
the phase space integration. Such an approach has been used recently
to compute the NLO QCD corrections to five jet production in $e^+e^-$
annihilation where one-loop amplitudes, calculated with generalized
unitarity, were interfaced with
{\sf Madevent/MadFKS}~\cite{Frederix:2010ne}.

We would like to walk the reader through the principal steps of the
implementation of $D$-dimensional generalized unitarity in a computer
code, but without discussing too many technical details.  To contrast
the numerical implementations of the diagrammatic and unitarity-based
calculations, we will first explain how ``fully automated''
diagrammatic calculations are performed. In order to keep this
discussion in line with the main theme of this review, we will assume
that the OPP method is employed for the reduction of tensor integrals
to the scalar basis, but a Passarino-Veltman reduction could also be
used.  We note that various implementations of diagrammatic
calculations
\cite{vanHameren:2009dr,Mastrolia:2010nb,Frederix:2010ne,Hirschi:2011pa}
can differ in the details -- for example, in how the diagrams are
grouped together before the numerical evaluation or in how the
rational parts are computed, but are otherwise very similar in spirit.

As the name suggests, diagrammatic calculations are based on Feynman
diagrams. The Feynman diagrams for a given process can be
automatically generated with programs like {\sf QGRAF}
\cite{Nogueira:1991ex} or {\sf FeynArts} \cite{Hahn:2000kx}. To do this, the
user needs to specify the allowed vertices and propagators, the
incoming and outgoing particles, and the number of loops.
Furthermore, the user can efficiently control the diagrams that are
generated by these programs and disregard one-particle reducible
diagrams such as tadpole diagrams or self-energy corrections to
external lines.

The package {\sf QGRAF}, for instance, generates all diagrams and produces
symbolic output for each of them, listing the vertices and the
propagators.  The momenta and color flow are automatically assigned to
the internal lines using the information on the momenta and color
configuration of the external particles.  For further computations,
the {\sf QGRAF} output needs to be turned into a form that allows one to
specify the ``quantum numbers'' of the external and internal
particles. For example, one needs to provide explicit Feynman rules
for the propagators and vertices so that a symbol for the quark
propagator with the momentum $p$ and the mass $m$ turns into $i
\delta_{ab} ( \s p - m)^{-1}$, the quark-gluon vertex turns into $-i
g_s/\sqrt{2}\; T^{a}_{ij} \gamma_\mu$ and so on.  It is
straightforward to accomplish this using computer algebra programs
such as {\sf Form} \cite{Vermaseren:2000nd}, {\sf Mathematica} or {\sf
  Maple}.
For example, the {\sf QGRAF} output can be turned into a {\sf Form}-readable
file that contains all Feynman diagrams listed one after the other.
The {\sf Form} program is then executed.  The program performs minimal
simplifications of the input.  It expresses color degrees of freedom
through the standardized color basis, contracts and removes all dummy
indices to an extent possible, and uses the equations of motion of the
external particles to simplify the output.  At the end of this
process, the {\sf Form} program writes the contribution of a given
Feynman diagram to different color-ordered amplitudes as sums of
products of kinematic functions with strings of gamma matrices
sandwiched between fermion spinors. The {\sf Form} program outputs the
resulting expression into a file that is further processed by a number
of symbolic manipulation routines.  Those routines shift the loop
momenta in each of the Feynman diagrams, to bring the denominators to
a standard form, \Eq(\ref{Scalarintegrals}), write out the remaining
contractions of dummy indices explicitly and produce {\sf Fortran}
programs that compute numerators and denominators of Feynman diagrams
for given value of the loop momentum.  In all these manipulations, the
number of dimensions is kept as a free parameter that can be passed to
the {\sf Fortran} routines.  After these manipulations are done, we
are in position to apply the OPP algorithm, as described in
Sect.~\ref{sect5}.

We now contrast this procedure with the calculation based on
generalized unitarity. As a first step, we analytically identify the
decomposition of the amplitude for the process of interest into
primitive amplitudes. For definiteness, we assume first that the
calculation involves only colored particles, in which case all
particles in each primitive amplitude are ordered. It follows that
there is just one parent diagram for each primitive amplitude and this
parent diagram can be described by the momenta, polarizations and
flavors of all external particles and flavors of all internal
propagators.
We note that, for programming purposes, it is often convenient to
assume that parent diagrams for amplitudes with $N$ external particles
contain $N$ propagators that depend on the loop momenta.  However,
when fermions are present, some primitive amplitudes contain a smaller
number of propagators in their parent diagrams. As we discussed in
Sect.~\ref{sect6}, to remedy the situation we introduce ``dummy
lines'' \cite{Ellis:2008qc} which do not correspond to propagators of
physical particles and therefore can not be ``cut'', but which are
very convenient for bookkeeping purposes.
For example, to describe a contribution to two-quark two-gluon
scattering amplitude which is proportional to the number of massless
fermions $n_f$, we can introduce the parent diagram with one dummy
line, while the parent diagram that describes $n_f$-dependent
contribution to a four-quark amplitude can be drawn using two dummy
lines.

\begin{figure}[t!]
\begin{center}
\includegraphics[angle=-90,scale=0.45]{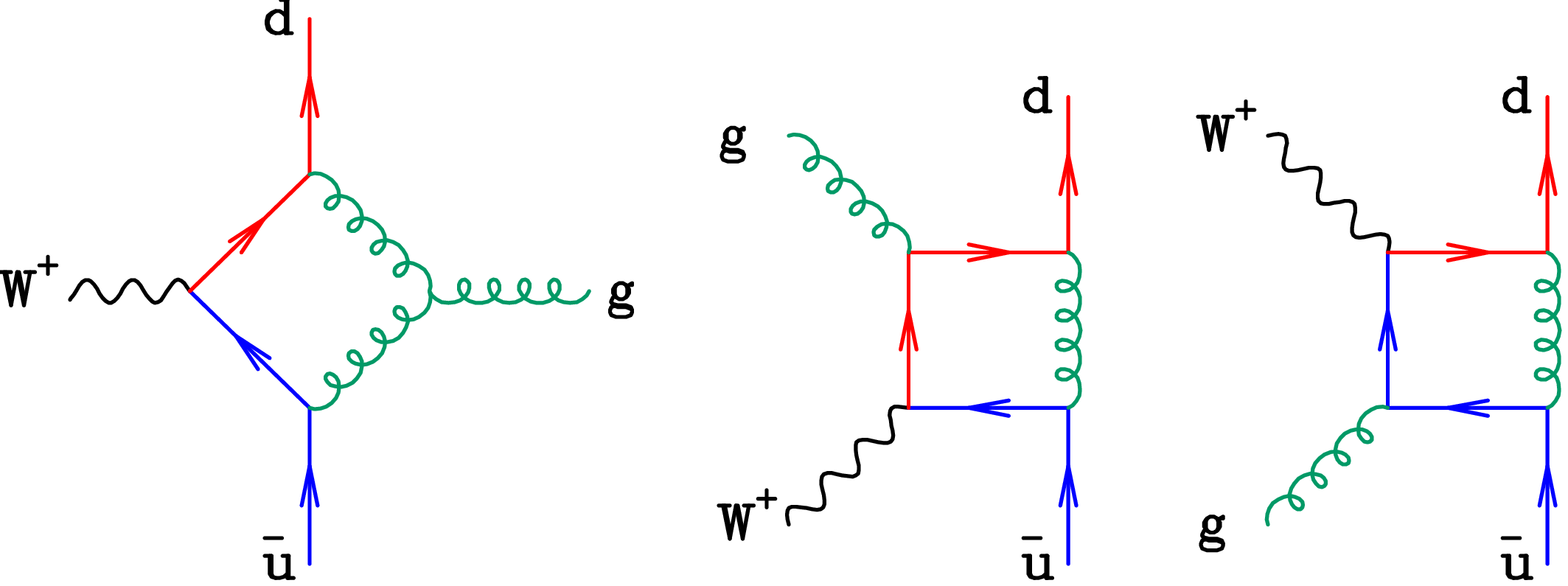}

\vspace{0.5cm}

\begin{picture}(0,0)(0,0)
\put(-90,0){(a)}
\put(18,0){(b)}
\put(103,0){(c)}
\end{picture}

\caption{
Parent diagrams for the $0\to \bar u d W^+ g$ amplitude.}
\end{center}
\label{fig3} 
\end{figure}

When colorless particles are involved, a primitive amplitude may have
more than one parent diagram. Those parent diagrams are obtained by
considering all possible permutations of the colorless particles with
respect to the external quarks and gluons. As an example, consider a
gluon-loop correction to the $0\to \bar u d W^+ g$ amplitude.  While
the $W^+$-boson is always attached to the fermion line, the gluon can
be emitted either by fermions or by the virtual gluon. These two
different possibilities correspond to two different primitive
amplitudes.  When the gluon is emitted by the virtual gluon there is
only one possible ordering, $\bar u W^+ d g$, so this primitive
amplitude involves only one parent diagram, see Fig.~\ref{fig3}a.
However, when the gluon is emitted from the fermion line, there are
two possible orderings of the $W^+$-boson with respect to the gluon,
$\bar u g W^+ d$ or $\bar u W^+ g d$. As a consequence, this primitive
amplitude has two parent diagrams shown in Figs.~\ref{fig3}b,c.

Following the OPP procedure, we have to consider all possible ``cuts''
of the parent diagram without cutting dummy lines. This means that we
have to identify all possible combinations of the inverse Feynman
propagators that can be set to zero simultaneously. The largest member
of the set contains five propagators, while the smallest member of the
set contains one propagator.  In principle, all possible combinations
of propagators must be included in this set. However, in practice,
sets with single massless propagators or sets with two massless
propagators with light-like reference momentum integrate to zero in
dimensional regularization and for this reason do not need to be
considered.

We study a member of a such a set with $k$ inverse propagators
$[d_{i_1}(l),d_{i_2}(l),$ $..,d_{i_k}(l)]$. By solving $k$ equations
$d_{i_j}(l) = 0$ with $j = 1, \ldots k$, we find the loop momentum $l_c$
that satisfies the cut conditions. Since for $l= l_c$ the ``internal''
particles $i_1, i_2, \ldots i_k$ are on-shell and since we know the
{\it external} particles that must appear between two consecutive
internal on-shell particles in the parent diagram, the on-shell
scattering amplitude that needs to be calculated is fixed.

In order to calculate the on-shell scattering amplitudes, we employ
the Berends-Giele recursion relations explained in
Sect.~\ref{bgampl}.  We can code up various recursion relations by
employing {\it recursive functions}, available in {\sf Fortran 90}.
The calculation of various currents can be sped up by avoiding the
re-calculation of the currents that have already been computed. This
can be done by assigning an integer index given by a different power
of two to each external particle. Once this is done, the sum of the
indices uniquely specifies a given current for fixed momenta and
helicities of the external particles.

We note that amplitudes with multiple fermion pairs or vector bosons
are complicated and that the knowledge of the quantum numbers of the
internal and external particles may be insufficient to fully specify
tree-level amplitudes that need to be computed once a particular cut
is considered.  As an example, consider the case of the four-quark
one-loop amplitude with a closed fermion loop. As we explained
earlier, the parent diagram involves two dummy lines and two virtual
fermion lines.  The double cut of the two fermion lines involves two
four-quark tree amplitudes. When the flavor of the internal and
external quarks is the same, these amplitudes involve $s$- and
$t$-channel contributions. However, one needs to forbid contributions
where {\it internal and external} quarks couple since these terms do
not give rise to a fermion loop contribution.  In a computer code,
such terms can be easily avoided by always assigning a different
flavor to the internal quark lines in amplitudes with closed 
fermion loops.

As another example we consider the one-loop amplitude for $0 \to Z q
\bar q g $. If we deal with color-ordered diagrams and use
color-stripped Feynman rules, a diagram exists where the $Z$-boson
mixes into an off-shell gluon that splits into quarks and gluons $g^*
\to \bar q q g$.  When color degrees of freedom are restored, such
diagram must vanish because of color conservation ${\rm Tr}(T^a) =
0$. A simple solution to this problem is to introduce special fermion
current for internal fermions and to disregard a contribution to this
current of the diagram $q \bar q \to g^* \to X$, provided that the
invariant mass of the two quarks equals to the mass of the $Z$-boson.
These examples illustrate that, occasionally, care is needed in the
recursive calculation of the tree-level amplitudes, when they are used
as an input for one-loop computations.  However, most of the time such
calculations are straightforward. We also note that if the OPP
reduction is applied to color-dressed amplitudes~\cite{Giele:2009ui}
these complications do not arise.

As we explained in Sect.~\ref{sect5}, tree on-shell amplitudes are
needed to find the OPP reduction coefficients.  To determine all the
coefficients uniquely, we must deal with spaces of various
dimensionalities and consider different embeddings of the loop momenta
into those higher-dimensional spaces.  The details of these choices
are explained in Sect.~\ref{sect5}.  To make computations efficient,
it is important to avoid repeating low-dimensional calculations when
higher-dimensional calculations are performed. For example, the
$(D-4)$-independent parts of the result are always computed with
four-dimensional momenta and polarization vectors.  Higher-dimensional
calculations are only needed to account for effects of
``extra-dimensional'' polarizations.

The OPP equations need to be solved once and for all.  The actual
solutions can be found in a variety of ways. By default, discrete
Fourier transforms \cite{Mastrolia:2008jb,Berger:2008sj} are employed
to compute the four-dimensional part, while other procedures, such as
e.g. direct Gauss elimination method, are used to calculate reduction
coefficients related to the rational parts or the cut-constructible
parts in singular kinematics.

\subsection{Checks and numerical instabilities}
\label{sec:checks}

When the on-shell/generalized unitarity framework is applied to a new
process, new tree-level primitive amplitudes and Berends-Giele
currents are in general required.  As the first check we verify that
when new currents are used with four-dimensional momenta and
polarization vectors, the tree-level amplitudes coincide with the ones
computed with e.g.  {\sf MadGraph} \cite{Alwall:2007st}.  Furthermore,
when coding up new one-loop amplitudes, it is useful to check that
coefficients of master integrals scale correctly when momenta and
masses of all external particles in the scattering amplitudes are
re-scaled by a constant factor. When we move from computing
higher-point coefficients to computing lower-point coefficients, we
require tree-level amplitudes with higher multiplicity. The re-scaling
test allows us to determine exactly the stage in the calculation when
something goes wrong which is very useful for debugging purposes.

Of course, the best check of any result is provided by an independent
calculation.  To this end, we use a Feynman-diagrammatic calculation
that employs the OPP method for the reduction of tensor integrals, as
described at the beginning of this Section.  The Feynman-diagrammatic
calculation does not require dealing with primitive amplitudes which
provides an important simplification and makes it quite independent
from unitarity computations.  Since it is sufficient to find agreement
between the Feynman-diagrammatic and the unitarity-based calculation
for a few selected points, the Feynman-diagrammatic calculation
neither needs to be particularly efficient nor does it need to have a
rescue system for points that turn out to be numerically unstable.

Typically, calculations are performed in standard double precision.
For some kinematic points this may be insufficient if numerical
instabilities develop because of divisions by small numbers or because
of large cancellations between separate terms.  For kinematic points
where this happens, it is necessary to repeat the calculation in
higher, for example quadruple, precision which is a built-in option in
{\sf Fortran 90}\footnote{Quadruple precision is not supported by
  {\sf gfortran} and {\sf g95} compilers but it is supported by the
  Intel Fortran Compiler ({\sf ifort}).}  or in multiple precision,
with packages like {\sf MPLib}~\cite{MP,MP1}.

However, because higher precision computations are expensive, it is
important to have a numerical procedure to detect unstable points, and
to repeat the higher-precision calculation for as limited a number of
contributions as possible.
There are several ways to control the numerical stability of the
calculation. For example, one can make use of the fact that
divergences of one-loop amplitudes are known \cite{Catani:2000ef} (see
also Section \ref{sec:singular}). When the numerical reduction is
performed and divergent parts of the master integrals are computed
with the use of the {\sf QCDLoop} program \cite{qcdloop}, the results
of the numerical calculation are compared with the predicted
\cite{Catani:2000ef} values for coefficients of $\ep^{-2}$ and
$\ep^{-1}$ terms.  Because two- and one-point functions are divergent,
the correct result for the $1/\ep$ pole of the amplitude suggests that
the reconstruction of the reduction coefficients works relatively well
all the way down to the lowest-point integrals. However, the rational
part of the amplitude can not be checked by comparing divergent terms
only.

Another way to detect instabilities is the so-called $N=N$ test.
Recall that the OPP method works by reconstructing the complete
parametric dependence of the integrand on the loop momentum.
Therefore, once the OPP coefficients in the decomposition of the
one-loop amplitude have been calculated, it is possible, for any
arbitrary loop-momentum, to compute a given residue of the amplitude
either using the OPP coefficients, or directly as a product of
tree-level amplitudes. The two results should coincide up to numerical
rounding errors. A larger difference between these two results is a
sign of a numerical instability.  We note that both, the $(D-4)$ part
of the parameterization of the residue of the cut and the $D=4$ part
are checked by this test. Moreover, by choosing either four- or
five-dimensional momentum on the cut, we can focus the test on either
cut-constructible or rational part of a particular amplitude.

However, the $N = N$ test has two unfortunate features.  First, since the test
needs to be performed for a randomly chosen  loop momentum, it may happen
that a bad choice of the loop momentum will not detect the numerical
instability.  Second, as we pointed out in Sect.~\ref{sect5}, many
of the coefficients in the OPP parameterization vanish after the
integration over the directions of respective transverse spaces which
implies that the precision with which they are reconstructed is
irrelevant for the precision of the final answer. Moreover, even those
coefficients that do not vanish under such integration may multiply
small or vanishing integrals so that, again, their precision is not
important.  Hence, we conclude that the $N = N$ test is typically too
strong except for unfortunate choices of the test momentum where it
can actually overestimate the accuracy of the calculation.

\begin{figure}[t!]
\begin{center}
\includegraphics[angle=-90,scale=0.25]{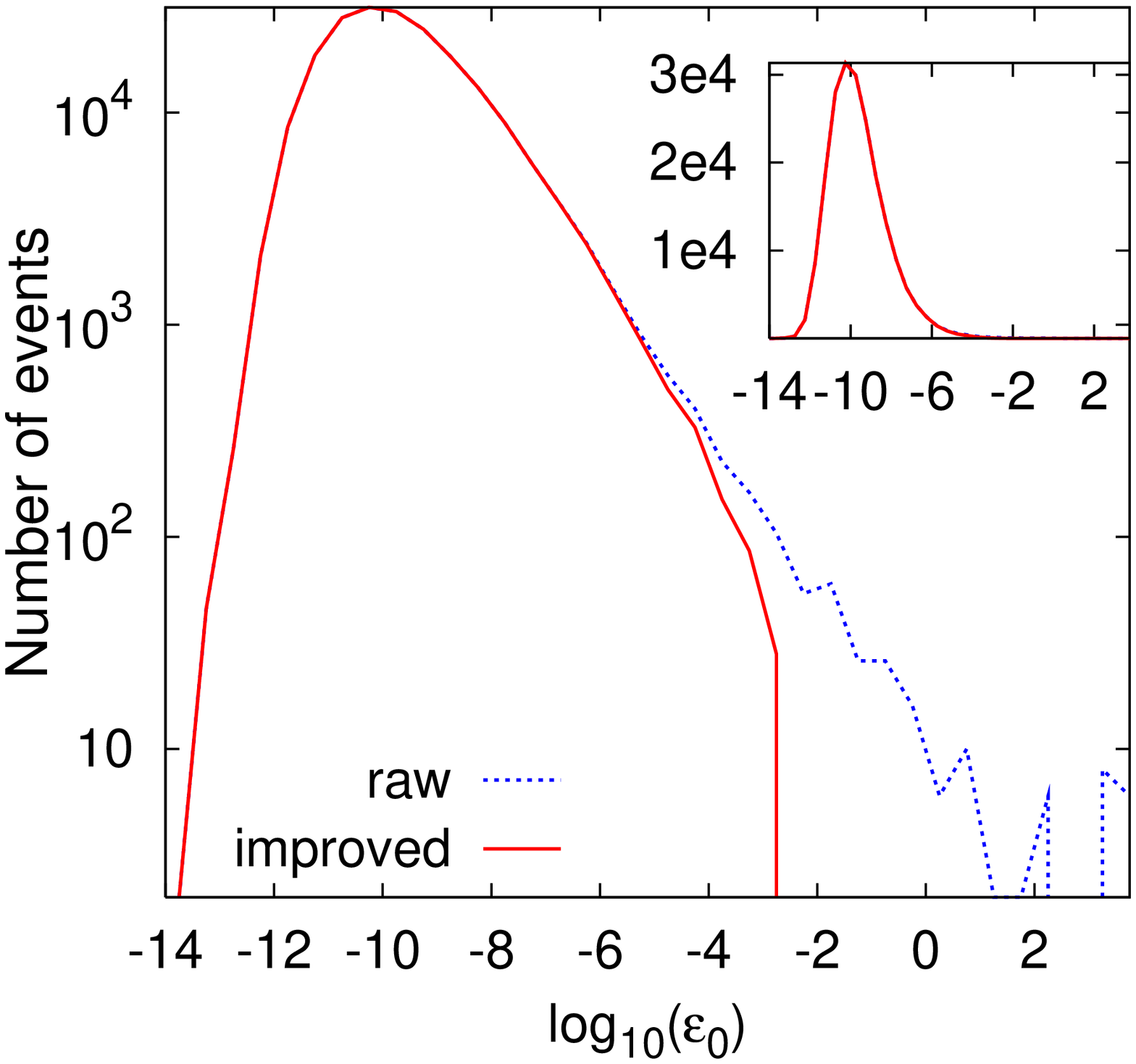}
\includegraphics[angle=-90,scale=0.25]{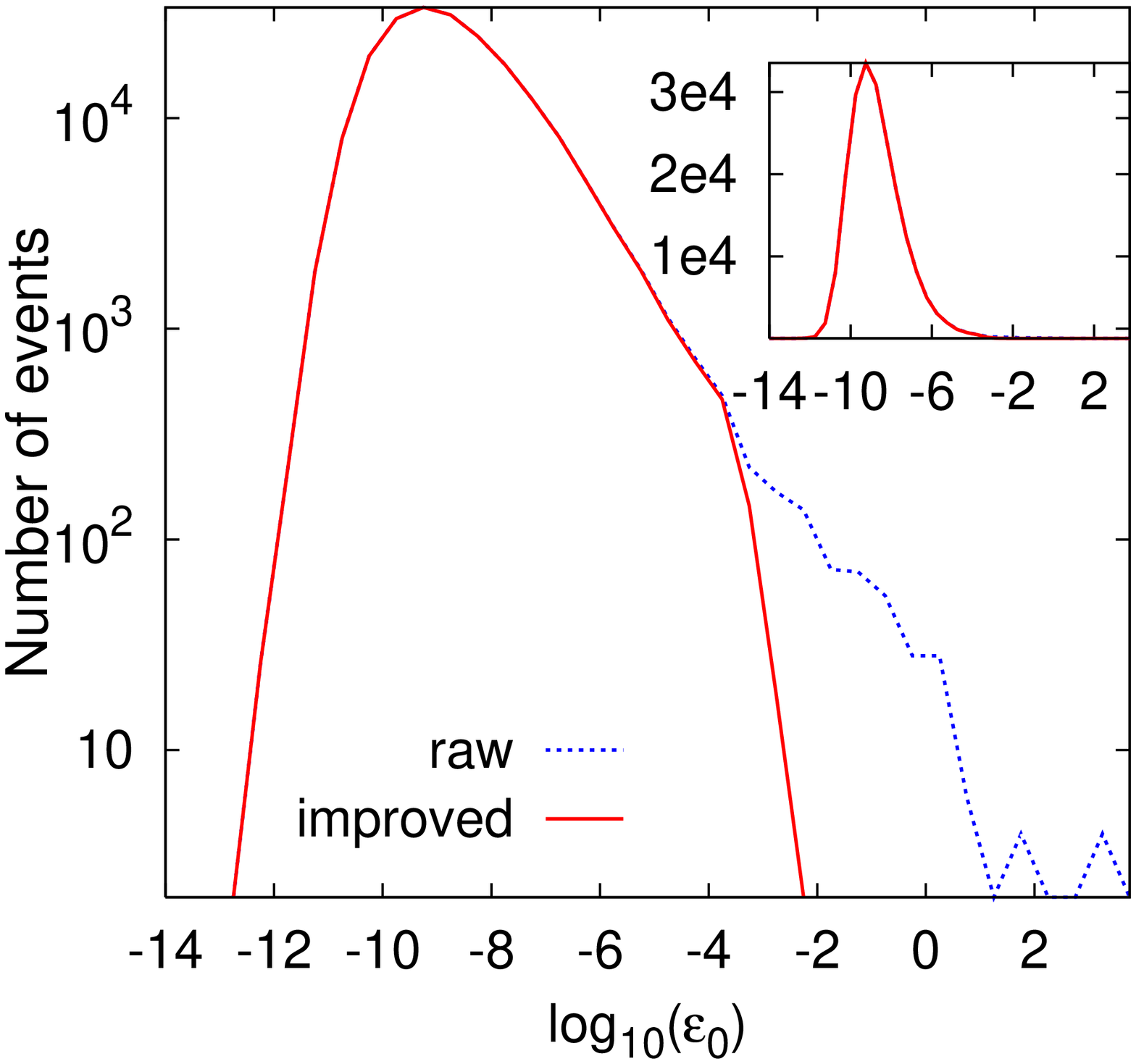}
\caption{Typical improvement in accuracy for some primitive amplitudes
  that describe $0 \to W^+ q \bar q + 3g$ process.  The accuracy of
  raw data is shown, in comparison with the results of numerical
  improvement. The inset shows the same plots in a linear scale. The
  figures are taken from Ref.~\cite{Ellis:2008qc}.}
\label{fig0}
\end{center}
\end{figure}

An interesting modification of the $N=N$ test that addresses the
shortcomings described above was recently suggested by R.~Pittau in
Ref.~\cite{Pittau:2010tk}. The idea is to perform the OPP reduction
twice, employing different ways to solve systems of linear equations.
Since this procedure gives two independent values for the required
one-loop amplitude, this is a good check but, if implemented naively,
it will simply double the computational time. The trick to avoid this
is to use {\it the reconstructed} numerator function, rather than the
actual one, to perform the second OPP reduction.  As a result, the
increase in the required computational time is modest.

There are other ways to estimate the accuracy of the calculation.  For
example, in Sect.~\ref{sect5} we explained that the parameterization of
the residues involves traceless tensors of restricted rank. This can
be checked numerically by performing a Fourier projection on terms that
{\it must} vanish, to verify that those terms are indeed compatible
with zero. This check can be performed at a moderate computational
cost. However, the interpretation of the outcome of the test might be
difficult, especially when large cancellations between different
coefficients take place.
Another -- rather simple-minded -- possibility is to check the size of
the one-loop virtual amplitude compared to the Born result.  We expect
the coefficient of $\alpha_s/(2\pi)$ in the virtual contribution to be
of the same order as the tree amplitude.  If this is not the case, the
phase-space point is considered suspicious and the calculation is
repeated in higher precision.  In order to find all unstable points,
one can combine those tests.
The power of these procedures to detect points that are computed
incorrectly is shown in Fig.~\ref{fig0}, taken from
Ref.~\cite{Ellis:2008qc}.  In a typical computation, less than one
percent of all phase-space points used to calculate the virtual
corrections, show numerical instabilities and are recomputed with
higher precision.

We also note that, instead of recurring to higher precision, one can
try to perform a small change in the kinematic configuration so that
numerical cancellations that, for the exceptional point, would occur
only beyond the double precision happen again in the double precision
regime.  For instance, one can consider replacing the one-loop result
at the exceptional point with the geometric average of the two results
obtained using two points in the phase-space that are close enough to
the exceptional point~\cite{Hirschi:2011pa}.

\subsection{Time dependence of the $D$-dimensional unitarity algorithm}

The development of new techniques for one-loop calculations is mainly
driven by the LHC physics, which forces us to study processes with a
large number of particles in the final state.  If we are to consider
processes with higher- and higher multiplicities, it is important to
understand how the computational time of the $D$-dimensional unitarity
algorithm scales with the number $N$ of external particles.

The calculation of one-loop amplitudes within $D$-dimensional
unitarity factorizes into the calculation of several
higher-dimensional tree amplitudes. The cost of computing one-loop
amplitudes is almost entirely dominated by the time needed for the
calculation of these tree-level amplitudes. In order to estimate how
the running time scales with the number of external particles
involved, it is thus sufficient to count the number of tree-level
amplitudes that have to be evaluated and to know how the computational
time of these building blocks scales with $N$.

For simplicity, let us first discuss the case of purely gluonic
amplitudes. An example involving a quark-pair, a vector boson, and an
arbitrary number of gluons will be considered at the end of this
Section. As long as only gluons are involved, primitive and
color-ordered amplitudes coincide, so that we do not need to
distinguish between them. Each color-ordered amplitude has only one
parent diagram with ordered external gluons.  The number of tree-level
amplitudes that one needs to compute to determine a single one-loop
color-ordered amplitude is given by
\beqn
\begin{split}
& n_{\rm tree} = \{(D_{s_1}-2)^2+(D_{s_2}-2)^2\} \\
& \times \left (
5\, c_{5} {N\choose 5}
+4\, c_{4} {N\choose 4}
+3\, c_{3} {N\choose 3}
+2\, c'_{2} \left[{N\choose 2}-N\right]
\right).
\label{eq:ntree}
\end{split}
\eeqn
The various terms in \Eq(\ref{eq:ntree}) have the following
origin:
\begin{itemize}
\item the two terms in the curly brackets take into account the sum
  over internal polarizations in $D_{s_1}$ and $D_{s_2}$
  dimensions. We note that \Eq(\ref{eq:ntree}) gives a conservative
  estimate since it is assumed there that the entire calculation is
  done in $D_{s_1}>4$ and $D_{s_2} >4$ dimensions. As already
  mentioned, it is instead convenient to compute the coefficients
  needed for the cut-constructible part in $D=4$. This decreases the
  running time for all $N$-gluon amplitudes by the same amount, so
  that this does not affect the way the running time scales with $N$;
\item the binomial coefficients denote the number of multiple cuts
  that a one-loop parent diagram, that is composed of massless
  particles and has $N$ external legs, can have.  In the case of
  double cuts, one can discard the $N$ cuts of self-energies on
  external legs, therefore one subtracts $N$ from the binomial
  coefficient;
\item the coefficients $5,4,3$, and $2$ count the number of tree-level
  amplitudes involved in pentuple, quadruple, triple, and double cuts,
  respectively;
\item finally, the coefficients $c_{i}$ denote the number of
  independent coefficients in the decomposition of the numerator
  function. Explicitly, we have $c_{5} = 1$, $c_{4} = 5$, $c_{3} =
  c_{2} = 10$. These coefficients also give the number of times one
  needs to evaluate the numerator function in order to fully solve the
  OPP system of equations.\footnote{We note that, for amplitudes with
    massless internal particles, tadpole diagrams do not need to be
    calculated.  Hence, out of ten OPP coefficients that contribute to
    each double cut, we need to know just {\it two} coefficients that
    give rise to non-vanishing contributions after integration over
    the transverse space is performed.  This is the reason that
    $c'_{2} = 2$ appears in \Eq(\ref{eq:ntree}).}
\end{itemize}
Altogether, \Eq(\ref{eq:ntree}) implies that, for large $N$, we need
to evaluate $\propto N^5$ tree-level amplitudes, to reconstruct a single
one-loop primitive amplitude, for one kinematic point.

As shown in \cite{Kleiss:1988ne} when Berends-Giele recursion
relations are implemented in a way that avoids recomputing the same
current appearing in different amplitudes, the time needed to compute
an $N$-gluon amplitude scales with the number of gluons as
\be
\tau_{\rm tree} = {N \choose 3} E_3 +{N \choose 4} E_4\,.
\label{eq:tau}
\ee
In \Eq(\ref{eq:tau}), $E_{3,4}$ denote times needed to evaluate a
single three- or four-gluon vertex and the binomial coefficients
simply count the number of such vertices.  Since vertices are never
recomputed, their {\it number} determines the time needed to compute a
tree-level amplitude.

To find the number of three-gluon vertices present in the amplitude
described by $N$ ordered gluons with momenta $k_1,\dots ,k_N$, we note
that such amplitude receives a unique contribution from a three-gluon
vertex that describes the interaction of three gluon currents with momenta
$k_{i_1}+k_{i_1+1}+...k_{i_2}$, $k_{i_2+1}+k_{i_2+2}+...k_{i_3}$,
$k_{i_3+1}+k_{i_3+2}+...k_{i_1-1}$, where all labels need to be
understood modulo $N$.  Because three numbers -- $i_1, i_2, i_3$ --
label the three-gluon vertex uniquely, the number of independent 
three-gluon vertices is given by the number of ways we can choose three
integers $i_1,i_2,i_3$ from a set of $N$ numbers. This is exactly the
binomial coefficient in the first term of \Eq(\ref{eq:tau}).  The same
reasoning holds for four-gluon vertices as well and explains the
binomial coefficient in the second term of \Eq(\ref{eq:tau}).

\begin{figure}[t!]
\begin{center}
\includegraphics[angle=270,scale=0.40]{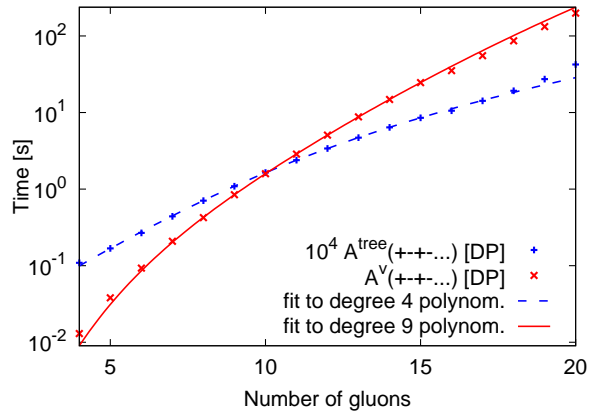}
\caption{
\label{fig:tvsN}
Time in seconds needed to compute tree amplitudes
(dashed (blue online)) and one-loop
ordered amplitudes (solid (red online))
with gluons of alternating helicity signs,
$A_N^{[1]} (+ - + - +\ldots)$, as a function
of the number of external gluons ranging between 4 to 20. The notation
[DP] refers to numerical
computations performed with double precision accuracy  in Fortran.
{\it Source}: Figure taken from Ref.~\cite{Giele:2008bc}.}
\end{center}
\end{figure}

It follows from \Eqs(\ref{eq:ntree},\ref{eq:tau}) that the overall
run-time for one-loop primitive amplitudes scales, for large $N$, as
\be \tau = \tau_{\rm tree} \cdot n_{\rm tree} \sim N^9\,.  \ee This is
illustrated in Fig.~\ref{fig:tvsN}, taken from
Ref.~\cite{Giele:2008bc} (see
also~\cite{Winter:2009kd,Lazopoulos:2008ex}), which shows the time
necessary to evaluate tree and one-loop gluon helicity amplitudes as a
function of the number of external gluons, as well as a fit to a
degree four and a degree nine polynomial.

\begin{figure}[t!]
\begin{center}
\includegraphics[angle=0,scale=0.72]{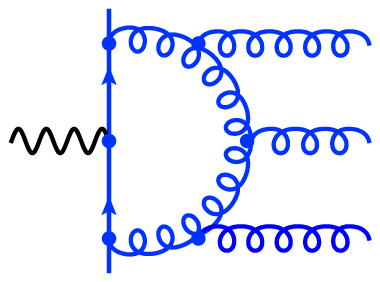}
\quad
\includegraphics[angle=0,scale=0.72]{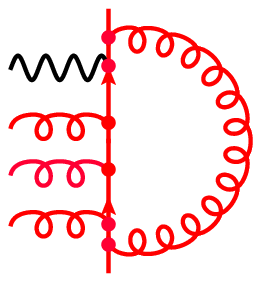}
\quad
\includegraphics[angle=0,scale=0.72]{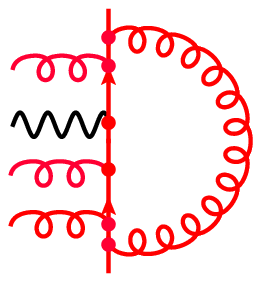}
\quad
\includegraphics[angle=0,scale=0.72]{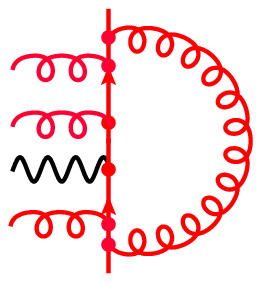}
\quad
\includegraphics[angle=0,scale=0.72]{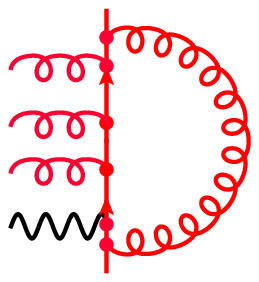}

\begin{picture}(0,0)(0,0)
\put(-140,-3){(a)}
\put(-65,-3){(b)}
\put(0,-3){(c)}
\put(68,-3){(d)}
\put(133,-3){(e)}
\end{picture}

\caption{\label{fig:qqWngprims} 
a) Parent diagram for the leading-color primitive amplitude
$A_v(\bar q,q,g_1,g_2,g_3)$; 
b-e) Parent diagrams for the sub-leading-color primitive amplitude
$A_v(\bar q,g_1,g_2,g_3,q)$. 
}
\end{center}
\end{figure}

We close this Section by commenting on what happens in more
complicated processes that involve quarks and/or vector bosons. In
such cases, for a fixed primitive, \Eq(\ref{eq:ntree}) needs to be
modified to include an additional factor $n_{\rm perm}$ which counts
the number of allowed permutations of colorless particles with respect
to quarks and gluons. \Eq(\ref{eq:tau}) still holds but $E_{3}$
denotes the average time to compute a three-point vertex, and for
four-point vertices $N$ denotes only the number of gluons. For
instance, the one-loop amplitude for the process $0\to \bar d u W^+ +
N$-gluons involves, among others, a leading-color primitive amplitude
$A_v(\bar q,q,g,\dots, g)$, where all external gluons are emitted from the
gluon in the loop, and a sub-leading color primitive amplitude $A_v(\bar q,
g, \dots g, q)$, where all gluons are emitted from the quark line. In
the case of the leading-color amplitude, no permutation of the $W$
boson is allowed, so that $n_{\rm perm}=1$. For $N = 3$ the
corresponding parent diagram is shown in
Fig. \ref{fig:qqWngprims}a. In contrast, in the case of the
sub-leading-color amplitude there are $n_{\rm perm} = n-2$ possible
attachments of the $W$ on the quark-line, where $n$ denotes the total
number of particles, i.e., $n= N +3$. For $N=3$ this primitive has
four different parent diagrams, as illustrated in
Figs. \ref{fig:qqWngprims}b-e.

\begin{figure}[!t]
\begin{center}
\includegraphics[angle=-90,scale=0.40]{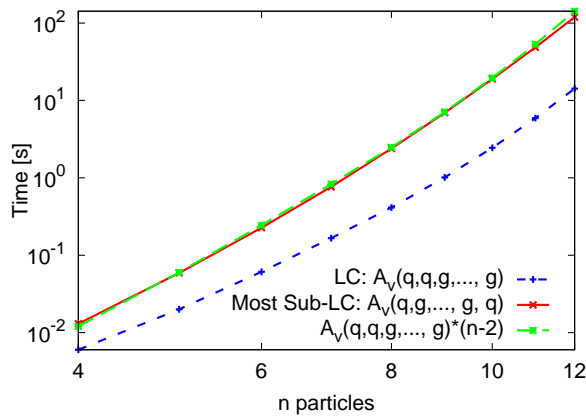}

\caption{\label{fig:timeqqWng} Time in seconds needed to compute the
  leading-color (LC) primitive amplitude $A_v(q,q,g\dots g)$ and the
  sub-leading-color primitive amplitude $A_v(q, g\dots g, q)$ for the
  process $0\to \bar u d W^+ (n-3) g$, as a function of the total
  number of particles $n$. The ratio of times required to compute 
sub-leading-color and leading color is given by $n-2$, with a very good 
approximation, so that it is hard to distinguish the solid and 
the dashed lines.}
\end{center}
\end{figure}

Fig.~\ref{fig:timeqqWng} shows the running time for the leading-color
amplitude, the sub-leading color one, as well as the latter one
rescaled by a factor of $n_{\rm perm} = n-2$. The running time for
both amplitudes is a polynomial in $n$ but, as expected, the
sub-leading color amplitude is $(n-2)$ times slower than the leading
color one.  Using the numerical dominance of the leading-color amplitudes,
one can organize the calculation in such a way that contributions of
sub-leading color amplitudes to cross-sections are computed with the
smaller relative accuracy than leading-color amplitudes.  Such
organization of the calculation ensures that not too much time is
spent on the calculation of pieces of the amplitude that give small
contributions to the final answer.  These ideas have been used in 
recent NLO calculation of some complicated processes
\cite{Berger:2009ep,Berger:2010zx}.

\section{Outlook}
\label{sec:conclu}

Perturbation theory plays a primary role in phenomenological
applications of quantum field theory. The first non-trivial step in
the perturbative expansion requires dealing with tree processes; this
is well-understood by now.  Indeed, there are several automated
computer programs
\cite{Mangano:2002ea,Gleisberg:2003xi,Alwall:2007st,Boos:2004kh} that
can, at leading order, calculate any process in the Standard Model and
its most popular extensions.  Given the scope of the LHC physics
program and the benefits of improving the reliability of theoretical
predictions, automating {\it one-loop computations} is considered to
be an important goal in phenomenologically-oriented high-energy theory
and it took about ten years of active research of many physicists to
get close to accomplishing this goal. Interestingly, arriving at this
point required achieving a deeper understanding of perturbation theory
in general and resulted in the development of a new 
framework for perturbative computations in quantum field theory that
goes beyond Feynman diagrams and traditional methods of
Passarino-Veltman reduction.

The subject of this review -- numerical $D$-dimensional unitarity --
is one of the results of multiple attempts, undertaken in recent years
by many theorists, to provide a flexible enough framework to serve as
a basis for automation.  Conceptually, $D$-dimensional unitarity is
very different from Feynman diagrams. It is an $S$-matrix approach where
locality is not manifest and loop amplitudes are bootstrapped from
tree amplitudes using unitarity.  The tree-level scattering amplitudes
need to be known for complex on-shell momenta and space-times of
higher dimensionality.  As we explained in this review, analytic
continuation of on-shell scattering amplitudes to higher-dimensional
space-times is easily accomplished by using Berends-Giele recursion
relations to construct tree amplitudes. Moreover, polarization states
of bosons and fermions can be extended to higher-dimensional space
times in a straightforward way if one does not require using
spinor-helicity methods in those calculations.

Generalized $D$-dimensional unitarity provides full amplitudes and
works for massless and massive particles, providing the level of
robustness that is required to address problems of high
phenomenological interest.  In this review, we tried to emphasize this
robustness by avoiding the spinor-helicity formalism in the
decomposition of the loop momenta entering the unitarity cuts.  An
intuitively simple decomposition of the loop momentum into physical
and transverse spaces that identifies very clearly the degrees of
freedom that are left unconstrained by unitarity cuts and a
geometrical picture of the transverse space allow us to choose a
parametrization of the transverse space in a most suitable and simple
way. We believe that, in doing so, many subtle features of the
unitarity-based approaches to one-loop computations are made very
transparent.

Before closing, we mention important research topics that we have not
discussed in this review. First, we only briefly mentioned the recent
developments in Passarino-Veltman reduction technology
\cite{Giele:2004iy,Denner:2005nn,Binoth:2006hk, Binoth:2008uq} that
lead to efficient and numerically-stable ways of computing multi-leg
one-loop Feynman diagrams \cite{Bredenstein:2009aj,Denner:2010jp}.
While these methods are traditional, the results that have been
achieved by applying them to realistic problems are spectacular and
the potential of traditional computational methods does not seem to be
anywhere near to being exhausted.

Second, in this review we did not talk about attempts to design a
framework where one-loop computations can be performed by pure
numerical methods \cite{Soper:2001hu,Passarino:2001wv,
  Ferroglia:2002mz,Nagy:2003qn, Anastasiou:2007qb,
  Lazopoulos:2007ix,Nagy:2006xy, Gong:2008ww,Catani:2008xa,
  Kilian:2009wy, 
Becker:2010ng,Becker:2011vg,
Catani:2008xa,
Bierenbaum:2010cy}. It will be
interesting to see how numerical methods will develop in the future,
given the robust semi-numerical nature of the OPP approach and recent
attempts to automate it
\cite{Berger:2010zx,Hirschi:2011pa}.

Third, we only briefly discussed  the spinor-helicity methods
\cite{Berends:1981rb} that play an important role in analytical
computations of tree amplitudes. In particular, those methods tend to
provide very compact expressions for tree amplitudes when external
particles are in definite helicity states.  Since tree amplitudes play
an important role in unitarity-based computations, one may expect that
spinor-helicity methods may be useful for boosting the efficiency of
one-loop computations of scattering amplitudes
\cite{Dixon:2010ik,Bourjaily:2010wh}.  Moreover, spinor-helicity
methods can be used to provide compact expressions for one-loop
scattering amplitudes as well, see e.g
Refs.~\cite{Bern:1994fz,Bern:1997sc,Badger:2010mg, Campbell:2010cz,
  Badger:2011yu}.  At the moment, it seems that the benefits of
analytic computations lie more in clarifying the generic properties of
the amplitudes, rather than in their phenomenological applications. In
particular, it is unclear if the benefits gained in the computing time
that follow from using analytic formulas in numerical codes are
sufficiently high, to outweigh the significant human efforts that are required
to obtain those analytic results.

Fourth, the ideas of unitarity overlap, at least partially, with many
exciting new developments in the field of computation of scattering
amplitudes in supersymmetric extensions of QCD and, in particular,
${\cal N}=4$ SUSY Yang-Mills. It is an interesting and non-trivial question
to what extent real needs of the physics program at the LHC can
benefit from those developments. While it appears that many of the
developments in that field rely on high degree of supersymmetry, a
number of ideas such as the recursion relations for the loop integrand
\cite{CaronHuot:2010zt,ArkaniHamed:2010kv,Boels:2010nw} may have more
general nature and, perhaps, can be successfully used in
non-supersymmetric quantum field theories, such as QCD and the
Standard Model.  Partially related to this is the barely explored
question of whether it is possible to apply unitarity-based ideas to
multi-loop computations.  On one hand, we do know that this is
possible in highly-supersymmetric theories, where calculations are
carried out to a very high order in the perturbative expansion. On the
other hand, in non-supersymmetric field theories, even the first step
-- the identification of a suitable basis of master integrals at
two-loops -- has been attempted only recently \cite{Gluza:2010ws}.
Hence, it will be interesting to watch in the future how new ideas
developed in the context of ${\cal N}=4$ SUSY Yang-Mills will penetrate into
more phenomenological research related to perturbative computations in
the Standard Model.

The field of perturbative computations for multi-particle processes
went through a remarkable transformation in the past few years. During
these years, the ability to perform specific computations that are of
importance for the Tevatron and the LHC physics program has increased
beyond the most optimistic expectations. The improvement in our
understanding of perturbative quantum field theory -- that is a
byproduct of these exciting developments -- gives us hope that the
momentum of the past several years can be carried forward, so that
even more complicated physics -- both in terms of the number of
external particles and in terms of the number of loops -- can be
addressed.

\section*{Acknowledgements} 

We would like to thank W.~Giele, T.~Melia, M.~Schulze, R.~R\"ontsch
for collaboration on various topics that we address in this review.
We are grateful to F.~Caola, W.~Giele and M.~Schulze for comments on
the manuscript.  
Z.K. would like to thank the Institute for the Physics and Mathematics
of the Universe (IPMU), University of Tokyo, Kashiwa, Japan, and the
Department for Theoretical Physics and the Cosmos, University of
Granada, Spain for hospitality where parts of this review have been
prepared.  This research is supported by the NSF under grant
PHY-0855365, by the British Science and Technology Council, by the
LHCPhenoNet network under the Grant Agreement PITN-GA-2010-264564, and
by the European Research and Training Network (RTN) grant Unification
in the LHC ERA under the Agreement PITN-GA-2009-237920. Fermilab is
operated by Fermi Research Alliance, LLC under Contract
No. DE-AC02-07CH11359 with the United States Department of Energy.

\appendix
\section{Details of Passarino-Veltman decomposition}
\label{app:PV}

The purpose of this appendix is to give a full description 
of the Passarino-Veltman decomposition~\cite{Passarino:1978jh}
for the bubble-, triangle-, and box- tensor integrals which appear in a renormalizable theory. 
Although these formulas 
have been given in the original paper it may be of use to present them here.
In presenting them we have converted 
the reduction formula to the Bjorken and Drell metric~\cite{Bjorken:1965zz}
$g_{\mu \nu} = {\rm diag}[1,-1,-1,-1]$, 
which is most commonly 
used. In addition, 
we believe that we have improved the notation so that the pattern of the
reduction is more apparent. Furthermore 
the original published paper contains typographical errors
(on line 2 and line 17 of page 205 of ref.~\cite{Passarino:1978jh}).   
We believe that the above remarks provide sufficient justification for 
presenting the Passarino-Veltman reduction  formulas again.

\subsubsection{ Appendix A.1. Two-point functions} 

We first recall (cf. Section~\ref{sec:tradoneloop}) 
our definitions of the form-factor expansions for the tensor integrals
\begin{eqnarray}
B^\mu &=& p_{1}^{\mu} B_1, \\
\label{Brank1}
B^{\mu \nu} &=& g^{\mu\nu} B_{00} + p_1^\mu p_1^\nu B_{11}.
\label{Brank2}
\end{eqnarray}
We shall refer to the coefficients $B_{i},B_{00},B_{11}$ as form factors.
The dependence of these form factors on the appropriate Lorentz
invariants, such as squares of the external momenta and masses 
of internal lines has been dropped for brevity.

By contracting through with $p_1$ and $g^{\mu \nu}$ 
the form factors can be expressed entirely in terms of scalar integrals.
(The singular case $p_1^2=0$ needs to be handled separately). The results are,
\begin{eqnarray}
B_{1}& = & \frac{1}{2 p_1^2} (f_1 B_{0}(1,2)+A_{0}(1)-A_{0}(2)), \nonumber \\
B_{11}&=&\frac{1}{2 p_1^2} (f_1 B_{1}(1,2)+A_{0}(2)-2 B_{00}(1,2)), \nonumber \\
B_{00} &= &\frac{1}{2 (D-1)} (2 m_1^2 B_{0}(1,2)+A_{0}(2)-f_1 B_{1}(1,2)) \, .
\label{Breduction}
\end{eqnarray}
The compact notation is as follows.
The propagators that occur in the scalar integrals, $A_0$ and $B_0$ are specified by $i$
where $m_i$ is the mass which is present in the propagator. Thus, for
instance, 
\beq
A_0(2) = \frac{1}{i \pi^{D/2}}\; \int d^Dl \; \frac{1}{l^2-m_2^2} \, .
 \eeq
The constant $f_1$ introduced in  \Eq(\ref{Breduction}) 
is part of 
the loop-momentum independent 
offsets that we will use in what follows 
\begin{eqnarray}
f_1 &=&            m_2^2-m_1^2-p_1^2 \, ,\nonumber \\
f_2 &=&        m_3^2-m_2^2-p_2^2 -2 p_1\cdot p_2  \, , \nonumber \\
f_3 &=&    m_4^2-m_3^2-p_3^2-2 p_3\cdot p_2-2 p_3\cdot p_1 \, .
\end{eqnarray}

\subsubsection{ Appendix A.2. Three-point functions} 

Turning now to the triangle integrals we have the form factor expansion
\begin{eqnarray}
C^\mu &=& p_{1}^{\mu} C_1 +p_{2}^{\mu} C_2, 
\label{Crank1} \\
C^{\mu \nu} 
&=& g^{\mu\nu} C_{00} + p_1^\mu p_1^\nu C_{11}+ p_2^\mu p_2^\nu C_{22}+ (p_1^\mu p_2^\nu +p_2^\mu p_1^\nu ) C_{12} 
\nonumber \\
            &=& g^{\mu\nu} C_{00} + \sum_{i,j=1}^2 p_i^\mu p_j^\nu C_{ij}\,, \;\;
\mbox{where}~C_{21}=C_{12}\;,
\label{Crank2} \\
C^{\mu \nu \alpha} 
&=& \sum_{i=1}^2  (g^{\mu \nu} p_i^\alpha+g^{\nu \alpha } p_i^\mu+g^{\alpha \mu} p_i^\nu) C_{00i}
+\sum_{i,j,k=1}^2 p_i^\mu p_j^\nu p_k^\alpha C_{ijk}\,.
\label{Crank3}
\end{eqnarray}
These too can be reduced to scalar integrals by contracting through with external momenta
and the metric tensor as explained in Section~\ref{Onelooptensorintegrals}. 
For example, for the two form factors
$C_1,C_2$ we have 
\begin{equation}
\left( \begin{array}{c} C_{1}\\
                        C_{2} \end{array}\right) = G_2^{-1}
\left( \begin{array}{c} R^{[c]}_1\\
                        R^{[c]}_2 \end{array}\right), 
\label{Cformfactoreq}
\end{equation}
where 
$G_2$ is the $2\times 2$ Gram matrix
\begin{equation}
G_2= \left( \begin{array}{cc} p_1\cdot p_1 & p_1 \cdot p_2 \\
                              p_1\cdot p_2 &  p_2 \cdot p_2\\
                         \end{array}\right).
\end{equation}
All the form factor pairs which satisfy equations 
of the form of \Eq(\ref{Cformfactoreq})
are shown in Table~\ref{Triangledoublets2}. Note that 
three of the coefficients in 
Table~\ref{Triangledoublets2} can be  
determined in more than one way. This provides an important
check of the reduction. The expressions for the right-hand sides 
listed in Table~\ref{Triangledoublets2} are,
\begin{table}[t]
\begin{center}
\begin{tabular}{|cc|c|}
\hline
\multicolumn{2}{|c|}{Form factors} & RHS \\
\hline
$C_{1}$     & $C_{2}$  & $R^{[c]}$ \\
\hline
$C_{11}$     & $C_{12}$  & $R^{[c1]}$ \\
$C_{12}$     & $C_{22}$  & $R^{[c2]}$ \\
\hline
$C_{001}$     & $C_{002}$  & $R^{[c00]}$ \\
$C_{111}$     & $C_{112}$  & $R^{[c11]}$ \\
$C_{112}$     & $C_{122}$  & $R^{[c12]}$ \\
$C_{122}$     & $C_{222}$  & $R^{[c22]}$ \\
\hline
\end{tabular}
\end{center}
\caption{Pairs of form factors which satisfy equations of the form
 \Eq(\ref{Cformfactoreq}).}
\label{Triangledoublets2}
\end{table}
\begin{equation}
\begin{split}
& R^{[c]}_{1}=\frac{1}{2} (f_1 C_0(1,2,3)+B_0(1,3)-B_0(2,3)),
\;\;\;\;\;\;\;\;\;\;\;\;\;\;\;\;\;\;\;\;\;\;\;\;\;\;\;\;\;\;\;\; 
\\
& R^{[c]}_{2}=\frac{1}{2} (f_2 C_0(1,2,3)+B_0(1,2)-B_0(1,3)). 
\end{split} 
\end{equation}
\begin{equation}
\begin{split}
& R^{[c1]}_{1}=\frac{1}{2} (f_1 C_{1}(1,2,3)+B_1(1,3)+B_0(2,3)
-2 C_{00}(1,2,3)), 
\;\;\;\;\;\;\;
\\ 
& R^{[c1]}_{2}=\frac{1}{2} (f_2 C_{1}(1,2,3)+B_1(1,2)-B_1(1,3)), 
\end{split} 
\end{equation}
\begin{equation}
\begin{split}
& R^{[c2]}_{1}=\frac{1}{2} (f_1 C_{2}(1,2,3)+B_1(1,3)-B_1(2,3)), 
\;\;\;\;\;\;\;\;\;\;\;\;\;\;\;\;\;\;\;\;\;\;\;\;\;\;\;\;\;\; 
\\ 
& R^{[c2]}_{2}=\frac{1}{2} (f_2 C_{2}(1,2,3)-B_1(1,3)-2 C_{00}(1,2,3)), 
\end{split}
\end{equation}
\begin{equation}
\begin{split} 
& R^{[c11]}_{1}=\frac{1}{2} (f_1 C_{11}(1,2,3)+B_{11}(1,3)-B_{0}(2,3)-4 C_{001}(1,2,3)),  
\\ 
& R^{[c11]}_{2}=\frac{1}{2} (f_2 C_{11}(1,2,3)+B_{11}(1,2)-B_{11}(1,3)),
\end{split} 
\end{equation}
\begin{equation}
\begin{split} 
& R^{[c22]}_{1}=\frac{1}{2} (f_1 C_{22}(1,2,3)+B_{11}(1,3)-B_{11}(2,3)), 
\;\;\;\;\;\;\;\;\;\;\;\;\;\;\;\;\;\;\;\;\;\;\;\; 
 \\ 
& R^{[c22]}_{2}=\frac{1}{2} (f_2 C_{22}(1,2,3)-B_{11}(1,3)-4 C_{002}(1,2,3)),
\label{SpecificR}
\end{split} 
\end{equation}
\begin{equation}
\begin{split} 
& R^{[c12]}_{1}=\frac{1}{2} (f_1 C_{12}(1,2,3)+B_{11}(1,3)+B_{1}(2,3)-2 C_{002}(1,2,3)), \\ 
& R^{[c12]}_{2}=\frac{1}{2} (f_2 C_{12}(1,2,3)-B_{11}(1,3)-2 C_{001}(1,2,3)), 
\end{split} 
\end{equation}
\begin{equation}
\begin{split} 
& R^{[c00]}_{1}=\frac{1}{2} (f_1 C_{00}(1,2,3)+B_{00}(1,3)
-B_{00}(2,3)),
\;\;\;\;\;\;\;\;\;\;\;\;\;\;\;\;\;\;\;\;\;\;\;\; 
\\
& R^{[c00]}_{2}=\frac{1}{2} (f_2 C_{00}(1,2,3)+B_{00}(1,2)-B_{00}(1,3)). 
\end{split} 
\end{equation}
To express our results we have introduced a compact notation.
Thus, for example, in \Eq(\ref{SpecificR}), $B_{11}(2,3)$ is defined as 
the $B_{11}$ coefficient of the integral
\beq
B^{\mu \nu}(p_2,m_2,m_3)=
\frac{1}{i \pi^{D/2}}\int \; d^D l  
\frac{l^\mu l^\nu} {(l^2-m_2^2) ((l+p_2)^2-m_3^2)}.
\label{CompactNotation}
\eeq
Note that in \Eq(\ref{CompactNotation})
the loop momentum $l$ has been shifted ($l \to l -p_1$) 
with respect to the defining equation for the triangle integrals, 
\be
\label{eq:Cint1}
\begin{split}
& C_0(p_1,p_2,m_1,m_2,m_3)= \\
&\frac{1}{i \pi^{D/2}}\int \; d^D l  \frac{1} {(l^2-m_1^2)((l+p_1)^2-m_2^2)((l+p_1+p_2)^2-m_3^2)} \, .
\end{split}
\ee

In addition we have the following relations,
\begin{eqnarray}
C_{00}(1,2,3)&=&\frac{1}{2 (D-2)}
   (2 m_1^2 C_0(1,2,3)-f_1 C_{1}(1,2,3) -f_2 C_{2}(1,2,3)\nonumber \\
&+&B_0(2,3)), \label{C00} \\
C_{001}(1,2,3)&=&
          \frac{1}{2 (D-1)} (2 m_1^2 C_{1}(1,2,3)
          - f_1 C_{11}(1,2,3)- f_2 C_{12}(1,2,3)\nonumber \\ 
          &-& B_0(2,3)), \label{C001}\\
C_{002}(1,2,3)&=& \frac{1}{2 (D-1)} (2 m_1^2 C_{2}(1,2,3) 
    -f_1 C_{12}(1,2,3)-f_2 C_{22}(1,2,3)\nonumber \\ 
&+&B_1(2,3)). 
\label{C002}
\end{eqnarray}
Note that \Eqs(\ref{C00},\ref{C001},\ref{C002}) are 
the only places where  the dependence on 
the dimensionality of space-time $D$
appears. Furthermore $C_{001},C_{002}$ are 
also determined by $D$-independent relations of the form of 
\Eq(\ref{Cformfactoreq}) as shown in Table~\ref{Triangledoublets2}.

\subsubsection{Appendix A.3. Four-point functions}

Turning now to the box coefficients we have the form factor expansion,
\begin{eqnarray}
 D^\mu &=& p_{1}^{\mu} D_1 +p_{2}^{\mu} D_2 +p_{3}^{\mu} D_3 \,,
\label{Drank1} \\
D^{\mu \nu}  &=& g^{\mu\nu} D_{00} + \sum_{i,j=1}^3 p_i^\mu p_j^\nu D_{ij}\,,
\label{Drank2} \\
D^{\mu \nu \alpha} 
&=& \sum_{i=1}^3  
g^{\{ \mu \nu} p_i^{\alpha \}}
D_{00i}
 + \sum_{i,j,k=1}^3 p_i^\mu p_j^\nu p_k^\alpha D_{ijk}\,,
\label{Drank3} \\
 D^{\mu \nu \alpha \beta} 
&=& 
g^{\{ \mu \nu} g^{\alpha \beta \} } 
D_{0000}
+\sum_{i,j=1}^3
g^{\{ \mu \nu} p_l^{\alpha} p_j^{\beta \}}
D_{00ij} 
\nn  \\
& +& \sum_{i,j,k,l=1}^3 p_i^\mu p_j^\nu p_k^\alpha p_l^\beta D_{ijkl}\, .
\label{Drank4}
\end{eqnarray}
The curly braces denote full symmetrization of the indices.
Many of these coefficients satisfy equations of the form
\begin{equation}
\left( \begin{array}{c} D_{1}\\
                        D_{2} \\
                        D_{3} \end{array}\right) = G_3^{-1}
\left( \begin{array}{c} R^{[d]}_1\\
                        R^{[d]}_2\\
                        R^{[d]}_3 \end{array}\right) \, ,
\label{Dformfactoreq}
\end{equation}
where $G_3$ is the $3\times 3$ Gram matrix
\begin{equation}
G_3= \left( \begin{array}{ccc} p_1\cdot p_1 & p_1 \cdot p_2 & p_1 \cdot p_3 \\
                              p_1\cdot p_2 &  p_2 \cdot p_2 &  p_2 \cdot p_3 \\
                              p_3\cdot p_1 &  p_3 \cdot p_2 &  p_3 \cdot p_3 \\
                         \end{array}\right) \, .
\end{equation}
All the form factor triplets which satisfy equations of the form \Eq(\ref{Dformfactoreq})
are shown in Table~\ref{Boxtriplets2}, where the notation for the right-hand sides of the corresponding 
linear equation is also defined.

\begin{table}[t]
\begin{center}
\begin{tabular}{|ccc|c|}
\hline
\multicolumn{3}{|c|}{Form factors} & RHS \\
\hline
$D_{1}$ & $D_{2}$ & $D_{3}$ & $R^{[d]}$ \\
\hline
$D_{11}$ & $D_{12}$ & $D_{13}$ & $R^{[d1]}$ \\
$D_{12}$ & $D_{22}$ & $D_{23}$ & $R^{[d2]}$ \\
$D_{13}$ & $D_{23}$ & $D_{33}$ & $R^{[d3]}$ \\
\hline
$D_{001}$ & $D_{002}$ & $D_{003}$ & $R^{[d00]}$ \\
$D_{112}$ & $D_{122}$ & $D_{123}$ & $R^{[d12]}$ \\
$D_{113}$ & $D_{123}$ & $D_{133}$ & $R^{[d13]}$ \\
$D_{123}$ & $D_{223}$ & $D_{233}$ & $R^{[d23]}$ \\
$D_{111}$ & $D_{112}$ & $D_{113}$ & $R^{[d11]}$ \\
$D_{122}$ & $D_{222}$ & $D_{223}$ & $R^{[d22]}$ \\
$D_{133}$ & $D_{233}$ & $D_{333}$ & $R^{[d33]}$ \\
\hline
$D_{0011}$ & $D_{0012}$ & $D_{0013}$ & $R^{[d001]}$ \\
$D_{0012}$ & $D_{0022}$ & $D_{0023}$ & $R^{[d002]}$ \\
$D_{0013}$ & $D_{0023}$ & $D_{0033}$ & $R^{[d003]}$ \\
$D_{1111}$ & $D_{1112}$ & $D_{1113}$ & $R^{[d111]}$ \\
$D_{1222}$ & $D_{2222}$ & $D_{2223}$ & $R^{[d222]}$ \\
$D_{1333}$ & $D_{2333}$ & $D_{3333}$ & $R^{[d333]}$ \\
$D_{1112}$ & $D_{1122}$ & $D_{1123}$ & $R^{[d112]}$ \\
$D_{1113}$ & $D_{1123}$ & $D_{1133}$ & $R^{[d113]}$ \\
$D_{1122}$ & $D_{1222}$ & $D_{1223}$ & $R^{[d122]}$ \\
$D_{1133}$ & $D_{1233}$ & $D_{1333}$ & $R^{[d133]}$ \\
$D_{1223}$ & $D_{2223}$ & $D_{2233}$ & $R^{[d223]}$ \\
$D_{1233}$ & $D_{2233}$ & $D_{2333}$ & $R^{[d233]}$ \\
$D_{1123}$ & $D_{1223}$ & $D_{1233}$ & $R^{[d123]}$ \\
\hline
\end{tabular}
\end{center}
\caption{Triplets of form factors which satisfy equations of the form,  \Eq(\ref{Dformfactoreq}).}
\label{Boxtriplets2}
\end{table}

We now give the definitions of the functions $R$ for the rank one boxes,
\begin{equation}
\begin{split}
& R^{[d]}_{1}=
  \frac{1}{2} (f_1 D_0(1,2,3,4)
  +C_{0}(1,3,4)-C_{0}(2,3,4)),
\\
& R^{[d]}_{2}=
  \frac{1}{2} (f_2 D_0(1,2,3,4)
  +C_{0}(1,2,4)-C_{0}(1,3,4)), 
\\
& R^{[d]}_{3}=
  \frac{1}{2} (f_3 D_0(1,2,3,4)
 +C_{0}(1,2,3)-C_{0}(1,2,4)).  
\end{split} 
\end{equation}
For the rank two boxes we obtain,
\begin{equation}
\begin{split} 
& R^{[d1]}_{1}=
 \frac{1}{2} (f_1 D_{1}(1,2,3,4)
 + C_{0}(2,3,4)
 + C_{1}(1,3,4)
 - 2 D_{00}(1,2,3,4)),
 \\
& R^{[d1]}_{2}=
 \frac{1}{2} (f_2 D_{1}(1,2,3,4)
 - C_{1}(1,3,4)
 + C_{1}(1,2,4)),
\\
& R^{[d1]}_{3} = 
 \frac{1}{2} (f_3 D_{1}(1,2,3,4)
 + C_{1}(1,2,3)
 - C_{1}(1,2,4)), 
 \\
& R^{[d2]}_{1}=
 \frac{1}{2} (f_1 D_{2}(1,2,3,4)
 - C_{1}(2,3,4)
 + C_{1}(1,3,4)),
\\
& R^{[d2]}_{2}=
 \frac{1}{2} (f_2 D_{2}(1,2,3,4)
 - C_{1}(1,3,4)
 + C_{2}(1,2,4)
 - 2 D_{00}(1,2,3,4)),
\\
& R^{[d2]}_{3}=
 \frac{1}{2} (f_3 D_{2}(1,2,3,4)
 + C_{2}(1,2,3)
 - C_{2}(1,2,4)), 
 \\
& R^{[d3]}_{1}=
 \frac{1}{2} (f_1 D_{3}(1,2,3,4)
 - C_{2}(2,3,4)
 + C_{2}(1,3,4)),
\\
& R^{[d3]}_{2}=
 \frac{1}{2} (f_2 D_{3}(1,2,3,4)
 - C_{2}(1,3,4)
 + C_{2}(1,2,4)),
\\
& R^{[d3]}_{3}=
 \frac{1}{2} (f_3 D_{3}(1,2,3,4)
 - C_{2}(1,2,4)
 - 2 D_{00}(1,2,3,4)). 
\end{split} 
\end{equation}
For the rank three boxes we have,
\begin{eqnarray}
R^{[d12]}_{1} & =&
 \frac{1}{2} (f_1 D_{12}(1,2,3,4)
 + C_{1}(2,3,4)
 + C_{11}(1,3,4)
- 2 D_{002}(1,2,3,4)), \nonumber \\
R^{[d12]}_{2} & =&
 \frac{1}{2} (f_2 D_{12}(1,2,3,4)
 + C_{12}(1,2,4)
 - C_{11}(1,3,4)
- 2 D_{001}(1,2,3,4)), \nonumber \\
R^{[d12]}_{3} &=& 
 \frac{1}{2} (f_3 D_{12}(1,2,3,4)
 + C_{12}(1,2,3)
 - C_{12}(1,2,4)).  
\end{eqnarray}

\begin{eqnarray}
R^{[d13]}_{1}&=&
 \frac{1}{2} (f_1 D_{13}(1,2,3,4)
 + C_{2}(2,3,4)
 + C_{12}(1,3,4)
 - 2 D_{003}(1,2,3,4)), \nonumber \\
R^{[d13]}_{2}&=&
 \frac{1}{2} (f_2 D_{13}(1,2,3,4)
 - C_{12}(1,3,4)
 + C_{12}(1,2,4)), \nonumber \\
R^{[d13]}_{3}&=&
 \frac{1}{2} (f_3 D_{13}(1,2,3,4)
 - C_{12}(1,2,4)
 - 2 D_{001}(1,2,3,4)).  
\end{eqnarray}
\begin{eqnarray}
R^{[d23]}_{1}&=& 
 \frac{1}{2} (f_1 D_{23}(1,2,3,4)
 - C_{12}(2,3,4)
 + C_{12}(1,3,4)), \nonumber \\
R^{[d23]}_{2}&=& 
 \frac{1}{2} (f_2 D_{23}(1,2,3,4)
 + C_{22}(1,2,4)
 - C_{12}(1,3,4)
 - 2 D_{003}(1,2,3,4)), \nonumber \\
R^{[d23]}_{3}&=& 
 \frac{1}{2} (f_3 D_{23}(1,2,3,4)
 - C_{22}(1,2,4)
 - 2 D_{002}(1,2,3,4)).  
\end{eqnarray}
\begin{eqnarray}
R^{[d11]}_{1}&=&
 \frac{1}{2} (f_1 D_{11}(1,2,3,4)
 - C_{0}(2,3,4)
 + C_{11}(1,3,4)
 - 4 D_{001}(1,2,3,4)), \nonumber \\
R^{[d11]}_{2}&=&
 \frac{1}{2} (f_2 D_{11}(1,2,3,4)
 - C_{11}(1,3,4)
 + C_{11}(1,2,4)), \nonumber \\
R^{[d11]}_{3}&=&
 \frac{1}{2} (f_3 D_{11}(1,2,3,4)
 + C_{11}(1,2,3)
 - C_{11}(1,2,4)).  
\end{eqnarray}
\begin{eqnarray}
R^{[d22]}_{1}&=&
 \frac{1}{2} (f_1 D_{22}(1,2,3,4)
 - C_{11}(2,3,4) 
 + C_{11}(1,3,4)), \nonumber \\
R^{[d22]}_{2}&=&
 \frac{1}{2} (f_2 D_{22}(1,2,3,4)
 - C_{11}(1,3,4)
 + C_{22}(1,2,4)
 - 4 D_{002}(1,2,3,4)), \nonumber \\
R^{[d22]}_{3}&=&
 \frac{1}{2} (f_3 D_{22}(1,2,3,4)
 + C_{22}(1,2,3)
 - C_{22}(1,2,4)).  
\end{eqnarray}
\begin{eqnarray}
R^{[d33]}_{1}&=&
 \frac{1}{2} (f_1 D_{33}(1,2,3,4)
 - C_{22}(2,3,4)
 + C_{22}(1,3,4)), \nonumber \\
R^{[d33]}_{2}&=&
 \frac{1}{2} (f_2 D_{33}(1,2,3,4)
 - C_{22}(1,3,4)
 + C_{22}(1,2,4)), \nonumber \\
R^{[d33]}_{3}&=&
 \frac{1}{2} (f_3 D_{33}(1,2,3,4)
 - C_{22}(1,2,4)
 - 4 D_{003}(1,2,3,4)).  
\end{eqnarray}
\begin{eqnarray}
R^{[d00]}_{1}&=&
 \frac{1}{2} (f_1 D_{00}(1,2,3,4)
 - C_{00}(2,3,4)
 + C_{00},1,3,4)), \nonumber \\
R^{[d00]}_{2}&=&
 \frac{1}{2} (f_2 D_{00}(1,2,3,4)
 - C_{00},1,3,4)
 + C_{00},1,2,4)), \nonumber \\
R^{[d00]}_{3}&=&
 \frac{1}{2} (f_3 D_{00}(1,2,3,4)
 + C_{00}(1,2,3)
 - C_{00},1,2,4)).  
\end{eqnarray}
For the rank four boxes we obtain
\begin{eqnarray}
R^{[d111]}_{1}&=&\frac{1}{2} (
            f_1 D_{111}(1,2,3,4)
          + C_{111}(1,3,4)
          + C_{0}(2,3,4)
          - 6 D_{0011}(1,2,3,4)), \nonumber \\
R^{[d111]}_{2}&=&\frac{1}{2} (
           f_2 D_{111}(1,2,3,4)
          - C_{111}(1,3,4)
          + C_{111}(1,2,4)), \nonumber \\
R^{[d111]}_{3}&=&\frac{1}{2} (
           f_3 D_{111}(1,2,3,4)
          - C_{111}(1,2,4)
          + C_{111}(1,2,3)).  
\end{eqnarray}
\begin{eqnarray}
R^{[d222]}_{1}&=&\frac{1}{2} (
           f_1 D_{222}(1,2,3,4)
          + C_{111}(1,3,4)
          - C_{111}(2,3,4)), \nonumber \\
R^{[d222]}_{2}&=&
          \frac{1}{2} (f_2 D_{222}(1,2,3,4)
          - C_{111}(1,3,4)
          + C_{222}(1,2,4)
          - 6 D_{0022}(1,2,3,4)), \nonumber \\
R^{[d222]}_{3}&=&
          \frac{1}{2} (f_3 D_{222}(1,2,3,4)
          - C_{222}(1,2,4)
          + C_{222}(1,2,3)).  
\end{eqnarray}
\begin{eqnarray}
R^{[d333]}_{1}&=&
         \frac{1}{2} (f_1 D_{333}(1,2,3,4)
          + C_{222}(1,3,4)
          - C_{222}(2,3,4)), \nonumber \\
R^{[d333]}_{2}&=&
         \frac{1}{2} (f_2 D_{333}(1,2,3,4)
          - C_{222}(1,3,4)
          + C_{222}(1,2,4)), \\
R^{[d333]}_{3}&=&
          \frac{1}{2} (f_3 D_{333}(1,2,3,4)
          - C_{222}(1,2,4)
          - 6 D_{0033}(1,2,3,4)). \nonumber  
\end{eqnarray}
\begin{eqnarray}
R^{[d112]}_{1}&=&\frac{1}{2} (f_1 D_{112}(1,2,3,4)
          + C_{111}(1,3,4)
          - C_{1}(2,3,4)
          - 4 D_{0012}(1,2,3,4)), \nonumber \\
R^{[d112]}_{2}&=&
          \frac{1}{2} (f_2 D_{112}(1,2,3,4)
          - C_{111}(1,3,4)
          + C_{112}(1,2,4)
          - 2 D_{0011}(1,2,3,4)), \nonumber \\
R^{[d112]}_{3}&=&
          \frac{1}{2} (f_3 D_{112}(1,2,3,4)
          - C_{112}(1,2,4)
          + C_{112}(1,2,3)).  
\end{eqnarray}
\begin{eqnarray}
R^{[d113]}_{1}&=&
          \frac{1}{2} (f_1 D_{113}(1,2,3,4)
          + C_{112}(1,3,4)
          - C_{2}(2,3,4)
          - 4 D_{0013}(1,2,3,4)), \nonumber \\
R^{[d113]}_{2}&=&
          \frac{1}{2} (f_2 D_{113}(1,2,3,4)
          - C_{112}(1,3,4)
          + C_{112}(1,2,4)), \nonumber \\
R^{[d113]}_{3}&=&
          \frac{1}{2} (f_3 D_{113}(1,2,3,4)
          - C_{112}(1,2,4)
          - 2 D_{0011}(1,2,3,4)).  
\end{eqnarray}
\begin{eqnarray}
R^{[d122]}_{1}&=&\frac{1}{2} (f_1 D_{122}(1,2,3,4)
          + C_{111}(1,3,4)
          + C_{11}(2,3,4)
          - 2 D_{0022}(1,2,3,4)), \nonumber \\
R^{[d122]}_{2}&=&
          \frac{1}{2} (f_2 D_{122}(1,2,3,4)
          - C_{111}(1,3,4)
          + C_{122}(1,2,4)
          - 4 D_{0012}(1,2,3,4)), \nonumber \\
R^{[d122]}_{3}&=&
          \frac{1}{2} (f_3 D_{122}(1,2,3,4)
          - C_{122}(1,2,4)
          + C_{122}(1,2,3)).  
\end{eqnarray}
\begin{eqnarray}
R^{[d133]}_{1}&=&
          \frac{1}{2} (f_1 D_{133}(1,2,3,4)
          + C_{122}(1,3,4)
          + C_{22}(2,3,4)
          - 2 D_{0033}(1,2,3,4)), \nonumber \\
R^{[d133]}_{2}&=&
         \frac{1}{2} (f_2 D_{133}(1,2,3,4)
          - C_{122}(1,3,4)
          + C_{122}(1,2,4)), \nonumber \\
R^{[d133]}_{3}&=&
          \frac{1}{2} (f_3 D_{133}(1,2,3,4)
          - C_{122}(1,2,4)
          - 4 D_{0013}(1,2,3,4)).  
\end{eqnarray}
\begin{eqnarray}
R^{[d223]}_{1}&=&\frac{1}{2} (f_1 D_{223}(1,2,3,4)
          + C_{112}(1,3,4)
          - C_{112}(2,3,4)), \nonumber \\
R^{[d223]}_{2}&=&
         \frac{1}{2} (f_2 D_{223}(1,2,3,4)
          - C_{112}(1,3,4)
          + C_{222}(1,2,4)
          - 4 D_{0023}(1,2,3,4)), \nonumber \\
R^{[d223]}_{3}&=&
          \frac{1}{2} (f_3 D_{223}(1,2,3,4)
          - C_{222}(1,2,4)
          - 2 D_{0022}(1,2,3,4)).  
\end{eqnarray}
\begin{eqnarray}
R^{[d233]}_{1}&=&\frac{1}{2} (f_1 D_{233}(1,2,3,4)
          + C_{122}(1,3,4)
          - C_{122}(2,3,4)), \nonumber \\
R^{[d233]}_{2}&=&
          \frac{1}{2} (f_2 D_{233}(1,2,3,4)
          - C_{122}(1,3,4)
          + C_{222}(1,2,4)
          - 2 D_{0033}(1,2,3,4)), \nonumber \\
R^{[d233]}_{3}&=&
          \frac{1}{2} (f_3 D_{233}(1,2,3,4)
          - C_{222}(1,2,4)
          - 4 D_{0023}(1,2,3,4)).  
\end{eqnarray}
\begin{eqnarray}
R^{[d123]}_{1}&=&
          \frac{1}{2} (f_1 D_{123}(1,2,3,4)
          + C_{112}(1,3,4)
          + C_{12}(2,3,4)
          - 2 D_{0023}(1,2,3,4)), \nonumber \\
R^{[d123]}_{2}&=&
          \frac{1}{2} (f_2 D_{123}(1,2,3,4)
          - C_{112}(1,3,4)
          + C_{122}(1,2,4)
          - 2 D_{0013}(1,2,3,4)), \nonumber \\
R^{[d123]}_{3}&=&
          \frac{1}{2} (f_3 D_{123}(1,2,3,4)
          - C_{122}(1,2,4)
          - 2 D_{0012}(1,2,3,4)).  
\end{eqnarray}
\begin{eqnarray}
R^{[d001]}_{1}&=&
          \frac{1}{2} (f_1 D_{001}(1,2,3,4)
          + C_{001}(1,3,4)
          + C_{00}(2,3,4)
          - 2 D_{0000}(1,2,3,4)), \nonumber \\
R^{[d001]}_{2}&=&
          \frac{1}{2} (f_2 D_{001}(1,2,3,4)
          - C_{001}(1,3,4)
          + C_{001}(1,2,4)), \nonumber \\
R^{[d001]}_{3}&=&
          \frac{1}{2} (f_3 D_{001}(1,2,3,4)
          - C_{001}(1,2,4)
          + C_{001}(1,2,3)).  
\end{eqnarray}
\begin{eqnarray}
R^{[d002]}_{1}&=&\frac{1}{2} (f_1 D_{002}(1,2,3,4)
          + C_{001}(1,3,4)
          - C_{001}(2,3,4)), \nonumber \\
R^{[d002]}_{2}&=&
          \frac{1}{2} (f_2 D_{002}(1,2,3,4)
          - C_{001}(1,3,4)
          + C_{002}(1,2,4)
          - 2 D_{0000}(1,2,3,4)), \nonumber \\
R^{[d002]}_{3}&=&
          \frac{1}{2} (f_3 D_{002}(1,2,3,4)
          - C_{002}(1,2,4)
          + C_{002}(1,2,3)). 
\end{eqnarray}
\begin{eqnarray}
R^{[d003]}_{1}&=&\frac{1}{2} (f_1 D_{003}(1,2,3,4)
          + C_{002}(1,3,4)
          - C_{002}(2,3,4)),\;\;\;\;\;\; \nonumber \\
R^{[d003]}_{2}&=&
          \frac{1}{2} (f_2 D_{003}(1,2,3,4)
          - C_{002}(1,3,4)
          + C_{002}(1,2,4)), \nonumber \\
R^{[d003]}_{3}&=&\frac{1}{2} (f_3 D_{003}(1,2,3,4)
          - C_{002}(1,2,4)
          - 2 D_{0000}(1,2,3,4)).\;\;\;\;\;\;
\end{eqnarray}

In addition to the relations detailed 
in Table~\ref{Boxtriplets2} we also have the following 
$D$-dependent relations, obtained by contracting with the metric tensor.
Of these only \Eqs(\ref{D0000reduction}) and (\ref{D00reduction}) 
are really necessary,
since the others are determined by 
the relations in Table~\ref{Boxtriplets2}. Nevertheless
the redundancy provides a good check of the 
implementation.
\begin{eqnarray}
&&D_{0000}(1,2,3,4) = \frac{1}{2(D-1)}(
          2 m_1^2 D_{00}(1,2,3,4)
          +C_{00}(2,3,4) \nonumber \\ 
        &-&f_1 D_{001}(1,2,3,4)
          -f_2 D_{002}(1,2,3,4)
          -f_3 D_{003}(1,2,3,4)
           ),  \label{D0000reduction}\\
&&D_{0011}(1,2,3,4) = \frac{1}{2(D-1)}(
          2 m_1^2 D_{11}(1,2,3,4)
          + C_{0}(2,3,4)\nonumber \\
        &-& f_1 D_{111}(1,2,3,4)
          - f_2 D_{112}(1,2,3,4)
          - f_3 D_{113}(1,2,3,4)
            ), \\
&&D_{0012}(1,2,3,4) = \frac{1}{2(D-1)}(
          2 m_1^2 D_{12}(1,2,3,4)
          -C_{1}(2,3,4) \nonumber \\
        &-&f_1 D_{112}(1,2,3,4)
          -f_2 D_{122}(1,2,3,4)
          -f_3 D_{123}(1,2,3,4)
           ),  \\
&&D_{0013}(1,2,3,4) = \frac{1}{2(D-1)}(
          2 m_1^2 D_{13}(1,2,3,4)
          -C_{2}(2,3,4)
          \nonumber \\
        &-&f_1 D_{113}(1,2,3,4)
          -f_2 D_{123}(1,2,3,4)
          -f_3 D_{133}(1,2,3,4)
           ),  \\
&&D_{0022}(1,2,3,4) = \frac{1}{2(D-1)}(
          2 m_1^2 D_{22}(1,2,3,4)
          +C_{11}(2,3,4)
          \nonumber \\
        &-&f_1 D_{122}(1,2,3,4)
          -f_2 D_{222}(1,2,3,4)
          -f_3 D_{223}(1,2,3,4)
          ),  \\
&&D_{0023}(1,2,3,4) = \frac{1}{2(D-1)}(
          2 m_1^2 D_{23}(1,2,3,4)
          +C_{12}(2,3,4)
          \nonumber \\
        &-&f_1 D_{123}(1,2,3,4)
          -f_2 D_{223}(1,2,3,4)
          -f_3 D_{233}(1,2,3,4)
          ),  \\
&&D_{0033}(1,2,3,4) = \frac{1}{2(D-1)}(
          2 m_1^2 D_{33}(1,2,3,4)
          +C_{22}(2,3,4)
          \nonumber \\
        &-&f_1 D_{133}(1,2,3,4)
          -f_2 D_{233}(1,2,3,4)
          -f_3 D_{333}(1,2,3,4)
          ),  \\
&&D_{001}(1,2,3,4) = \frac{1}{2(D-2)}(
          2 m_1^2 D_{1}(1,2,3,4)
          -C_{0}(2,3,4)
          \nonumber \\
        &-&f_1 D_{11}(1,2,3,4)
          -f_2 D_{12}(1,2,3,4)
          -f_3 D_{13}(1,2,3,4)
           ),\\
&&D_{002}(1,2,3,4) = \frac{1}{2(D-2)}(
          2 m_1^2 D_{2}(1,2,3,4)
          +C_{1}(2,3,4)
          \nonumber \\
        &-&f_1 D_{12}(1,2,3,4)
          -f_2 D_{22}(1,2,3,4)
          -f_3 D_{23}(1,2,3,4)
          ),  \\
&&D_{003}(1,2,3,4) = \frac{1}{2(D-2)}(
          2 m_1^2 D_{3}(1,2,3,4)
          +C_{2}(2,3,4)
           \nonumber \\
         &-&f_1 D_{13}(1,2,3,4)
          - f_2 D_{23}(1,2,3,4)
          - f_3 D_{33}(1,2,3,4)
           ),  \\
&&D_{00}(1,2,3,4) = \frac{1}{2(D-3)}(
     2 m_1^2 D_0(1,2,3,4)
   +C_{0}(2,3,4)
\nonumber \\ 
  &-&f_1 D_{1}(1,2,3,4)
  - f_2 D_{2}(1,2,3,4)
  - f_3 D_{3}(1,2,3,4)
    ). \label{D00reduction}
\end{eqnarray}

\section{Rational terms of specific tensor integrals}
\label{app:RT}

As an illustration, we present results of the computation 
of  rational parts of  
some of the tensor integrals. We begin with the rank-two two-point 
function. To calculate  the rational part, we perform the reduction of 
the integral, along the lines described above, and trace  
the part of the reduction that depends on the $(D-4)$-component 
of the loop momentum. Considering the rank-two two-point integral 
\be
I_2^{\mu \nu} = 
\int \frac{{\rm d}^Dl }{(2\pi)^D}
\frac{l^\mu l^\nu}{d_0 d_1},
\ee
where $d_i = (l + q_i)^2 - m_i^2$ 
we find  its rational part 
\be
\label{ra2}
{\cal R}[I_2]_{\mu \nu} = 
i c_\Gamma \left ( 
\frac{\omega_{\mu \nu}}{3} 
+ \frac{g^{-2\ep}_{\mu \nu}}{2\ep}
\right ) F_{01}.
\ee
In \Eq(\ref{ra2}),  $\omega^{\mu \nu}$ is 
the metric tensor of the transverse space 
$\omega_{\mu \nu} (q_1 - q_0)^\mu = 0$ 
and $g^{-2\ep}_{\mu \nu}$ is the metric tensor of the 
$(D-4)$-dimensional space. Also,  we introduced
the short-hand notation  
\be
F_{ij} = \frac{m_i^2+m_j^2}{2}
- \frac{(q_i-q_j)^2}{6},
\ee
and the loop factor $\cg$ reads,
\begin{equation}
\cg = 
\frac{1}{(4\pi)^{2-\ep}}
\frac{\Gamma(1+\ep)\Gamma^2(1-\ep)}{\Gamma(1-2\ep)} = 
\frac{(4 \pi)^\ep }{16 \pi^2} \frac{1}{\Gamma(1-\ep)}+\cO(\ep^3)\ .
\end{equation}
We note that the presence of the $\epsilon$-dependent part of the 
metric tensor in the result is the  reflection of the fact that the 
rank-two two-point function is divergent. This term does not contribute 
when $I_{\mu \nu}$ is contracted with the four-dimensional 
vectors, as it is often the case, but when $I_{\mu \nu}$ is contracted 
with the full $D$-dimensional metric tensor,  it does contribute.
In fact, the two terms shown in the right hand side of 
\Eq(\ref{ra2}) can be identified with the two contributions 
to the rational part ${\cal R}_{1,2}$, discussed at the end of 
Section~\ref{sec_rational}.

For the three-point functions, we need to consider the rank-two and 
the rank-three tensor integrals. For the rank-two three-point 
function  
\be
I_3^{\mu \nu} = 
\int \frac{{\rm d}^Dl }{(2\pi)^D}
\frac{l^\mu l^\nu}{d_0 d_1 d_2},
\ee
the rational part is remarkably simple. It reads 
\be
{\cal R}[I_3]_{\mu \nu} = 
i c_\Gamma 
\left ( 
\frac{\omega_{\mu \nu}(q_1,q_2)}{4} 
+ \frac{g^{-2\ep}_{\mu \nu}}{4\ep}
\right ).
\label{Remarkablysimple}
\ee
However, 
the rational part becomes significantly more complex for the rank-three 
three-point function  
\be
I_3^{\mu \nu \alpha} = 
\int \frac{{\rm d}^Dl }{(2\pi)^D}
\frac{l^\mu l^\nu l^\alpha}{d_0 d_1 d_2}.
\ee
We find 
\beqa
{\cal R}[I_3]^{\mu \nu \alpha} 
&= i c_{\Gamma}\left(
\sum \limits_{i=1}^{2}  v_i^\mu v_i^\nu v_i^\alpha c_{iii} 
+ \sigma_2^\mu \sigma_2^\nu \sigma_2^\alpha c_{000} \right . \nonumber \\
&\left. + \sum \limits_{i=1}^{2} v_i^{\{\alpha} c_{\perp,i}^{\mu \nu \}}
+ \sigma_2^{\{\alpha} c_{\perp,0}^{\mu \nu \}}
+ t_{3,1}^{\{ \mu \nu} v^{\alpha \}}\right), 
\eeqa
and the curly braces $\{ \ldots \}$ 
indicate symmetrization over the three cyclic permutations of the indices.
As usual $v_i^\mu$ are the 
basis vectors of the physical space in van Neerven - Vermaseren basis, 
$\sigma_n = \sum \limits_{i=1}^{n} v_i$ and 
\be
\begin{split}  
& c_{111} = \frac{\Delta_2 F_{02}}{6 q_2^2},\;\;\;c_{222} = \frac{\Delta_2 F_{01}}{6 q_1^2},\;\;\;
c_{000} = -\frac{\Delta_2 F_{12}}{6 (q_2 - q_1)^2},
\\
& c_{\perp,1}^{\alpha \beta} = t_{3,2}^{\alpha \beta} \frac{F_{02}}{2},\;\;\; 
  c_{\perp,2}^{\alpha \beta} = t_{3,2}^{\alpha \beta} \frac{F_{01}}{2},\;\;\; 
c_{\perp,0}^{\alpha \beta} = - t_{3,2}^{\alpha \beta} \frac{F_{12}}{2}.
\end{split}
\ee
The Kronecker delta contracted with the 
two momenta is denoted by 
$\Delta_2 = \Delta(q_1,q_2) = \delta^{q_1 q_2}_{q_1 q_2}$, 
see Section~\ref{sec3} for details. 
Also, we introduced 
\be
t_{3,1}^{\alpha \beta} 
= \frac{\omega^{\alpha \beta}(q_1,q_2)}{4}
+ \frac{g_{-2\ep}^{\mu \nu} }{4 \epsilon},
\;\;\;\;
t_{3,2}^{\alpha \beta} 
= 
\frac{\omega^{\alpha \beta}(q_1,q_2)}{3}
+ \frac{g_{-2\ep}^{\mu \nu} }{2 \epsilon},
\ee
and the auxiliary vector 
\be
v^\mu = \frac{1}{2} 
\sum \limits_{i=1}^{2} v_i^\mu \left ( m_i^2 - m_0^2 - q_i^2 \right ).
\ee

A similar situation occurs with the four-point functions. 
For the rational part of the rank-three four-point function  
\be
I_4^{\mu \nu \alpha} = 
\int \frac{{\rm d}^Dl }{(2\pi)^D} 
\frac{l^\mu l^\nu l^\alpha}{d_0 d_1 d_2 d_3},
\ee
we obtain 
\be
{\cal R}[I_4]^{\mu \nu \alpha} 
= i c_{\Gamma}\left(\sum \limits_{i=1}^{3} v_i^\mu v_i^\nu v_i^\alpha d_{iii} 
+ \sigma_3^\mu \sigma_3^\nu \sigma_3^\alpha d_{000}\right),
\label{r34pt}
\ee
where 
\be
d_{iii} = \frac{\Delta_3}{8Q_{ii}},\;\;\; d_{000} = - \frac{\Delta_3}{8 Q}.
\ee
Here, we use the following general notation. Consider an $n+1$-point function  
with  momenta $q_i$. The Gram-determinant matrix 
is given by $G_{ij} = (q_i \cdot q_j)$. We   denote the elements 
of the inverse Gram matrix by 
\be
G^{-1}_{ij} = \frac{Q_{ij}}{\Delta_n},
\ee
where $\Delta_n = \delta^{q_1...q_n}_{q_1...q_n}$.
With this notation, $Q_{ij}$ equals to   $(-1)^{i+j}$ times the determinant 
of a minor of the Gram matrix obtained by removing the $i$-th row and 
the $j$-th column. These minors can be obtained from the scalar 
product of the van Neerven-Vermaseren basis vectors 
\be
Q_{ij} = \Delta_n \; v_i v_j .
\ee 
We also employ the short-hand notation 
\be
Q_i = \sum \limits_{j=1}^{n} Q_{ij},\;\;\; Q = \sum \limits_{i=1}^{n} Q_i.
\ee

We note that, while the rational part of the 
rank-three four-point function and 
\Eq(\ref{r34pt})
is remarkably compact,  a similar 
formula for the rank-four four-point is much more involved 
and we do not present it here.

\section{Cutkosky rules}
\label{app:Cut}
Scattering  amplitudes in quantum field theories are functions of 
scalar products of four-momenta of 
scattered particles. It is possible 
to complexify these scalar products  and study how scattering 
amplitudes depend on them. 
In particular, a $2 \to 2$  scattering process 
of particles of mass $m$
can be fully described 
by two independent Mandelstam variables $s$ and 
$t = -(s-4m^2)/2\; (1-\cos \theta_{\rm cms})$, where $\theta_{\rm cms}$ is the 
scattering angle in the center-of-mass frame. 
It is useful to consider the scattering amplitude as a function of 
$s$, keeping $\theta_{\rm cms}$ fixed. It follows from the   
unitarity of the scattering matrix $S$ that  
the scattering amplitude is an analytic
function in the complex $s$-plane with possible single-particle poles
 in non-physical regions
and branch cuts corresponding to multi-particle thresholds.
Indeed, since 
$S^\dagger S=1,\;\;S=1+iT$,
\be
2 \, {\rm Im}\;T=-i(T-T^{\dagger})=T^\dagger T=T T^\dagger.
\label{eq_un}
\ee
Taking the  matrix element of both sides of the above equation 
between two-particle states and using the completeness relation, we 
find
\be
2\, {\rm
  Im}\, \bra{p_3,p_4}T\aket{p_1,p_2}=
\sum_n{\bra{n}T |{p_3,p_4}\rangle}^*\bra{n}T\aket{p_1,p_2}.
\label{eq_uni}
\ee
The scattering amplitude $M(s,t)$ is defined by factoring the energy-momentum 
conserving $\delta$-function from the matrix element of the $T$-matrix,
$\bra{p_3,p_4}T\aket{p_1,p_2}=(2\pi)^4\delta(p_1+p_2-p_3-p_4)M(s,t)$.
As the scattering energy increases, the imaginary part of the
scattering amplitude receives contributions from intermediate 
states of higher multiplicities.  As a result $M(s,t)$
has branch cuts in the complex $s$-plane at $(n m)^2$, with 
$n = (2,3, 4,...)$. The discontinuity of the amplitude 
$M(s,t)$ 
at these 
cuts is given by \Eq(\ref{eq_uni}).  The analysis of
the analytic properties of multi-particle amplitudes based on
unitarity and crossing symmetry becomes more and more cumbersome
with increasing number of  external particles.  In perturbation
theory, however, the Cutkosky formula \cite{Cutkosky:1960sp}
$1/(p^2 - m^2 + i \delta) \to (-2\pi i) \delta(p^2 - m^2) \theta(p_0)$ 
provides us with a simple recipe 
for calculating discontinuities across branch cuts
of multi-loop, multi-leg  Feynman diagrams.  In this Appendix, 
we discuss the origin of the Cutkosky rule using the 
example of a two-point function. Our presentation follows 
Ref.~\cite{QFTLL} where further details can be found.  

\begin{figure}[t]
\begin{center}
\includegraphics[angle=0,scale=0.4]{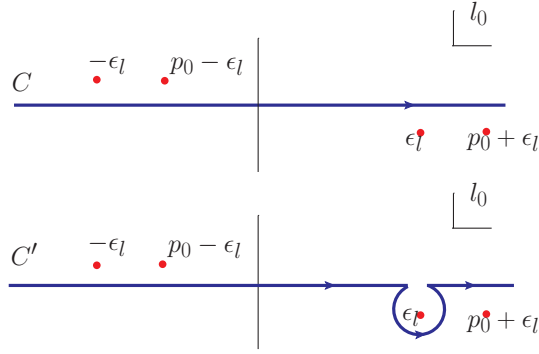}
\caption{Integration contours in the $l_0$ plane. See text for details.}
\label{figC31}
\end{center}
\end{figure}

Consider a diagram that contributes to 
a correlation function of two scalar fields in $\varphi^3$-theory
\be
\label{twopoint}
\begin{split} 
& {\cal I}^{(2)}(p^2,m^2)=\int \frac{{\rm d}^4l}{(2\pi)^4} 
\frac{1}{D_1D_2},
\\
&   D_1=l^2-m^2+ i0 ,\;\;\;  D_2=(p-l)^2-m^2+ i 0\,.
\end{split} 
\ee
This diagram is an analytic function of $p^2$, with a cut along the positive 
real axis, starting at $p^2 = 4 m^2$.  To understand the origin of 
the discontinuity across the cut, we start  computing  the integral 
in \Eq(\ref{twopoint}) by integrating over $l_0$.  We do so by using the 
residue theorem.  We work in a reference frame where $p = (p_0, \vec 0)$ 
and $p_0 > 0$.  There are four poles that we need to consider: 
\be
\label{eq_poles}
\begin{split} 
& D_1:~~~~~a_1)~l_0 = \epsilon_{l} - i0,~~~~~~~~~~~~~~~~
b_1)~l_0 = -\epsilon_{l} +i0,
\\ 
& D_2:~~~~~a_2)~l_0 = p_0 + \epsilon_{\l} - i0,~~~~~
~~~~b_2)~l_0 =  p_0 - \epsilon_{l} + i0,
\end{split} 
\ee
where $\epsilon_{l} = \sqrt{\vec l^2 + m^2}$. From \Eq(\ref{eq_poles}), 
it follows that two $l_0$-poles are located 
above and two $l_0$-poles are located  below 
the real axis, see the upper inset in Fig.~\ref{figC31}.
 We can also compute the distance between the poles in 
\Eq(\ref{eq_poles}). We find that the distance between the poles $b_2$ and 
$a_1$ can vanish, while distances between all other poles are larger  than zero. 
This observation has important consequences. Suppose we start with a $p_0$ value 
such that $ 0 < p_0 < 2 m$. In that case poles $b_{1,2}$ are above the negative real 
axis and poles $a_{1,2}$ are below the positive real axis. It is clear that 
the integration contour can be deformed to complex $l_0$-values such that 
${\rm Re}(l_0) < 0, {\rm Im}(l_0) < 0$ and ${\rm Re}(l_0) > 0, {\rm Im}(l_0) > 0$. 
Once the integration contour is deformed, it becomes clear 
that the integral can not develop 
a discontinuity in $p^2$.  Indeed, the discontinuity is the result of 
differing values of the integral when computed with $p_{0} \to p_{0} \pm i \delta$, 
$\delta \to 0$. The location of the poles changes when $p_0$ is 
substituted by $p_0 \pm i \delta$. 
However, because the integration contour is far away from the poles, 
the result of the integration is not affected. 

This  argument fails if the integration contour {\it cannot} 
be moved  away from the poles. In our example,  this happens if the integration 
contour is pinched between the two poles, so that they are on top of each other. 
This can only happen for the two poles  $b_2$ and $a_1$ in \Eq(\ref{eq_poles}).
 The distance between the poles $b_2$ and $a_1$ vanishes if  
$p_0 = 2 \epsilon_{l}$; this  can only happen if $p_0 \ge 2 m$. The 
Lorentz-invariant  form of this condition 
is $p^2 \ge 4m^2$, which is the point in the complex $p^2$-plane
where the two-particle 
cut starts. 

To compute the discontinuity, we go back to \Eq(\ref{twopoint}) 
and write it as an integral over the contour $C$ shown in 
the upper inset in Fig.~\ref{figC31} 
\be
\label{twopointI_1}
{\cal I}^{(2)}(p^2,m^2)=\int \frac{{\rm d}^3 \vec l}{(2\pi)^3}
\int \limits_{C}^{} \frac{{\rm d} l_0 }{(2\pi) } 
\frac{1}{(l_0^2 - \epsilon_{l}^2)( (l_0 - p_0)^2 - \epsilon_l^2) }.
\ee
We can write this integral as the sum of two terms: the integral over  
a contour $C'$ (see the lower inset in Fig.~\ref{figC31} )
 where pinching does not occur and therefore
no discontinuity is present,
and the residue at the pole 
$a_1$
\be
\label{twopointI_2}
{\cal I}^{(2)}=\int \frac{{\rm d}^3 \vec l}{(2\pi)^3}
\left [ 
\int \limits_{C'}^{} \frac{{\rm d} l_0}{(2\pi)}... 
+  \oint \limits_{\gamma_{a_1}}^{} \frac{{\rm d} l_0}{(2\pi)}...  \right ].
\ee
The integral over $\gamma_{a_1} $ can be written as 
\be
\oint \limits_{\gamma_{a_1}}^{} \frac{{\rm d} l_0 }{(2 \pi)}
\frac{1}{(l_0^2 - \epsilon_{l}^2)( (l_0 - p_0)^2 - \epsilon_l^2) }
= \int  \frac{{\rm d} l_0 }{(2 \pi)} 
\frac{(-2\pi i) \delta_+(l^2 - m^2)}{( (l_0 - p_0)^2 - \epsilon_l^2) },
\label{eq_c5}
\ee
where $\delta_+(l^2 - m^2) = \delta(l^2 - m^2) \theta(l_0)$.
The discontinuity of the integral in \Eq(\ref{eq_c5}) can be computed 
directly. For our choice of the reference frame, $p^2 \pm i \delta \equiv 
p_0^2 \pm i \delta$. Therefore, 
\be
\begin{split} 
 {\rm Disc} \left [ {\cal I}_2 \right ] &  = 
 {\cal I}_2 (p^2 + i\delta ) - {\cal I}_2( p^2 - i \delta)  
\\
&
=\int \frac{{\rm d}^4 l }{(2\pi)^4 }
(-2 \pi i) \delta_+(l^2 - m^2)
\times \left [ 
\frac{1}{(l_0 - p_0 - i\delta)^2 - \epsilon_l^2  }
- {\rm c.c}
\right ]
\\
& 
= \int \frac{{\rm d}^4 l }{(2\pi)^4 }
(-2 \pi i) \delta_+(l^2 - m^2)
\left [ 
\frac{1}{(l-p)^2 -m^2 + i \delta  }
- {\rm c.c.}
\right ].
\end{split} 
\ee
As the final step, we use 
\be
\frac{1}{x+i \delta} = {\rm P} \left [ \frac{1}{x} \right ]- i \pi \delta (x).
\ee
to simplify the difference of the two terms in square brackets and find 
\be 
 {\rm Disc} \left [{\cal I}_2 \right ]  
= \int \frac{{\rm d}^4 l }{(2\pi)^4 }
(-2 \pi i) \delta_+(l^2 - m^2)\;
(-2\pi i) \delta_+((l-p)^2 - m^2).
\ee
Hence, the discontinuity of the integral ${\cal I}_2$ is obtained by replacing 
both propagators with $\delta$-functions, in agreement with the Cutkosky rule.

We now discuss how this result generalizes. We begin with the formulation 
of a more thorough condition for the appearance of the discontinuities 
in one-loop Feynman diagrams.  As follows from the above discussion, 
discontinuities appear when the integration contour becomes trapped between 
poles of the integrand of a given Feynman diagram.   To streamline a discussion 
of when this happens, we go back to our example and combine 
denominators using Feynman parameters 
\be
{\cal I}^{(2)}(p^2,m^2)=\int \frac{{\rm d}^4l}{(2\pi)^4} 
\frac{1}{D_1D_2} 
= 
\int \frac{{\rm d}^4l}{(2\pi)^4}  \int {\rm d} \alpha_1 {\rm d}\alpha_2 
\frac{\delta (1 - \sum \alpha_i)}{D^2},
\ee
where $D = \alpha_1 D_1 + \alpha_2 D_2$.  Singularities in the above expression appear, if 
$D = 0$.  The {\it necessary condition} for that is $D_1(l^*) = 0, D_2(l^*) = 0$.
Since $D$ is a quadratic polynomial 
 in the loop momentum $l$, equation $D(l) = 0$ contains two solutions
which, in general, are  different. In such a situation, 
the integration contour can be  deformed, the pinch singularity can be avoided and no discontinuity 
appears.  This does not occur if the two solutions 
of the quadratic equation  coincide. Suppose we write 
\be
D(l) = (l_0 - a(\vec l,p,m))(l_0 - b(\vec l,p,m)),
\ee
where $a$ and $b$ are 
 functions of the three-momentum $\vec l$, external momentum $p$ 
and the mass $m$.  The condition that the poles 
coincide  is 
\be
\left. \frac{\partial D}{\partial l_0 } \right |_{l_0 = a~{\rm or}~b} = 0.
\ee
The covariant generalization of this equation is straightforward
\be
\left. \frac{\partial D}{\partial l^\mu } \right |_{l = l^*} = 0.
\ee
Because $D(l) = \alpha_1 D_1 + \alpha_2 D_2$ and $D_{1,2} = (l+q_{1,2})^2 - m^2$, 
we find 
\be
\left. \frac{\partial D}{\partial l_\mu } \right |_{l = l^*} = 0 \Rightarrow 
\alpha_1 l_1^* + \alpha_2 l_2^* = 0, 
\label{eq_c_15}
\ee
where $l_{1,2}^* = l^* + q_{1,2}$ and $l^*$ is such that $D_1(l^*) 
= D_2(l^*) = 0$. 
If \Eq(\ref{eq_c_15})  has solutions for real values 
of the loop momentum {\it and} for physical values of Feynman parameters 
$0 \le \alpha_{1,2} \le 1$,  there is a pinch 
singularity and, correspondingly,  
a discontinuity in a given Feynman diagram.

The generalization of the 
above discussion leads to Landau equations 
\cite{Landau:1959fi} 
(see Ref.~\cite{eden:1966ev} for a  pedagogical introduction).  
To present these equations, we consider an $L$-loop 
Feynman integral with $N$ propagators $D_i=q^2_i-m_i^2, i \in N$
\be
\begin{split}
\label{feynintmulti}
{\cal I}(\{p_i\},\{m_i\})=&\int 
\prod \limits_{k=1}^{L}
\frac{d^4l_k}{(2\pi)^4} \;
\frac{{\cal N}(\{l_i\}\,\{p_i\})}{\prod \limits_{i=1}^ND_i}\,. 
\end{split}
\ee
The set of equations that determines the positions of the pinch 
singularities  of the integral~(\ref{feynintmulti}) that 
lead to a discontinuity across an $r$-particle cut is 
\ba
&D_i=q_i^2-m_i^2=0,\quad (i=1,2,...,r \le 4 L), 
\label{Landau1} \\
&\sum_{i \subset {\rm loop\ j}} \alpha_i^{(j)}q_i=0\,
\ \  {\rm for\  every\  loop}.
\label{Landau2}
\ea
The first set of Landau equations~\Eq(\ref{Landau1}) 
selects a finite number of vanishing propagators, i.e. defines a cut.
This is a necessary condition for a Feynman integral to develop a singularity and 
to have a discontinuity in a particular channel.  As we saw in the explicit 
example, the second set of equations~(\ref{Landau2})  
is a condition that the singularities 
trap the integration contour. 

The Landau equations provide us with sets of homogeneous linear equations
for the coefficients $\alpha_i^{(j)}$.  The condition to have
non-trivial solutions requires that the corresponding determinants
vanish.  The equations given by the vanishing determinants allow us to
calculate the positions of the branch point 
singularities associated with a particular channel.

To illustrate how the Landau equations work, we go back to our example -- the integral ${\cal I}_2$. 
We slightly generalize 
it by considering two different masses in the inverse Feynman 
propagators $D_1$ and $D_2$.  The position of the singularity for the 
double cut is given by the  Landau equations  
\be
 l^2-m_1^2=0\,,\quad (l-p)^2-m_2^2=0\,,\quad
(\alpha_1+\alpha_2)l^{\mu}-\alpha_2 p^{\mu}=0. 
\ee
Contracting the last equation with $l$ and $p$, we obtain a linear system of equations
for $\alpha_{1,2}$ 
\be
\begin{split} 
\label{linearseteq}
(\alpha_1+\alpha_2)m_1^2-\alpha_2 (l \cdot p)&=0,\\
(\alpha_1+\alpha_2)(l \cdot p)-\alpha_2 p^2&=0.
\end{split}
\ee
These equations   have 
non-trivial solutions for $\alpha_1,\alpha_2$ provided that 
the corresponding determinant vanishes. Hence, 
\be
m_1^2 p^2-(l \cdot p)^2=0\,,\quad {\rm and }\quad l \cdot p=\frac{1}{2}(m_1^2+p^2-m_2^2).
\ee
Solving this equation for $p^2$ we obtain  
\be 
p^2_{\pm}=(m_1\pm m_2)^2.
\ee
In spite of the two solutions, the physical two-particle threshold is at $p^2=p^2_{+}$
because we need to satisfy additional constraints $l^{0} > 0$ and $p^0-l^0 > 0$. This
leaves only one solution for the branch cut point  $p^2 = p_+^2=(m_1+m_2)^2$.

Having discussed the examples of how Cutkosky rules are used, the reasons 
discontinuities appear and 
equations that can be used to find locations of branch cuts, we are
now in a position 
to write down a general formula for  the discontinuity  of a 
Feynman diagram in a particular 
$r$-particle channel. We note that a complete  proof of Cutkosky 
rules is non-trivial; details can be found in 
\cite{Cutkosky:1960sp,Nishijima:1962}.

To be specific, we imagine that in a general Feynman 
diagram \Eq(\ref{feynintmulti}), 
a discontinuity appears if the first $r$ propagators go on the mass-shell. 
The discontinuity across this cut can be  computed 
using Cutkosky rules \cite{Cutkosky:1960sp} 
\be
{\rm Disc}\left [ {\cal I} \right ] 
=\int \prod \limits_{k=1}^{L} \frac{d^4l_k}{(2\pi)^4} \;
  \frac{ {\cal N}(\{l_i\}\,\{p_i\}) \prod \limits_{i=1}^{r} (-2\pi i) \delta^{(+)}(q_i^2-m_i^2)}
{\prod \limits_{j=r+1}^{N} (q_j^2 - m_j^2)}.
\label{full_form}
\ee
The above equation is the main result of this Appendix.  We hope to have 
sufficiently motivated it with the above discussion. 
We finish this Appendix 
with a few remarks.
\begin{itemize}
\item[a)]
As shown in \Eq(\ref{eq_un}), unitarity implies 
non-linear 
relations between scattering amplitudes. Those relations 
can be used to compute the discontinuities of scattering  amplitudes
at a given order in the perturbative expansion, in terms of  amplitudes 
at lower
orders.
\item[b)]
Unitarity is built into perturbation theory in an even more detailed
manner. 
It is not necessary for the external legs to  have on-shell values. 
If the Landau equations have
solutions, the Cutkosky formula provides the correct singularities also
in the case  of off-shell legs. Only the cut lines need be on-shell.
\item[c)]
The diagrammatic version of the Cutkosky formula \Eq(\ref{full_form}) 
can be applied to scattering amplitudes, leading to {\it generalized unitarity relations}. 
Consider the triangle singularity.
It  appears in many diagrams. Specializing to 
one loop and summing
up all diagrams with such a discontinuity, we obtain 
\be
\begin{split} 
&{\rm Disc}_{\l\to l_{n_1n_2n_3}}\; \left [ A_n^{\rm 1-loop} \right ] = 
\sum_{\rm states}\int \frac{{\rm d}^4 l}{(2\pi)^4} 
\prod \limits_{i=1}^{3} (-2\pi i) \delta_+(l_i^2- m_i^2) \\
&~~~~~~~~~ \times 
A^{\rm tree}_{n_1}(-l_3,l_1)
A^{\rm tree}_{n_2}(-l_1,l_2)A^{\rm tree}_{n_3}(-l_2,l_3),
\end{split}
\ee
where $A_n^{\rm 1-loop}$ and $A^{\rm tree}$ are one-loop and tree scattering 
on-shell scattering amplitudes. 

\item[d)]
In $D$-dimensions the maximal possible number of  cut propagators 
of an $L$-loop Feynman diagram 
is
$D \times L$ since in that case
all components of the loop momenta  are fixed by on-shell constraints. 
Obviously, when the maximal cut is evaluated, no integration over the loop 
momentum is needed since 
the integral reduces to a sum over discrete solutions.
For one-loop computations in $D$-dimensions, with $D > 4$, at most 
five cuts are required, since the extra components of the 
loop momenta decouple from the four-dimensional kinematics. 
\end{itemize}

\section{Spinor Helicity methods}
\label{App:SH}
The main thrust of this review has been the description of methods
that are applicable to one-loop amplitudes containing both massless 
and massive particles. However QCD contains massless gluons and, at energies
much larger than their masses, quarks can also be considered massless. 
Considerable simplification can be achieved in massless amplitudes by
the use of spinor helicity methods~\cite{Berends:1981rb,Xu:1986xb}. 
These methods, as they apply to tree diagrams, have been extensively reviewed 
in refs.~\cite{Mangano:1990by,Dixon:1996wi,Peskin:2011in}.
In the main text, Section \ref{Sec:Oneloopexample},
spinor-helicity methods are used in the calculation of a simple one-loop diagram.
The purpose of this appendix is to give a short review of these methods 
to elucidate the examples given in the main text. 
Reviews of the application of spinor-helicity methods to loop diagrams can also 
be found in refs.~\cite{Bern:2007dw,Berger:2009zb,Britto:2010xq}.

\subsection{Spinor solutions for massless fermions}
We choose to work in the Weyl representation for the gamma matrices, 
\Eq(\ref{eq:gammamatrices}). In this representation upper and 
lower components have different helicities
\beq \label{eq:gammaCHIRAL}
\gamma_R\ 
\ =\frac{1}{2} (1+\gamma_5) =
\ \left(\begin{matrix}{\bf 1}&{\bf 0}\cr{\bf 0}&{\bf
      0}\cr \end{matrix}\right)\ ,\quad 
\gamma_L\ 
\ =\frac{1}{2} (1-\gamma_5) =
\ =\ \left(\begin{matrix}{\bf 0}&{\bf 0}\cr{\bf 0}& {\bf 1}\cr \end{matrix}\right)\ .
\eeq

The massless spinor solutions of the Dirac equations are 
(c.f.\ \Eq(\ref{eq:Diracsolutions}))
\beq \label{eq:explicitspinor}
u_{+}(p) =
  \left[ \begin{matrix} \sqrt{p^+} \cr 
   \sqrt{p^-} e^{i\varphi_p} \cr
                  0 \cr 
   0 \cr  \end{matrix}\right] , \hskip3mm
u_{-}(p) =
  \left[ \begin{matrix} 0 \cr 
                  0 \cr
                \sqrt{p^-} e^{-i\varphi_p} \cr 
                                 -\sqrt{p^+} \cr  \end{matrix}\right] , 
\eeq
where 
\beq \label{eq:phasekdef}
e^{\pm i\varphi_p}\ \equiv\ 
  \frac{ p^x \pm ip^y }{ \sqrt{(p^x)^2+(p^y)^2} }
\ =\  \frac{ p^x \pm ip^y }{ \sqrt{p^+p^-} }\ ,
\qquad p^\pm\ =\ E \pm p^z.  
\eeq
In this representation the Dirac conjugate spinors are
\be
\begin{split}
\label{eq:explicitspinorconjg}
& \overline{u_+}(p) \equiv  u_+^\dagger(p) \gamma^0 =
  \left[ 0, 0, \begin{matrix} \sqrt{p^+} , 
   \sqrt{p^-} e^{-i\varphi_p}  \cr  \end{matrix}\right], 
\\
& \overline{u_-}(p) = 
  \left[ \sqrt{p^-} e^{i\varphi_p}, -\sqrt{p^+}, 0,0 \right]\,,  
\end{split}
\ee
and the spinors are normalized such that $u^\dagger_{\pm} u_\pm = 2 E$. 
In the massless limit the antiparticle spinors obey the same Dirac equation
and we may choose the phase such that particle spinors $u(p)$ and
antiparticle spinors $v(p)$ satisfy 
\beq
u_{\pm}(p)= v_{\mp}(p)\,.
\eeq

We now introduce a bra and ket notation for spinors corresponding to
(massless) momenta $p_i$, $i=1,2,\ldots,n$ labeled by the index $i$
\be
\begin{split}
\label{eq:spinorshort}
& |i+\rangle\ \equiv\ |i\rangle\ \equiv\ |p_i+\rangle \ 
\equiv\   u_{+}(p_i)\ =\ v_{-}(p_i),  \\
& |i-\rangle\ \equiv\ |i]   \    \equiv |p_i -\rangle \
\equiv\  u_{-}(p_i)\ =\ v_{+}(p_i),  \\
& \langle i+| \equiv\ [i|    \    \equiv   \langle p_i+|
\equiv\  \overline{u_+}(p_i) \ =\ \overline{v_-}(p_i),  \\
& \langle i-| \equiv \ \langle i |\ \equiv \langle p_i-| 
\equiv\  \overline{u_-}(p_i) \ =\ \overline{v_+}(p_i).
\end{split}
\ee
We further define the basic spinor products by
{ \begin{eqnarray} \label{eq:basicspinordef}
&&\langle i j \rangle \ \equiv\ \langle i-|j+\rangle
  \ =\ \overline{u_-}(p_i) u_+(p_j), \\
&&\;\; [ij]\ \equiv\ \langle i+|j-\rangle
  \ =\ \overline{u_+}(p_i) u_-(p_j). 
\end{eqnarray} }
The helicity projection implies that products like 
$\langle i+|j+ \rangle$ vanish
\begin{equation}
\langle i+| j+ \rangle = [i j \rangle = \langle i-|j- \rangle = \langle i j ] = 0.
\end{equation}
It is also straightforward to verify that the spinor products satisfy
the following relations 
\begin{equation}
\langle ij \rangle = -\langle ji \rangle \qquad 
[ ij ] = -[ ji ]\,,
\end{equation}
which imply
\be
\langle ii \rangle = [ii] = 0\,. 
\ee

We will use the first two notations in \Eq(\ref{eq:spinorshort})
interchangeably. Thus we may write
\be
\begin{split} 
& \langle i - | \gamma^\mu | j- \rangle \equiv  \langle i | \gamma^\mu | j]\,,   \\
& \langle i + | \gamma^\mu | j+ \rangle \equiv  [ i | \gamma^\mu | j \rangle\,.  
\end{split} 
\ee

\subsection{Spinor products}
For the case where both energies are positive, $p_i^0 >0,\, p_j^0
> 0$, we can write the spinor products explicitly as 
\be
\label{eqma}
\begin{split} 
& \spa{i}.{j} = \sqrt{p_i^- p_j^+} e^{i\varphi_{p_i}}
              - \sqrt{p_i^+ p_j^-} e^{i\varphi_{p_j}}
\ =\ \sqrt{|s_{ij}|} e^{i\phi_{ij}}, 
\\
& \spb{i}.{j} = \sqrt{p_i^+ p_j^-} e^{-i\varphi_{p_j}}
 -\sqrt{p_i^- p_j^+} e^{-i\varphi_{p_i}}
\ =\ -\sqrt{|s_{ij}|} e^{-i\phi_{ij}}\,,
\end{split}
\ee
where  $s_{ij}\ =\ (p_i+p_j)^2\ =\ 2 p_i\cdot p_j$, and  
\begin{equation} \label{eq:phiijdef}
\cos\phi_{ij}\ =\ { p_i^x p_j^+ - p_j^x p_i^+ 
             \over \sqrt{|s_{ij}| p_i^+ p_j^+} }
\ , \qquad
\sin\phi_{ij}\ =\ { p_i^y p_j^+ - p_j^y p_i^+ 
             \over \sqrt{|s_{ij}| p_i^+ p_j^+} }
\ .
\end{equation}
Thus, the spinor products are, up to a phase, square roots of Lorentz
products.  {\it For real momenta} with positive energy components 
we have that $\spa i.j^* = \spb j. i$.  
Note, however, that for complex momenta this is no longer true.

By explicit construction one can show that 
\begin{equation}
 \label{schoutenderive}
| b+\rangle \langle c-| \>-\> | c+\rangle \langle b-| = \langle c- | b+ \rangle \gamma_R \,.
\end{equation}
Contracting this equation with $\langle a-|$ from the left and with $|d+\rangle $ from the 
right,  we obtain the Schouten identity
\begin{equation}
\langle a-| b+\rangle \langle c-|d+\rangle - \langle a-| c+\rangle \langle b-|d+\rangle = 
\langle c- | b+ \rangle \langle a- | d+ \rangle\ \;.
\end{equation}
We can write it more compactly as 
\be \label{schouten}
\begin{split} 
& \spa a.b  \spa c.d  - \spa a. c \spa b.d-\spa c.b  \spa a.d =0 \,, \\
& \spb a.b  \spb c.d  - \spb a. c \spb b.d-\spb c.b  \spb a.d =0 \,.
\end{split} 
\ee
Note that the expressions on the left of \Eq(\ref{schouten}) are totally 
antisymmetric under the exchange of $a,b,c$.
The Schouten identity follows because the totally antisymmetric product of three 
 two-component objects is equal to zero.

The importance of spinor products  for describing on-shell scattering amplitudes in gauge 
theories is related to their  natural connection to square roots of four-momenta 
scalar products $s_{ij}$, 
cf. \Eq(\ref{eqma}). Since gauge scattering amplitudes have square-root singularities in $s_{ij}$, 
they become simple functions when written in terms of spinor products. 


\begin{table}[t]
\begin{eqnarray}
&& \langle pq \rangle = \langle p-| q+ \rangle ,\;\; [pq] =  \langle p+|q- \rangle \nn \\
&&\langle p \pm| \gamma_\mu | p \pm \rangle = 2 p_\mu \nn \\
&&\langle p+| q+ \rangle = \langle p-|q- \rangle = \langle pp \rangle = [pp] = 0\\ 
&&\langle pq \rangle = -\langle qp \rangle, \; \;
[ pq ] = -[ qp ] \nn \\
&& 2 | p \pm \rangle \langle q \pm | =
{\scriptstyle \frac{1}{2}} (1 \pm \gamma_5) \gamma^\mu 
\langle q\pm| \gamma_\mu | p\pm\rangle \nn \\
&&\langle  p q \rangle^* = - {\rm sign}(p\cdot q) [pq] = 
{\rm sign}(p\cdot q) [qp] \nn \\
&&|\langle  p q \rangle|^2 = \langle  p q \rangle \langle  p q \rangle^* = 
2 | p\cdot q | \equiv |s_{pq}| \nn \\
&&\langle  p q \rangle [q p]  = 2  p\cdot q  \equiv s_{pq} \nn \\
&&\langle p\pm| \gamma_{\mu_1} \ldots \gamma_{\mu_{2n+1}} | q\pm\rangle =
\langle q\mp|\gamma_{\mu_{2n+1}}  \ldots \gamma_{\mu_1}| p\mp\rangle \nn \\
&&\langle p\pm| \gamma_{\mu_1} \ldots \gamma_{\mu_{2n}} | q\mp\rangle =
- \langle q\pm|\gamma_{\mu_{2n}}  \ldots \gamma_{\mu_1}| p\mp\rangle \nn \\
&&\langle  a+ | \gamma_\mu | b+ \rangle \langle  c- | \gamma^\mu | d- \rangle =
 2  [ ad ]  \langle  cb \rangle,\;\;\; ({\rm Fierz}) \nn \\
&&
\langle  a\pm | \gamma^\mu | b\pm \rangle \;  \gamma_\mu = 
2 \; \bigg[ \; | a \mp \rangle \langle b \mp |
       +| b \pm \rangle \langle a \pm | \; \bigg],\;\;\; 
({\rm Fierz}+{\rm Charge~conjugation}) \nn \\
&&\langle  ab \rangle \langle  cd \rangle =
  \langle  ad \rangle \langle  cb \rangle 
+ \langle  ac \rangle \langle  bd \rangle.\;\;\; ({\rm Schouten}) \nn 
\end{eqnarray}
\caption{A summary of relations valid for massless spinors.}
\label{Relations}
\end{table}

We will now explain how spinors are used to simplify the description of 
massless gauge vector bosons. 
We  consider 
a gluon with momentum $k$ and gauge vector $b$.  Its  polarization
vector can be written as 
\begin{equation} \label{gluonpolarization}
\varepsilon_\mu^\pm(k,b) = \pm
\frac{ \langle k \pm | \gamma_\mu | b\pm \rangle}
{\sqrt{2} \langle  b \mp | k \pm\rangle }.
  \end{equation}
Using Fierz and charge-conjugation equations, see Table~\ref{Relations},
we find 
\be
\label{gamepplusminus}
\begin{split} 
& \gamma^\mu \varepsilon_\mu^{+}(k,b) = 
\frac {\sqrt{2} \Big[ 
|  k - \rangle \langle   b-| \>+\> | b + \rangle \langle   k+| \Big]}
{\langle  b  k  \rangle }, \\
&
\gamma^\mu \varepsilon_\mu^{-}(k,b) = 
\frac{\sqrt{2} 
\Big[  | k + \rangle \langle   b+ |
\> +\> |  b- \rangle \langle   k- | \Big]}
{ [ k  b  ] }.
\end{split}
\ee 
Different choices of the vector $b$ correspond to different gauge choices; the corresponding 
polarization vectors differ by an amount  proportional to the gluon momentum.
Specifically, 
\begin{eqnarray}
&&\varepsilon_\mu^{+}(k,b) - \varepsilon_\mu^{+}(k,c) = 
\frac{ \langle k + | \gamma_\mu | b+ \rangle}
{\sqrt{2} \langle  b  k  \rangle }
-\frac{ \langle k + | \gamma_\mu | c+ \rangle}
{\sqrt{2} \langle  c  k  \rangle } \nn \\
&=&
\frac{1}
{\sqrt{2} \langle  b  k  \rangle \langle  c  k  \rangle } 
\Big[ 
  \langle k + | \gamma_\mu | b+ \rangle  \langle c k \rangle
 -\langle k + | \gamma_\mu | c+ \rangle  \langle b k \rangle 
 \Big] 
\nn \\
&=&
\frac{1}
{\sqrt{2} \langle  b  k  \rangle \langle  c  k  \rangle }
\Big[ \langle k + | \gamma_\mu | k+ \rangle  \langle c b \rangle \Big]
=
\frac{\sqrt{2} \langle c b \rangle}
{\langle  b  k  \rangle \langle  c  k  \rangle } k_\mu
\end{eqnarray}
where we have used \Eq(\ref{schoutenderive}).
\subsubsection{Three point vertices}
\label{App:qqg}
We will now calculate the amplitudes for the scattering of three massless partons; these will be important
building blocks for the calculation of loop diagrams using spinor helicity techniques.
By direct insertion of the color-ordered Feynman rules, Fig.~\ref{fig6.1}, we have
\beqn \label{qbqgmpp}
-i \ m_3({\qb}_1^-,q_2^+,g_3^+)&=&-\frac{\spb 2.3 \spa b. 1}{\spa b.3}
                      = -\frac{{\spb 2.3}^2}{\spb 1.2}, \nn \\
-i \ m_3({\qb}_1^-,q_2^+,g_3^-)&=&-\frac{\spb 2.b \spa 3. 1}{\spb 3.b}
                      =     - \frac{{\spa 3.1}^2}{\spa 1.2}. 
\eeqn
The last steps follow from the momentum conservation which allows us 
to write  
\beq \label{Momcon}
\spa{b}.{1} \spb1.2 + \spa{b}.{3} \spb3.2 = 0
\eeq
and obtain $ \spa{b}.{1}/\spa{b}.{3} = \spb2.3/\spb1.2$. 
Using \Eq(\ref{Momcon}) and similar equations
allows to remove all the dependence on the auxiliary vector $b$ in \Eq(\ref{qbqgmpp}).

In a similar fashion for the $ggg$ amplitudes we have that, (choosing the gauge vector $b$ the same for all 
polarizations),
\beqn \label{MHV1}
- i \ m_3(g_1^-,g_2^+,g_3^+)&=&
\sqrt{2} \Big[\varepsilon_1^- \cdot \varepsilon_2^+ \  \varepsilon_3^+ \cdot p_1-\varepsilon_3^+ \cdot \varepsilon_1^-  \ \varepsilon_2^+ \cdot p_1\Big]
\nn \\
&=&-\frac{\spa1.b^2 \spb2.3 }{\spa b.2 \spa b.3} 
=-\frac{\spb2.3^3 }{\spb1.2 \spb 3.1 }, 
\eeqn
\beqn \label{MHV2}
-i \ m_3(g_1^-,g_2^+,g_3^-)&=&
\sqrt{2} \Big[\varepsilon_3^- \cdot \varepsilon_2^+  \ \varepsilon_1^- \cdot p_2-\varepsilon_1^- \cdot \varepsilon_2^+ \  \varepsilon_3^- \cdot p_1\Big]
\nn \\
&=&\frac{\spb2.b^2 }{\spb 1.b \spb b.2 \spb 3.b } 
\Big[ \spa 1.2 \spa 3.b -\spa 1.b \spa 3.2 \Big]
\nn \\
&=&-\frac{\spb2.b^2 \spa 1.3 }{\spb 1.b \spb 3.b }
= \frac{ \spa3.1^3 }{\spa1.2 \spa 2.3 }.
\eeqn
In deriving these results we have used the expressions for the polarization vectors in terms of
spinors, \Eqs(\ref{gluonpolarization}-\ref{gamepplusminus}), the Schouten identity, \Eq(\ref{schouten}), 
and momentum conservation.
Note that \Eqs(\ref{MHV1},\ref{MHV2}) have the characteristic form of the maximal-helicity-violating 
(MHV) amplitudes~\cite{Mangano:1990by}. 
These three-point vertices can only be defined for complex momenta, since we have 
\beq \label{mass-shell}
p_1 \cdot p_2 =p_2 \cdot p_3 =p_3 \cdot p_1 =0 \; .
\eeq
Thus, for example, we have that in the first equation in \Eq(\ref{qbqgmpp}), $\spa1.2=\spa2.3=\spa3.1=0$ so that 
\Eq(\ref{mass-shell}) is satisfied.
For complex momenta this does not imply that $\spb1.2,\spb2.3$ or $\spb3.1$ are equal to zero.

Note also that there are simple relations between the $\qb q g$ and $ggg$ amplitudes that can be derived using
supersymmetry relations~\cite{Mangano:1990by},
\beqn
&&m_3(g_1^-,g_2^+,g_3^-)=-m_3({\qb}_1^-,q_2^+,g_3^-) \frac{\spa 3.1 }{\spa 2.3}, \nn \\
&&m_3(g_1^-,g_2^+,g_3^+)=+m_3({\qb}_1^-,q_2^+,g_3^+) \frac{\spb 2.3 }{\spb 3.1} \, .
\eeqn

\subsubsection{The quark-gluon scattering process}
\label{qqbggsec}

\begin{figure}[t]
\begin{center}
\includegraphics[angle=270,scale=0.7]{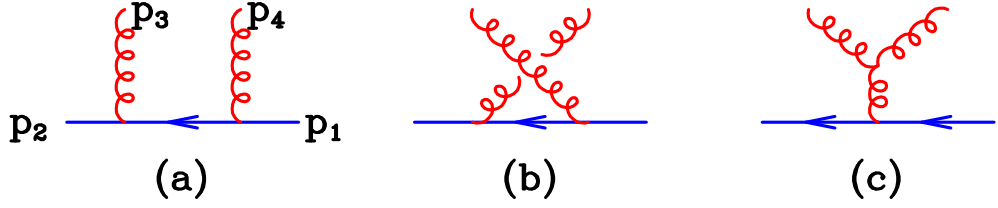}
\caption{Feynman diagrams for the process $0 \rightarrow \bar qqgg$.}
\label{qqgg}
\end{center}
\end{figure}

To further illustrate how spinor techniques should be used 
at tree-level, we consider 
the process shown in Fig.~(\ref{qqgg}), $\bar q  q gg$ scattering.
The momenta are labelled according to,
\beq
\label{sec34e1}
0  \rightarrow \bar{q}(p_1)+q(p_2)+ g(p_3)+g(p_4) \;.
\eeq

We first decompose the amplitude in terms of color ordered 
sub-amplitudes that are separately gauge invariant,
\be
\begin{split} 
{\cal A}^{\rm tree}_4({\qb}_1^{h_1}, q_2^{h_2}, g_3^{h_3},g_4^{h_4})
& = g_s^2 \left[ \left( T^{a_3} T^{a_4} \right)_{i_2 \ib_1}  m_4(\qb_1^{h_1},q_2^{h_2},g_3^{h_3},g_4^{h_4}) \right.
\\ 
& \left.       + \left( T^{a_4} T^{a_3} \right)_{i_2 \ib_1}  m_4(\qb_1^{h_1},q_2^{h_2},g_4^{h_4},g_3^{h_3})\right] \; . 
\end{split}
\ee
This decomposition is obvious  for the (non-abelian) diagrams ($a$) and ($b$) in Fig.~(\ref{qqgg}).
The diagram containing the triple-gluon vertex contributes to both $m_4(\qb_1,q_2,g_3,g_4)$ and $m_4(\qb_1,q_2,g_4,g_3)$
(with opposite signs)
due to the color algebra relation,
\beq
f^{abc} T^c = -\frac{i}{\sqrt{2}} \left[ T^a, T^b \right] \;.
\eeq

We take a negative helicity
quark line and compute $m_4(\qb_1,q_2,g_3,g_4)$ only. Because of the color factor, diagram ($b$)
does not contribute to this color stripped amplitude and the other two
diagrams are given by
\be
 \label{mqbqgga}
\begin{split}
&    m^{(a)}_4 =  \frac{-i}{2} \langle 2 | \slsh{\varepsilon_3} 
\frac{(\slsh{p_2}+ \slsh{p_3})}
{\spa3.2 \spb2.3} 
\slsh{\varepsilon_4}  | 1 ], \\
&    m_4^{(c)}= 
 \frac{-i}{\spa3.4 \spb4.3}
 { \Big[ \varepsilon_3 \cdot \varepsilon_4 \;
\langle 2| \slsh{p_4} | 1 ]} 
+  \varepsilon_4 \cdot p_3 \;
\langle 2| \slsh{\varepsilon_3} | 1 ] 
- \varepsilon_3 \cdot p_4 \;
\langle 2| \slsh{\varepsilon_4} | 1 ] \Big].
\end{split} 
\ee
At this point the calculation can be greatly simplified by an astute choice of
gauge vectors $b_3$ and $b_4$, cf. \Eq(\ref{gluonpolarization}).
When the helicities of the two gluons are the same, we shall choose 
the two reference momenta $b_3,b_4$ to be the same; it then follows that 
${ \varepsilon_3 \cdot \varepsilon_4 =0}$. 
For the positive helicity case we choose ${ b_3=b_4=p_2}$ so that,
\beq
\langle 2 | \slsh{\varepsilon_3}^+ =
\langle 2 | \slsh{\varepsilon_4}^+ = 0 \;.
\eeq
We thus see that when  both gluons have positive helicity, the amplitude vanishes
${ m_4(\qb_1^+,q_2^-,g_3^+,g_4^+)} = 0$. 
Similarly it is easy to show that 
${ m_4(\qb_1^+,q_2^-,g_3^-,g_4^-)} = 0$ by choosing 
${ b_3=b_4=p_1}$.

The remaining helicity combination, when the gluons have opposite helicities, is most simple
to compute by choosing $b_3=p_4$ and $b_4=p_3$. In that case we have the simplification,
\beq
{ \varepsilon_3 \cdot \varepsilon_4=\varepsilon_3 \cdot p_4=\varepsilon_4 \cdot p_3=0} \;.
\eeq
We again find that the contribution from diagram ($c$) vanishes and only diagram ($a$) remains.
This is a remarkable result: we have computed the 
quark gluon scattering matrix element in a non-Abelian theory, with
no net contribution from the diagram involving the three gluon vertex.
Its effect is completely fixed by gauge invariance. Completing the calculation we find
from \Eq(\ref{mqbqgga}),
\beq
-i \   m_4(\qb_1^+,q_2^-, g_3^+,g_4^-) =
  \frac{\spa2.4^2 \spb3.1}{\spa3.2 \spa4.3 \spb4.3} \;.
\label{eq:treeqggqb}
\eeq
By multiplying top and bottom by $\spb3.1^2$ and using momentum conservation we can put this
in a simpler form,
\beq
{ -i \  m_4(\qb_1^+,q_2^-, g_3^+,g_4^-)} =  
\frac{\spb3.1^3}{\spb3.4 \spb4.1 \spb1.2} \;.
\label{qqggppmm}
\eeq
Similarly, the result for the opposite helicity choice is,
\beq
{ -i \  m_4(\qb_1^+,q_2^-,g_3^-,g_4^+)} = 
{ -  \frac{[41]^2 [24]}{[23][34][21]}} \;.
\eeq
Finally, the non-zero amplitudes for $m_4(\qb_1^+,q_2^-,g_4,g_3)$ can be obtained by Bose symmetry 
(interchanging 3 and 4),
\be
\begin{split}
& { -i \  m_4(\qb_1^+,q_2^-, g_4^+,g_3^-)} =  \frac{\spb4.1^3}{\spb4.3 \spb3.1 \spb1.2} \;, \\
& { -i \  m_4(\qb_1^+,q_2^-,g_4^-,g_3^+)} = { - \frac{[31]^2 [23]}{[24][43][21]}} \;,
\end{split}
\ee
and parity invariance of the strong interactions means that,
\be
  m_4(\qb_1^{h_1},q_2^{h_2},g_3^{h_3},g_4^{h_4})=m_4^{*}(\qb_1^{-h_1},q_2^{-h_2},g_3^{-h_3},g_4^{-h_4}) \;.
\ee
\subsubsection{Amplitude for four quark scattering}
In a similar way, the amplitude for four-quark scattering can easily be written down,
\beq
\B^\tree (\qb_1,q_2,\Qb_3,Q_4) =  g_s^2 
\Bigg[\delta_{i_2 \ib_1}\delta_{i_4 \ib_3} -\frac{1}{N_c}\delta_{i_2 \ib_1}\delta_{i_4 \ib_3}\Bigg] 
m_4(\qb_1, q_2,\Qb_3,Q_4) \, ,
\eeq
where
\be
\label{qbQqQb}
\begin{split}
  -i m_4(\qb_1^+, q_2^-,\Qb_3^+,Q_4^-) 
 &  = -\frac{\spab{4}.{\gamma^\mu}.{3} \spab{2}.{\gamma_\mu}.{1}}{2 s_{12}} 
  = -\frac{\spa4.2 \spb1.3}{\spa1.2 \spb2.1 } \\
& =  -\frac{\spa4.2 \spb1.3^2}{\spb1.3 \spa1.2\spb2.1} 
= \frac{\spb1.3^2}{\spb4.3 \spb2.1}\,. 
\end{split} 
\ee
In deriving the last line of \Eq(\ref{qbQqQb}) we have used momentum conservation
$\spb1.3 \spa1.2 = -\spab{2}.{\slsh{1}}.{3}= \spab{2}.{\slsh{4}}.{3}= -\spb4.3 \spa4.2$.

\section{Results for selected scalar integrals}
\label{App:int}

In this Appendix, we present scalar integrals that are 
needed for the computation of $q \bar q gg$ primitive amplitude 
discussed in Section~\ref{Sec:Oneloopexample}. In all integrals, we neglect
${\cal O}(\ep)$ terms. 

The result for the box integral with all external lines light-like is,
\be 
\label{eq:I4}
\begin{split} 
&I_4(0,0,0,0;\sud,\sdt;0,0,0,0)
=\frac{\mu^{2 \e}}{\sud \sdt}  \\
&\times \Big[\frac{2}{\e^2} 
\Big((-\sud-i \varepsilon)^{-\e}
     +(-\sdt-i \varepsilon)^{-\e}\Big)
-\ln^2\Big(\frac{-\sud-i \varepsilon}{-\sdt-i \varepsilon}\Big) 
- \pi^2 \Big]+\cO(\e).
\end{split} 
\ee
The result for a triangle integral with two massless external lines is,
\be 
\label{eq:I3}
\begin{split} 
& I_3(0,0,p^2;0,0,0)=
\frac{\mu^{2 \e}}{\e^2} \Bigg(\frac{(-p^2-i \varepsilon)^{-\e}}{p^2}\Bigg) \\
& =\frac{1}{p^2} \Bigg(\frac{1}{\e^2}
+\frac{1}{\e} \ln \Big( \frac{\mu^2}{-p^2-i\varepsilon}\Big)
+\frac{1}{2} \ln^2 \Big( \frac{\mu^2}{-p^2-i\varepsilon}\Big)\Bigg)+\cO(\e). 
\end{split}
\ee
Lastly, the result for the bubble integral is
\beqn \label{eq:I2}
I_2(p^2;0,0)&=&
\Big(\frac{\mu^{2}}{-p^2-i \varepsilon }\Big)^\epsilon 
\Bigg(\frac{1}{\epsilon}+2\Bigg) \nn \\
&=& \frac{1}{\e}
+\ln \Big( \frac{\mu^2}{-p^2-i\varepsilon}\Big) +2 +\cO(\e).
\eeqn

\addcontentsline{toc}{section}{Bibliography}

\newpage
\providecommand{\href}[2]{#2}\begingroup\raggedright\endgroup


\begin{thebibliography}{100}

\bibitem{Bern:2008ef}
{\bf NLO Multileg Working Group} Collaboration, Z.~Bern {\em et.~al.}, {\it
  {The NLO multileg working group: Summary report}},
  \href{http://xxx.lanl.gov/abs/0803.0494}{{\tt 0803.0494}}.

\bibitem{Bredenstein:2009aj}
A.~Bredenstein, A.~Denner, S.~Dittmaier, and S.~Pozzorini, {\it {NLO QCD
  corrections to pp $\to$ t anti-t b anti-b + X at the LHC}},  {\em Phys. Rev.
  Lett.} {\bf 103} (2009) 012002,
  [\href{http://xxx.lanl.gov/abs/0905.0110}{{\tt 0905.0110}}].

\bibitem{Bredenstein:2010rs}
A.~Bredenstein, A.~Denner, S.~Dittmaier, and S.~Pozzorini, {\it {NLO QCD
  corrections to top anti-top bottom anti-bottom production at the LHC: 2. full
  hadronic results}},  {\em JHEP} {\bf 1003} (2010) 021,
  [\href{http://xxx.lanl.gov/abs/1001.4006}{{\tt 1001.4006}}].

\bibitem{Bevilacqua:2009zn}
G.~Bevilacqua, M.~Czakon, C.~G. Papadopoulos, R.~Pittau, and M.~Worek, {\it
  {Assault on the NLO Wishlist: pp $\to$ tt bb}},  {\em JHEP} {\bf 09} (2009)
  109, [\href{http://xxx.lanl.gov/abs/0907.4723}{{\tt 0907.4723}}].

\bibitem{Berger:2009zg}
C.~Berger, Z.~Bern, L.~J. Dixon, F.~Febres~Cordero, D.~Forde, {\em et.~al.},
  {\it {Precise Predictions for $W$ + 3 Jet Production at Hadron Colliders}},
  {\em Phys.Rev.Lett.} {\bf 102} (2009) 222001,
  [\href{http://xxx.lanl.gov/abs/0902.2760}{{\tt 0902.2760}}].

\bibitem{KeithEllis:2009bu}
R.~K. Ellis, K.~Melnikov, and G.~Zanderighi, {\it {W+3 jet production at the
  Tevatron}},  {\em Phys. Rev.} {\bf D80} (2009) 094002,
  [\href{http://xxx.lanl.gov/abs/0906.1445}{{\tt 0906.1445}}].

\bibitem{Berger:2009ep}
C.~Berger, Z.~Bern, L.~J. Dixon, F.~Febres~Cordero, D.~Forde, {\em et.~al.},
  {\it {Next-to-Leading Order QCD Predictions for W+3-Jet Distributions at
  Hadron Colliders}},  {\em Phys.Rev.} {\bf D80} (2009) 074036,
  [\href{http://xxx.lanl.gov/abs/0907.1984}{{\tt 0907.1984}}].

\bibitem{Melnikov:2009wh}
K.~Melnikov and G.~Zanderighi, {\it {W+3 jet production at the LHC as a signal
  or background}},  {\em Phys. Rev.} {\bf D81} (2010) 074025,
  [\href{http://xxx.lanl.gov/abs/0910.3671}{{\tt 0910.3671}}].

\bibitem{Berger:2010vm}
C.~Berger, Z.~Bern, L.~J. Dixon, F.~Cordero, D.~Forde, {\em et.~al.}, {\it
  {Next-to-Leading Order QCD Predictions for $Z,\gamma^*$+3-Jet Distributions at
  the Tevatron}},  {\em Phys.Rev.} {\bf D82} (2010) 074002,
  [\href{http://xxx.lanl.gov/abs/1004.1659}{{\tt 1004.1659}}].

\bibitem{Frederix:2010ne}
R.~Frederix, S.~Frixione, K.~Melnikov, and G.~Zanderighi, {\it {NLO QCD
  corrections to five-jet production at LEP and the extraction of alphas(MZ)}},
   {\em JHEP} {\bf 11} (2010) 050,
  [\href{http://xxx.lanl.gov/abs/1008.5313}{{\tt 1008.5313}}].

\bibitem{Melia:2010bm}
T.~Melia, K.~Melnikov, R.~Rontsch, and G.~Zanderighi, {\it {Next-to-leading
  order QCD predictions for $W^+W^+jj$ production at the LHC}},  {\em JHEP}
  {\bf 1012} (2010) 053, [\href{http://xxx.lanl.gov/abs/1007.5313}{{\tt
  1007.5313}}].

\bibitem{Denner:2010jp}
A.~Denner, S.~Dittmaier, S.~Kallweit, and S.~Pozzorini, {\it {NLO QCD
  corrections to WWbb production at hadron colliders}},  {\em Phys.Rev.Lett.}
  {\bf 106} (2011) 052001, [\href{http://xxx.lanl.gov/abs/1012.3975}{{\tt
  1012.3975}}].

\bibitem{Bevilacqua:2010ve}
G.~Bevilacqua, M.~Czakon, C.~Papadopoulos, and M.~Worek, {\it {Dominant QCD
  Backgrounds in Higgs Boson Analyses at the LHC: A Study of pp $\to$ t anti-t
  + 2 jets at Next-To-Leading Order}},  {\em Phys.Rev.Lett.} {\bf 104} (2010)
  162002, [\href{http://xxx.lanl.gov/abs/1002.4009}{{\tt 1002.4009}}].

\bibitem{Bevilacqua:2010qb}
G.~Bevilacqua, M.~Czakon, A.~van Hameren, C.~G. Papadopoulos, and M.~Worek,
  {\it {Complete off-shell effects in top quark pair hadroproduction with
  leptonic decay at next-to-leading order}},  {\em JHEP} {\bf 1102} (2011) 083,
  [\href{http://xxx.lanl.gov/abs/1012.4230}{{\tt 1012.4230}}].

\bibitem{Melia:2011dw}
T.~Melia, K.~Melnikov, R.~Rontsch, and G.~Zanderighi, {\it {NLO QCD corrections
  for $W^+W^-$ pair production in association with two jets at hadron
  colliders}},  \href{http://xxx.lanl.gov/abs/1104.2327}{{\tt 1104.2327}}.

\bibitem{Greiner:2011mp}
N.~Greiner, A.~Guffanti, T.~Reiter, and J.~Reuter, {\it {NLO QCD corrections to
  the production of two bottom-antibottom pairs at the LHC}},
  \href{http://xxx.lanl.gov/abs/1105.3624}{{\tt 1105.3624}}.

\bibitem{Berger:2010zx}
C.~Berger, Z.~Bern, L.~J. Dixon, F.~Cordero, D.~Forde, {\em et.~al.}, {\it
  {Precise Predictions for W + 4 Jet Production at the Large Hadron Collider}},
   {\em Phys.Rev.Lett.} {\bf 106} (2011) 092001,
  [\href{http://xxx.lanl.gov/abs/1009.2338}{{\tt 1009.2338}}].

\bibitem{Ita:2011wn}
  H.~Ita, Z.~Bern, L.~J.~Dixon, F.~Febres Cordero, D.~A.~Kosower and D.~Maitre,
  Phys.\ Rev.\ D {\bf 85}, 031501 (2012)
  [\href{http://xxx.lanl.gov/abs/1108.2229}{{\tt 1108.2229}}].

\bibitem{Bern:1997sc}
Z.~Bern, L.~J. Dixon, and D.~A. Kosower, {\it {One-loop amplitudes for e+ e- to
  four partons}},  {\em Nucl. Phys.} {\bf B513} (1998) 3--86,
  [\href{http://xxx.lanl.gov/abs/hep-ph/9708239}{{\tt hep-ph/9708239}}].

\bibitem{Britto:2004nc}
R.~Britto, F.~Cachazo, and B.~Feng, {\it {Generalized unitarity and one-loop
  amplitudes in N = 4 super-Yang-Mills}},  {\em Nucl. Phys.} {\bf B725} (2005)
  275--305, [\href{http://xxx.lanl.gov/abs/hep-th/0412103}{{\tt
  hep-th/0412103}}].

\bibitem{Ossola:2006us}
G.~Ossola, C.~G. Papadopoulos, and R.~Pittau, {\it {Reducing full one-loop
  amplitudes to scalar integrals at the integrand level}},  {\em Nucl. Phys.}
  {\bf B763} (2007) 147--169,
  [\href{http://xxx.lanl.gov/abs/hep-ph/0609007}{{\tt hep-ph/0609007}}].

\bibitem{Ellis:2007br}
R.~K. Ellis, W.~Giele, and Z.~Kunszt, {\it {A Numerical Unitarity Formalism for
  Evaluating One-Loop Amplitudes}},  {\em JHEP} {\bf 0803} (2008) 003,
  [\href{http://xxx.lanl.gov/abs/0708.2398}{{\tt 0708.2398}}].

\bibitem{Giele:2008ve}
W.~T. Giele, Z.~Kunszt, and K.~Melnikov, {\it {Full one-loop amplitudes from
  tree amplitudes}},  {\em JHEP} {\bf 04} (2008) 049,
  [\href{http://xxx.lanl.gov/abs/0801.2237}{{\tt 0801.2237}}].

\bibitem{Ellis:2008ir}
R.~K. Ellis, W.~T. Giele, Z.~Kunszt, and K.~Melnikov, {\it {Masses, fermions
  and generalized D-dimensional unitarity}},  {\em Nucl.Phys.} {\bf B822}
  (2009) 270--282, [\href{http://xxx.lanl.gov/abs/0806.3467}{{\tt 0806.3467}}].

\bibitem{Berger:2009zb}
C.~F. Berger and D.~Forde, {\it {Multi-Parton Scattering Amplitudes via
  On-Shell Methods}},  {\em Ann.Rev.Nucl.Part.Sci.} (2009)
  [\href{http://xxx.lanl.gov/abs/0912.3534}{{\tt 0912.3534}}].

\bibitem{Britto:2010xq}
R.~Britto, {\it {Loop amplitudes in gauge theories: modern analytic
  approaches}},  \href{http://xxx.lanl.gov/abs/1012.4493}{{\tt 1012.4493}}.

\bibitem{Schwinger:1962tp}
J.~S. Schwinger, {\it {Gauge Invariance and Mass. 2.}},  {\em Phys.Rev.} {\bf
  128} (1962) 2425--2429.

\bibitem{Berends:1987cv}
F.~A. Berends and W.~Giele, {\it {The Six Gluon Process as an Example of
  Weyl-Van Der Waerden Spinor Calculus}},  {\em Nucl. Phys.} {\bf B294} (1987)
  700.

\bibitem{Forde:2007mi}
D.~Forde, {\it {Direct extraction of one-loop integral coefficients}},  {\em
  Phys. Rev.} {\bf D75} (2007) 125019,
  [\href{http://xxx.lanl.gov/abs/0704.1835}{{\tt 0704.1835}}].

\bibitem{Badger:2008cm}
S.~Badger, {\it {Direct Extraction Of One Loop Rational Terms}},  {\em JHEP}
  {\bf 0901} (2009) 049, [\href{http://xxx.lanl.gov/abs/0806.4600}{{\tt
  0806.4600}}].

\bibitem{Mastrolia:2009dr}
P.~Mastrolia, {\it {Double-Cut of Scattering Amplitudes and Stokes' Theorem}},
  {\em Phys.Lett.} {\bf B678} (2009) 246--249,
  [\href{http://xxx.lanl.gov/abs/0905.2909}{{\tt 0905.2909}}].

\bibitem{Badger:2008rn}
S.~Badger, N.~Bjerrum-Bohr, and P.~Vanhove, {\it {Simplicity in the Structure
  of QED and Gravity Amplitudes}},  {\em JHEP} {\bf 0902} (2009) 038,
  [\href{http://xxx.lanl.gov/abs/0811.3405}{{\tt 0811.3405}}].

\bibitem{Soper:2001hu}
D.~E. Soper, {\it {Choosing integration points for QCD calculations by
  numerical integration}},  {\em Phys.Rev.} {\bf D64} (2001) 034018,
  [\href{http://xxx.lanl.gov/abs/hep-ph/0103262}{{\tt hep-ph/0103262}}].

\bibitem{Passarino:2001wv}
G.~Passarino, {\it {An Approach toward the numerical evaluation of multiloop
  Feynman diagrams}},  {\em Nucl.Phys.} {\bf B619} (2001) 257--312,
  [\href{http://xxx.lanl.gov/abs/hep-ph/0108252}{{\tt hep-ph/0108252}}].

\bibitem{Ferroglia:2002mz}
A.~Ferroglia, M.~Passera, G.~Passarino, and S.~Uccirati, {\it {All purpose
  numerical evaluation of one loop multileg Feynman diagrams}},  {\em
  Nucl.Phys.} {\bf B650} (2003) 162--228,
  [\href{http://xxx.lanl.gov/abs/hep-ph/0209219}{{\tt hep-ph/0209219}}].

\bibitem{Nagy:2003qn}
Z.~Nagy and D.~E. Soper, {\it {General subtraction method for numerical
  calculation of one loop QCD matrix elements}},  {\em JHEP} {\bf 0309} (2003)
  055, [\href{http://xxx.lanl.gov/abs/hep-ph/0308127}{{\tt hep-ph/0308127}}].

\bibitem{Anastasiou:2007qb}
C.~Anastasiou, S.~Beerli, and A.~Daleo, {\it {Evaluating multi-loop Feynman
  diagrams with infrared and threshold singularities numerically}},  {\em JHEP}
  {\bf 0705} (2007) 071, [\href{http://xxx.lanl.gov/abs/hep-ph/0703282}{{\tt
  hep-ph/0703282}}].

\bibitem{Lazopoulos:2007ix}
A.~Lazopoulos, K.~Melnikov, and F.~Petriello, {\it {QCD corrections to
  tri-boson production}},  {\em Phys.Rev.} {\bf D76} (2007) 014001,
  [\href{http://xxx.lanl.gov/abs/hep-ph/0703273}{{\tt hep-ph/0703273}}].

\bibitem{Nagy:2006xy}
Z.~Nagy and D.~E. Soper, {\it {Numerical integration of one-loop Feynman
  diagrams for N-photon amplitudes}},  {\em Phys.Rev.} {\bf D74} (2006) 093006,
  [\href{http://xxx.lanl.gov/abs/hep-ph/0610028}{{\tt hep-ph/0610028}}].

\bibitem{Gong:2008ww}
W.~Gong, Z.~Nagy, and D.~E. Soper, {\it {Direct numerical integration of
  one-loop Feynman diagrams for N-photon amplitudes}},  {\em Phys.Rev.} {\bf
  D79} (2009) 033005, [\href{http://xxx.lanl.gov/abs/0812.3686}{{\tt
  0812.3686}}].

\bibitem{Kilian:2009wy}
W.~Kilian and T.~Kleinschmidt, {\it {Numerical Evaluation of Feynman Loop
  Integrals by Reduction to Tree Graphs}},
  \href{http://xxx.lanl.gov/abs/0912.3495}{{\tt 0912.3495}}.

\bibitem{Becker:2010ng}
S.~Becker, C.~Reuschle, and S.~Weinzierl, {\it {Numerical NLO QCD
  calculations}},  {\em JHEP} {\bf 1012} (2010) 013,
  [\href{http://xxx.lanl.gov/abs/1010.4187}{{\tt 1010.4187}}].

\bibitem{Becker:2011vg}
S.~Becker, D.~Goetz, C.~Reuschle, C.~Schwan, and S.~Weinzierl, {\it {NLO
  results for five, six and seven jets in electron-positron annihilation}},
  \href{http://xxx.lanl.gov/abs/1111.1733}{{\tt 1111.1733}}.

\bibitem{Catani:2008xa}
S.~Catani, T.~Gleisberg, F.~Krauss, G.~Rodrigo, and J.-C. Winter, {\it {From
  loops to trees by-passing Feynman's theorem}},  {\em JHEP} {\bf 09} (2008)
  065, [\href{http://xxx.lanl.gov/abs/0804.3170}{{\tt 0804.3170}}].

\bibitem{tHooft:1972fi}
G.~'t~Hooft and M.~J.~G. Veltman, {\it {Regularization and Renormalization of
  Gauge Fields}},  {\em Nucl. Phys.} {\bf B44} (1972) 189--213.

\bibitem{Landau:1959fi}
L.~D. Landau, {\it {On analytic properties of vertex parts in quantum field
  theory}},  {\em Nucl. Phys.} {\bf 13} (1959) 181--192.

\bibitem{Kinoshita:1962ur}
T.~Kinoshita, {\it {Mass singularities of Feynman amplitudes}},  {\em J. Math.
  Phys.} {\bf 3} (1962) 650--677.

\bibitem{Marciano:1975de}
W.~J. Marciano, {\it {Dimensional Regularization and Mass Singularities}},
  {\em Phys. Rev.} {\bf D12} (1975) 3861.

\bibitem{Melrose:1965kb}
D.~B. Melrose, {\it {Reduction of Feynman diagrams}},  {\em Nuovo Cim.} {\bf
  40} (1965) 181--213.

\bibitem{vanNeerven:1983vr}
W.~L. van Neerven and J.~A.~M. Vermaseren, {\it {Large loop integrals}},  {\em
  Phys. Lett.} {\bf B137} (1984) 241.

\bibitem{Bern:1992em}
Z.~Bern, L.~J. Dixon, and D.~A. Kosower, {\it {Dimensionally Regulated One-Loop
  Integrals}},  {\em Phys. Lett.} {\bf B302} (1993) 299--308,
  [\href{http://xxx.lanl.gov/abs/hep-ph/9212308}{{\tt hep-ph/9212308}}].

\bibitem{tHooft:1978xw}
G.~'t~Hooft and M.~Veltman, {\it {Scalar One Loop Integrals}},  {\em
  Nucl.Phys.} {\bf B153} (1979) 365--401.

\bibitem{Denner:1991qq}
A.~Denner, U.~Nierste, and R.~Scharf, {\it {A Compact expression for the scalar
  one loop four point function}},  {\em Nucl. Phys.} {\bf B367} (1991)
  637--656.

\bibitem{Ellis:2007qk}
R.~K. Ellis and G.~Zanderighi, {\it {Scalar one-loop integrals for QCD}},  {\em
  JHEP} {\bf 02} (2008) 002, [\href{http://xxx.lanl.gov/abs/0712.1851}{{\tt
  0712.1851}}].

\bibitem{qcdloop}
R.~K. Ellis and G.~Zanderighi, ``{QCDLoop}.'' http://qcdloop.fnal.gov.

\bibitem{Denner:2010tr}
A.~Denner and S.~Dittmaier, {\it {Scalar one-loop 4-point integrals}},  {\em
  Nucl. Phys.} {\bf B844} (2011) 199--242,
  [\href{http://xxx.lanl.gov/abs/1005.2076}{{\tt 1005.2076}}].

\bibitem{vanHameren:2010cp}
A.~van Hameren, {\it {OneLOop: for the evaluation of one-loop scalar
  functions}},  \href{http://xxx.lanl.gov/abs/1007.4716}{{\tt 1007.4716}}.

\bibitem{Passarino:1978jh}
G.~Passarino and M.~J.~G. Veltman, {\it {One Loop Corrections for $e^+ e^-$
  Annihilation Into $\mu^+ \mu^-$ in the Weinberg Model}},  {\em Nucl. Phys.} {\bf
  B160} (1979) 151.

\bibitem{Nogueira:1991ex}
P.~Nogueira, {\it {Automatic Feynman graph generation}},  {\em J. Comput.
  Phys.} {\bf 105} (1993) 279--289.

\bibitem{Hahn:2000kx}
T.~Hahn, {\it {Generating Feynman diagrams and amplitudes with FeynArts 3}},
  {\em Comput. Phys. Commun.} {\bf 140} (2001) 418--431,
  [\href{http://xxx.lanl.gov/abs/hep-ph/0012260}{{\tt hep-ph/0012260}}].

\bibitem{Binoth:2008uq}
T.~Binoth, J.-P. Guillet, G.~Heinrich, E.~Pilon, and T.~Reiter, {\it {Golem95:
  A Numerical program to calculate one-loop tensor integrals with up to six
  external legs}},  {\em Comput.Phys.Commun.} {\bf 180} (2009) 2317--2330,
  [\href{http://xxx.lanl.gov/abs/0810.0992}{{\tt 0810.0992}}].

\bibitem{Cullen:2010hz}
G.~Cullen {\em et.~al.}, {\it {Recent Progress in the Golem Project}},  {\em
  Nucl. Phys. Proc. Suppl.} {\bf 205-206} (2010) 67--73,
  [\href{http://xxx.lanl.gov/abs/1007.3580}{{\tt 1007.3580}}].

\bibitem{Denner:2005nn}
A.~Denner and S.~Dittmaier, {\it {Reduction schemes for one-loop tensor
  integrals}},  {\em Nucl. Phys.} {\bf B734} (2006) 62--115,
  [\href{http://xxx.lanl.gov/abs/hep-ph/0509141}{{\tt hep-ph/0509141}}].

\bibitem{Giele:2004ub}
W.~Giele, E.~Glover, and G.~Zanderighi, {\it {Numerical evaluation of one-loop
  diagrams near exceptional momentum configurations}},  {\em
  Nucl.Phys.Proc.Suppl.} {\bf 135} (2004) 275--279,
  [\href{http://xxx.lanl.gov/abs/hep-ph/0407016}{{\tt hep-ph/0407016}}].

\bibitem{vanOldenborgh:1989wn}
G.~J. van Oldenborgh and J.~A.~M. Vermaseren, {\it {New Algorithms for One Loop
  Integrals}},  {\em Z. Phys.} {\bf C46} (1990) 425--438.

\bibitem{Kajantie}
E.~Byckling and K.~Kajantie, {\em {Particle kinematics}}.
\newblock {Wiley}, London and New York, 1973.

\bibitem{Ellis:2008qc}
R.~K. Ellis, W.~T. Giele, Z.~Kunszt, K.~Melnikov, and G.~Zanderighi, {\it
  {One-loop amplitudes for $W^+$ 3 jet production in hadron collisions}},  {\em
  JHEP} {\bf 01} (2009) 012, [\href{http://xxx.lanl.gov/abs/0810.2762}{{\tt
  0810.2762}}].

\bibitem{Lazopoulos:2009zn}
A.~Lazopoulos, {\it {Precise evaluation of the Cut Constructible part of one
  loop amplitudes within D-dimensional unitarity}},
  \href{http://xxx.lanl.gov/abs/0911.5241}{{\tt 0911.5241}}.

\bibitem{Mastrolia:2008jb}
P.~Mastrolia, G.~Ossola, C.~G. Papadopoulos, and R.~Pittau, {\it {Optimizing
  the Reduction of One-Loop Amplitudes}},  {\em JHEP} {\bf 06} (2008) 030,
  [\href{http://xxx.lanl.gov/abs/0803.3964}{{\tt 0803.3964}}].

\bibitem{Berger:2008sj}
C.~F. Berger {\em et.~al.}, {\it {An Automated Implementation of On-Shell
  Methods for One- Loop Amplitudes}},  {\em Phys. Rev.} {\bf D78} (2008)
  036003, [\href{http://xxx.lanl.gov/abs/0803.4180}{{\tt 0803.4180}}].

\bibitem{Melnikov:2010iu}
K.~Melnikov and M.~Schulze, {\it {NLO QCD corrections to top quark pair
  production in association with one hard jet at hadron colliders}},  {\em
  Nucl.Phys.} {\bf B840} (2010) 129--159,
  [\href{http://xxx.lanl.gov/abs/1004.3284}{{\tt 1004.3284}}].

\bibitem{Melnikov:2009dn}
K.~Melnikov and M.~Schulze, {\it {NLO QCD corrections to top quark pair
  production and decay at hadron colliders}},  {\em JHEP} {\bf 08} (2009) 049,
  [\href{http://xxx.lanl.gov/abs/0907.3090}{{\tt 0907.3090}}].

\bibitem{Bern:1994cg}
Z.~Bern, L.~J. Dixon, D.~C. Dunbar, and D.~A. Kosower, {\it {Fusing gauge
  theory tree amplitudes into loop amplitudes}},  {\em Nucl. Phys.} {\bf B435}
  (1995) 59--101, [\href{http://xxx.lanl.gov/abs/hep-ph/9409265}{{\tt
  hep-ph/9409265}}].

\bibitem{Ossola:2008xq}
G.~Ossola, C.~G. Papadopoulos, and R.~Pittau, {\it {On the Rational Terms of
  the one-loop amplitudes}},  {\em JHEP} {\bf 05} (2008) 004,
  [\href{http://xxx.lanl.gov/abs/0802.1876}{{\tt 0802.1876}}].

\bibitem{Bern:2002zk}
Z.~Bern, A.~De~Freitas, L.~J. Dixon, and H.~L. Wong, {\it {Supersymmetric
  regularization, two-loop QCD amplitudes and coupling shifts}},  {\em Phys.
  Rev.} {\bf D66} (2002) 085002,
  [\href{http://xxx.lanl.gov/abs/hep-ph/0202271}{{\tt hep-ph/0202271}}].

\bibitem{Garzelli:2010qm}
M.~Garzelli, I.~Malamos, and R.~Pittau, {\it {Feynman rules for the rational
  part of the Electroweak 1-loop amplitudes in the $R_{\xi}$ gauge and in the
  Unitary gauge}},  \href{http://xxx.lanl.gov/abs/1009.4302}{{\tt 1009.4302}}.

\bibitem{Garzelli:2009is}
M.~Garzelli, I.~Malamos, and R.~Pittau, {\it {Feynman rules for the rational
  part of the Electroweak 1-loop amplitudes}},  {\em JHEP} {\bf 1001} (2010)
  040, [\href{http://xxx.lanl.gov/abs/0910.3130}{{\tt 0910.3130}}].

\bibitem{Draggiotis:2009yb}
P.~Draggiotis, M.~Garzelli, C.~Papadopoulos, and R.~Pittau, {\it {Feynman Rules
  for the Rational Part of the QCD 1-loop amplitudes}},  {\em JHEP} {\bf 0904}
  (2009) 072, [\href{http://xxx.lanl.gov/abs/0903.0356}{{\tt 0903.0356}}].

\bibitem{Bern:2005cq}
Z.~Bern, L.~J. Dixon, and D.~A. Kosower, {\it {Bootstrapping multi-parton loop
  amplitudes in QCD}},  {\em Phys.Rev.} {\bf D73} (2006) 065013,
  [\href{http://xxx.lanl.gov/abs/hep-ph/0507005}{{\tt hep-ph/0507005}}].

\bibitem{Berger:2006ci}
C.~F. Berger, Z.~Bern, L.~J. Dixon, D.~Forde, and D.~A. Kosower, {\it
  {Bootstrapping one-loop QCD amplitudes with general helicities}},  {\em Phys.
  Rev.} {\bf D74} (2006) 036009,
  [\href{http://xxx.lanl.gov/abs/hep-ph/0604195}{{\tt hep-ph/0604195}}].

\bibitem{Mangano:1988kk}
M.~L. Mangano, {\it {The Color Structure of Gluon Emission}},  {\em Nucl.
  Phys.} {\bf B309} (1988) 461.

\bibitem{Mangano:1990by}
M.~L. Mangano and S.~J. Parke, {\it {Multiparton amplitudes in gauge
  theories}},  {\em Phys.Rept.} {\bf 200} (1991) 301--367,
  [\href{http://xxx.lanl.gov/abs/hep-th/0509223}{{\tt hep-th/0509223}}].

\bibitem{Dixon:1996wi}
L.~J. Dixon, {\it {Calculating scattering amplitudes efficiently}},
  \href{http://xxx.lanl.gov/abs/hep-ph/9601359}{{\tt hep-ph/9601359}}.

\bibitem{Bern:1994fz}
Z.~Bern, L.~J. Dixon, and D.~A. Kosower, {\it {One loop corrections to two
  quark three gluon amplitudes}},  {\em Nucl.Phys.} {\bf B437} (1995) 259--304,
  [\href{http://xxx.lanl.gov/abs/hep-ph/9409393}{{\tt hep-ph/9409393}}].

\bibitem{Bern:1996je}
Z.~Bern, L.~J. Dixon, and D.~A. Kosower, {\it {Progress in one loop QCD
  computations}},  {\em Ann.Rev.Nucl.Part.Sci.} {\bf 46} (1996) 109--148,
  [\href{http://xxx.lanl.gov/abs/hep-ph/9602280}{{\tt hep-ph/9602280}}].

\bibitem{Berends:1988zn}
F.~A. Berends and W.~T. Giele, {\it {Multiple Soft Gluon Radiation in Parton
  Processes}},  {\em Nucl. Phys.} {\bf B313} (1989) 595.

\bibitem{Kleiss:1988ne}
R.~Kleiss and H.~Kuijf, {\it {Multi - gluon cross-sections and five jet
  production at hadron colliders}},  {\em Nucl. Phys.} {\bf B312} (1989) 616.

\bibitem{DelDuca:1999rs}
V.~Del~Duca, L.~J. Dixon, and F.~Maltoni, {\it {New color decompositions for
  gauge amplitudes at tree and loop level}},  {\em Nucl. Phys.} {\bf B571}
  (2000) 51--70, [\href{http://xxx.lanl.gov/abs/hep-ph/9910563}{{\tt
  hep-ph/9910563}}].

\bibitem{GieleColor}
W.~Giele, private~communication.

\bibitem{Bern:2008qj}
Z.~Bern, J.~J.~M. Carrasco, and H.~Johansson, {\it {New Relations for
  Gauge-Theory Amplitudes}},  {\em Phys. Rev.} {\bf D78} (2008) 085011,
  [\href{http://xxx.lanl.gov/abs/0805.3993}{{\tt 0805.3993}}].

\bibitem{BjerrumBohr:2010zs}
N.~E.~J. Bjerrum-Bohr, P.~H. Damgaard, T.~Sondergaard, and P.~Vanhove, {\it
  {Monodromy and Jacobi-like Relations for Color-Ordered Amplitudes}},  {\em
  JHEP} {\bf 06} (2010) 003, [\href{http://xxx.lanl.gov/abs/1003.2403}{{\tt
  1003.2403}}].

\bibitem{Feng:2010my}
B.~Feng, R.~Huang, and Y.~Jia, {\it {Gauge Amplitude Identities by On-shell
  Recursion Relation in S-matrix Program}},  {\em Phys. Lett.} {\bf B695}
  (2011) 350--353, [\href{http://xxx.lanl.gov/abs/1004.3417}{{\tt 1004.3417}}].

\bibitem{Chen:2011jx}
Y.-X. Chen, Y.-J. Du, and B.~Feng, {\it {A Proof of the Explicit Minimal-basis
  Expansion of Tree Amplitudes in Gauge Field Theory}},  {\em JHEP} {\bf 02}
  (2011) 112, [\href{http://xxx.lanl.gov/abs/1101.0009}{{\tt 1101.0009}}].

\bibitem{Giele:2009ui}
W.~Giele, Z.~Kunszt, and J.~Winter, {\it {Efficient Color-Dressed Calculation
  of Virtual Corrections}},  {\em Nucl. Phys.} {\bf B840} (2010) 214--270,
  [\href{http://xxx.lanl.gov/abs/0911.1962}{{\tt 0911.1962}}].

\bibitem{Gleisberg:2008fv}
T.~Gleisberg and S.~Hoche, {\it {Comix, a new matrix element generator}},  {\em
  JHEP} {\bf 12} (2008) 039, [\href{http://xxx.lanl.gov/abs/0808.3674}{{\tt
  0808.3674}}].

\bibitem{Kunszt:1994mc}
Z.~Kunszt, A.~Signer, and Z.~Trocsanyi, {\it {Singular terms of helicity
  amplitudes at one loop in QCD and the soft limit of the cross-sections of
  multiparton processes}},  {\em Nucl. Phys.} {\bf B420} (1994) 550--564,
  [\href{http://xxx.lanl.gov/abs/hep-ph/9401294}{{\tt hep-ph/9401294}}].

\bibitem{Catani:1998bh}
S.~Catani, {\it {The singular behaviour of {QCD} amplitudes at two-loop
  order}},  {\em Phys. Lett.} {\bf B427} (1998) 161--171,
  [\href{http://xxx.lanl.gov/abs/hep-ph/9802439}{{\tt hep-ph/9802439}}].

\bibitem{Catani:2000ef}
S.~Catani, S.~Dittmaier, and Z.~Trocsanyi, {\it {One loop singular behavior of
  QCD and SUSY QCD amplitudes with massive partons}},  {\em Phys.Lett.} {\bf
  B500} (2001) 149--160, [\href{http://xxx.lanl.gov/abs/hep-ph/0011222}{{\tt
  hep-ph/0011222}}].

\bibitem{Cheung:2009dc}
C.~Cheung and D.~O'Connell, {\it {Amplitudes and Spinor-Helicity in Six
  Dimensions}},  {\em JHEP} {\bf 0907} (2009) 075,
  [\href{http://xxx.lanl.gov/abs/0902.0981}{{\tt 0902.0981}}].

\bibitem{CaronHuot:2010rj}
S.~Caron-Huot and D.~O'Connell, {\it {Spinor Helicity and Dual Conformal
  Symmetry in Ten Dimensions}},  \href{http://xxx.lanl.gov/abs/1010.5487}{{\tt
  1010.5487}}.

\bibitem{collins}
J.~Collins, {\em {Renormalization}}.
\newblock Cambridge University Press, Cambridge, 1984.

\bibitem{BG}
F.~Berends and W.~Giele, {\it {Recursive calculations for Processes with n
  gluons}},  {\em Nucl. Phys.} {\bf B306} (1988) 759.

\bibitem{Melnikov:2011ta}
K.~Melnikov, M.~Schulze, and A.~Scharf, {\it {QCD corrections to top quark pair
  production in association with a photon at hadron colliders}},  {\em
  Phys.Rev.} {\bf D83} (2011) 074013,
  [\href{http://xxx.lanl.gov/abs/1102.1967}{{\tt 1102.1967}}].

\bibitem{Britto:2005fq}
R.~Britto, F.~Cachazo, B.~Feng, and E.~Witten, {\it {Direct Proof Of Tree-Level
  Recursion Relation In Yang- Mills Theory}},  {\em Phys. Rev. Lett.} {\bf 94}
  (2005) 181602, [\href{http://xxx.lanl.gov/abs/hep-th/0501052}{{\tt
  hep-th/0501052}}].

\bibitem{ArkaniHamed:2008yf}
N.~Arkani-Hamed and J.~Kaplan, {\it {On Tree Amplitudes in Gauge Theory and
  Gravity}},  {\em JHEP} {\bf 04} (2008) 076,
  [\href{http://xxx.lanl.gov/abs/0801.2385}{{\tt 0801.2385}}].

\bibitem{Badger:2011yu}
S.~Badger, R.~Sattler, and V.~Yundin, {\it {One-Loop Helicity Amplitudes for
  ttbar Production at Hadron Colliders}},
  \href{http://xxx.lanl.gov/abs/1101.5947}{{\tt 1101.5947}}.

\bibitem{Broadhurst:1991fy}
D.~J. Broadhurst, N.~Gray, and K.~Schilcher, {\it {Gauge invariant on-shell
  Z(2) in QED, QCD and the effective field theory of a static quark}},  {\em Z.
  Phys.} {\bf C52} (1991) 111--122.

\bibitem{ArkaniHamed:2008gz}
N.~Arkani-Hamed, F.~Cachazo, and J.~Kaplan, {\it {What is the Simplest Quantum
  Field Theory?}},  {\em JHEP} {\bf 1009} (2010) 016,
  [\href{http://xxx.lanl.gov/abs/0808.1446}{{\tt 0808.1446}}].

\bibitem{Kilgore:2007qr}
W.~B. Kilgore, {\it {One-loop Integral Coefficients from Generalized
  Unitarity}},  \href{http://xxx.lanl.gov/abs/0711.5015}{{\tt 0711.5015}}.

\bibitem{Parke:1986gb}
S.~J. Parke and T.~R. Taylor, {\it {An Amplitude for n Gluon Scattering}},
  {\em Phys. Rev. Lett.} {\bf 56} (1986) 2459.

\bibitem{Kunszt:1993sd}
Z.~Kunszt, A.~Signer, and Z.~Trocsanyi, {\it {One loop helicity amplitudes for
  all 2 to 2 processes in QCD and N=1 supersymmetric Yang-Mills theory}},  {\em
  Nucl.Phys.} {\bf B411} (1994) 397--442,
  [\href{http://xxx.lanl.gov/abs/hep-ph/9305239}{{\tt hep-ph/9305239}}].

\bibitem{Peskin:2011in}
M.~E. Peskin, {\it {Simplifying Multi-Jet QCD Computation}},
  \href{http://xxx.lanl.gov/abs/1101.2414}{{\tt 1101.2414}}.

\bibitem{Drummond:2010ep}
J.~M. Drummond, {\it {Hidden Simplicity of Gauge Theory Amplitudes}},  {\em
  Class. Quant. Grav.} {\bf 27} (2010) 214001,
  [\href{http://xxx.lanl.gov/abs/1010.2418}{{\tt 1010.2418}}].

\bibitem{NigelGlover:2008ur}
E.~N. Glover and C.~Williams, {\it {One-Loop Gluonic Amplitudes from Single
  Unitarity Cuts}},  {\em JHEP} {\bf 0812} (2008) 067,
  [\href{http://xxx.lanl.gov/abs/0810.2964}{{\tt 0810.2964}}].

\bibitem{Lee:1977eg}
B.~W. Lee, C.~Quigg, and H.~Thacker, {\it {Weak Interactions at Very
  High-Energies: The Role of the Higgs Boson Mass}},  {\em Phys.Rev.} {\bf D16}
  (1977) 1519.

\bibitem{Chanowitz:1985hj}
M.~S. Chanowitz and M.~K. Gaillard, {\it {The TeV Physics of Strongly
  Interacting W's and Z's}},  {\em Nucl.Phys.} {\bf B261} (1985) 379.

\bibitem{Ellis:1975ap}
J.~R. Ellis, M.~K. Gaillard, and D.~V. Nanopoulos, {\it {A Phenomenological
  Profile of the Higgs Boson}},  {\em Nucl.Phys.} {\bf B106} (1976) 292.

\bibitem{Ellis:1991qj}
R.~K. Ellis, W.~J. Stirling, and B.~R. Webber, {\it {QCD and collider
  physics}},  {\em Camb. Monogr. Part. Phys. Nucl. Phys. Cosmol.} {\bf 8}
  (1996) 1--435.

\bibitem{Binoth:2006hk}
T.~Binoth, J.~Guillet, and G.~Heinrich, {\it {Algebraic evaluation of rational
  polynomials in one-loop amplitudes}},  {\em JHEP} {\bf 0702} (2007) 013,
  [\href{http://xxx.lanl.gov/abs/hep-ph/0609054}{{\tt hep-ph/0609054}}].

\bibitem{Adler:1969gk}
S.~L. Adler, {\it {Axial vector vertex in spinor electrodynamics}},  {\em Phys.
  Rev.} {\bf 177} (1969) 2426--2438.

\bibitem{Bell:1969ts}
J.~S. Bell and R.~Jackiw, {\it {A PCAC puzzle: pi0 $\to$ gamma gamma in the
  sigma model}},  {\em Nuovo Cim.} {\bf A60} (1969) 47--61.

\bibitem{Pittau:2011qp}
R.~Pittau, {\it {Primary Feynman rules to calculate the epsilon-dimensional
  integrand of any 1-loop amplitude}},
  \href{http://xxx.lanl.gov/abs/1111.4965}{{\tt 1111.4965}}.

\bibitem{Binoth:2010xt}
T.~Binoth {\em et.~al.}, {\it {A proposal for a standard interface between
  Monte Carlo tools and one-loop programs}},  {\em Comput. Phys. Commun.} {\bf
  181} (2010) 1612--1622, [\href{http://xxx.lanl.gov/abs/1001.1307}{{\tt
  1001.1307}}].

\bibitem{Hirschi:2011pa}
V.~Hirschi {\em et.~al.}, {\it {Automation of one-loop QCD corrections}},
  \href{http://xxx.lanl.gov/abs/1103.0621}{{\tt 1103.0621}}.

\bibitem{vanHameren:2009dr}
A.~van Hameren, C.~G. Papadopoulos, and R.~Pittau, {\it {Automated one-loop
  calculations: a proof of concept}},  {\em JHEP} {\bf 09} (2009) 106,
  [\href{http://xxx.lanl.gov/abs/0903.4665}{{\tt 0903.4665}}].

\bibitem{Mastrolia:2010nb}
P.~Mastrolia, G.~Ossola, T.~Reiter, and F.~Tramontano, {\it {Scattering
  AMplitudes from Unitarity-based Reduction Algorithm at the Integrand-level}},
   {\em JHEP} {\bf 08} (2010) 080,
  [\href{http://xxx.lanl.gov/abs/1006.0710}{{\tt 1006.0710}}].

\bibitem{Vermaseren:2000nd}
J.~Vermaseren, {\it {New features of FORM}},
  \href{http://xxx.lanl.gov/abs/math-ph/0010025}{{\tt math-ph/0010025}}.

\bibitem{Alwall:2007st}
J.~Alwall {\em et.~al.}, {\it {MadGraph/MadEvent v4: The New Web Generation}},
  {\em JHEP} {\bf 09} (2007) 028,
  [\href{http://xxx.lanl.gov/abs/0706.2334}{{\tt 0706.2334}}].

\bibitem{MP}
D.~H. Bailey, ``{A Portable High Performance Multiprecision Package}.'' NASA
  Ames RNR Technical Report RNR-90-022, 1990.

\bibitem{MP1}
D.~H. Bailey, ``{A Fortran-90 Based Multiprecision System}.'' RNR Technical
  Report RNR-94-013, 1994.

\bibitem{Pittau:2010tk}
R.~Pittau, {\it {Testing and improving the numerical accuracy of the NLO
  predictions}},  {\em Comput. Phys. Commun.} {\bf 181} (2010) 1941--1946,
  [\href{http://xxx.lanl.gov/abs/1006.3773}{{\tt 1006.3773}}].

\bibitem{Giele:2008bc}
W.~T. Giele and G.~Zanderighi, {\it {On the Numerical Evaluation of One-Loop
  Amplitudes: The Gluonic Case}},  {\em JHEP} {\bf 06} (2008) 038,
  [\href{http://xxx.lanl.gov/abs/0805.2152}{{\tt 0805.2152}}].

\bibitem{Winter:2009kd}
J.-C. Winter and W.~T. Giele, {\it {Calculating gluon one-loop amplitudes
  numerically}},  \href{http://xxx.lanl.gov/abs/0902.0094}{{\tt 0902.0094}}.

\bibitem{Lazopoulos:2008ex}
A.~Lazopoulos, {\it {Multi-gluon one-loop amplitudes numerically}},
  \href{http://xxx.lanl.gov/abs/0812.2998}{{\tt 0812.2998}}.

\bibitem{Mangano:2002ea}
M.~L. Mangano, M.~Moretti, F.~Piccinini, R.~Pittau, and A.~D. Polosa, {\it
  {ALPGEN, a generator for hard multiparton processes in hadronic collisions}},
   {\em JHEP} {\bf 07} (2003) 001,
  [\href{http://xxx.lanl.gov/abs/hep-ph/0206293}{{\tt hep-ph/0206293}}].

\bibitem{Gleisberg:2003xi}
T.~Gleisberg {\em et.~al.}, {\it {SHERPA 1.alpha, a proof-of-concept version}},
   {\em JHEP} {\bf 02} (2004) 056,
  [\href{http://xxx.lanl.gov/abs/hep-ph/0311263}{{\tt hep-ph/0311263}}].

\bibitem{Boos:2004kh}
{\bf CompHEP} Collaboration, E.~Boos {\em et.~al.}, {\it {CompHEP 4.4:
  Automatic computations from Lagrangians to events}},  {\em Nucl. Instrum.
  Meth.} {\bf A534} (2004) 250--259,
  [\href{http://xxx.lanl.gov/abs/hep-ph/0403113}{{\tt hep-ph/0403113}}].

\bibitem{Giele:2004iy}
W.~T. Giele and E.~W.~N. Glover, {\it {A calculational formalism for one-loop
  integrals}},  {\em JHEP} {\bf 04} (2004) 029,
  [\href{http://xxx.lanl.gov/abs/hep-ph/0402152}{{\tt hep-ph/0402152}}].

\bibitem{Bierenbaum:2010cy}
I.~Bierenbaum, S.~Catani, P.~Draggiotis, and G.~Rodrigo, {\it {A Tree-Loop
  Duality Relation at Two Loops and Beyond}},  {\em JHEP} {\bf 10} (2010) 073,
  [\href{http://xxx.lanl.gov/abs/1007.0194}{{\tt 1007.0194}}].

\bibitem{Berends:1981rb}
F.~A. Berends, R.~Kleiss, P.~De~Causmaecker, R.~Gastmans, and T.~T. Wu, {\it
  {Single Bremsstrahlung Processes in Gauge Theories}},  {\em Phys.Lett.} {\bf
  B103} (1981) 124.

\bibitem{Dixon:2010ik}
L.~J. Dixon, J.~M. Henn, J.~Plefka, and T.~Schuster, {\it {All tree-level
  amplitudes in massless QCD}},  {\em JHEP} {\bf 01} (2011) 035,
  [\href{http://xxx.lanl.gov/abs/1010.3991}{{\tt 1010.3991}}].

\bibitem{Bourjaily:2010wh}
J.~L. Bourjaily, {\it {Efficient Tree-Amplitudes in N=4: Automatic BCFW
  Recursion in Mathematica}},  \href{http://xxx.lanl.gov/abs/1011.2447}{{\tt
  1011.2447}}.

\bibitem{Badger:2010mg}
S.~Badger, J.~M. Campbell, and R.~K. Ellis, {\it {QCD corrections to the
  hadronic production of a heavy quark pair and a W-boson including decay
  correlations}},  {\em JHEP} {\bf 1103} (2011) 027,
  [\href{http://xxx.lanl.gov/abs/1011.6647}{{\tt 1011.6647}}].

\bibitem{Campbell:2010cz}
J.~M. Campbell, R.~K. Ellis, and C.~Williams, {\it {Hadronic production of a
  Higgs boson and two jets at next-to-leading order}},  {\em Phys.Rev.} {\bf
  D81} (2010) 074023, [\href{http://xxx.lanl.gov/abs/1001.4495}{{\tt
  1001.4495}}].

\bibitem{CaronHuot:2010zt}
S.~Caron-Huot, {\it {Loops and trees}},
  \href{http://xxx.lanl.gov/abs/1007.3224}{{\tt 1007.3224}}.

\bibitem{ArkaniHamed:2010kv}
N.~Arkani-Hamed, J.~L. Bourjaily, F.~Cachazo, S.~Caron-Huot, and J.~Trnka, {\it
  {The All-Loop Integrand For Scattering Amplitudes in Planar N=4 SYM}},  {\em
  JHEP} {\bf 1101} (2011) 041, [\href{http://xxx.lanl.gov/abs/1008.2958}{{\tt
  1008.2958}}].

\bibitem{Boels:2010nw}
R.~H. Boels, {\it {On BCFW shifts of integrands and integrals}},  {\em JHEP}
  {\bf 11} (2010) 113, [\href{http://xxx.lanl.gov/abs/1008.3101}{{\tt
  1008.3101}}].

\bibitem{Gluza:2010ws}
J.~Gluza, K.~Kajda, and D.~A. Kosower, {\it {Towards a Basis for Planar
  Two-Loop Integrals}},  {\em Phys.Rev.} {\bf D83} (2011) 045012,
  [\href{http://xxx.lanl.gov/abs/1009.0472}{{\tt 1009.0472}}].

\bibitem{Bjorken:1965zz}
J.~D. Bjorken and S.~D. Drell, {\em {Relativistic quantum mechanics}}.
\newblock McGraw-Hill, 1965.

\bibitem{Cutkosky:1960sp}
R.~E. Cutkosky, {\it {Singularities and discontinuities of Feynman
  amplitudes}},  {\em J. Math. Phys.} {\bf 1} (1960) 429--433.

\bibitem{QFTLL}
V.~Berestetskii, E.~Lifschitz, and L.~Pitaevsky, {\em {Quantum
  Electrodynamics}}.
\newblock Butterworth-Heineman, 1982.

\bibitem{eden:1966ev}
R.~J. Eden, P.~V. Landshoff, D.~I. Olive, and J.~C. Polkinghorne, {\em {The
  Analytic S-Matrix}}.
\newblock Cambridge at the University Press, 1966.

\bibitem{Nishijima:1962}
K.~Nishijima, {\it {Unitarity Condition and Anomalous Vertex Functions}},  {\em
  Phys. Rev.} {\bf 126} (1962) 852--860.

\bibitem{Xu:1986xb}
Z.~Xu, D.-H. Zhang, and L.~Chang, {\it {Helicity Amplitudes for Multiple
  Bremsstrahlung in Massless Nonabelian Gauge Theories}},  {\em Nucl. Phys.}
  {\bf B291} (1987) 392.

\bibitem{Bern:2007dw}
Z.~Bern, L.~J. Dixon, and D.~A. Kosower, {\it {On-Shell Methods in Perturbative
  QCD}},  {\em Annals Phys.} {\bf 322} (2007) 1587--1634,
  [\href{http://xxx.lanl.gov/abs/0704.2798}{{\tt 0704.2798}}].

\end{thebibliography}
\end{document}